# Plasma waveguides for high-intensity laser pulses

J.E. Shrock, B. Miao, E. Rockafellow, and H.M. Milchberg*

*Institute for Research in Electronics and Applied Physics and Dept. of Physics, University of Maryland, College Park, MD 20742, USA*

**Abstract**

Fundamental to many applications of laser pulses in science and technology is an extended interaction length with matter that significantly exceeds the distance over which the pulse would normally diffract and transversely spread. At low intensity, the interaction could simply be the linear refraction provided by a glass optical fiber to keep the pulse from spreading. At higher pulse intensities, however, the goal extends beyond diffraction-free pulse transport: an extended interaction length of high intensity light can give rise to bright secondary sources of photons and, at relativistic intensities, beams of high energy charged particles. Since these applications require laser intensities well above the threshold for ionization of atoms, new methods for overcoming pulse diffraction in a plasma have been developed. Chief among them are plasma waveguides – optical fibers composed of plasma, with characteristic mode structure. This article reviews the methods and theory of plasma waveguides, highlighting the recent development of meter-scale plasma waveguides that have been instrumental to the laser acceleration of high-charge electron beams to energies ~10 GeV.

## I. INTRODUCTION

### A. Extending the range of high intensity laser pulse propagation

Over the past 35 years, the attainable peak intensity of ultrashort laser pulses has increased by approximately 6 orders of magnitude to $\sim 10^{23}\ \mathrm{W/cm^2}$ (J. W. Yoon *et al.* 2021; ICUIL 2024). Many of the applications of pulses throughout this intensity range are enhanced by extending the high intensity interaction length in some material medium, generally owing to the favorable length scaling of an associated nonlinear process. Laser-driven secondary sources of electromagnetic radiation or high energy charged particles are the most well-known applications. One extensively studied secondary source of energetic photons is high harmonic generation (HHG) in gases (McPherson *et al.* 1987; X. F. Li *et al.* 1989; Winterfeldt, Spielmann, and Gerber 2008; Krausz and Ivanov 2009) where efficiency can be increased by extending the interaction length, provided that the intensity remains high and the process avoids dephasing (Milchberg, Durfee, and McIlrath 1995; Rundquist *et al.* 1998; Zepf *et al.* 2007). Another well-studied secondary photon source is x-ray lasers (Matthews *et al.* 1985; Suckewer *et al.* 1985; Rocca *et al.* 1994; Milchberg, Durfee, and Lynch 1995; Butler *et al.* 2003; Chou *et al.* 2007; Depresseux *et al.* 2015), where an increased interaction length leads to laser pumping of an axially extended population inversion, increasing the gain. A source of high energy charged particles is laser wakefield acceleration (LWFA) of electrons in plasmas, a highly active research area since the concept was first proposed (Tajima and Dawson 1979). Here, an increase in electron energy gain requires maintenance of high laser intensity with an extended laser interaction length, and mitigation of both dephasing and laser energy depletion (E. Esarey, Schroeder, and Leemans 2009). The accelerated electrons are also a source of forward-directed short wavelength "betatron radiation" from their transverse oscillation during the acceleration process (Corde *et al.* 2013).

* milch@umd.edu

Before we discuss plasma-based methods to counter diffraction and spreading of high intensity pulses, we first consider the diffractive limit on interaction length. The lowest order beamlike solution to the electromagnetic wave equation is the Gaussian beam (Yariv 1989)

$$E(r,z) = E_0 \frac{w_0}{w(z)} e^{-r^2/w^2(z)} e^{i\left(kz - tan^{-1}\frac{z}{z_0} + \frac{kr^2}{2R(z)}\right)}, \quad (1)$$

for a field $E$ with center wavenumber $k = 2\pi/\lambda$ propagating along z in a uniform medium of refractive index $n$, where $k_0 = \omega/c = k/n$ and $\lambda_0 = n\lambda$ are the vacuum wavenumber and wavelength, $\omega$ is the light angular frequency, and $c$ is the speed of light in vacuum. Here, the interaction length $L_{int}$ is limited to approximately $2z_0$ centered on the narrowest and highest intensity location of the beam at $z = 0$, the "beam waist". In Eq. (1), $z_0 = n\pi w_0^2/\lambda_0$ is the Rayleigh range, the distance over which the peak intensity drops by a factor of 2 owing to diffractive spreading, $w_0$ and $w(z) = w_0(1 + z^2/z_0^2)^{1/2}$ are the $1/e$ field radius (spot size) at the waist ($z = 0$) and at general $z$, and $R(z) = z(1 + z_0^2/z^2)$ is the phase front radius of curvature. For a laser pulse energy $\varepsilon$ and pulse duration $\tau_{pulse}$, the diffraction-limited product of intensity and interaction length is

$$IL_{int} \sim \varepsilon n/\lambda_0 \tau_{pulse} \quad . \quad (2)$$

Extending the intensity-interaction length product beyond the diffraction limit demands that the medium in which the nonlinear process is driven should support some form of optical guiding, either nonlinear guiding enabled by the pulse itself ("self-guiding"), or a preformed guiding structure.

Nonlinear self-guiding originates from optical self-focusing (Boyd 2008). We consider the nonrelativistic ($a_0 \ll 1$) and relativistic intensity ($a_0 > \sim 1$) regimes, delineated by the dimensionless vector potential

$$a_0 = eA_0/mc^2 = eE_L/mc\omega \approx 0.86\, \lambda_0[\mu m]\sqrt{I[10^{18}\, W/cm^2]} \ , \quad (3)$$

where $A_0$ is the peak laser vector potential, $E_L$ is the peak amplitude of the laser field, $e$ is the electron charge, and $m$ is the electron rest mass. Because the "quiver" momentum of the electron in the laser field is $p = eA_0/c$, $a_0 = p/mc$ is the ratio of the electron momentum to $mc$. Note that in Eq. (3) we have used Gaussian units for $a_0$ and will do so for physics expressions elsewhere in this paper. However, as seen on the right side of Eq. (3), for numerical values of intensity, we use mixed SI-Gaussian units ($W/cm^2$) as this usage is widespread in the field.

At very low laser intensity, the induced dipole moment of an atom or molecule in the propagation medium is linear in the laser electric field. With increasing intensity, the lowest order nonlinear correction (in a near-instantaneously responding centrosymmetric medium such as an atomic gas or plasma) is quadratic in the electric field **E**, so that the refractive index is modified, at lowest order, to $n = n_0 + n_2'|\mathbf{E}|^2$ (Boyd 2008), where the coefficient $n_2'$ is the *nonlinear index of refraction*. This is more commonly written in terms of the beam intensity $I$ as $n = n_0 + n_2 I$, where $n_2 = n_2' 8\pi/c$ (in Gaussian units). Portions of phase fronts in the high intensity regions, near the beam center, will therefore lag those in regions of lower intensity at the beam edge, inducing inward phase front curvature. When the inward phase curvature barely exceeds the outward phase curvature imposed by diffraction, self-focusing of a Gaussian beam occurs at a critical peak power $P_c \sim 3.77\lambda_0^2/8\pi n_0 n_2$ (Boyd 2008).

For $a_0 \ll 1$, the nonlinear dipole moment is that of a neutral atom, and $n_2 = 12\pi^2\chi^{(3)}/n_0^2 c$ (Gaussian units), where $\chi^{(3)}$ is the third order nonlinear scalar susceptibility of the medium. For example, for 1 atm of argon, $n_2 = 9.7 \times 10^{-20}$ $\mathrm{cm^2/W}$ (Wahlstrand, Cheng, and Milchberg 2012) at $\lambda = 800\ nm$, giving $P_c \sim 10$ GW, easily achieved by a modest 1 mJ, 100 fs laser pulse. For laser power $P > P_c$ , the self-focusing beam undergoes collapse – a runaway shrinking of spot size and increase of intensity – until it reaches the ionization threshold of the propagation medium and generates a defocusing plasma which arrests the collapse. The competition between self-focusing and plasma defocusing gives rise to self-guiding and filamentation (Couairon and Mysyrowicz 2007; Bergé *et al.* 2007; Milchberg *et al.* 2014).

The utility of a nonrelativistic filament—especially for femtosecond laser pulses, where the energy requirement is modest— stems from the greatly extended axial range $L_{fil}$ of the narrow high intensity "core", or central region of the filament transverse intensity profile. Here $L_{fil} \gg \pi d_{core}^2/4\lambda$, where $d_{core}$ is the filament core diameter; typically, $d_{core} \sim 200$ μm in atmospheric pressure gases (Jhajj *et al.* 2016). Taking $L_{int} = L_{fil}$, the product of peak intensity and interaction length is $IL_{int} \gg \varepsilon_{core}/\lambda\tau$ , much greater than the diffraction-imposed limit of Eq. (2), where $\varepsilon_{core}$ is the energy in the filament core. In general, $\varepsilon_{core}/\varepsilon \ll 1$ for nonrelativistic filamentation; in atmospheric pressure gases, $\varepsilon_{core}/\varepsilon \sim 0.1$, so that most of the pulse energy lies outside the core in the much wider, lower intensity beam periphery or "reservoir", which continuously exchanges energy with the core (Mlejnek, Wright, and Moloney 1998; Le *et al.* 2024). While filamentary self-guiding in gases yields useful light sources such as supercontinuum generation (Couairon and Mysyrowicz 2007; Bergé *et al.* 2007) and HHG (Tosa *et al.* 2003; Steingrube *et al.* 2011), and has important applications such as laser-induced breakdown spectroscopy (Skrodzki, Burger, and Jovanovic 2017), LIDAR (Kasparian *et al.* 2003), lightning capture (Houard *et al.* 2023), and generation of air waveguides (Jhajj *et al.* 2014; Rosenthal *et al.* 2014; Goffin, Tartaro, and Milchberg 2023; Goffin *et al.* 2023), efficiency is limited by the low energy in the filament core.

For relativistic intensities with $a_0 > \sim 1$, a gas propagation medium is tunnel-ionized very early in the pulse (Augst *et al.* 1989) so that the remaining pulse envelope interacts entirely with plasma. In this case, the free electrons provide the nonlinear dipole response, with the nonlinearity originating from the light speed limitation on the laser-driven quiver velocity. The nonlinear refractive index coefficient associated with the quadratic term in the free electron response is $n_2' = N_e \pi e^4/m^3 c^2 \omega^4$, where $N_e$ is the electron density, giving the critical power for relativistic self-focusing $P_c = 17.4\,(\omega/\omega_p)^2 = 17.4\,N_{cr}/N_e$ (GW) (Schmidt and Horton 1985; Sun *et al.* 1987), where $N_{cr} = m\omega^2/4\pi e^2$ is the critical electron density (the highest plasma density for which light at frequency $\omega$ has a real propagation wavenumber) and $\omega_p = (4\pi N_e e^2/m)^{1/2}$ is the plasma frequency. For $\lambda = 800$ nm ($N_{cr} \sim 1.74 \times 10^{21}$ $\mathrm{cm^{-3}}$), 1 atm of fully ionized helium plasma ($N_e \sim 5 \times 10^{19}$ $\mathrm{cm^{-3}}$) gives $P_c \sim 1$ TW.

For the lower densities consistent with multi-GeV laser plasma acceleration ($N_e \sim 10^{17}$ $\mathrm{cm^{-3}}$, see Sec. II.F), $P_c \sim 0.3$ PW. In the relativistic case, self-focusing collapse is arrested when the rapidly increasing laser ponderomotive pressure gradient expels electron density from the beam axis region. The resulting density depression (of transverse dimension $d_{core} \sim \lambda_p = 2\pi c/\omega_p$, the plasma wavelength) is a self-generated refractive index structure that traps and self-guides the pulse. As a result of ponderomotive trapping, $\varepsilon_{core}/\varepsilon$ can be much larger than in the case of nonrelativistic filamentation, approaching unity. As in the nonrelativistic case, $IL_{\text{int}} \gg \varepsilon_{core}/\lambda\tau$, the diffraction-imposed limit. For example, at $N_e = 5 \times 10^{19}$ cm$^{-3}$, a relativistically self-focused pulse will collapse and then self-guide in a channel of diameter $d_{core} \sim \lambda_p \sim 5$ μm for a distance much greater than the diffraction limit of $\pi d_{core}^2/2\lambda \sim 50\ \mu m$, limited mainly by pulse energy loss due to plasma wave excitation (E. Esarey, Schroeder, and Leemans 2009).

The main application of relativistic self-guiding has been LWFA of electrons, which can be greatly enhanced by sustaining relativistically intense laser fields ($a_0 > 1$) over propagation distances in plasma much longer than the Rayleigh range. An early example leading to GeV-level electron beams (Clayton *et al.* 2010) used λ=800nm, 250 TW, 60 fs laser pulse to relativistically self-guide in plasma with the relatively high density $N_e = 1.5 \times 10^{18}$ cm$^{-3}$, resulting in limitation of electron energy gain by dephasing (E. Esarey, Schroeder, and Leemans 2009) (see Sec. II.F) rather than diffraction. Another example, leading to ~2 GeV electron beams, is (X. Wang *et al.* 2013), which used relativistic self-guiding of λ=1.057μm, 120 J, 150 fs pulses (0.8 PW) over a 7 cm long helium gas cell. More recently, (Constantin Aniculaesei *et al.* 2023) used self-guiding of 130 J, 135 fs laser pulses through a 10 cm helium gas cell to demonstrate nanoparticle-assisted electron acceleration to ~10 GeV. Other PW-class self-guided results include a ~3-GeV dual-stage experiment (H. T. Kim *et al.* 2013) and stable 2-GeV operation obtained with waveform-controlled pulses (H. T. Kim *et al.* 2017).

A useful short wavelength photon source that has accompanied relativistically self-guided LWFA of electrons is betatron emission (Rousse *et al.* 2004; Kiselev, Pukhov, and Kostyukov 2004; Corde *et al.* 2013; Albert *et al.* 2014; Albert *et al.* 2017; Hojbota *et al.* 2023), where accelerated electrons in the wakefield oscillate transversely with respect to the near-stationary positive ion background and emit forward-directed synchrotron-like radiation. The peak betatron emission frequency scales as $\gamma^2$, or the square of the accelerated electron energy, where $\gamma$ is the Lorentz factor.

While relativistic self-guiding provides a relatively straightforward route to extending the intensity-interaction length product for laser-driven nonlinear processes, it has two significant drawbacks: (1) a laser power threshold for self-focusing beam collapse and filamentation, and (2) lack of independent control of the high intensity beam profile and its axial variation. For (1), the power threshold for self-focusing and its self-guided spot size may not be commensurate with the peak intensity and spot size demanded by a particular application. And (2) is a disadvantage in applications where control of the profile and the axial extent of high intensity propagation are essential for optimizing the secondary source, such as in phase matching of HHG (Milchberg, Durfee, and McIlrath 1995; Zepf *et al.* 2007) or controlling the laser mode structure and resulting plasma wake in LWFA (Shrock *et al.* 2024).

Another approach for extending the interaction length uses a focal-intensity maximum whose axial velocity is set by spatiotemporal structure. (Sainte-Marie, Gobert, and Quéré 2017) showed that the velocity of an ultrashort intensity peak in vacuum can be controlled through spatiotemporal coupling. One important realization, described by (Froula *et al.* 2018), sends a chirped pulse through a chromatic focusing system; other ultrashort implementations use pulse-front shaping, axiparabola-echelon pairs, or related spatiotemporal optics (Ambat *et al.* 2023; J.J. Pigeon *et al.* 2024). These arrangements can sustain a high-intensity focus over distances far beyond a conventional Rayleigh range and prescribe forward, backward, subluminal, or superluminal focal motion. Such beams require high total beam energy to achieve high intensity throughout the focal region because the peak fluence at each axial location is formed by a subsample of the beam aperture. Controlled focal velocity has been proposed for dephasingless LWFA (Palastro *et al.* 2020; Miller *et al.* 2023), while (C. Caizergues *et al.* 2020) proposed phase-locked acceleration using a superluminal laser-intensity structure. Recently, proof-of-principle demonstrations showed that acceleration is possible, and energy gain improves in the super-luminal cases, although overall laser-electron efficiency is still limited (Liberman *et al.* 2026; Arrowsmith *et al.* 2026). Because the optical principle does not consider propagation through an ionizing plasma, the coupled laser-plasma evolution must be incorporated into the experiment design.

### B. Preformed plasma waveguides

A preformed waveguiding structure would bypass the laser power threshold for relativistic self-guiding and enable independent control of the transverse intensity profile and the axial extent of high intensity propagation. A guiding structure must mitigate diffractive beam spreading by imposing inward phase front curvature to compensate the outward curvature from diffraction; this requires a larger effective refractive index near the beam center than at its periphery. At low laser intensities, step-index and graded-index silica optical fibers (Snyder and Love 1984) and photonic crystal fibers (Russell 2003) satisfy this requirement, and have been used for nonlinear generation of harmonics (Canagasabey *et al.* 2009; Heckl *et al.* 2009) and supercontinuum emission (Dudley, Genty, and Coen 2006; Granzow *et al.* 2011). Gas filled hollow core fibers have also provided the extended interaction length used for generation of supercontinuum (Nisoli, De Silvestri, and Svelto 1996) and high harmonics (Rundquist *et al.* 1998), although this guiding method depends on a narrow range of acceptance angles for efficient grazing incidence reflection from the inner wall, and is thus highly sensitive to misalignment. Plasma-based applications of hollow core fibers are discussed in Sec. IV.

As pulse intensity is increased, the performance of optical fibers is degraded by stimulated Raman scattering in the bulk silica (Agrawal 2019) followed at even higher intensity by damage from multiphoton-ionization-seeded avalanche breakdown. For $\sim 100$ fs pulses at $\lambda = 800$ nm, the intensity threshold for bulk breakdown damage is $\sim 10^{12} - 10^{13}\ \mathrm{W/cm^2}$ (Schaffer, Brodeur, and Mazur 2001) ($a_0 \sim 10^{-4} - 10^{-3}$), and even lower for fiber entrance surface damage. This is an intensity level more than 6 orders of magnitude lower than needed for LWFA, and $\sim 1 - 2$ orders lower than needed for HHG. At higher intensity, damage to hollow core fibers occurs from beam misalignment and refraction into the fiber wall from gas ionization by the laser. At the intensities needed for these applications in LWFA and HHG, solid materials would be quickly converted to dense plasmas, with strong pulse absorption and then reflection occurring for plasma density exceeding $N_{cr}$.

An optical guiding structure that enables extended propagation and is immune to high intensity optical damage is an optical fiber made of plasma at density below $N_{cr}$. This is the regime of gas density plasmas, best-suited for applications in HHG and LWFA. The particular structure that forms such a plasma fiber or waveguide along an axis $z$ is determined by the plasma refractive index. An extensive treatment of plasma guiding structures is presented in Sec. II; here we motivate the initial discussion by writing down the real refractive index of a fully ionized, collisionless plasma, $n(\mathbf{r}_\perp, \omega) = (1 - N_e(\mathbf{r}_\perp)/N_{cr})^{1/2} \approx 1 - N_e(\mathbf{r}_\perp)/2N_{cr}$ (for $N_e(\mathbf{r}_\perp)/N_{cr} \ll 1$), where $\mathbf{r}_\perp$ is a vector perpendicular to the $z$ axis (Durfee, Lynch, and Milchberg 1995). The negative polarizability of free electrons is responsible for $n < 1$, with the consequence that $n$ increases with decreasing $N_e$. Thus, a plasma guiding structure requires a plasma density minimum on the beam axis, as first discussed in (L. C. Steinhauer 1970; Loren C. Steinhauer and Ahlstrom 1971a). Using the familiar terms for the higher and lower refractive index regions of an optical fiber, the near-axis region of the plasma waveguide, where most of the transmitted energy propagates, acts as the *core*, and the increasing plasma density in the wings provides the *cladding*, which supplies transverse refractive confinement. These are functional labels and do not imply a step-index profile; both regions may vary continuously in electron density and neutral composition.

There have been two main approaches for producing plasma guiding structures, with their characteristic plasma density and associated refractive index profiles illustrated in Fig. 1. One is the generation of freestanding plasma waveguides by short laser pulses in gas targets, first demonstrated in (Durfee and Milchberg 1993; Durfee, Lynch, and Milchberg 1994; Durfee, Lynch, and Milchberg 1995), where the cylindrical hydrodynamic shock expansion into the background gas gives rise to the waveguide profile structure in Fig. 1(a). The central region, with minimum density $N_{e0}$, forms the waveguide core. The shock wall, peaking at some maximum density $N_{e,max}$, forms the waveguide cladding. This type of plasma waveguide was the first to exhibit tunable optical mode structure, with $L_{int}/z_0 \gg 1$.

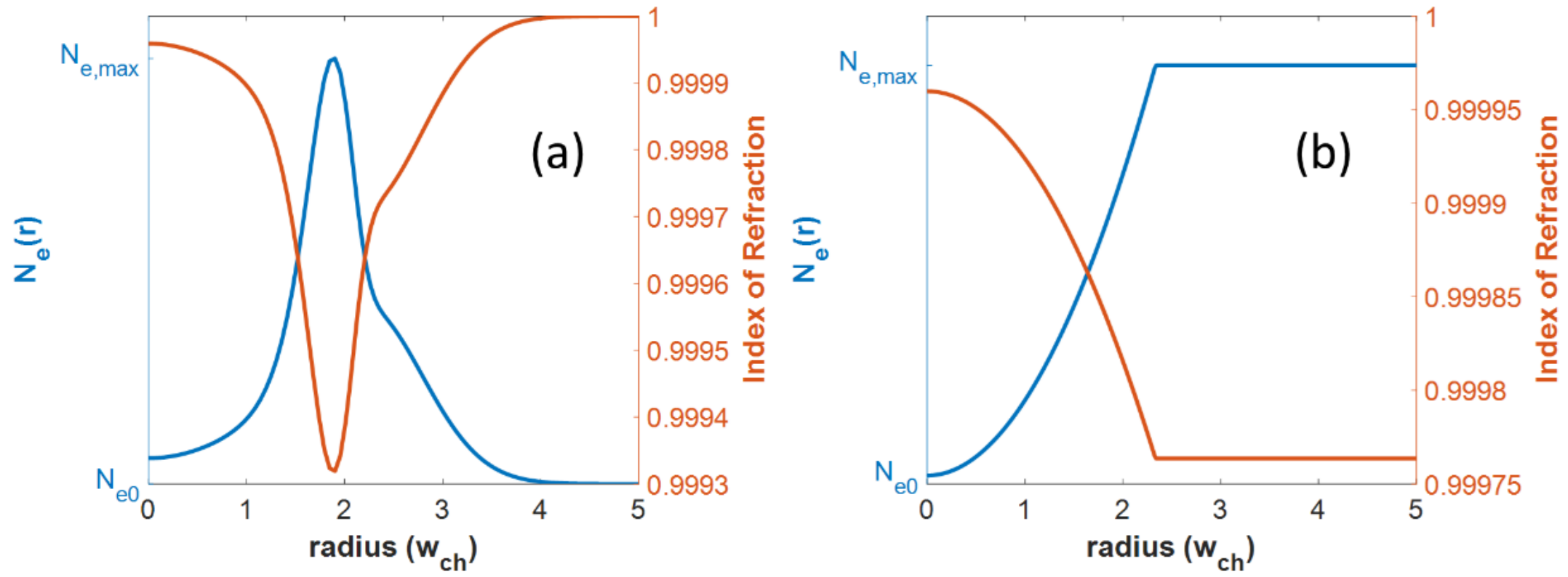

**Figure 1.** Typical plasma density and associated refractive index profiles for **(a)** "leaky" and **(b)** "finite" plasma waveguides. The profiles in (a) are characteristic of hydrodynamic plasma waveguides (Durfee and Milchberg 1993; Durfee, Lynch, and Milchberg 1995) and those in (b) characteristic of capillary discharge waveguides (Zigler *et al.* 1996; Ehrlich *et al.* 1996; Butler, Spence, and Hooker 2002; Spence, Butler, and Hooker 2003).

The other approach to plasma waveguides is based on electric discharges in long, narrow capillaries, first demonstrated using surface ablative plasma in (Zigler *et al.* 1996; Ehrlich *et al.* 1996; Ehrlich *et al.* 1998), and later using a gas fill in (Spence and Hooker 2000b; Spence, Butler, and Hooker 2001; Butler, Spence, and Hooker 2002; Spence, Butler, and Hooker 2003), where the thermal profile of the resulting quasi-steady state plasma forms the guiding structure. Such a profile can be modeled as in Fig. 1(b), with plasma density increasing from $N_{e0}$ to some maximum $N_{e,max}$ and then held constant. With sufficiently high $N_{e,max}$, the infinite parabolic profile is a good model for the discharge capillary waveguide.

### C. Outline of the review

The organization of this review follows the physical questions that arise in extending the propagation of intense laser pulses: What plasma structures can confine the light, how can these structures be created and controlled, and how does a pulse evolve within them? These questions link the theory of guided propagation to the formation, diagnosis, and practical use of plasma waveguides. Section II develops the theoretical framework, beginning with the refractive index profiles needed for guiding and their radiating, bound, and leaky modes. It then examines coupling of an incident laser pulse into these modes and how their interference, leakage, and differing group velocities determine subsequent propagation. Modal, envelope, and mean-radius descriptions provide complementary views of this evolution. The section extends the discussion to relativistically intense pulses, of special interest for laser wakefield acceleration, for which the plasma response couples transverse confinement to wake excitation, pulse reshaping, and depletion, with consequences for electron injection and acceleration. The section closes with the density and length requirements for LWFA. Section III reviews laser-generated hydrodynamic waveguides, tracing their development through collisional heating of gases, femtosecond laser heating of clusters, and optical field ionization. It examines how the initial energy deposition and subsequent expansion determine the guiding structure, including the role of the surrounding neutral shock and methods for generating confining plasma cladding at low density. Section IV covers hollow dielectric capillaries and discharge-generated plasma waveguides, explaining how wall ablation, gas heating, and heat transport establish the guiding conditions, and how an additional heater laser can modify a discharge channel.

Section V describes experimental methods for producing and diagnosing the laser-generated and discharge capillary waveguides reviewed in Secs. III and IV, together with models of their formation and performance. It begins with diagnostics of plasma and neutral-density profiles, guided-pulse transmission and mode content, and longitudinal gas-density variation. For laser-generated guides, it describes the generation and propagation of Bessel and Bessel-like beams, along with the extended gas jets and gas cells in which these beams form plasma columns. For discharge capillary guides, it discusses capillary construction, gas delivery, and discharge operation. The section concludes with hydrodynamic modeling of laser-generated waveguides, including how laser ionization and heating establish the initial conditions for plasma expansion. A brief closing discussion links discharge-capillary modeling to the fuller treatment in Sec. IV.B.3. Throughout the review, experimental examples relate waveguide formation and pulse propagation to applications, especially laser wakefield acceleration. Section VI looks ahead to how longer plasma waveguides and greater control of their structure can extend the reach of intense laser–matter interactions, enabling brighter radiation sources, higher electron energies from laser wakefield acceleration, and new opportunities in relativistic nonlinear optics.

## II. THEORY OF PLASMA WAVEGUIDES

In this section, we review the principal approaches used to describe optical propagation in plasma waveguides. The common physical questions addressed are coupling and mode content, matched and mismatched propagation, spot-size and centroid motion, modal group delay, nonlinear modification of the plasma profile, depletion, and computational scaling with propagation length.

We describe three complementary levels of description: Modal and leaky-eigenmode treatments resolve linear coupling, leakage, interference, and group delay for an arbitrary specified refractive index profile. Envelope, variational, effective-radius, and centroid models provide descriptions of matched propagation, mismatch oscillations, stability, and nonlinear evolution. Full-wave, particle-in-cell (PIC), and quasistatic calculations are used when ionization, wake excitation, pulse depletion, kinetic response, or strongly nonlinear evolution must be modeled self-consistently.

The first linear modal treatment of plasma waveguides was developed by (Durfee, Lynch, and Milchberg 1994; Durfee, Lynch, and Milchberg 1995), and was extended into general treatments of coupling and the radiating, bound, and leaky spectra of finite plasma structures by (T. R. Clark and Milchberg 1998a; T. R. Clark and Milchberg 2000). The linear theory is readily applicable to plasma waveguides generated through the experimental techniques reviewed in this paper and provides insight into propagation effects even for relativistic intensity pulses, which can modify the waveguide structure (P. Sprangle *et al.* 1992; Benedetti *et al.* 2012; Shrock *et al.* 2024).

In parallel, a useful hierarchy was developed of quasistatic, envelope, source-dependent-expansion, and mean-radius models for intense-pulse propagation (P. Sprangle *et al.* 1992; E. Esarey *et al.* 1993; Eric Esarey, Krall, and Sprangle 1994; Krall *et al.* 1994; E Esarey and Leemans 1999). Those descriptions reduce the propagating field to a slowly varying envelope or a small set of transverse parameters and are discussed after the modal treatment.

We first focus on the modal theory in Secs. II.A–C and II.D.1. The bound mode discussion of Sec. II.B applies to plasma waveguides of all types. The parts of Section II.B on radiating and quasibound modes are most appropriate for free-standing plasma waveguides generated in a gas background by a short pulse laser.

### A. Refractive index profiles for optical guiding

The wave nature of light ensures that light beams undergo diffractive spreading as illustrated by the lowest order Gaussian beam of Eq. (1), where the diffraction-limited interaction length is $2z_0$. However, many applications of intense laser pulses require high intensity to be maintained over distances far beyond the Rayleigh range. This motivates the use of a transverse refractive index profile that counteracts diffractive spreading.

For Gaussian beam propagation over a small distance $z = \delta z$ from the beam waist, Eq. (1) shows that the diffractive phase advance of the field from wavefront curvature at beam center ($r = 0$) is $\Delta\Phi_d = \tan^{-1}(z/z_0) \cong \delta z/z_0$. Countering diffraction requires a compensating on-axis phase retardation $\Delta\Phi_f = k_0 \Delta n \delta z$, where $\Delta n = n(r=0) - n(r=w_0)$ is the on-axis index contrast with respect to the beam periphery. Setting $\Delta\Phi_f = \Delta\Phi_d$ effectively flattens the phase fronts and eliminates diffraction, giving

$$\Delta n = \frac{1}{k_0 z_0} \quad . \tag{4}$$

A refractive index profile $n(\mathbf{r}_\perp, \omega)$ satisfying Eq. (4) uniformly along $z$ forms a waveguide, where we assume $\omega$-dependence for generality. For a specific $n(\mathbf{r}_\perp, \omega)$, the guided beam profiles with planar phase fronts comprise the linear *bound modes* of the waveguide. Such modes are fully trapped, with no transverse energy leakage out of the waveguide as they propagate. In this review, we consider guiding structures for which linear modes can be classified as bound, quasibound, or free, as we will discuss below.

In applying the above discussion to plasma waveguides, we consider the Drude model plasma dielectric response $\varepsilon(\mathbf{r}_\perp, \omega) = 1 - \omega_p^2/\omega(\omega + i\nu) + 4\pi\chi(\mathbf{r}_\perp, \omega) = n^2(\mathbf{r}_\perp, \omega)/\mu$, where $\nu$ is the electron collision frequency, $\mu$ ($= 1$) is the magnetic permeability, and $\chi(\mathbf{r}_\perp, \omega)$ is the combined complex susceptibility of any ions and neutrals present (T. R. Clark and Milchberg 2000). Thus,

$$n^2(\mathbf{r}_\perp, \omega) = 1 - \omega_p^2(\mathbf{r}_\perp)/\omega(\omega + i\nu) + 4\pi\chi(\mathbf{r}_\perp, \omega) = (n_r + in_i)^2 \ . \tag{5}$$

where $n_r$ and $n_i$ are the real and imaginary parts of the refractive index.

As discussed earlier, major applications of plasma waveguides operate in the gas density plasma regime, where $N_e/N_{cr} \ll 1$. For ultrashort laser pulses in this density regime, the plasma can be taken as collisionless ($\nu/\omega \sim 0$). For example, for a laser wavelength $\lambda = 0.8$ μm, typical of ultrashort pulse Ti:Sapphire lasers, $N_e/N_{cr}$ ranges from $\sim 10^{-5}$ to $10^{-2}$ for applications ranging from plasma-based x-ray lasers, which favor higher plasma densities for greater energy storage and gain (Butler *et al.* 2003), to multi-GeV laser-plasma wakefield accelerators, which favor lower densities to mitigate dephasing (see Sec. II.F) between the electron bunch and the wakefield (E. Esarey, Schroeder, and Leemans 2009). For $\omega$ far from any atomic or ionic resonances in a gas density plasma, $\chi(\mathbf{r}_\perp, \omega) \ll 1$ and is real. Thus, an excellent approximation for the refractive index everywhere inside and outside the waveguide is

$$n(\mathbf{r}_\perp, \omega) = (1 - N_e(\mathbf{r}_\perp)/N_{cr} + 4\pi\chi(\mathbf{r}_\perp, \omega))^{1/2} \approx 1 - \tfrac{1}{2}N_e(\mathbf{r}_\perp)/N_{cr} + 2\pi\chi(\mathbf{r}_\perp, \omega) \ , \tag{6}$$

where the $\chi$ term can be neglected inside the guide for the case of full ionization there; the population of neutrals and the susceptibility of the ions will both be negligible. The negative polarizability of free electrons is responsible for the negative sign in Eqs. (5) and (6), so that higher refractive index corresponds to lower plasma density; plasma waveguides therefore require density minima for trapping of light. If the plasma density increases parabolically from the beam axis according to $N_e(r) = N_{e0} + r^2/\pi r_e w_{ch}^4$, where $r_e = e^2/m_e c^2$ is the classical electron radius, the lowest order bound mode (or "matched" mode), is a Gaussian with $1/e^2$ intensity spot radius $w_{ch}$ (see Sec. II.B.2). Conversely, this can be viewed as a necessary condition on plasma density contrast $\Delta N_e$ for guiding of a lowest order mode of spot size $w_{ch}$:

$$\Delta N_e = N_e(r = w_{ch}) - N_{e0} \geq 1/\pi r_e w_{ch}^2 \ . \tag{7}$$

Alternatively, using Eq. (4) together with Eq. (6) gives the same result.

### B. Radiating, bound, and leaky modes of plasma waveguides

The Fourier transformed wave equation for a given laser field $\tilde{E}(\mathbf{r}_\perp, z, \omega) = \int_{-\infty}^{\infty} dt\, e^{i\omega t} E(\mathbf{r}_\perp, z, t)$ propagating in a waveguide with index of refraction profile $n(\mathbf{r}_\perp, z, \omega)$ is

$$\nabla^2 \tilde{E} + \frac{\omega^2}{c^2} n^2 \tilde{E} = \nabla\left(-\frac{1}{n^2} \tilde{E} \cdot \nabla n^2\right) , \tag{8}$$

where $z$ is the propagation axis. The polarization-dependent term on the right side of Eq. (8) is negligible for $|\Delta(n^2)/n^2| \ll |\Delta\tilde{E}/\tilde{E}|$, where $|\Delta(n^2)|$ and $|\Delta\tilde{E}|$ are the absolute changes in $n^2$ and $\tilde{E}$ across a transverse scale width $w_{ch}$. This weak-guidance condition is satisfied for the smooth, underdense profiles considered here, where $|\Delta(n^2)/n^2| \sim \Delta N_e/N_{cr} \ll 1$, and $|\Delta\tilde{E}/\tilde{E}| \sim 1$. Thus, to this accuracy, solutions for axially invariant waveguides are polarization-independent transverse modes of the form $\tilde{E}(\mathbf{r}_\perp, z, \omega) = u(\mathbf{r}_\perp, \omega) e^{i\beta z}$, where $\beta$ is the longitudinal propagation wavenumber. Equation (8) then becomes the Helmholtz equation,

$$\nabla_\perp^2 u(\mathbf{r}_\perp, \omega) + \kappa^2 u(\mathbf{r}_\perp, \omega) = 0 , \tag{9}$$

where $\nabla_\perp^2$ is the transverse Laplacian and $\kappa^2 \equiv k_\perp^2 = (n\omega/c)^2 - \beta^2$ is the square of the transverse (or perpendicular) wavenumber $\boldsymbol{\kappa} = \mathbf{k}_\perp$. This gives

$$\kappa^2(\mathbf{r}_\perp, \omega) = k_0^2\left(1 - \frac{\beta^2}{k_0^2} - \frac{N_e(\mathbf{r}_\perp)}{N_{cr}} + 4\pi\chi(\mathbf{r}_\perp, \omega)\right), \tag{10}$$

where $\chi(\mathbf{r}_\perp, \omega)$ is real.

In the case of freestanding plasma waveguides in a gas background, propagating wave solutions exist both inside and outside the waveguiding structure and one must consider three types of propagating solution to Eq. (9) based on the spatial dependence of $\kappa^2$. In general, allowed solutions must have some region of $\mathbf{r}_\perp$ where $\kappa^2 > 0$ (T. R. Clark and Milchberg 2000). Among allowed solutions are the cases where:

(1) $\kappa^2 > 0$ for all $\mathbf{r}_\perp$: $\kappa$ is real everywhere and the solutions are freely propagating *radiating modes*.

(2) $\kappa^2 > 0$ for $|\mathbf{r}_\perp| < r_b(\varphi)$ and $\kappa^2 < 0$ for $|\mathbf{r}_\perp| > r_b(\varphi)$, where $r_b(\varphi)$ is a boundary varying with azimuthal angle $\varphi$: these idealized solutions are *bound modes*.

(3) $\kappa^2 < 0$ for $r_{b,in}(\varphi) < |\mathbf{r}_\perp| < r_{b,out}(\varphi)$, between an inner and an outer boundary, and $\kappa^2 > 0$ elsewhere: these solutions are called *leaky* or *quasibound* modes because the freely propagating solutions for $|\mathbf{r}_\perp| > r_{b,out}$ originate from tunneling or leakage from the waveguide interior.

Laser-produced hydrodynamic plasma waveguides (see Sec. III) are of type (3), where $r_{b,in}$ and $r_{b,out}$ correspond to radial locations at the inner and outer walls of the cylindrical hydrodynamic shock generated in these structures; see Fig. 1(a). Such guides strictly support only radiating and leaky modes. However, in practice, plasma waveguide parameters are chosen so that guided pulses are well-confined; under such conditions, well-bound leaky modes are practically indistinguishable from true bound modes. Here we further specify the cladding as the plasma waveguide region outside $r_b$ or $r_{b,in}$ and the region internal to that as the core.

For the remainder of this section, we will consider plasma waveguides that are azimuthally symmetric—though we note that asymmetric waveguides have quasibound and bound solutions to Eq. (9) and have been observed experimentally (R. J. Shalloo *et al.* 2019; B. Miao *et al.* 2020). We consider solutions separable in $r$ and $\phi$, $u(\mathbf{r}_\perp,\omega) = u(r,\phi,\omega) = u(r,\omega)e^{im\varphi}$, where $m$ is an integer, ensuring $u$ is single-valued in $\phi$. Then Eq. (9) becomes

$$\frac{d^2u}{ds^2} + \frac{1}{s}\frac{du}{ds} + \left(n^2(s,\omega) - \frac{\beta^2}{k_0^2} - \frac{m^2}{s^2}\right)u = 0 \ , \tag{11}$$

where $s = k_0 r$, and $n(s)$ is the plasma refractive index (Eq. (5)).

Figure 2 plots examples of radiating, bound, and quasibound modes for $m = 0$ superimposed on the waveguide index profiles giving rise to each type of mode.

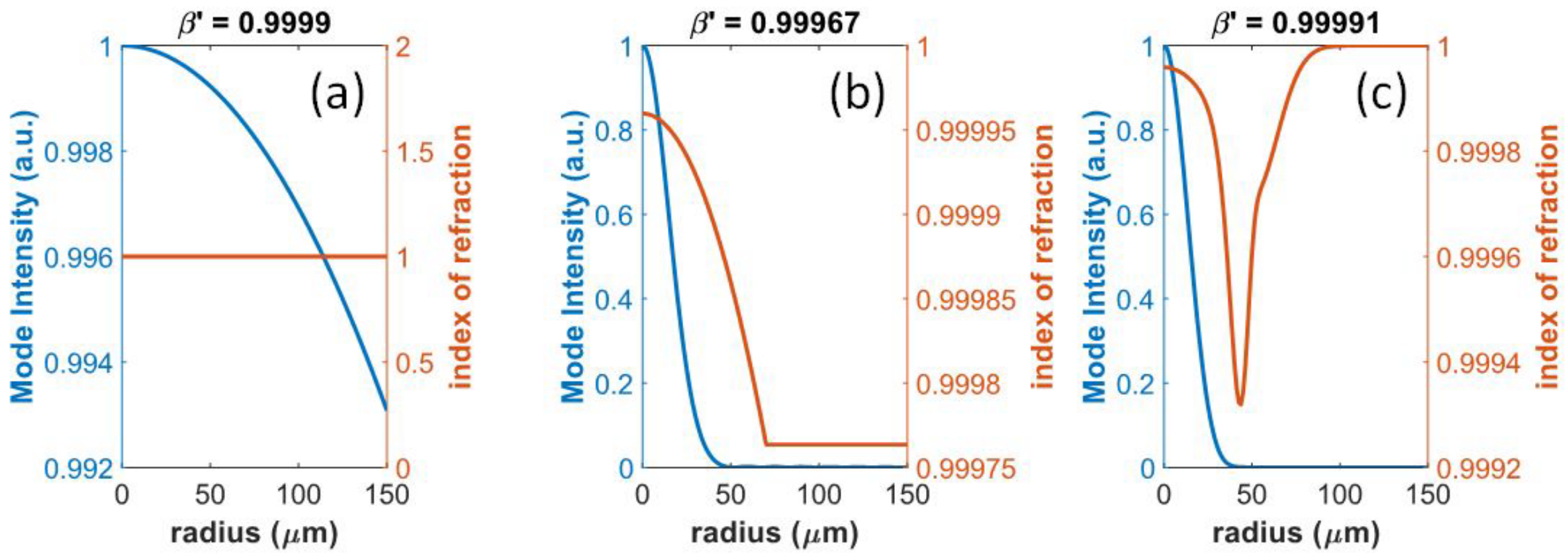

**Figure 2.** Radiating, bound, and quasibound modes. For the $\beta' = \beta/k_0$ values above each panel, the corresponding mode and waveguide index of refraction are plotted for **(a)** a radiating mode, **(b)** a bound mode of the finite parabolic index profile in Fig. 1(b), and **(c)** a quasibound mode of the 'leaky' profile in Fig. 1(a).

### 1. Radiating modes

For wave propagation along $\pm z$, $\beta$ is real. For $\kappa^2 > 0$ everywhere, the components of the transverse wavenumber $\boldsymbol{\kappa}$ are also real and there is no trapping or wave confinement with respect to the propagation axis. A mode whose vector wavenumber $\mathbf{k} = (\boldsymbol{\kappa}, \beta)$ has all real components is a *radiating mode*. As long as the light frequency satisfies $\omega > \omega_p$, continuous real values of $\beta$ are allowed. For the case of a uniform medium with $\nu = 0$, solutions to Eq. (11) give one such mode that is of particular importance,

$$\tilde{E}(\mathbf{r}_\perp, z, \omega) \propto J_{|m|}(\kappa r)e^{im\varphi}e^{i\beta z} \ , \tag{12}$$

which we call a $q^{th}$ order Bessel beam (where $q = |m|$), or a "$J_q$ beam", where $J_q$ is a $q^{th}$ order Bessel function. Bessel beams appear in the discussion of Sec. II.B.3 and are reviewed in Sec. V.B.

## 2. Bound modes

If a modal solution $\tilde{E}(\mathbf{r}_\perp, z, \omega)$ of Eq. (9) is bound, then $\int d^2\mathbf{r}_\perp |\tilde{E}(\mathbf{r}_\perp, z, \omega)|^2$ is finite for all $z$, and allowed values of $\beta$ are discrete. As the plasma density in the "finite profile" waveguide of Fig. 1(b) increases with radius to a finite level, the waveguide has a finite number of bound modes.

However, we first consider an idealized plasma waveguide where $N_e(\mathbf{r}_\perp)$ increases without limit for $|\mathbf{r}_\perp| > r_b(\varphi)$, and *all* modes are bound. In particular, an infinite number of bound mode solutions can be found for the "infinite" parabolic profile $N_e(r) = N_{e0} + r^2/\pi r_e w_{ch}^4$, with no limit on $r$, where $\kappa^2$ becomes increasingly negative for $r > r_b$. The bound normalized solutions of Eq. (11) for $\nu = 0$ are the Laguerre-Gauss modes

$$u_{pm}(r, \varphi, \omega) = \sqrt{\frac{2p!}{\pi(p + |m|)!}}\, w_{ch}^{-1}(2r^2/w_{ch}^2)^{|m|/2} e^{-r^2/w_{ch}^2} L_p^{|m|}(2r^2/w_{ch}^2) e^{im\varphi} \,. \tag{13}$$

Here $p\ (= 0,1,2,\dots)$ and $m\ (= 0, \pm 1, \pm 2, \dots)$ are radial and azimuthal mode indices, $w_{ch}$ is the $1/e$ amplitude radius of the lowest order or matched $(p, m) = (0,0)$ mode solution, and $L_p^{|m|}$ is an associated Laguerre polynomial (T. R. Clark and Milchberg 2000). The propagation wavenumber associated with these solutions is

$$\beta_{pm} = k_0(1 - N_{e0}/2N_{cr} - 2(2p + |m| + 1)/k_0^2 w_{ch}^2). \tag{14}$$

for $N_{e0}/N_{cr} \ll 1$ and $(2p + |m| + 1)/k_0^2 w_{ch}^2 \ll 1$, which are well satisfied in low density guides with low order modes. The modal group velocity is then

$$v_g = \left(\frac{\partial \beta_{pm}}{\partial \omega}\right)^{-1} \approx c(1 - N_{e0}/2N_{cr} - 2(2p + |m| + 1)/k_0^2 w_{ch}^2) \tag{15}$$

(Milchberg, Durfee, and McIlrath 1995; T. R. Clark and Milchberg 2000; Schroeder *et al.* 2011; L. Feder *et al.* 2020), which displays the plasma dispersion contribution (2nd term) and the mode-dependent waveguide structure contribution (3rd term), where $v_g \to c(1 - N_{e0}/2N_{cr})$ for a uniform plasma ($k_0 w_{ch} \to \infty$). Higher order modes for a given waveguide have lower group velocity, and a mode with given $(p, m)$ has lower group velocity in a waveguide with smaller $w_{ch}$ and higher $N_{e0}$. This means that if a short pulse couples into multiple transverse modes at the guide entrance, they can longitudinally separate (Shrock *et al.* 2024), especially in the long waveguides discussed later (see Secs. II.D, II.E, III.C.4). Even if the waveguide length and mode size do not result in clear mode separation, the modal dispersion can result in pulse lengthening and delay, as shown using nonparaxial envelope analyses in an infinite parabolic channel (E Esarey and Leemans 1999; Schroeder *et al.* 2011). While the infinite parabolic profile is an idealization of an actual physical guide, its modes and associated wavenumbers are very useful for calculations involving quasibound modes.

For the finite profile waveguide of the type shown in Fig. 1(b), the number of bound modes depends on the guide depth $N_{e,max} - N_{e0}$. We define $r_{\max}$ as the radius at which the plasma density reaches $N_{e,max}$. For a cylindrically symmetric guide $N_e(r) = N_{e0} + (N_{e,max} - N_{e0})(r/r_{max})^2$ for $r \le r_{max}$ and $N_e(r) = N_{e,max}$ for $r > r_{max}$, complementary expressions for $(p, m)$ mode cutoff in terms of channel depth and radius are (Durfee, Lynch, and Milchberg 1994):

$$N_{e,max} - N_{e0} = \Delta N_e(r_{max}) \gtrsim (2p + |m| + 1)^2/\pi r_e r_{max}^2 \tag{16a}$$

$$\frac{r_{max}}{w_{ch}} \gtrsim \sqrt{2p + |m| + 1}\ . \tag{16b}$$

An important property of plasma waveguides is that bound transverse mode shapes are wavelength-independent. This is exhibited by the Laguerre-Gauss solutions to the infinite waveguide (Eq. (13)) and can be seen more generally: Eq. (9) can be rewritten $[\nabla_\perp^2 - 4\pi r_e N_e(\mathbf{r}_\perp)]u(\mathbf{r}_\perp, \omega) = -\alpha^2 u(\mathbf{r}_\perp, \omega)$, where $\alpha^2 = k_0^2 - \beta^2$. The left side operator has no dependence on $\omega$, so that the eigenvalues $-\alpha^2$ and eigenmodes $u$ are wavelength independent (T. R. Clark and Milchberg 2000). Wavelength independence of the mode structure is of particular importance for using plasma waveguides for higher order frequency conversion in (Milchberg, Durfee, and McIlrath 1995).

### 3. Leaky or quasibound modes

For leaky or quasibound modes, $\beta$ is no longer discrete, and a continuum of values is allowed. As mentioned, all laser-generated hydrodynamic waveguides are of this type. A useful and illuminating analysis of the leaky plasma waveguide treats it as a scattering structure. Here, the pulse propagating in the waveguide is considered to be the superposition of an incoming and an outgoing conical wave. For a cylindrically symmetric hydrodynamic plasma waveguide whose outer boundary (at $r = r_{b,out}$) is surrounded by neutral gas, the solution to Eq. (11) for $r > r_{b,out}$ is

$$\tilde{E}(\mathbf{r}_\perp, z, \omega) = \tfrac{1}{2} e^{i\beta z}[a_+ H_m^{(1)}(\kappa_0 r) + a_- H_m^{(2)}(\kappa_0 r)]e^{im\varphi}, \tag{17}$$

where $H_m^{(1.2)}$ are $m$th order Hankel functions of the first and second kind, $a_+$ and $a_-$ are complex coefficients whose ratio $a_+/\,a_-$ depends on the plasma structure, and $\kappa_0 = k_0(1 - \beta^2/k_0^2 + 4\pi\chi_0)^{1/2}$, where $\chi_0 = \chi(r > r_b)$ represents the contribution of neutral gas outside the waveguide. For $\nu/\omega = 0$ (no plasma absorption), $|a_+/a_-| = 1$ and for $\nu/\omega > 0$, $|a_+/a_-| < 1$ (T. R. Clark and Milchberg 2000).

For $\kappa_0 r \gg 1$, well outside the waveguide, the asymptotic form of Eq. (17) is

$$\begin{aligned}\tilde{E}(\mathbf{r}_\perp, z, \omega) &\approx (2\pi\kappa_0 r)^{-1/2} e^{i\beta z}\big(a_+ e^{-i(\pi/4 + m\pi/2)} e^{i\kappa_0 r} + a_- e^{i(\pi/4 + m\pi/2)} e^{-i\kappa_0 r}\big)e^{im\varphi} \\ &= \tilde{E}_{out} + \tilde{E}_{in}\ ,\end{aligned} \tag{18}$$

which we identify as the superposition of outward and inward propagating conical waves, where $\tilde{E}_{out}$ is the wave scattered by the plasma structure in response to the incident wave $\tilde{E}_{in}$. Thus, any field $\tilde{E}(\mathbf{r}_\perp, z, \omega) = u(r, \omega)e^{im\varphi}e^{i\beta z}$ inside the plasma waveguide, whether a radiating mode or quasibound mode, can be viewed as the result of superposition of incoming and outgoing conical waves, each propagating at the approach angle

$$\gamma = \tan^{-1}(\kappa_0/\beta). \tag{19}$$

with respect to the $z$-axis (T. R. Clark and Milchberg 2000). Note that for a uniform medium without a plasma structure, $a_+/a_- = 1$, and $\tilde{E}(\mathbf{r}_\perp, z, \omega) \propto J_{|m|}(\kappa r)e^{im\varphi}e^{i\beta z}$, the same radiating mode result as Eq. (12). The Bessel beam can therefore be viewed as the superposition of incoming and outgoing conical waves of equal amplitude.

The scattering model can be used to numerically solve for the leaky or quasibound modes of general cylindrically symmetric plasma waveguides. Because for quasibound modes the normalized propagation wavenumber $\beta' = \beta/k_0$ is a continuous parameter, it is useful to replace $u(r,\omega)$ above by $u(r,\beta')$, showing $\beta'$ as an explicit argument, with $\omega$ as an implicit dependence. For a given azimuthal mode number $m$ and normalized wavenumber $\beta'$, we find $u(r,\beta')$ numerically by integrating Eq. (11) from $r = 0$ to $r = r_{b,out}$ and then matching the numerical solution to the analytic solution (Eq. (17)) for $r \geq r_{b,out}$, subject to continuity of the electric and magnetic fields at $r = r_{b,out}$. An unnormalized measure of the degree of coupling of the incoming conical wave field $E_{in}$ to the field in the plasma waveguide $\tilde{E}(\mathbf{r}_\perp, z, \omega) = u(r,\beta')e^{im\varphi}e^{i\beta z}$ is the resonance function

$$\eta(\beta') \equiv \frac{\int_A \left|\tilde{E}(\mathbf{r}_\perp, z, \omega)\right|^2 dA}{\left|\tilde{E}_{in}\right|^2 A} = \frac{\int_A |u(r,\beta')|^2 dA}{\left|\tilde{E}_{in}\right|^2 A} \propto \int_A |u(r,\beta')|^2 dA, \tag{20}$$

where $A$ is the area for $r < r_{b,out}$ and we take $\left|\tilde{E}_{in}\right| = 1$. The function $\eta(\beta')$ is constructed by varying $\beta'$ in small increments in Eq. (11) and repeatedly solving it for $u(r,\beta')$.

The continuous resonance function $\eta(\beta')$ maps out the mode structure of the plasma waveguide. For *radiation modes*, $\eta{\sim}1$ for a continuous range of $\beta'$, because the conical wave is not trapped by the channel. For weak trapping associated with mild damping in the cladding ($\int_{r_{b,in}}^{r_{b,out}} \kappa^2 dr \lesssim 0$), $\eta > 1$. For increasing damping in the cladding, with $\int_{r_{b,in}}^{r_{b,out}} \kappa^2 dr$ increasingly negative, $\eta \gg 1$ for narrow ranges of $\beta'$ centered about particular discrete values $\beta'_j$. These values identify the *quasibound modes* of the waveguide and can be viewed as quasi-resonances $\kappa_j = k_{\perp,j} = k_0(1-(\beta'_j)^2)^{1/2}$ of a cylindrical Fabry-Perot resonator (T. R. Clark and Milchberg 2000). As damping in the cladding continues to increase, $\eta(\beta'_j) \to \infty$ and the range of quasibound modes distributed around $\beta'_j$ converges to a *bound mode* precisely at $\beta'_j$ . For a given $m$, the values of $j$ are mapped to the radial mode index $p$, so that each continuous set of quasibound modes centered around a particular $\beta'_j$ is associated with a specific bound mode indexed by $(p,m)$. We therefore can use analytic results for bound modes, such as $\beta'_{pm} = \beta_{pm}/k_0$ and $v_g$ from Eqs. (14) and (15), as good approximations for corresponding quasibound modes. In general, for a given $m$, the quasibound mode with the largest quasiresonant value of $\beta'$ corresponds to the $p = 0$ radial mode.

An example of this mode identification procedure is shown in Figure 3. Fig. 3(a) plots $\eta(\beta')$ for the leaky plasma density profile in Fig. 1(a) for $m = 0$ and $m = 1$, where the $p$ resonances are identified. The associated $p = 0$ and $p = 1$ leaky modes $u(r,\beta')$ are plotted in Fig. 3(b). Figure 3(c) shows that if $N_e(r)$ in Fig. 1(a) is halved, resonances in $\eta(\beta')$ are wider, and the amplitude of $u(r,\beta')$ for $r > r_{b,out}$ is increased (green curve in panel (b)). This corresponds to greater leakage. The (0,0) lowest order quasibound mode is also called the "quasi-matched" mode.

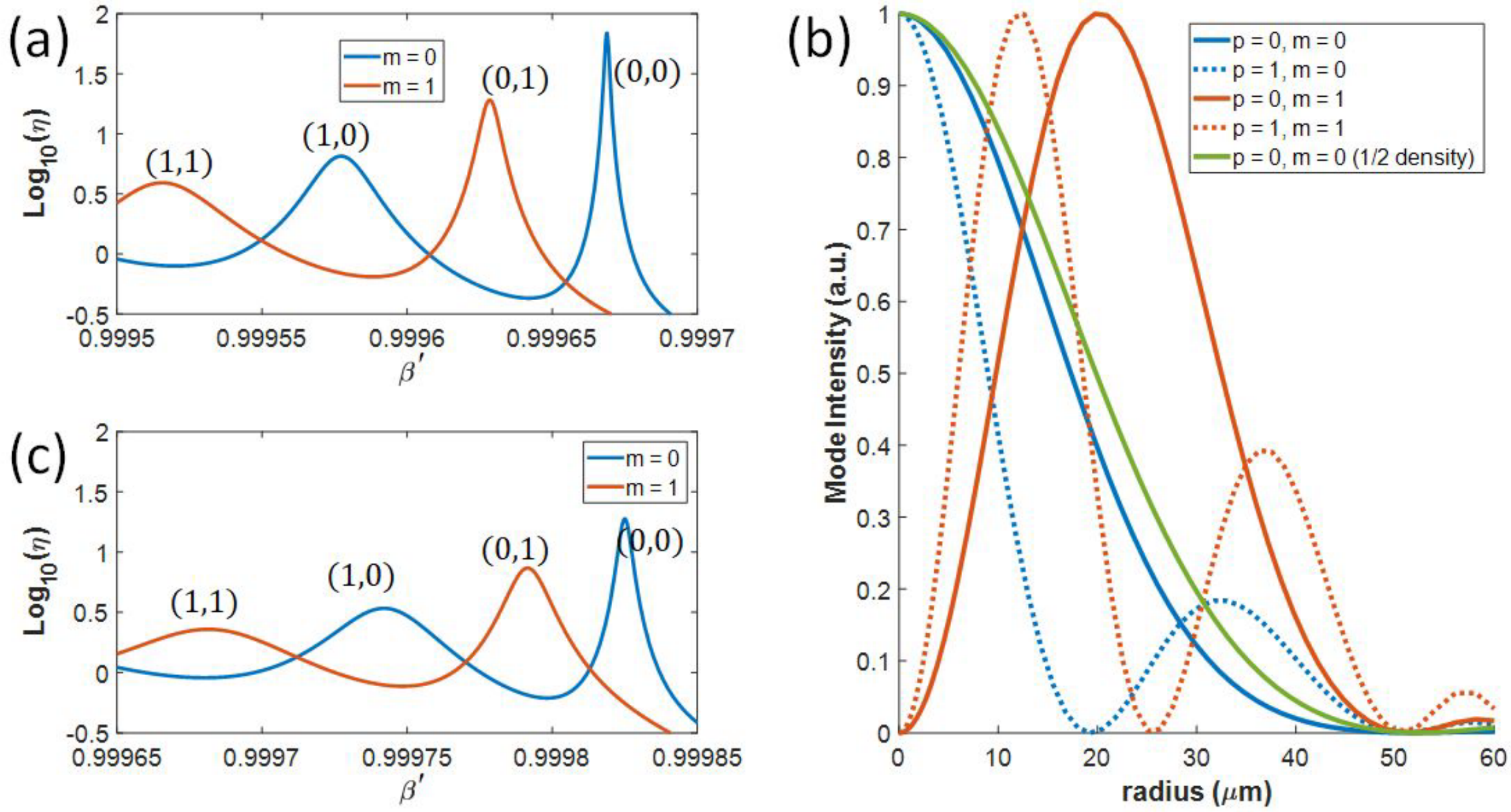

**Figure 3.** Calculated modes of the leaky plasma density profile $N_e(r)$ in Fig. 1(a). **(a)** $\eta(\beta')$ curves for $m = 0$ and $m = 1$; **(b)** Calculated mode intensity plots for $m = 0$ ($p = 0$ and $p = 1$) and $m = 1$ ($p = 0$ and $p = 1$). Also included is the $p = 0, m = 0$ mode with $N_e(r)$ in Fig. 1(a) halved. **(c)** $\eta(\beta')$ curve ($m = 0$ and $m = 1$) for $N_e(r)$ in Fig. 1(a) halved.

A description of leaky modes that accounts for the quasiresonant behaviour of $\eta(\beta)$ is obtained by considering complex propagation wave numbers $\beta_{c,j} = \beta_j + i\alpha_j$, where $\alpha_j$ describes attenuation along the guide resulting from tunneling loss of field energy into the outgoing conical field $\tilde{E}_{\mathrm{out}}$ (T. R. Clark and Milchberg 2000). The purely outgoing radiation condition makes the transverse eigenvalue problem non-Hermitian, with the quasibound modes corresponding to resonance poles at complex propagation wave numbers (Snyder and Love 1984; Leung, Liu, and Young 1994; Ching *et al.* 1998). For the propagation factor $e^{i\beta_{c,j}z}$, a forward propagating leaky mode has $\alpha_j > 0$ and an axial dependence $e^{i\beta_j z - \alpha_j z}$.

If the $j^{\mathrm{th}}$ quasiresonance is sufficiently narrow and isolated, a distinguishable transverse profile $u_j(r)e^{im\phi}$ can be associated with it, and we write for the propagating mode $\tilde{E}_j(\mathbf{r}_\perp, z, \omega) = u_j(r)e^{im\phi}q_j(z)$, where $q_j(z) = A_j e^{i\beta_j z - \alpha_j z}$ for $z > 0$. The distribution of real propagation wave numbers $\beta$ associated with this axial amplitude is described by the Fourier transform pair

$$q_j(z) = \frac{1}{\sqrt{2\pi}} \int_{-\infty}^{\infty} \tilde{q}_j(\beta) e^{i\beta z}\, d\beta,$$
$$\tilde{q}_j(\beta) = \frac{1}{\sqrt{2\pi}} \int_0^{\infty} q_j(z) e^{-i\beta z}\, dz = \frac{A_j}{\sqrt{2\pi}\left[\alpha_j + i(\beta - \beta_j)\right]} \quad . \tag{21}$$

The lower limit in the second integral corresponds to taking $z > 0$ propagation of the launched mode to begin at $z = 0$ so that $q_j(z) = 0$ for $z < 0$. Near the quasiresonance, the waveguide core field produced by an incoming conical wave $\tilde{E}_{in}$ has a resonant contribution proportional to $\left(\beta - \beta_j - i\alpha_j\right)^{-1}$. This is the resonant loading of the guide, analogous, as mentioned, to a cylindrical Fabry–Perot resonator, with $\tilde{E}_{out}$ as the leakage wave. When the incident wave coupling and transverse profile vary slowly across the resonance and the non-resonant field is negligible, the enhancement defined in Eq. (20) is proportional to the power spectrum in $\beta$,

$$\eta_j(\beta) \simeq C_j\left|\tilde{q}_j(\beta)\right|^2 = \frac{C_j\left|A_j\right|^2}{2\pi\left[\alpha_j^2 + \left(\beta - \beta_j\right)^2\right]} \quad , C_j > 0. \tag{22a}$$

The constant $C_j$ accounts for the normalization of the incident field and the transverse integral over the core. This Lorentzian has a full width at half maximum $\Delta\beta = 2\alpha_j$. The power propagating in the guide varies as $P_j(z) = P_j(0)e^{-2\alpha_j z}$, giving the $1/e$ power attenuation length

$$L_{1/e} = \frac{1}{2\alpha_j} \simeq \frac{1}{\Delta\beta} = \frac{1}{k_0\Delta\beta'} \quad , \qquad \beta' = \frac{\beta}{k_0}. \tag{22b}$$

A plasma waveguide of length $L_{guide}$ is adequately confining only if $L_{1/e} > L_{guide}$, with "highly confining" applied to guides with $L_{1/e} \gg L_{guide}$.

An approximate cutoff condition for quasibound modes can be found by applying the result from the finite parabolic waveguide (Eq. (16a)), where $r_{\max}$ is identified with the radius of the density maximum in the cladding, at the shock peak. The condition $\kappa^2(r_{max}) < 0$ for an evanescent region at the cladding peak yields the approximate relation,

$$\Delta N_e(r_{max}) \gtrsim \frac{(2p + |m| + 1)^2}{\pi r_e r_{max}^2} \ , \tag{23}$$

as an estimate of the density contrast required to form a confining barrier for a $(p, m)$ mode. The attenuation length depends on the width and shape of this barrier.

Experiments have shown in some cases that hydrodynamic plasma waveguides at long delays resemble step-index fibers (B. Miao *et al.* 2020; L. Feder *et al.* 2020), with a relatively flat core and sharply rising cladding. In this case, we can use a step-index fiber analysis, for which the fundamental channel mode radius is given by (Marcuse 1977) as $w_{ch} \approx a(0.6484 + 1.619V^{-3/2} + 2.879V^{-6} + \cdots)$, where $a$ is the core radius and the step-index fiber parameter (Snyder and Love 1984) is $V = k_0 a(n_{core}^2 - n_{cladding}^2)^{1/2} \approx a(4\pi r_e \Delta N_e)^{1/2}$ , where $\Delta N_e = N_{e,max} - N_{e0}$. For a well-bound leaky mode of the step-like guide, approximate solutions to Eq. (11) for $r < a$ are $u_{pm}(r, \varphi, \omega) \propto J_{|m|}(\kappa_{pm} r)e^{im\varphi}$ ($m = 0, \pm1, \pm2, \ldots$and $p = 0,1,2 \ldots$ ) where $\kappa_{pm} a = \alpha_{pm}$ is the $p^{th}$ root of $J_{|m|}\left(\kappa_{pm} a\right) = 0$, and $\alpha_{pm} = \alpha_{p|m|}$. Note that here $p$ is a counting index, where we label the first root as $p = 0$ to make the fundamental mode label $(p, m) = (0,0)$ similar to the LG case. For $N_{e0}/N_{cr} \ll 1$, the associated propagation wavenumber for the step index plasma waveguide is

$$\beta_{pm} = k_0\left(1 - N_{e0}/2N_{cr} - \alpha_{pm}^2/2k_0^2a^2\right) \quad , \tag{24}$$

and an approximate condition for efficient guiding of a $(p, m)$ mode in a step-index plasma fiber is $V > \sim\alpha_{pm}$ or

$$\Delta N_e(r_{max}) > \sim \frac{\alpha_{pm}^2}{4\pi r_e a^2} \ . \tag{25}$$

For example, efficient guiding of a (0,0) mode in a step-index plasma waveguide ($\alpha_{00} = 2.405$) requires $\Delta N_e(r_{max}) > 4 \times 10^{17}$ cm$^{-3}$ for a core radius $a = 20$ µm.

### C. Linear coupling into plasma waveguides

The end-coupling efficiency of an incident pulse into the modes of a plasma waveguide can be calculated by decomposing the transverse incident field $\tilde{E}_i(r, \varphi, \omega)$ at the entrance of the waveguide into the quasibound modes and the radiation field (T. R. Clark and Milchberg 2000; Thomas R. Clark Jr. 1998),

$$\tilde{E}_i(r, \varphi, \omega) = \sum_{p,m} a_{pm} u_{pm}(r, \varphi, \omega) + \tilde{E}_{rad}(r, \varphi, \omega), \tag{26}$$

where $a_{pm}$ is the amplitude coefficient of the $(p, m)$ quasibound mode $u_{pm}$ (associated with wavenumber $\beta_{pm}$), and $\tilde{E}_{rad}(r, \phi, \omega)$ is the radiation field, which is non-negligible only for $r > r_b$. The $a_{pm}$ coefficients, for unit-normalized and approximately orthogonal modes in the well-confined limit, are determined by

$$a_{pm} = \int_A u_{pm}^* \tilde{E}_i dA \ , \tag{27}$$

where the integral is over the waveguide cross section. The power $P_{pm}$ linearly coupled into a given mode, and the coupling efficiency $\epsilon_{pm}$ are

$$P_{pm} = \frac{c}{8\pi}\left|a_{pm}\right|^2, \quad \epsilon_{pm} = \frac{\left|a_{pm}\right|^2}{\int_A \left|\tilde{E}_i\right|^2 dA} \ . \tag{28}$$

For an azimuthally symmetric waveguide injected on axis by an azimuthally symmetric pulse, coupling occurs only to the $m = 0$ waveguide modes ($a_{p,m} = 0$ for $m \neq 0$) so that only ($p \geq 0, m = 0$) modes survive. Coupling into $|m| > 0$ modes requires breaking of azimuthal symmetry (T. R. Clark and Milchberg 1998a), accomplished with an incident beam that is asymmetric or transversely offset, or whose axis is tilted with respect to the guide axis. In general, the coupling coefficients $a_{pm}$ can be calculated numerically. An instructive analytic result is obtained for a fundamental Gaussian beam of waist $w_0$ injected at a small angle $\theta$ with respect to the axis of an infinite parabolic guide, where the waveguide modes $u_{pm}$ are given by Eq. (13), and coupling efficiency into the $(0, m)$ and $(p, 0)$ modes from Eq. (28) (T. R. Clark and Milchberg 2000) is

$$\epsilon_{0m} = \frac{P_{0m}}{P_i} = \frac{4\,(k\theta w_0)^{2|m|}}{2^{|m|}|m|!}\left(\frac{r}{1+r^2}\right)^{2|m|+2} \exp\left[-\frac{(k\theta w_0)^2}{2(1+r^2)}\right] \tag{29a}$$

$$\epsilon_{00} = \frac{P_{00}}{P_i} = 4\left(\frac{r}{1+r^2}\right)^2 \exp\left[-\frac{(k\theta w_0)^2}{2(1+r^2)}\right] \tag{29b}$$

$$\epsilon_{p0}(\theta=0) = \frac{P_{p0}}{P_i} = 4\left(\frac{r}{1+r^2}\right)^2 \left(\frac{1-r^2}{1+r^2}\right)^{2p} \tag{29c}$$

Here $r = w_0/w_{ch}$ and Eq. (29c) is for $\theta = 0$. As $\theta \to 0$, $\epsilon_{0m} \to 0$ (for $m \neq 0$) while $\epsilon_{00} \to 1$, provided that the input Gaussian beam is well-matched to the channel ($r = 1$). As $\theta$ increases, $\epsilon_{00}$ decreases exponentially, with more energy coupled into the $|m| > 0$ modes. The optimum tilt angle for coupling into the $(0, m)$ mode is obtained from Eq. (29a) as

$$\theta_{opt} = \frac{\sqrt{2|m|}}{kw_0}(1 + w_0^2/w_{ch}^2)^{1/2} \ . \tag{30}$$

At fixed tilt angle $\theta$ and spot-size ratio $r$, the coupling in Eq. (29a) varies with $|m|$ as $A^{|m|}/|m|!$. The most strongly excited $p = 0$ mode is therefore

$$|m|_{opt} = \text{int}(A), \quad A = \frac{(k\theta w_0)^2 r^2}{2(1+r^2)^2} \ , \tag{31}$$

where $\text{int}(A)$ denotes the largest integer not exceeding $A$. If $A$ is an integer $\geq 1$, the adjacent modes $|m| = A - 1$ and $|m| = A$ have equal maximum coupling. For $r \gg 1$, $A \approx (k\theta w_{ch})^2/2$, independent of the input spot size. For $r \ll 1$, $A \approx (k\theta w_0)^2 r^2/2$, so excitation of nonzero azimuthal orders is suppressed as the input spot underfills the guide. For $\theta = 0$, coupling into $p > 0$ modes can occur when the injected beam waist underfills or overfills the waveguide, as seen from Eq. (29c), or is located away from the waveguide entrance (Durfee, Lynch, and Milchberg 1994). Coupling into $p > 0$ modes also occurs when the radial dependence of the incident field is different than that of $u_{00}(r, \beta_{00})$, such as would occur when coupling a Gaussian beam into a step-index waveguide, or a Jinc-like beam into a parabolic channel (Leemans *et al.* 2014; Djordjević *et al.* 2018). In any of these conditions, slight transverse offsets between injected beam and guide axes are sufficient to break cylindrical symmetry and lead to $m \neq 0$ modes. At sufficiently low waveguide depth $\Delta N_e(r_{max})$, only the (0,0) mode will survive, as seen from Eq. (16a), although one might have $L_{1/e} < L_{guide}$.

End-coupling into experimentally generated waveguides can be complicated by the presence of density ramps (gradients) at the entrance of the waveguide. These are of particular concern for guiding high power pulses, where ionization-induced refraction may hinder coupling (Durfee, Lynch, and Milchberg 1994; Nikitin *et al.* 1997; Nikitin *et al.* 1999; Dorchies *et al.* 1999). Particularly for hydrodynamically generated waveguides, reduced heat deposition at lower densities may locally reduce the mode size, causing coupling losses for a pulse matched to the waveguide size after the ramp. The entrance ramp also changes the beam radius and wavefront curvature before the pulse reaches the main guiding channel. Consequently, matching the vacuum focal spot to the channel mode does not ensure matched propagation. (Giacone *et al.* 2005) and (Dimitrov *et al.* 2007) examined this entrance evolution using PIC simulations. (Dimitrov *et al.* 2007) additionally used a spot-size evolution equation to select the incident waist and vacuum focus position, accounting for focusing within the ramp. The resulting reduction of spot-size oscillations and transverse leakage shows how the entrance profile and incident beam can be matched together.

The introduction of a "funnel mouth" at the waveguide entrance can mitigate the effects of end-taper. For hydrodynamic plasma waveguides, the plasma funnel has been shown to act as a lens and improve coupling efficiency (K. Y. Kim *et al.* 2002; Cooley *et al.* 2003; Wu *et al.* 2005). This implementation required an additional laser pulse for funnel formation. A more recent technique demonstrated in OFI-generated guides (Tripathi *et al.* 2025; Sloss *et al.* 2026) forms the funnel and waveguide simultaneously with a specially tailored Bessel-like beam (see Sec. V.B).

Experimental assessment of coupling requires separating the energy launched into guided modes from losses accumulated during propagation. (Spence, Butler, and Hooker 2003) inferred these contributions separately in gas-filled discharge capillaries, obtaining coupling losses below 4% and a propagation-loss coefficient of $\sim 7\ \mathrm{m}^{-1}$. The transverse distribution of transmitted energy is also important: (Geltner, Ping, and Suckewer 2003) observed substantial transmission through discharge-ablated LiF microcapillaries, while only a small fraction of the incident energy emerged in the narrow guided mode. These measurements illustrate the need to resolve both transmission losses and the spatial distribution of the transmitted field when assessing coupling.

One of the features of free-standing plasma waveguides, with finite cladding thickness, is the ability to side-couple into them. An experimental demonstration of this process was presented in (T. R. Clark and Milchberg 1998a), which reported frequency selective side coupling of incoming conical waves (provided by Bessel beams) into higher order quasibound modes. As discussed in Sec. II.B.3, a quasibound mode of the waveguide can be considered as resulting from conical waves incoming and outgoing from the plasma structure. From Eq. (19), these conical waves couple most efficiently to the $j^{th}$ quasibound mode at a frequency-dependent approach angle

$$\gamma(\omega) = \tan^{-1}\big(\kappa_0(\omega)/\beta_j\big) \ . \tag{32}$$

Side-coupling of an incoming conical wave is the inverse of the leakage process: the incident light is tuned for deliberate excitation of a specific quasibound mode, which it excites by tunneling through the waveguide cladding into the core. For the $(p, m)$ quasibound mode of a leaky parabolic guide, Eq. (32) becomes, for small $\gamma$,

$$k_0 \tan\gamma \approx \left(4\pi r_e N_{e0} + \frac{4}{w_{ch}^2}(2p + |m| + 1) + 4\pi k_0^2 \chi_0\right)^{1/2} \ , \tag{33}$$

showing that side coupling to the $(p, m)$ mode of a given waveguide is optimized by adjustment of the input angle or wavelength.

An experimental configuration employed in (T. R. Clark and Milchberg 1998a) to demonstrate wavelength-dependent side coupling is shown in Fig. 4(a). The plasma channel was produced through the hydrodynamic expansion of an inverse Bremsstrahlung (IB)-heated plasma column (Sec. III.A) generated by a $J_0$ Bessel beam produced by passing a laser pulse through an axicon, a conical glass lens (Sec. V.B.1). A counter-propagating supercontinuum Bessel beam probe was focused along the channel by an axicon, enabling preferential excitation of specific higher order modes by variation of $k_0$ with a fixed $\gamma$, as predicted by Eq. (33). Figure 4(b) shows increasingly higher order modes excited by expanding the waveguide size with pump-probe delay. Note that side-coupling preferentially excites higher order modes. Because these are the leakiest modes, they are the easiest to side-couple to.

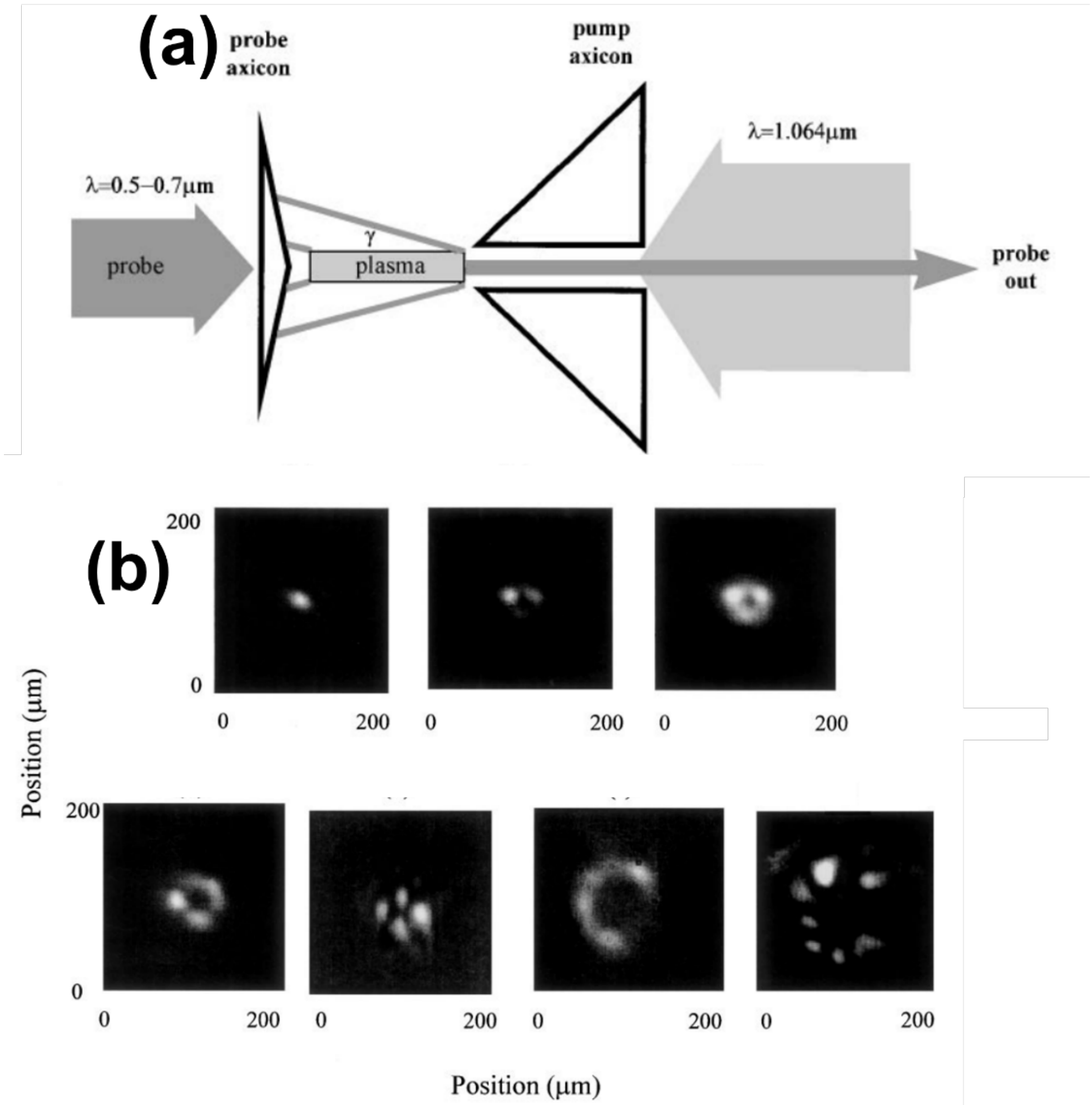

**Figure 4.** Higher order mode excitation via side coupling of a supercontinuum probe pulse (both panels adapted from (T. R. Clark and Milchberg 1998a)). **(a)** Experimental layout. A supercontinuum probe beam (entering from the left) is focused by an axicon, tunneling into the channel generated by a counterpropagating Bessel beam. **(b)** Increasingly higher order modes excited by increasing the waveguide size with pump-probe delay. Note that side-coupling preferentially excites higher order modes. Because these are the leakiest modes, they are the easiest to side-couple to.

### D. Multimode propagation in plasma waveguides

#### 1. Linear modal description

In the linear regime ($a_0 \ll 1$), pulse propagation dynamics in plasma waveguides is dominated by modal effects. Energy coupled into two co-propagating modes with wavenumbers $\beta_{p_1 m_1}$, $\beta_{p_2 m_2}$ will interfere as they propagate through the guide with a spatial frequency and period

$$\Delta\beta = \left|\beta_{p_1m_1} - \beta_{p_2m_2}\right|, \qquad \Lambda = \frac{2\pi}{\Delta\beta} \ . \tag{34}$$

This interference is called *mode beating* and results in both longitudinal and transverse deformations of the guided field. The relative amplitude of the interference effects is determined by the relative coupling into the different modes. For coupling coefficients $a_{p_1m_1}, a_{p_2m_2}$ (Eq. (26)), the local intensity ratio at a specified transverse position $r$ is

$$\mathcal{R} = \left|\frac{a_{p_1m_1}u_{p_1m_1}(r)}{a_{p_2m_2}u_{p_2m_2}(r)}\right|^2, \tag{35}$$

with the numerator chosen to be the larger local intensity. The spatial mode functions are evaluated at the observation point; $\left|a_{p_1m_1}/a_{p_2m_2}\right|^2$ gives the power ratio for unity-normalized modes. For two modes of the same polarization, the local modulation contrast or visibility is then (L. Feder *et al.* 2020)

$$v = \frac{I_{max} - I_{min}}{I_{max} + I_{min}} = \frac{2\sqrt{\mathcal{R}}}{\mathcal{R} + 1} \ . \tag{36}$$

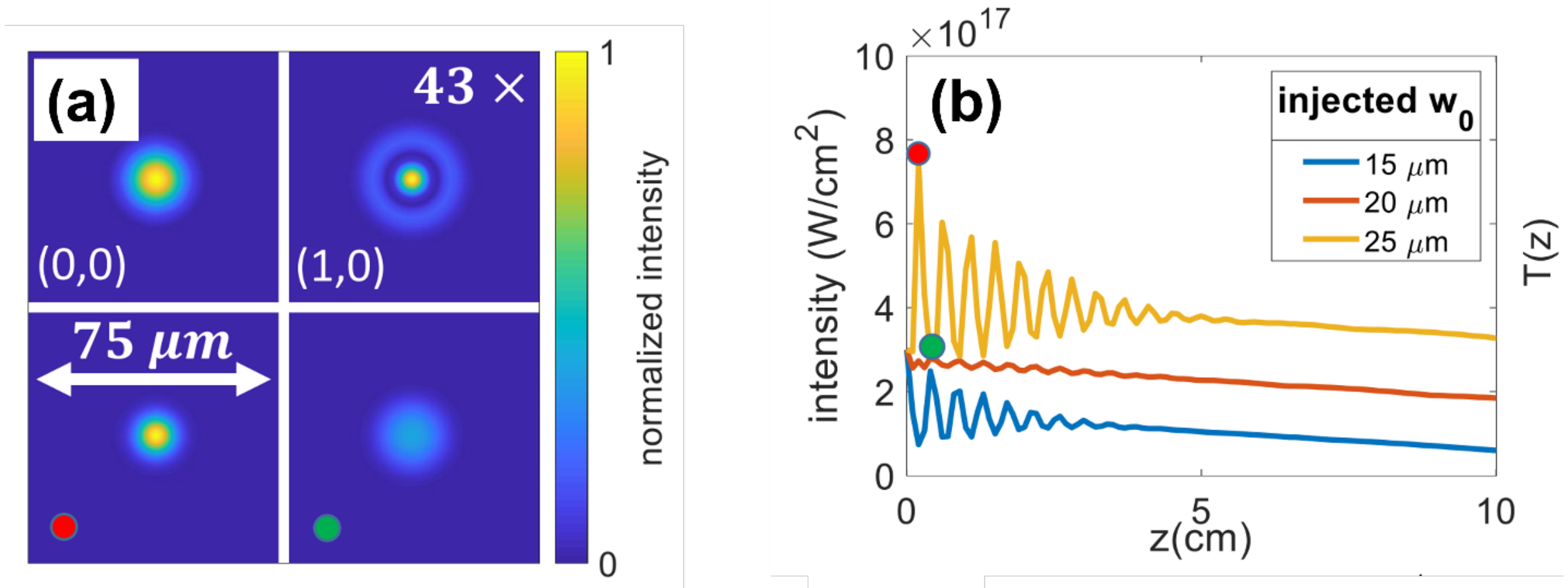

**Figure 5.** Effects of (0,0)-(1,0) mode beating in leaky plasma waveguides observed in YAPPE ("yet another pulse propagation effort") simulations (L. Feder *et al.* 2020) using the unidirectional pulse propagation algorithm (Kolesik and Moloney 2004) (both panels adapted from (L. Feder *et al.* 2020)). **(a)** Top panels: calculated $(0,0), (1,0)$ relative intensity profiles for the two lowest order modes of an experimentally characterized waveguide ($w_{ch}$~$20\ \mu m$) and an injected Gaussian pulse with $w_0 = 25\ \mu m$. Bottom panels: calculated intensity profiles showing the effects of beating at different points during propagation. **(b)** Intensity vs. propagation length for pulses with different $w_0$ injected into the guide. The red and green dots on the gold curve correspond to the bottom panels in (a). The gold, red, and blue curves are for overfilled, matched, and underfilled injection, respectively.

An example of higher order mode excitation and beating is provided in Figure 5 (L. Feder *et al.* 2020). The top panels of Fig. 5(a) plot the calculated intensity profiles of $(0,0)$ and $(1,0)$ modes for an experimentally generated channel with $w_{ch}$~$20\ \mu m$. For an injected Gaussian pulse with $w_0 = 25\ \mu m$ (overfilled injection), the calculated quasibound mode excitation in the actual channel profile gives an on-axis intensity ratio $\mathcal{R} \approx 43$. Beam propagation simulations (L. Feder *et al.* 2020) in Fig. 5(b) show beating between these modes (gold curve), with the red and green dots marking the propagation positions corresponding to the images in the bottom two panels of Fig. 5(a). The other curves are for underfilled injection ($w_0 = 15\ \mu m$, blue curve) and matched injection ($w_0 = 20\ \mu m$, red curve). For all curves, the decaying amplitude is due to guide leakage. At fixed input peak intensity, as in these simulations, the ideal-parabolic coupling estimate in Eq. (29b), together with the change in beam area, explains why, when $w_0 > w_{ch}$ $(r > 1)$, the intensity of the guided $(0,0)$ mode is higher than that of the matched mode: the oversized spot is squeezed into the narrower channel, increasing the intensity. But when $w_0 < w_{ch}$, the intensity is reduced.

While mode beating can be minimized by optimizing coupling into the fundamental mode of the waveguide (Durfee, Lynch, and Milchberg 1994; E Esarey and Leemans 1999; T. R. Clark and Milchberg 2000), initial linear matching does not generally eliminate oscillations once an intense pulse modifies the channel. In long, low-density waveguides, sustained beating of the lowest-order modes of the modified channel can periodically enhance ionization injection and shape the accelerated electron spectrum (Shrock *et al.* 2024); see Sec. II.E and Fig. 7.

For linear propagation, there are two primary mechanisms for the reduction of mode beating: mode slippage (or walkoff) and leakage. As discussed earlier in the context of Eq. (15), higher order quasibound modes $p > 0$ or $m \neq 0$ have lower group velocity and can undergo group velocity walkoff, lagging the fundamental $(0,0)$ mode. Higher order modes are progressively slower, with narrower guides (smaller $w_{ch}$) enhancing the walkoff effect.

For short ($< \sim 100$ fs) pulses propagating in long plasma waveguides (tens to hundreds of Rayleigh ranges), such as those used for multi-GeV LWFA (B. Miao, Shrock, *et al.* 2022; Picksley *et al.* 2024), higher order modes increasingly lag the fundamental mode to the point where their temporal envelopes no longer overlap the fundamental, and beating ceases (Shrock *et al.* 2024). For example, a 40 fs FWHM pulse with $w_0 = 30$ μm couples into the $(p = 1,2,3, \ldots, m = 0)$ modes of a waveguide with $w_{ch} \approx 27$ μm. All of these modes walk off and trail the fundamental (0,0) mode after just 5 cm of propagation, and beating ceases. This is shown in Fig. 6, which uses the particle-in-cell code WarpX (Vay *et al.* 2021) to compute the laser propagation in a 20 cm long plasma waveguide. The early phase of propagation (Fig. 6(a)) shows clear effects of mode beating on the laser intensity envelope. After ~18 mm of propagation, higher order structure begins to detach from the leading part of the pulse. By 50 mm (panel (b)), the effect of group velocity walkoff is clearly seen, with evenly spaced pulselets of increasingly higher order structure trailing the pulse head. This means that wake excitation is primarily caused by energy propagating in the fundamental (0,0) mode of the channel.

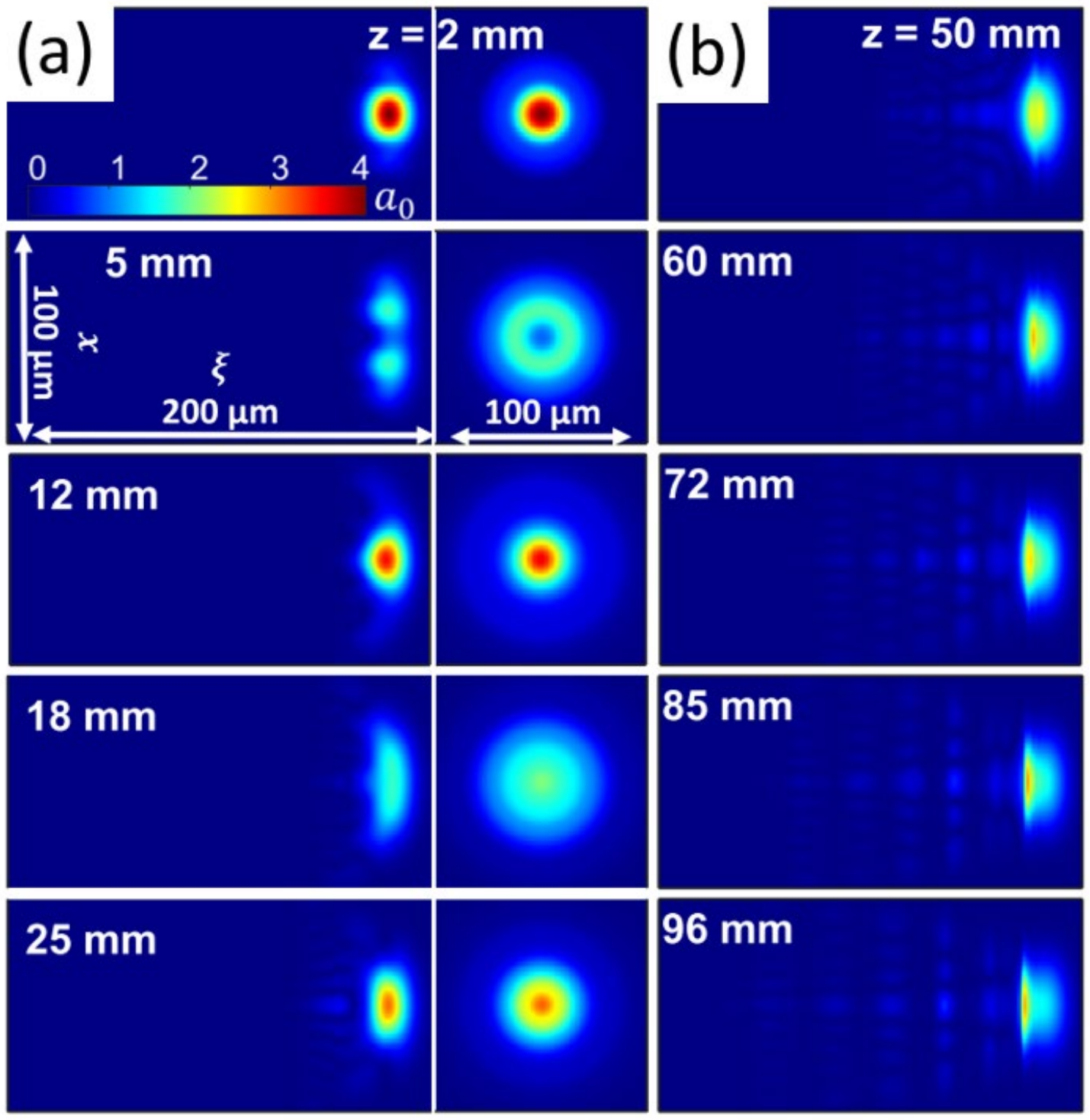

**Figure 6.** Example of group velocity walkoff of higher order modes in a low-density plasma waveguide (partially adapted from (Shrock *et al.* 2024)). Pulse propagation is left to right. Here $x$ is a transverse coordinate and $\xi = v_g t - z$ is a space coordinate in the moving frame of the pulse. **(a)** *left column*: $x\xi$ slices of a mismatched high intensity pulse. *Right column*: $\xi$-integrated mode. **(b)** $x\xi$ slices from the next 50 mm of propagation showing continued walkoff of higher order modes and self-steepening.

For longer pulses, incomplete walkoff may reduce the beating modulation contrast and cause effective pulse lengthening and delay of the energy centroid (E Esarey and Leemans 1999; Schroeder *et al.* 2011); in those cases, elimination of higher-order mode content from the pulse envelope requires leakage of the higher order modes. In highly confining guides, sufficient leakage to mitigate beating can require meters of propagation (T. R. Clark and Milchberg 2000; B. Miao *et al.* 2020; L. Feder *et al.* 2020; Spence, Butler, and Hooker 2003; Gonsalves *et al.* 2019). Lower index contrast plasma waveguides can act as mode-filters by preferentially leaking higher order modes (Antonsen and Mora 1995; T. R. Clark and Milchberg 2000; Djordjević *et al.* 2018). For ultrashort ($< \sim 100$ fs) pulses, group velocity walkoff can mitigate mode beating in both high and low contrast waveguides (Shrock *et al.* 2024).

## 2. Envelope and mean-radius descriptions

The modal description makes explicit how interference and relative delay among transverse modes determine the evolving pulse profile. A complementary approach follows the pulse envelope or its transverse moments, providing a direct way to examine how diffraction, plasma focusing, and the intensity-dependent plasma response govern the beam radius and its evolution. Early work by (P. Sprangle *et al.* 1992) derived and numerically solved a two-dimensional axisymmetric relativistic Maxwell–fluid model, using quasistatic, cycle-averaged, and paraxial approximations to follow intense pulses over many Rayleigh lengths. For analytic interpretation, (Eric Esarey, Krall, and Sprangle 1994) used a source-dependent expansion (SDE) in Laguerre–Gaussian functions but closed the description by retaining an approximately Gaussian lowest-order profile, thereby obtaining an evolution equation for the spot radius. The full SDE permits higher-order components; the Gaussian closure simplifies the transverse dynamics. That equation and related effective-radius models describe the competition among diffraction, channel focusing, relativistic self-focusing, and wake-induced focusing, and were used to derive matching conditions and self-modulation thresholds and growth rates. Related work treated radius tailoring and hose or self-modulation (Krall *et al.* 1994; Phillip Sprangle, Krall, and Esarey 1994).

Subsequent work carried these reduced descriptions into the guided ultrashort pulse and accelerator regimes, typically assuming a cylindrical, highly confining waveguide with no leakage, a good model for capillary discharge waveguides (Secs. IV.B.3 and IV.B.4). In (E Esarey and Leemans 1999), the derived nonparaxial corrections and equations for spot size and pulse centroid evolution in an ideal parabolic plasma channel showed pulse length broadening via modal group velocity dispersion and the effects of mode beating (described there as "betatron oscillation" of the spot size). They also derived damping of these oscillations through phase mixing associated with the pulse's finite bandwidth. (Schroeder *et al.* 2011) solved the corresponding nonparaxial envelope problem for arbitrary Laguerre–Gaussian order and used centroid and spot-size evolution to quantify the group velocity and pulse lengthening of mismatched pulses. These results mostly assume bound modes. WKB-based leakage rates may also be incorporated into reduced models applied to truncated parabolic leaky profiles (Volfbeyn, Esarey, and Leemans 1999; Djordjević *et al.* 2018).

These reduced models are useful for estimating the matched spot size and the oscillations caused by imperfect matching, understanding how relativistic and wake-induced focusing modify the balance between diffraction and channel focusing, and determining when perturbations to the pulse grow during propagation, provided that the phase front curvature is well-defined. Their predictions can also be used to infer waveguide properties from measured beam evolution, assuming the transverse channel structure is known. For example, (Gonsalves *et al.* 2010) inferred the matched spot size of a capillary discharge waveguide from output beam radius and centroid measurements, relating these propagation parameters to the channel's transverse focusing strength. The modest computational requirements of reduced models also make them useful for design scans and for supplying estimates or initial conditions for more complete simulations. When the actual transverse profile, entrance coupling, bound- or leaky-mode cutoff, and multimode content are important, the modal treatment supplies a direct physical picture; for leaky profiles it also gives the attenuation of individual quasibound modes. When the pulse strongly modifies the channel or drives a large-amplitude wake, the field must be evolved with a self-consistent plasma response, as discussed in Sec. II.E.

### E. Propagation of relativistic intensity pulses

For pulses with field strengths approaching and exceeding $a_0 \sim 1$, the laser ponderomotive force, $\mathbf{F}_p = -mc^2\nabla\bar{\gamma}$, can modify the plasma density profile forming the waveguide. Here $\bar{\gamma} = (1 + a_0^2/2)^{1/2}$ is the laser cycle-averaged Lorentz factor for an electron in the laser field; the gradient in $\bar{\gamma}$ originates from the spatial dependence of $a_0$ in the ponderomotively modified plasma waveguide. Perturbation of the plasma via the ponderomotive force results in the head and tail of a guided pulse experiencing different local plasma structure, resulting in highly complex pulse evolution. This is compounded over the extended propagation distances at the low plasma density required for high-energy LWFA, as discussed in Sec. II.F. Resolving the extreme pulse distortion and plasma modification, as well as electron trapping and the accelerated particle distribution, requires a kinetic description, provided by particle-in-cell (PIC) (Birdsall and Langdon 1991; Fonseca *et al.* 2002; Arber *et al.* 2015), boosted-frame PIC (Vay 2007), and quasistatic PIC simulations (Mora and Antonsen 1997; Huang *et al.* 2006), even with symmetry assumptions (Rémi Lehe *et al.* 2016).

Before PIC simulations could routinely resolve this complexity, reduced analytic models established much of the physical understanding used to design relativistic channel-guided experiments. In the weakly relativistic regime, (E. Esarey *et al.* 2000) retained the leading nonparaxial correction, allowing energy to move longitudinally within the pulse. They showed how the resulting power-dependent group velocity steepens the pulse envelope and analyzed the coupling between forward Raman scattering and self-modulation driven by spot-size variations. (P. Sprangle, Hafizi, and Peñano 2000) calculated modulation growth while accounting for the finite longitudinal extent of the initial disturbance and the transverse structure of the laser pulse.

In (Benedetti *et al.* 2012) a transverse-moment approach was used to determine how the spot size should vary along an intense pulse to minimize spot-size oscillations during propagation in a plasma channel. This tailoring accounts for the variation of the plasma's focusing response along the pulse. Earlier work by (Krall *et al.* 1994) demonstrated extended propagation of radius-tailored pulses in an initially uniform plasma. The quasi-matched solutions of (Benedetti *et al.* 2012) apply to general radial channel profiles and include the plasma response, treated analytically in the weakly relativistic limit and numerically at higher intensities. The initial spot size of each pulse slice is chosen to balance diffraction against plasma focusing, keeping its intensity-weighted mean-square radius constant during propagation. This quasi-matching condition is derived neglecting self-steepening and pulse depletion, which can change the slice power and plasma response and thereby alter the required matching. A later model (Benedetti *et al.* 2015) relaxed the assumption of Gaussian profiles to numerically extract a "supermatched" pulse shape from coupled equations for laser field and plasma density evolution. Supermatching suppresses transverse mismatch over distances short compared with the depletion length. Experimental investigation at high power requires control of the transverse profile along the pulse; emerging techniques such as coherent combining of fiber lasers may enable more complex mode matching in the future (Zhou *et al.* 2015; R. Lehe *et al.* 2025; Stark *et al.* 2023; Kling *et al.* 2024; Rainville *et al.* 2024).

In LWFA applications of plasma waveguides, longitudinal pulse evolution can limit the useful acceleration distance. (Shadwick, Schroeder, and Esarey 2009) related laser energy depletion to photon red-shifting and pulse reshaping, with initial compression followed by lengthening and loss of resonant wake excitation, while (Benedetti *et al.* 2015) showed that nonlinear pulse evolution can lower the wake phase velocity below the laser group velocity, thereby changing trapping and dephasing. For weakly relativistic pulses in plasma channels, (Benedetti *et al.* 2015) obtained analytical expressions for depletion, self-steepening, and intensity centroid motion, and related this evolution to the wake phase velocity through comparison with numerical solutions.

Over long propagation distances, transverse confinement and longitudinal pulse evolution become closely coupled. The pulse modifies the plasma density and hence its own focusing environment, while energy transfer to the wake reshapes the pulse and changes the intensity driving that response. A pulse initially matched to the channel can therefore develop oscillations as it propagates, with consequences for wake excitation, electron injection, and acceleration. Following this feedback requires self-consistent calculations of the evolving laser field and plasma response. Fluid models can follow the coupled evolution of the laser field and plasma density within the limits of their approximations. Kinetic PIC calculations, including formulations that use the quasistatic approximation, also follow particle trapping and the accelerated electron distribution. Recent PIC studies relate this coupled evolution to measurements in long, low-density plasma waveguides (B. Miao, Shrock, *et al.* 2022; Shrock *et al.* 2024; Picksley *et al.* 2024; Ludwig *et al.* 2025).

To interpret this coupled evolution, we first consider how excitation and interference of waveguide modes shape the propagating pulse. This heuristic approach uses estimates from linear theory to explain key features of recent simulations and experiments in meter-scale, low-density plasma waveguides, with reasonable quantitative agreement (Shrock *et al.* 2024). Here, the simple picture is applied to the results of PIC simulations, plotted in Fig. 7, which model propagation of relativistic intensity pulses in a 20 cm long plasma waveguide. Upon injection of a mode-mismatched pulse with $a_0 > 1$ into a plasma waveguide, pulse evolution occurs in three stages: (I) mode beating and higher order mode group velocity walkoff; (II) mode beating of the (0,0) and (1,0) modes of the ponderomotively-modified waveguide and pulse erosion; (III) pulse depletion, stretching, and collapse.

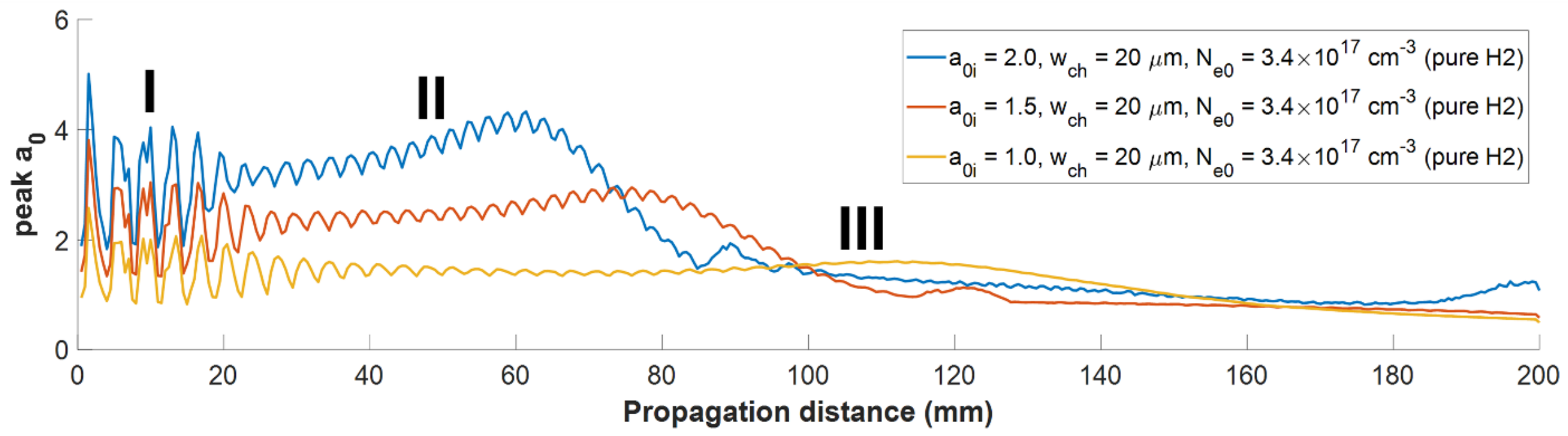

**Figure 7.** Examples of three stages of pulse evolution (adapted from online supplement of (Shrock *et al.* 2024)).

Stage I is understood from linear mode beating, as discussed in Sec. II.D. A linearly mismatched or misaligned incident pulse couples into multiple higher order modes of the plasma waveguide, resulting in mode beating that is sustained until leakage of higher order modes and/or group velocity walkoff causes the beating to decay. Beating is potentially deleterious to LWFA, for example, by inducing distortions of the plasma wakefield. Mode beating can be minimized by linearly matching the injected pulse mode to the fundamental mode of the waveguide, and ensuring alignment of the beam and waveguide axes in transverse position and pointing. As discussed above, group velocity walkoff is more pronounced for smaller diameter guides. For the wider waveguides proposed for implementation of >1 m waveguides for LWFA to tens of GeV ($w_{ch}$~100 μm) (Ludwig *et al.* 2025), significantly longer propagation distances are needed for group velocity walkoff of higher order modes. For transversely offset pulses, (Xia *et al.* 2026) developed an analytical description of centroid-oscillation damping by higher-order mode leakage and group-velocity walkoff. At relativistic intensity, they identified an additional mechanism: channel modification and red-shifting cause different longitudinal slices of the pulse to oscillate at different rates, producing phase mixing that reduces the oscillation of the pulse-averaged centroid.

Stage II begins as the initial mode beating has ceased. With the initial higher-order content reduced, the intense fundamental mode continues to drive a wake and modifies the channel; there is a longitudinal dependence of the guided mode size as a result of ponderomotive modification of the channel by the pulse. As the pulse propagates through the guide, pulse front erosion and red-shifting from wake excitation (Shadwick, Schroeder, and Esarey 2009), along with self-steepening (E. Esarey *et al.* 2000) at the back of the pulse, act to continuously feed energy from the leading edge of the pulse, where $w_{ch} = w_{ch,0}$ (the unperturbed guide mode size), to the cavitated region where the pulse centroid is located and the perturbed mode size $w_{ch,p} < w_{ch,0}$. This effective spot size mismatch couples dominantly to the $(0,0)$ and $(1,0)$ modes of the ponderomotively modified channel, leading to a beat period $\Lambda = 2\pi/|\beta_{10} - \beta_{00}| = \pi^2\lambda_0(w_{ch,p}/\lambda_0)^2$. For the initially Gaussian pulses studied, these stage II oscillations occur in the long, low density plasma waveguides appropriate for multi-GeV LWFA, whether or not the injected pulse is linearly mode matched to the guide (Shrock *et al.* 2024). For working gases doped for ionization injection (Rowlands-Rees *et al.* 2008; Pak *et al.* 2010; McGuffey *et al.* 2010; Clayton *et al.* 2010; M. Chen *et al.* 2012) into the laser wakefield, these oscillations are correlated with enhancement and suppression of injection (Shrock *et al.* 2024), leading to modulated electron energy spectra.

Eventually, the accumulated nonlinear effects result in depletion of the drive pulse, extreme red-shifting and broadening, and rapid decay of the peak intensity; this is stage III. The depletion and red-shifting reduce the group velocity of the drive pulse, and cause early onset of depletion induced dephasing (Shadwick, Schroeder, and Esarey 2009; B. Miao, Shrock, *et al.* 2022), which can dramatically alter the accelerated bunch properties, including reduction of the overall accelerated energy and loss of any unique energy structure. Significant modification of the pulse through nonlinear plasma interaction is a key concern for implementation of meter-scale plasma waveguides. In general, lower plasma densities and lower intensity drive pulses can mitigate these effects, but with the push towards longer guides for LWFA, the accumulated effects can become a concern, even in the quasilinear regime.

### F. Plasma waveguide density and length requirements for LWFA

In laser wakefield acceleration, the useful acceleration distance is limited by both laser depletion and dephasing of the accelerated electrons relative to the plasma wake (E. Esarey, Schroeder, and Leemans 2009). A simple dephasing estimate illustrates why increasing the target electron energy requires lower plasma densities and longer waveguides. Dephasing occurs as the velocity of the accelerated bunch $v_e$ increases and surpasses the phase velocity $v_\varphi$ of the laser-driven plasma wakefield; the electrons can then move from an accelerating to a decelerating phase of the plasma wave, corresponding to passage across the accelerating half-cycle of a linear plasma wake. With $v_e \approx c$ and $v_\varphi \approx v_g$, the laser pulse group velocity (Eq. (15)), the corresponding distance over which this occurs is the dephasing length $L_d \approx \frac{1}{2}\lambda_p(1 - v_g/c)^{-1}$. For a density-based estimate we retain only the plasma-density term in $v_g$, neglecting the waveguide contribution to the group delay. We also neglect nonlinear changes to the wake velocity, including pulse-front etching as laser energy is transferred to the wake (Decker et al. 1996). This gives $L_d \sim \lambda(N_{e0}/N_{cr})^{-3/2}$, showing why lower plasma density permits longer acceleration distances.

Depletion imposes a second requirement on the guiding length: the pulse must retain enough energy to sustain the accelerating wake over that distance. Energy transfer to the wake red-shifts the laser spectrum, while pulse reshaping changes the rate of energy loss. Eventual pulse lengthening can weaken wake excitation before the laser energy is exhausted (Shadwick, Schroeder, and Esarey 2009). An energy-balance estimate compares the laser energy with the energy deposited per unit length in the wake. In the one-dimensional, weakly relativistic limit, with the pulse duration $\tau_{pulse}$ chosen for resonant wake excitation, $\tau_{pulse} \sim \pi/\omega_{p0} \propto N_{e0}^{-1/2}$, the characteristic pump-depletion length scales as $L_{pd} \propto \lambda_p^3 \lambda^{-2} a_0^{-2}$ (E. Esarey, Schroeder, and Leemans 2009). At fixed $a_0$ and $\lambda$, both $L_d$ and $L_{pd}$ scale as $N_{e0}^{-3/2}$, while requiring a longer pulse to remain resonant. At relativistic intensities, depletion also depends on pulse shape and its nonlinear evolution (Shadwick, Schroeder, and Esarey 2009). Thus, exploiting a meter-scale guide requires sufficient driver energy and an appropriate pulse duration to sustain the wake over the intended acceleration length; the coupled evolution discussed in Sec. II.E must be considered alongside the dephasing estimate.

A particular waveguide application other than LWFA may have its own criterion for "low density", which also depends on the driver laser wavelength. For this paper, we use an LWFA-based criterion—acceleration of an electron bunch to $\sim$10 GeV using a $\lambda = 800$ nm drive laser pulse. From the scalings in (W. Lu *et al.* 2007), LWFA of electrons to $\sim$10 GeV requires meter-scale acceleration with a laser pulse with $a_0 \sim 2$ driving the plasma wave. For $L_d \sim 1$ m, this gives $N_{eo}/N_{cr} \sim (\lambda/L_d)^{2/3} \sim 10^{-4}$ so that $N_{eo} \sim 2 \times 10^{17}\ cm^{-3}$. In the remainder of this paper, we will take this as our approximate criterion for a "low-density" plasma waveguide.

## III. LASER-GENERATED HYDRODYNAMIC PLASMA WAVEGUIDES

The earliest plasma waveguides–exhibiting low order optical mode structure and guiding of pulses over many Rayleigh ranges–were generated using short pulse laser ionization and heating of gases. The explosive hydrodynamic expansion of the laser-heated plasma into the background gas produced narrow electron density profiles with a minimum on axis, the desired profile for optical guiding. The technique was developed at the University of Maryland in the 1990s (Durfee and Milchberg 1993; Durfee, Lynch, and Milchberg 1994; Durfee, Lynch, and Milchberg 1995; T. R. Clark and Milchberg 1997; Fan, Clark, and Milchberg 1998; Nikitin *et al.* 1997), along with simulation and theoretical tools to model waveguide evolution and properties (Durfee and Milchberg 1993; T. R. Clark and Milchberg 2000), and diagnostics for their measurement (T. R. Clark and Milchberg 1997; Nikitin *et al.* 1999). These waveguides were formed from the hydrodynamic evolution of elongated gas density plasmas generated by $\sim$100 ps pulses with an extended focus generated by Bessel beams (Secs. II.B.1 and V.B). The use of such relatively short pulses was essential to ensure that the duration of laser heating was much shorter than the ~nanosecond acoustic timescale of the gas plasma evolution. Such impulsive laser heating enabled generation of optical fiber-like waveguides capable of guiding lowest order transverse modes with radii as low as a few microns. A quantitative discussion of relevant laser heating and plasma response timescales is found in the next section.

Early theoretical work from the 1970s recognized that a laser-heated plasma could induce a wide on-axis density depression, forming a refractive index structure to guide laser pulses modeled with ray optics (L. C. Steinhauer 1970; Loren C. Steinhauer and Ahlstrom 1971a). That and related work (Loren C. Steinhauer and Ahlstrom 1971b) were motivated by axially extending the interaction length for $CO_2$ laser-heated nuclear fusion (Dawson *et al.* 1971). Early experiments used laser pulses tens of nanoseconds long to generate breakdown sparks in gases, leading to several-mm-wide axial depressions in plasma density. For example, an electron beam injected into a spark produced with a 30-J, 30-ns Nd:glass laser pulse ($\lambda$=1.05 μm) exhibited reduced scattering due to the decreased plasma density in the beam path (Askar'yan and Tarasova 1974). Later work (Chu and Johnson 1975) measured the shock wave profile created during the interaction of a 30 J, 150-ns $CO_2$ laser pulse ($\lambda$=10.6 μm) with 30 Torr of helium. Refraction of the $CO_2$ beam was interpreted as lensing of the trailing part of the pulse by the plasma density profile created earlier in the pulse. In other early work, it was established that plasma density minima generated using pulsed power could refract a beam, using either an electrical discharge (Molen, Kristiansen, and Hagler 1973) or a theta pinch (Amherd and Vlases 1974). In these experiments, the wide plasmas were probed with large diameter beams whose Rayleigh range was greater than the plasma length; no increase in the high intensity propagation distance was demonstrated. In all of these early calculations and experiments, the generated plasma structures were at least hundreds of wavelengths wide, well modeled by a ray optics analysis, but much too wide for any distinguishable transverse optical modes.

### A. Laser generation of plasma waveguides by collisional ionization and heating of gases

The 1990s Maryland experiments served as a proof-of-principle demonstration that laser pulses short compared to the plasma acoustic timescale could generate narrow plasma waveguides capable of guiding low order optical modes over tens of Rayleigh ranges, akin to glass optical fibers. These new developments in optics and plasma dynamics motivated the theoretical and computational modeling of plasma waveguides, including the simulation of laser-induced hydrodynamics in gas plasmas (Durfee and Milchberg 1993; Durfee, Lynch, and Milchberg 1995; T. R. Clark and Milchberg 1997) and the scattering model for quasibound modes presented in Sec. II (T. R. Clark and Milchberg 1998a; T. R. Clark and Milchberg 2000).

In these early experiments, a zeroth order Bessel beam ($J_0$), produced by a 100 ps FWHM pulse incident on a conical glass axicon (Sec. V.B.1), generated a $< 2$ cm long line focus in either a gas-backfilled chamber or in a cm-scale gas jet (Nikitin *et al.* 1999), using Ar, $N_2$, Xe, and $N_2O$. The pulse was of sufficient peak intensity ($\sim 10^{13} - 10^{14}\ \mathrm{W/cm^2}$ in the $J_0$ beam central peak of radius $r_p = 2.405/k \sin\gamma$ ) and duration to seed initial gas ionization via multiphoton ionization and then drive further ionization and avalanche breakdown by collisional or inverse-Bremsstrahlung heating (Durfee, Lynch, and Milchberg 1995). Similar conical axicon-based experiments generated fully ionized waveguides in backfill helium gas (Gaul *et al.* 2000). In all experiments, sufficient initial gas densities were needed for efficient avalanche breakdown to full ionization or to a closed ionic shell (for example, heliumlike nitrogen, $N^{5+}$, in nitrogen gas) so that a subsequent guided pulse would do little additional ionization. The requirement of efficient avalanche ionization set the minimum initial plasma column density to be $> \sim 10^{19}$ cm$^{-3}$, with resulting electron temperature peaked on axis to tens of eV.

Once the initial plasma is generated, its radial pressure gradient launches a single-cycle acoustic wave that propagates as a cylindrical shock or blast wave moving at the plasma sound speed $c_s = (\gamma_c \bar{Z} k_B T_e / m_i)^{1/2}$ into the ambient gas, where $\gamma_c, \bar{Z}, k_B, T_e, m_i$, are the specific heat ratio, mean ionization state, Boltzmann constant, electron temperature, and ion mass. In its very early evolution, the shock consists of peripheral neutral gas compressed by the expanding plasma; it quickly becomes ionized by radial thermal conduction and UV radiation from the plasma, provided there is sufficient thermal energy in the plasma and the electron collisional mean free path does not exceed the shock width. The ionized shock becomes the effective cladding of the hydrodynamic plasma waveguide. For the collisionally heated waveguides considered here, the initial plasma is typically sufficiently hot and dense ($\sim 100$ eV, $\sim 10^{19}$ cm$^{-3}$) to continuously supply the thermal energy needed to rapidly ionize the fresh neutral gas compressed by the expanding shock. Moreover, the ion-ion collisional mean free path is typically smaller than the initial plasma radius, $\lambda_{ii}/r_p < 1$ (where $\lambda_{ii}$ is a low end estimate of the width of the ionized shock), so a well-defined ionized shock always develops (Durfee, Lynch, and Milchberg 1995; T. R. Clark and Milchberg 1997). In low-density OFI-heated plasmas, the conditions for shock formation and ionization are considerably more restrictive, as discussed in Sec. III.C.

As the shock propagates radially, the on-axis mass density drops and, owing to quasi-neutrality, the plasma density drops by the same fraction. The lower density near-axis plasma forms the waveguide core; the ionized cladding is formed by outward thermal conduction to the fresh neutral gas compressed by the outward propagating shock. Figure 8(a)-(c) shows early interferometric (see Sec. V.A.1) measurements of this process, with analysis of the shock expansion plotted in panels (d) and (e) (T. R. Clark and Milchberg 1997). The result is a waveguide profile similar to that sketched in Fig. 1(a), with a higher refractive index (lower plasma density) in the core region radially inside the shock, and a lower refractive index (higher plasma density) at the shock region. As seen in Fig. 8(d)(e), the radial location of the density peak of the expanding shock is well fit by the expression for a cylindrical blast wave (Lin 1954; Zel'dovich and Raizer 1967).

$$R(t) = \sigma_0 \left(\frac{E_{th}}{\rho_0}\right)^{1/4} t^{1/2} \,, \tag{37}$$

where $\sigma_0 \sim 1$ is a dimensionless parameter depending on the specific heat ratio $\gamma_c$, $E_{th}$ is the initial thermal energy per unit length deposited by the laser, and $\rho_0$ is the initial mass density of the gas.

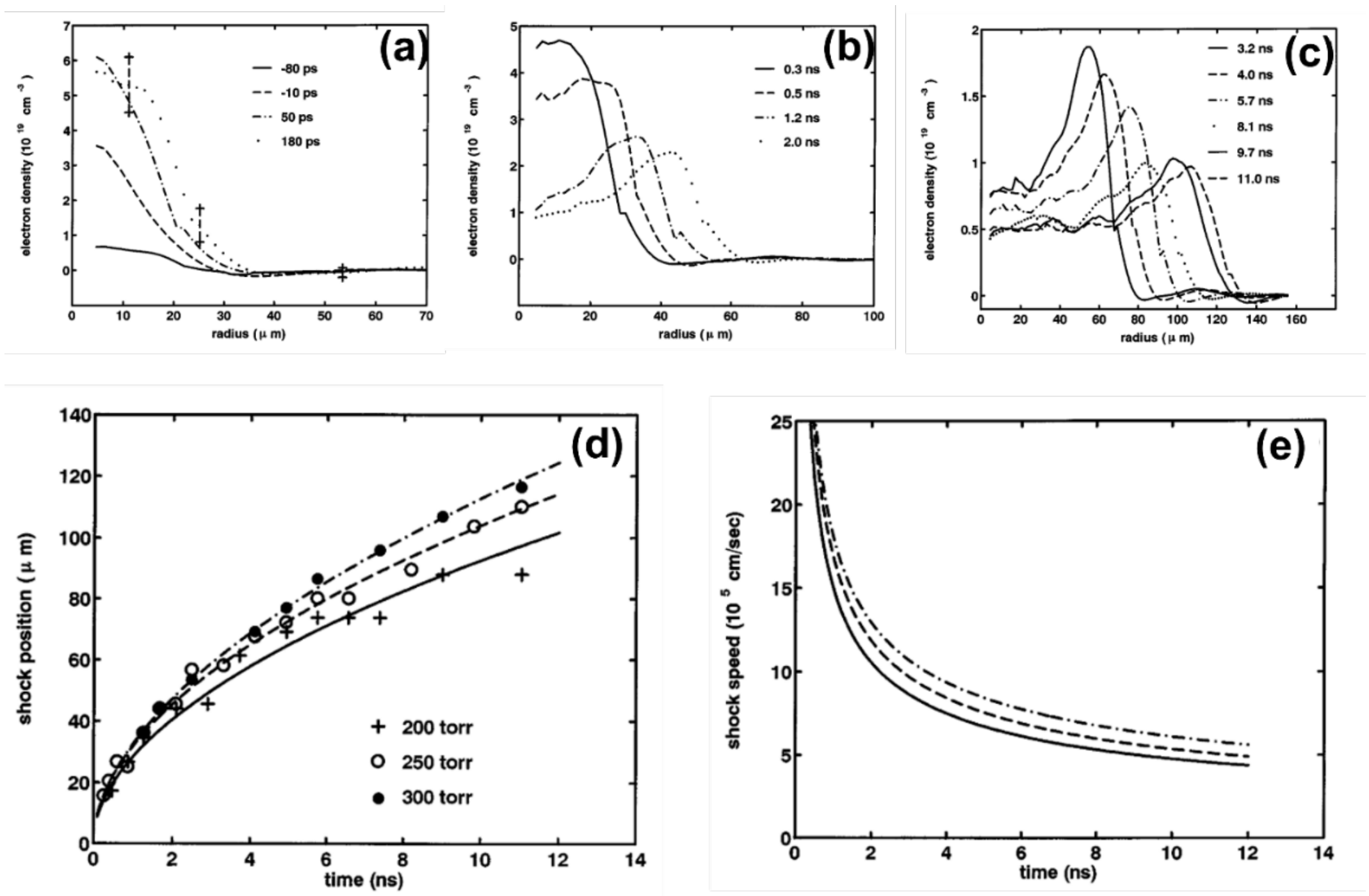

**Figure 8.** Early characterization of hydrodynamic expansion of plasma channels (all panels adapted from (T. R. Clark and Milchberg 1997)). **(a)-(c)** Expanding plasma profiles reconstructed from transverse interferometry measurements (see Sec. V.A.1). **(d)** Expanding shock position and **(e)** speed plotted vs. time extracted from interferometric measurements of the expanding plasma column.

Two timescales govern the evolution of the initial plasma column into a plasma waveguide (assuming the expanding shock is readily ionized): (1) the time to develop an initial density dip on axis, which coincides with the shock development timescale $\tau_s \sim \lambda_{ii}/c_s$, and (2) the overall plasma channel evolution timescale $\tau_{ch} \sim r_p/c_s$, where in general $\tau_s < \tau_{ch}$. For plasma densities $\sim 10^{19}\ \mathrm{cm}^{-3}$, the timescales are $\tau_s \sim 50\ \mathrm{ps}$ and $\tau_{ch} \sim 1$ ns; these timescales are borne out in Fig. 8.

In practice, it is preferable that the waveguide generation pulsewidth $\tau_{pulse}$ be shorter than $\tau_s$ so that ionization and heating of the plasma is impulsive on an acoustic timescale and the laser ceases to heat the plasma as the waveguide begins to form. Even the 100 ps Bessel beam pulse (where $\tau > \tau_s$) of the early experiments was long enough to side-couple into its own self-generated waveguide (Fan, Parra, and Milchberg 2000; Fan *et al.* 2002). Pulses with $\tau > \tau_{ch}$ will continue to heat the evolving plasma, excessively widening any waveguide structure and possibly driving instabilities and additional shocks, including shock implosions (T. R. Clark and Milchberg 1998b).

Support of tunable optical mode structure by plasma waveguides was first demonstrated in (Durfee and Milchberg 1993; Durfee, Lynch, and Milchberg 1994; Durfee, Lynch, and Milchberg 1995; T. R. Clark and Milchberg 1998a). Since the waveguide expands as the shock propagates radially outward (until it stagnates when the temperature drops too low from expansion work), the most efficient coupling to the guide's lowest order or fundamental mode is obtained when the injected pulse is delayed with respect to the $J_0$ channel-forming pulse until the waveguide expands to $w_{ch} \approx w_0$ (Eq. (29b)), with the injected beam waist located at the channel entrance. For a sufficiently low waveguide contrast $\Delta N_e$ (see Eq. (23)) the guide will support *only* the lowest order $(0{,}0)$ mode. Figures 9(a) and 9(b) show the effect on the guided output mode of such a waveguide from varying the input coupling of a probe beam (by moving its waist axially back from the guide entrance). It is seen that only the coupling efficiency is affected (Fig. 9(a)), but the mode shape remains the same and near-Gaussian (Fig. 9(b)). For a higher contrast guide, Eq. (23) predicts the trapping of higher order modes. For Fig. 9(c)-(h), the probe pulse waist was placed ~1 mm in advance of the guide entrance so that the beam was diverging as it entered the channel and could excite higher order modes. With increasing delay of the probe pulse, it is seen that mixes of azimuthal modes ($m > 0$) appear ((c)-(e)), followed by the onset of radial modes ($p > 0$) ((f)-(h)).

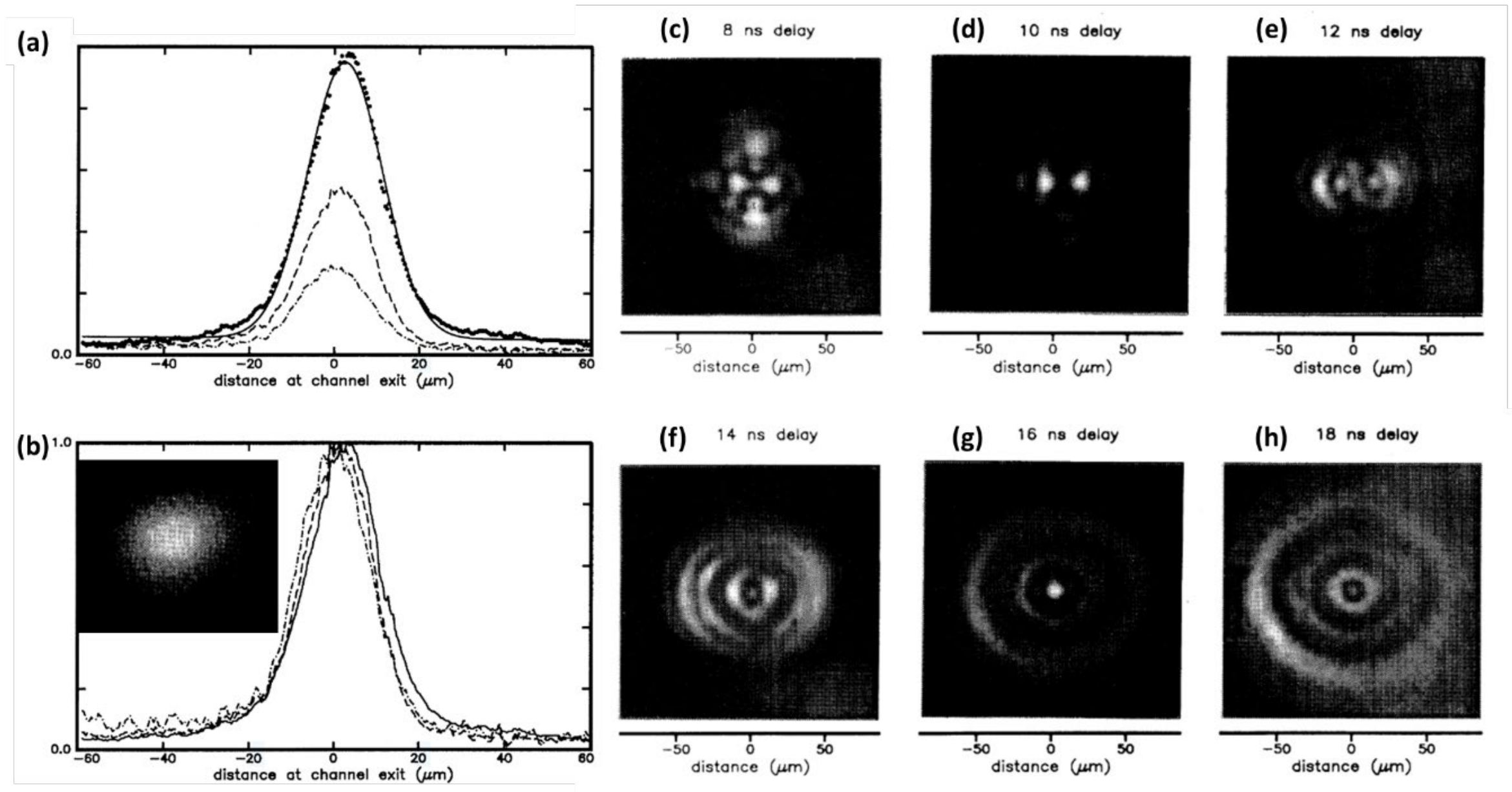

**Figure 9.** Demonstration of transverse mode properties of plasma waveguides. For a waveguide with density contrast $\Delta N_e$ sufficient for confinement of only the fundamental mode (see Eq. (23)): (a) Sequence of lineouts of waveguide exit mode for varying the input coupling of a probe beam by moving its waist axially back from the guide entrance; (b) Curves of (a) normalized to their peak values, with the inset mode image. For a waveguide with higher density contrast, (c)-(h) are exit mode images for increasing probe injection delay after the $J_0$ pulse. Figure adapted from (Durfee, Lynch, and Milchberg 1995; Durfee, Lynch, and Milchberg 1994).

Plasma waveguides generated by collisional ionization and heating have several drawbacks. The most significant, as mentioned, is the necessity of high gas densities ($> 10^{19}\ \mathrm{cm}^{-3}$) to ensure efficient breakdown and heating, resulting in waveguides with on-axis plasma densities $>$ $\sim 2 \times 10^{18}\ \mathrm{cm}^{-3}$ (after shock expansion), an order of magnitude higher than optimal for multi-GeV LWFA. In addition, avalanche breakdown requires seed electrons produced through multiphoton ionization; this restricted target gases to those with relatively low ionization potential (IP), which is why Xe and $N_2O$ were among the low IP gases used in the earliest experiments (Durfee, Lynch, and Milchberg 1995; Fan, Clark, and Milchberg 1998).

The strong local density dependence in the heating and ionization of avalanche-generated waveguides makes them sensitive to density variation in gas jets or axial variations in the Bessel beam intensity, which can produce channels with axially varying density and cross-sectional profile. Gas density falloff at the ends of a gas jet can cause waveguide tapering at its ends, limiting coupling (Nikitin *et al.* 1999), as can axial variation of the Bessel beam intensity in uniform gas density targets (Nikitin *et al.* 1997). One solution to mitigate channel tapering at the guide entrance was the use of an auxiliary laser pulse to supplement the heating at the end and widen the waveguide to form an "end funnel" (K. Y. Kim *et al.* 2002; K.-Y. Kim 2003; Cooley *et al.* 2003; Wu *et al.* 2005).

One approach for ensuring the generation of a high population of avalanche seed electrons was the igniter-heater method (Volfbeyn, Esarey, and Leemans 1999), enabling the use of target gases with higher ionization potentials than in the original hydrodynamic waveguide experiments of (Durfee and Milchberg 1993; Durfee, Lynch, and Milchberg 1994; Durfee, Lynch, and Milchberg 1995). This two-pulse method uses a subpicosecond pulse ($\sim 10^{15}\,\mathrm{W/cm^2}$, the "igniter") to generate the seed electrons via optical field ionization, followed by a longer pulse ($> \sim 100$ ps, the "heater") to drive collisional ionization and inverse bremsstrahlung heating. In the approach detailed in (Volfbeyn, Esarey, and Leemans 1999), the two pulses used cylindrical lens focusing to produce elongated waveguides. A schematic of the setup as well as shadowgraphy showing the geometries of the different generated plasma structures is shown in Fig. 10. The igniter-heater method for hydrodynamic waveguides was employed in the first demonstration of LWFA using a plasma waveguide (Geddes *et al.* 2004; Geddes *et al.* 2005). In these experiments, a $\sim 9$ TW, $w_0 = 8\,\mu\mathrm{m}$ drive pulse was guided in 2.2 mm plasma waveguides with on-axis densities $\sim 2 \times 10^{19}\,\mathrm{cm^{-3}}$, accelerating electrons to ~100 MeV. Under similar laser and plasma density conditions without a waveguide, the beam divergence, charge, and energy gain were inferior.

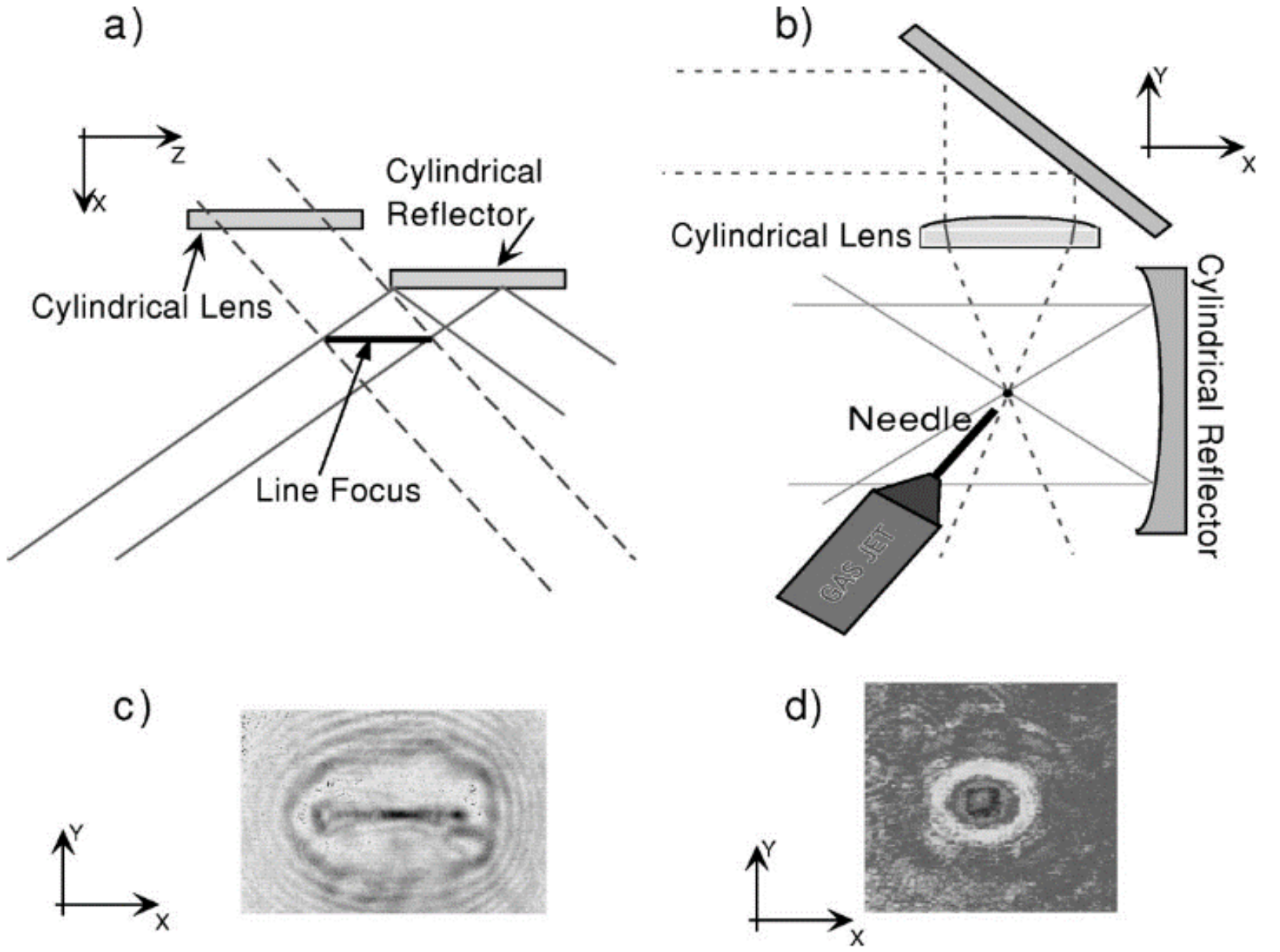

**Figure 10.** Illustration of ignitor-heater waveguide formation method (adapted from (Volfbeyn, Esarey, and Leemans 1999)). **(a)(b)** Experimental schematics for generating slab (a) and cylindrical (b) waveguides. **(c)(d)** Shadowgraphy showing slab (c) and cylindrical (d) waveguide structures.

Another plasma waveguide scheme originally demonstrated using long pulse Bessel beams was the hollow channel. This scheme did not rely on hydrodynamic expansion to form the core and cladding; the cladding was directly generated with a 5th order Bessel beam ($J_5$) (Fan *et al.* 2000), leaving the central region un-ionized; this scheme was aimed at applications such as high order nonlinear optics (Milchberg, Durfee, and McIlrath 1995) or generation of hollow plasma channels for positron acceleration (Kimura *et al.* 2011). Hollow plasma channels generated by femtosecond pulses for positron acceleration were demonstrated in (S. Gessner *et al.* 2016).

### B. Plasma waveguide generation by femtosecond laser heating of clustered gases

Use of clustered gas targets for waveguide generation enables both higher efficiency laser heating and lower density plasma waveguides. If a pulsed gas jet is operated with sufficiently high backing pressure and/or low gas reservoir temperature, and an appropriately shaped nozzle, rapid adiabatic expansion cooling of the flowing gas occurs, and atoms or molecules are susceptible to efficient agglomeration or clustering through van der Waals forces (Hagena and Obert 1972). Noble gases such as Kr and Xe easily form clusters without auxiliary cryogenic cooling of the gas jet valve and nozzle, while clustering of species such as $H_2$, Ar, $N_2$ is improved with cooling (Kumarappan, Kim, and Milchberg 2005). Clusters can contain as many as $10^6$–$10^7$ atoms/molecules and range up to several hundred Å in diameter. Internally, a cluster has the density of a solid, even though the volume average density of a gas of clusters can be as low as an unclustered gas. In the irradiation of a puff of clusters from a nozzle by an intense laser pulse, the microscopic interaction occurs initially with a locally solid material, which is rapidly turned into dense plasma, so that efficient high-density collisional processes are dominant in the cluster ionization and IB heating.

Plasma waveguide generation in clustered gases has used both short ($< 1$ ps) and longer ($\sim 100$ ps) pulses. Each regime relies on the same key idea: the dense clusters, $\sim 1000\times$ the average gas density, are efficiently ionized and heated by the laser pulse, independent of the overall gas density or the number of clusters (Milchberg, McNaught, and Parra 2001; K. Y. Kim *et al.* 2003; Kumarappan, Kim, and Milchberg 2005). This means that waveguide density can be controlled over multiple orders of magnitude without substantial changes in heating laser requirements. Clusters are ionized and heated even more efficiently than solid targets, as they are fully immersed in the optical field (Milchberg, McNaught, and Parra 2001).

The evolution to plasma waveguide formation from short pulse interaction with clusters occurs as follows. The leading edge of the short laser pulse converts a cluster to a near-solid density plasma "ball", which explodes on a timescale $\tau \sim R/c_s < \sim 1$ ps, where $R < 0.1$ μm is the cluster radius and $c_s \sim 10^7$ cm/s is the typical expansion speed, approximately the plasma sound speed (K. Y. Kim *et al.* 2003). Within $10-100$ ps (Ditmire *et al.* 1997; Ditmire, Smith, and Hutchinson 1998; Milchberg, McNaught, and Parra 2001), the exploding cluster plasmas merge, along with plasma originating from residual unclustered gas, to form a hot, locally uniform plasma in the heating laser focal volume. The hot plasma then expands radially, driving a shock wave in the surrounding clusters and gas. Provided there is sufficient thermal transport from the initial plasma, the shock is also ionized; this is the case for collisional heating of both conventional gas targets (Sec. III.A) and clustered gas targets, where the electron temperature can be tens to hundreds of eV. The ionized cylindrical shock then forms the waveguide cladding, similar to Sec. III.A. Even with pulses much longer than $R/c_s$, enhanced laser absorption up to 35% has been measured, which is $\sim 5-10\times$ more efficient than in an unclustered gas of the same average density (Milchberg *et al.* 2006). This effect is caused by the rapid cooling of the exploding cluster plasma, causing subsequent efficient absorption by a longer pulse. As an example, 100 ps long Bessel beams ($\gg R/c_s$) have been demonstrated to efficiently ionize and heat elongated argon and nitrogen cluster jets to form plasma waveguides (Sheng *et al.* 2005).

In a clustered gas, an extended plasma column can be generated by a Bessel beam generated by an axicon (Sheng *et al.* 2005). A conventional lens-focused pump pulse may also generate an extended plasma column through self-focusing and self-guiding in clustered gas (Milchberg, McNaught, and Parra 2001; Alexeev *et al.* 2003; Kumarappan, Kim, and Milchberg 2005). Unlike in non-clustered gases, where ionization-induced refraction quickly defocuses an ionizing pulse, the presence of exploding clusters produces a focusing index of refraction in the plasma (Milchberg, McNaught, and Parra 2001; Alexeev *et al.* 2003). This effect is unrelated to relativistic self-focusing and occurs at 5 orders of magnitude lower laser intensity (Alexeev *et al.* 2003); it has been used to generate cm-scale plasma waveguides in clustered gases and demonstrate guiding in them (Kumarappan, Kim, and Milchberg 2005). A sketch of the experimental setup, evolving waveguide profiles, and a typical guided mode are shown in Fig. 11.

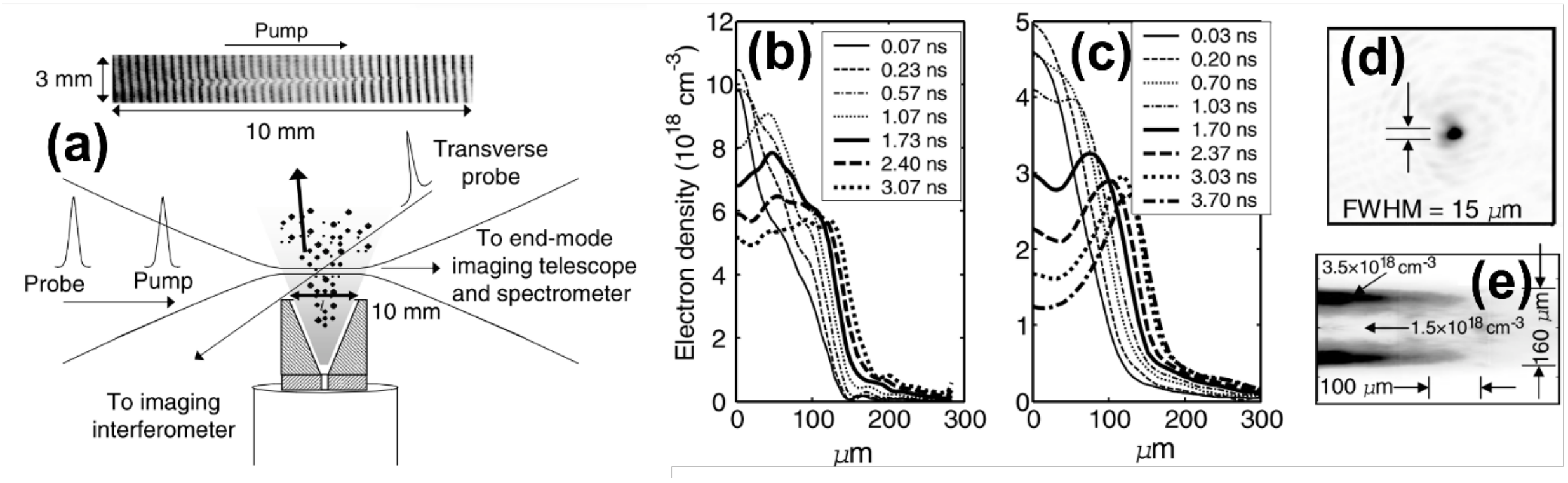

**Figure 11.** Plasma waveguide generation in clustered gases (adapted from (Kumarappan, Kim, and Milchberg 2005)). **(a)** Schematic of experimental setup. **(b)(c)** Channel formation under different gas conditions. **(d)** Example guided mode. **(e)** Example electron density profile at waveguide entrance.

An important feature of clustered gas flow is that it can be sculpted using both optical and mechanical techniques to form axially modulated plasma waveguides whose transverse refractive index structure varies with $z$. Such waveguides have been proposed for phase matched nonlinear generation of light down to the THz range (Antonsen, Palastro, and Milchberg 2007), quasi-phase-matched LWFA (S. J. Yoon, Palastro, and Milchberg 2014), and phase matched direct laser acceleration (York *et al.* 2008). In the latter application, the modulated waveguide acts as a "slow wave structure" to accomplish mitigation of dephasing between the electron beam and the laser field (Layer *et al.* 2007). For example, well-defined axial density modulations can be induced in sheet cluster flows from jets using mechanical blockages such as fine wires, without generating deleterious shockwaves that can introduce shot-to-shot density spikes in supersonic gas flow (S. J. Yoon *et al.* 2013). Axially modulated plasma waveguides were demonstrated using this technique in (Layer *et al.* 2009), with the setup and some results shown in Fig. 12. For wire blockages, shocks are eliminated because the cluster collisional mean free path $\lambda_{mfp}$ becomes much larger than the wire diameter $d_{wire}$, where the opposite is the case for conventional gas. For example, in (Layer *et al.* 2009), $\lambda_{mfp}$~1 mm for argon clusters, and $d_{wire} = 25 - 50$ μm, so that the cluster encounter with the wire is essentially ballistic.

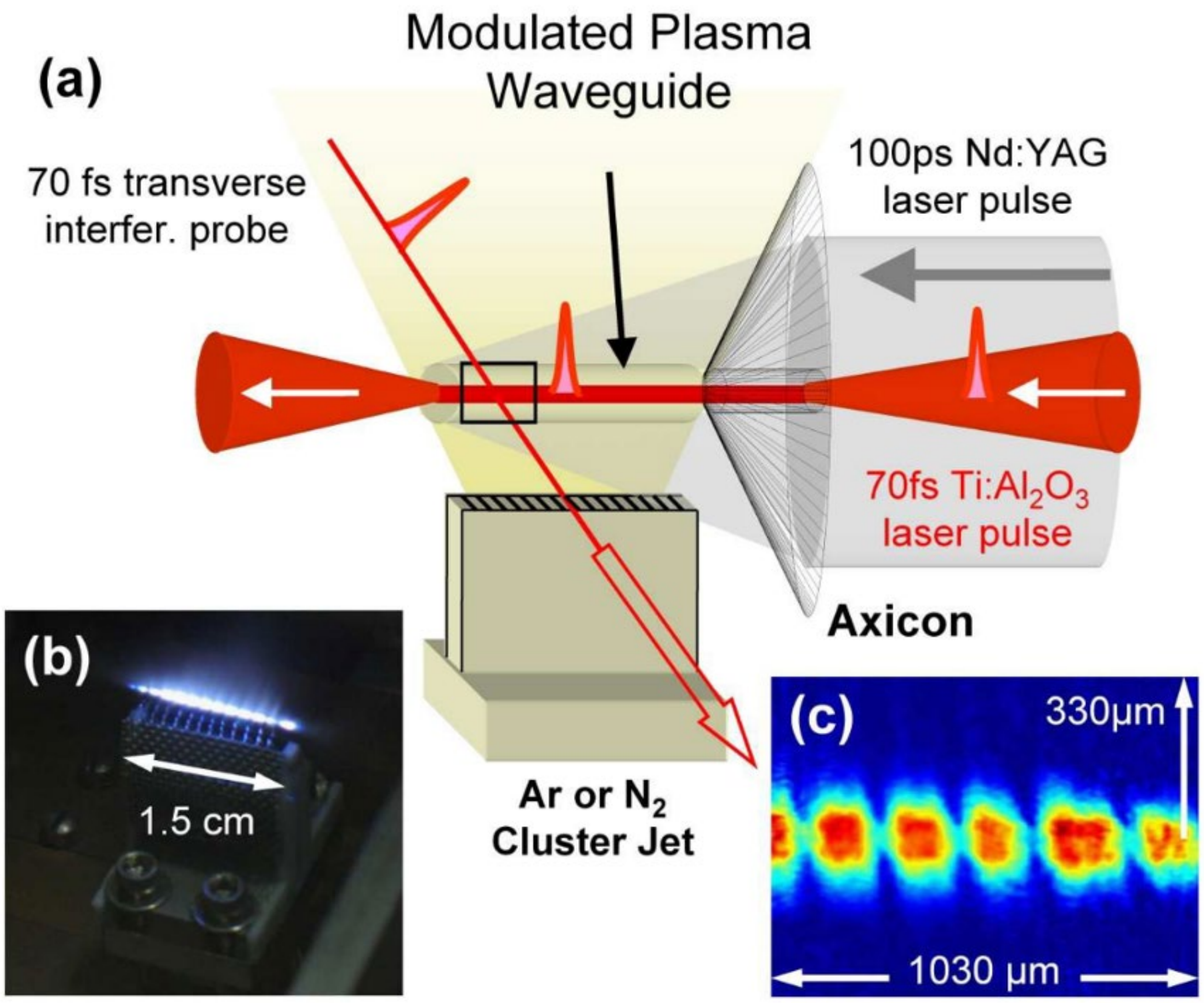

**Figure 12.** Modulated plasma waveguide generation (adapted from (Layer *et al.* 2009)). **(a)** Schematic of experimental setup. **(b)** Photograph of modulated plasma fluorescence above 1.5 cm jet. **(c)** Measured phase shift from axially modulated plasma column.

A more flexible method for producing an axially modulated plasma waveguide is to use an axially modulated Bessel beam to ionize and heat a uniform cluster sheet flow, as demonstrated in (Layer *et al.* 2007) using a fixed phase mask (ring grating) and in (Hine *et al.* 2016) using a spatial light modulator (SLM). In the latter, the modulated Bessel beam is generated by interfering a Gaussian beam with a much weaker beam that has been radially phase modulated by a SLM, producing radial intensity modulations with a controllable period. The radially modulated beam is then focused through an axicon to generate a $J_0$ Bessel beam line focus with axially modulated intensity. The $z$-dependent ionization and heating causes a $z$-dependent transverse refractive index structure whose period is controlled by the SLM. Figure 13 shows the optical setup and measurements of axially modulated waveguides.

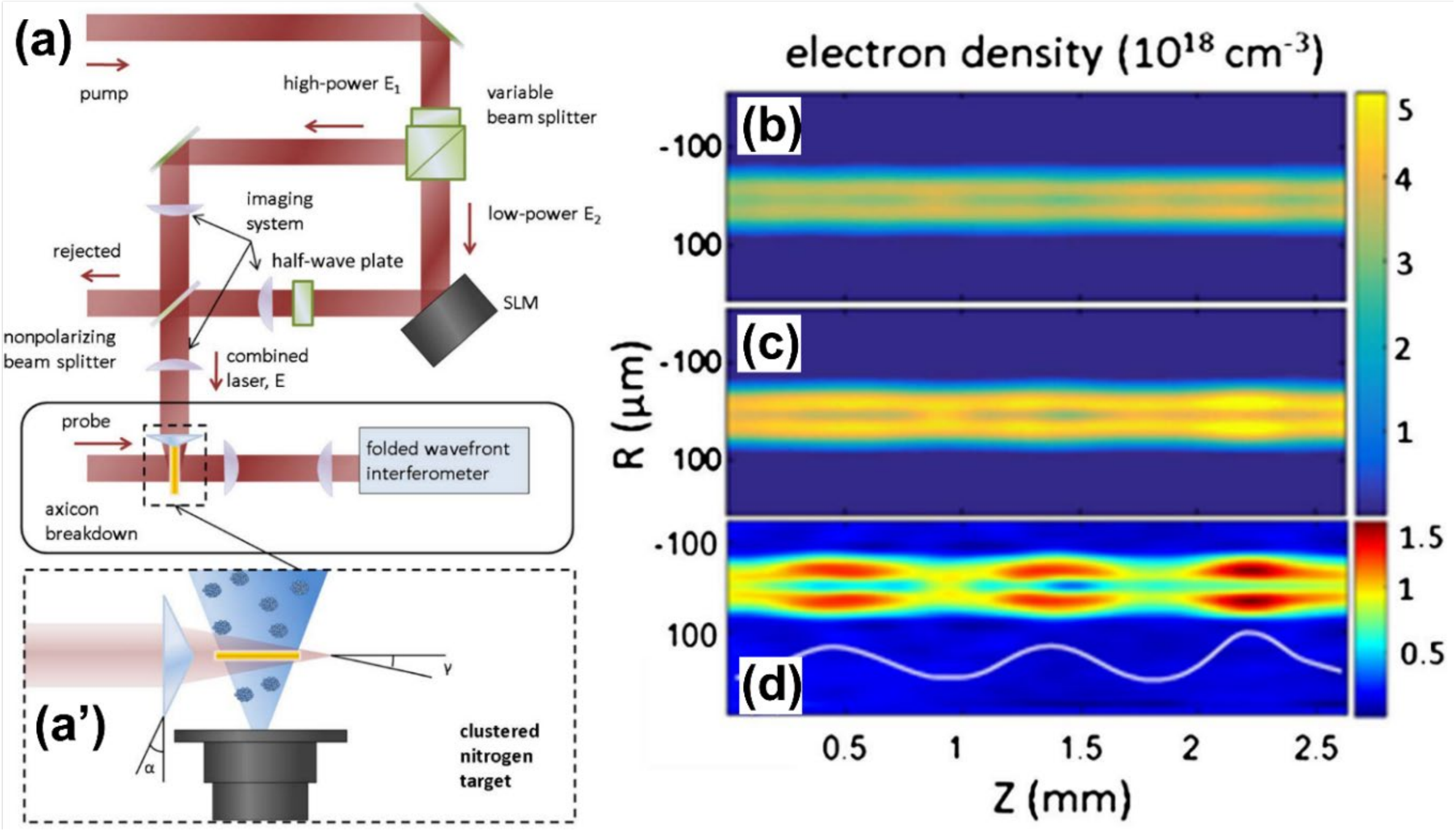

**Figure 13.** Modulated waveguide formation with an SLM (adapted from (Hine *et al.* 2016)). **(a)** Schematic of experimental setup with inset **(a′)** of clustering gas jet. **(b)(c)** Electron density profiles for plasma waveguides formed without (b) and with (c) axial modulation. **(d)** difference between (b) and (c).

Waveguide formation in clustered gas can also occur without reliance on hydrodynamic channel expansion, as shown in (Ditmire, Smith, and Hutchinson 1998). This technique relies on the fact that ions in laser-heated clusters can reach much higher ionization states than in dilute gases (Ditmire *et al.* 1996; Milchberg, McNaught, and Parra 2001). A weak ($10^{12}$ W/cm$^2$) prepulse ~200 ps in advance of a $10^{15}$ W/cm$^2$ pump pulse dissociates clusters near the optical axis with minimal ionization, leaving a central region of uniform gas surrounded by clustered gas. The wings of the pump pulse are sufficiently intense to cause cluster heating and explosion away from the optical axis, forming the cladding of a plasma waveguide. The central gas region is still susceptible to ionization by a guided high intensity pulse, potentially causing deleterious refraction or beam distortion that could affect some applications.

### C. Plasma waveguide generation using optical field ionization (OFI)

The last few years have seen the extension of the laser-induced hydrodynamic waveguide technique to meter-scale lengths and the low ($< \sim 2 \times 10^{17}$ cm$^{-3}$) on-axis densities required for multi-GeV acceleration. The recent results, outlined below, derive from the original hydrodynamic waveguide concept (Durfee and Milchberg 1993), including the use of Bessel and Bessel-like beams, but rely on the hydrodynamics being driven by a comparatively lower temperature plasma generated through direct liberation of bound electrons into the continuum by the optical field (Corkum, Burnett, and Brunel 1989). Given the ~10 × density reduction from hydrodynamic expansion as measured in (L. Feder *et al.* 2020; B. Miao *et al.* 2024), the low-density criterion of Sec. II.F corresponds to an initial density $\sim 10^{18}$ $cm^{-3}$ of fully ionized hydrogen gas.

Unlike collisional ionization and heating, OFI-based heating depends on the single atom/molecule interaction with the laser field, and thus is density-independent (other than through density-dependent Bessel beam propagation effects (B. Miao *et al.* 2024)). This enables significantly more control over the waveguide longitudinal and transverse structure, including independent generation of the waveguide cladding, when needed (Sec. III.C.4). We will discuss the possible need for these techniques for OFI waveguides in Sec. III.C.2. With these auxiliary techniques, the key waveguide parameters, $\Delta N_e$ and $w_{ch}$, can be controlled largely independently of the on-axis density $N_{e0}$. A major limitation in LWFA applications of capillary discharge waveguides (see Sec. IV) is the inability to straightforwardly decouple $\Delta N_e$ and $N_{e0}$.

While collisional heating requires appreciable laser energy to sustain moderate laser intensity ($> \sim 10^{13} - 10^{14}$ W/cm$^2$) over ~100 ps pulses, OFI demands only sufficient *peak* intensity, which can be delivered with ultrashort, lower energy pulses. This reduces the laser energy required per unit length of waveguide by more than an order of magnitude. For example, the pulse energy requirement for collisional heating is ~50-500 mJ/cm (Durfee, Lynch, and Milchberg 1995; Sheng *et al.* 2005), depending on the gas species and whether it is clustered or unclustered, and on the Bessel beam parameters. For OFI heating of hydrogen, the requirement is $\sim 5 - 10$ mJ/cm (Lemos, Grismayer, Cardoso, Figueira, *et al.* 2013; R. J. Shalloo *et al.* 2019; Picksley, Alejo, Cowley, *et al.* 2020; Shrock *et al.* 2022; B. Miao, Shrock, *et al.* 2022; L. Feder *et al.* 2020; B. Miao *et al.* 2024). For the OFI heating of low-density $H_2$ ($< \sim 10^{18}$ cm$^{-3}$) appropriate for multi-GeV LWFA, the Bessel beam absorption is a few percent at most; most of the laser energy transmits through the OFI-generated plasma as an outgoing conical wave.

An estimate for the electron temperature in OFI plasmas is the electron ponderomotive energy $U_p = e^2E^2/4m\omega^2$, giving $U_p$[eV]~9.33 $I(10^{14}$ W/cm$^2)\lambda^2(\mu$m$^2)$ at the time and location of ionization. For hydrogen, where the ionization threshold is $\sim 10^{14}$ W/cm$^2$, $k_BT_e \sim U_p \sim 10$ eV for a linearly polarized laser pulse at central wavelength λ=800nm, typical of high peak power Ti:Sapphire lasers (Kiani *et al.* 2023). More accurately, the temperature is determined as an average weighted by the OFI ionization rate over the pulse history, $k_BT_e(r,z) = (2/3)[\int d\xi\, (dN_e/d\xi)]^{-1} \int d\xi\, (dN_e/d\xi)\, |\mathbf{p}|^2/2m$, where $\xi = v_g t - z$ is the local longitudinal coordinate in a frame moving at the group velocity $v_g$ of the generating pulse, $dN_e/d\xi$ is the OFI ionization rate, $\mathbf{A}$ is the laser pulse vector potential, and $\mathbf{p} = e\mathbf{A}(r,\xi;z)/c$ is the residual drift momentum associated with the vector potential at the time of ionization. The integration is taken over the full laser pulse envelope. Simulations of optical field ionization and heating by a propagating intense Bessel beam are shown in Sec. V.B.3. It is important to note that for pulses with peak laser intensities exceeding the OFI threshold for the species of interest, an additional rise in $k_BT_e$ will not occur—ionization events yielding free electrons with the same initial 'birth' momenta and temperature will simply move earlier in the pulse. However, a higher intensity pulse can widen the transverse extent of OFI (B. Miao *et al.* 2020), giving rise to an increased energy deposition per unit length of waveguide, which will drive faster hydrodynamic expansion, as shown by Eq. (37).

The first OFI-generated hydrodynamic waveguides were demonstrated by (Lemos, Grismayer, Cardoso, Figueira, *et al.* 2013; Lemos, Grismayer, Cardoso, Geada, *et al.* 2013), using lens-based heating, followed by a demonstration of guiding of a low intensity probe pulse (Lemos *et al.* 2018).

These guides, generated in hydrogen and helium and up to 8 mm in length, were in the plasma density range $10^{18} - 10^{19}\, cm^{-3}$. Low-density hydrodynamic OFI (HOFI) waveguides in hydrogen were subsequently demonstrated at Oxford by (R. J. Shalloo *et al.* 2018), who used lens-based heating in a hydrogen gas cell to generate the first HOFI structures with on-axis densities in the range $\sim 10^{17}\, cm^{-3}$ relevant to multi-GeV laser-wakefield acceleration. This result was followed by demonstration of 16-mm-long OFI waveguide generation in a hydrogen gas cell using an axicon lens (R. J. Shalloo *et al.* 2019). Guiding of $\sim 10^{17}\, W/cm^2$ pulses over $\sim 5\, cm - 20 cm$ in hydrogen gas cells (Picksley, Alejo, Cowley, *et al.* 2020; Picksley, Alejo, Shalloo, *et al.* 2020) was demonstrated at Oxford and in extended low density gas jets (B. Miao *et al.* 2020; L. Feder *et al.* 2020) at Maryland. These and related results are discussed quantitatively in the subsections below.

The first application of OFI-heated hydrodynamic plasma waveguides to LWFA (Oubrerie, Leblanc, *et al.* 2022) combined a 15-mm-long guide (at $N_{e0} \sim 1.4 \times 10^{18}\, cm^{-3}$) with controlled injection to produce ~1-GeV electron beams using 50 TW injected pulses. The guides were generated in a gas jet using an axiparabola (see Sec. V.B.2). In the low-density ($N_{e0} \sim 10^{17} cm^{-3}$) OFI regime, (B. Miao, Shrock, *et al.* 2022) then reported the first multi-GeV LWFA in a 20-cm-long OFI plasma waveguide, producing quasi-monoenergetic peaks to 5 GeV using 300 TW injected pulses. This experiment introduced the use of meter-scale supersonic gas jets and diffractive transmission axicons for plasma waveguide generation. Also at low density, (Picksley *et al.* 2023) generated 1.2-GeV bunches using truncated-channel injection in an 11 cm gas cell; (Shrock *et al.* 2024) used ionization injection to produce quasi-monoenergetic multi-GeV peaks and examined multi-modal propagation effects. At the highest energies from guided LWFA to date, a LBNL-Maryland collaboration (Picksley *et al.* 2024) used injection of a 500 TW pulse into an axicon-generated guide in a 30-cm-long hydrogen gas jet, along with nitrogen ionization injection, to produce a 9.2-GeV peak with charge beyond 10 GeV. (Rockafellow *et al.* 2025) used a diffractive axicon-generated guide in a 30 cm hydrogen gas jet to generate LWFA electron bunches with energies to 10 GeV and nC-level charge above 1 GeV using $< 400$ TW drive laser pulses. (Lahaye *et al.* 2025) later reported ~2-GeV acceleration in a 6-cm-long axiparabola-generated helium guide using 200 TW pulses.

### 1. Conditions for neutral-shock formation in impulsive OFI heating

As in the case of collisional heating (Sec. III.A), OFI heating is impulsive on the shock-development timescale. The subsequent shock dynamics, however, differ significantly. In the low-density gases to which OFI heating is applied, the initial plasma cannot be assumed to supply sufficient thermal energy to ionize the fresh neutral gas compressed by the expanding shock. Thus, two distinct conditions must be satisfied: formation of a well-defined neutral shock, and its subsequent ionization by the OFI-heated plasma. We consider these requirements separately, beginning with neutral-shock formation.

For the typical <100 fs pulsewidths used in OFI heating, $\tau_{pulse} \ll \tau_s$. We first consider well-defined *neutral* shock generation in hydrogen. After OFI heating of hydrogen to $T_{e0} \sim 10$ eV in a column of initial radius $r_p$, the plasma pressure is initially $\sim 400$ times that of the surrounding room temperature hydrogen. A supersonic plasma expansion thus occurs. For a well-defined neutral shock to form on the expanding plasma periphery, the mean free path of neutral-neutral collisions must be smaller than the size of the plasma driver, or $\lambda_{nn}/r_p < 1$. Using $\lambda_{nn} = (N_g\sigma_{nn})^{-1}$, where $\sigma_{nn}$ is the neutral-neutral collisional cross section, gives $N_g > (\sigma_{nn}r_p)^{-1}$ as the minimum hydrogen density needed to form a shock. Using $\sigma_{nn} \sim 3 \times 10^{-15} cm^2$ (Poling, Prausnitz, and O'Connell 2020) for $H_2$-$H_2$ collisions at room temperature gives $N_g[cm^{-3}] > 3 \times 10^{18}/r_p\ [\mu m]$. So for the case of an initial plasma radius $r_p \sim 10\ \mu m$ (B. Miao *et al.* 2024), a rough initial gas density threshold for neutral shock generation is $N_g \sim 3 \times 10^{17}\ cm^{-3}$. This may pose an interesting challenge to the development of hydrodynamic plasma waveguides with $N_{e0} \sim 10^{16}\ cm^{-3}$, as are envisioned for 100 GeV-class single-stage accelerators (Ludwig *et al.* 2025). In practice, fractional pre-ionization of the plasma periphery by short wavelength radiation from the plasma (Zel'dovich and Raizer 1967; Drake 2018)—a radiative precursor—can further reduce the low density threshold in $N_g$ by adding neutral–ion collisions to the momentum transfer rate.

## 2. Conditions for ionization of a neutral shock by an OFI-heated plasma

Once a neutral shock is generated, two conditions must be satisfied for shock wall ionization by the initial OFI plasma generated in the background gas. The first and most important is that the thermal flux $q$ from the plasma be sufficient to supply the ionization power demand of the neutrals effectively flowing through the shock layer as it propagates outward, $q > q_{ioniz}$, where $q_{ioniz} = N_g c_s \varepsilon_{req}$, accounting for mass flux conservation across the shock. Here, $N_g$ is the molecular gas density, $c_s \sim (k_B T_{e0}/m_i)^{1/2}$ is the approximate shock speed. The energy expenditure per incident $H_2$ molecule $\varepsilon_{req}$ includes dissociation, ionization of both atoms, and the kinetic energy retained by the two liberated electrons (neglecting ion heating and additional excitation or radiative losses). Taking the electron mean kinetic energy to be $\langle E_e \rangle \sim 10$ eV, we estimate $\varepsilon_{req} \simeq D_0(H_2) + 2I_H + 2\langle E_e \rangle \simeq 4.48 + 2(13.60) + 2(10) \simeq 52$ eV. Here, $D_0(H_2)$ is the molecular dissociation energy (Liu et al. 2009), and $I_H$ is the atomic ionization energy (Kramida et al. 2024).

The maximum thermal flux that can be supplied by the initial plasma is set by the free streaming limit, $q_{max} = f N_{e,init} k_B T_{e0} v_{th}$, where $f \sim 0.1$ is a typical flux limiting factor in gas density plasmas (Cowie and McKee 1977; Drake 2018), $T_{e0}$ is the OFI plasma initial temperature and $v_{th} = (8k_B T_{e0}/\pi m)^{1/2}$ is the electron thermal speed. For fully ionized $H_2$, the initial peak electron density on axis is $N_{e,init} = 2N_g$. If a fraction $\alpha < 1$ of the heat flow contributes to heating and ionizing the inner shock wall (Regan *et al.* 2007; Fabbro, Max, and Fabre 1985), while the rest is lost to radiation, geometric effects and electrostatic sheath potentials, ionization of the wall requires $\alpha q_{max} > q_{ioniz}$. Here, we take $\alpha = 0.5$ for moderately efficient coupling.

If $\alpha q_{\max} > q_{\text{ioniz}}$ is satisfied, gradient-based conduction can supply the ionization demand if the coupled conductive flux is sufficient, $\alpha q_{grad} \gtrsim q_{ioniz}$, where $q_{grad} \sim \kappa_e |\nabla T_e| \sim \kappa_e T_{e0}/r_p$ and $\kappa_e$ is the electron thermal conductivity (Braginskii 1965; Spitzer 2006). Collisions establish this heat flow when the initial plasma radius spans at least a few electron collisional mean free paths, $r_p > r_{p,min} \sim n_{\text{mfp}}\, \lambda_{e,i+n}$. Here $\lambda_{e,i+n} = v_{th}/(\nu_{ei} + \nu_{en})$, and $\nu_{ei}$ and $\nu_{en}$ are the electron-ion (Braginskii 1965; Spitzer 2006) and electron-neutral (Raĭzer 1997) collision frequencies. At the partially ionized edge of the expanding plasma, we approximate $\nu_{ei} \sim \nu_{en}$, giving $\lambda_{e,i+n} \sim \lambda_{ei}/2$.

In evaluating $\lambda_{ei}$ for Eq. (38), we further take the local singly ionized hydrogen density to be $N_i \sim N_g$ for the partially ionized edge; complete dissociation and ionization at the original gas density would give $N_i = 2N_g$. For the case $r_p < r_{p,min}$, the edge electrons are more likely to ballistically stream through the shock and not heat it. A rough estimate for the minimum initial plasma radius to permit conductive ionization of the inner shock wall is therefore (in Gaussian units),

$$r_{p,\mathrm{min}} \sim n_{\mathrm{mfp}} \frac{3(k_B T_{e0})^2}{2\pi Z^2 e^4 N_g \ln\Lambda} \; ; \quad r_{p,\mathrm{min}}[\mu\mathrm{m}] \sim 2.3 \times 10^{17}\, n_{\mathrm{mfp}} \frac{(T_{e0}[\mathrm{eV}])^2}{Z^2 \ln\Lambda\, N_g[\mathrm{cm}^{-3}]} . \quad (38)$$

In Fig. 14, we plot $r_{p,min}$ vs. $N_g$ for a range of initial plasma temperatures, taking $n_{\mathrm{mfp}} = 3$, $Z = 1$, and the Coulomb logarithm (Spitzer 2006) $\ln\Lambda \sim 7$ at sub-atmospheric density. The boundary between the solid and dashed curves is set by $\alpha q_{max} = q_{ioniz}$, which gives $k_B T_{e0,cut} = (\alpha f)^{-1} \varepsilon_{req} (N_g/N_{e,init})(\pi m_e/(8 m_i))^{1/2}$. For hydrogen, with $m_i = m_{proton}$, $f = 0.1$, $\alpha = 0.5$, and $N_g/N_{e,init} = \frac{1}{2}$ for the fully ionized initial column, the estimated energy expenditure $\varepsilon_{req} \simeq$ 52 eV per molecule gives $\varepsilon_{req}(N_g/N_{e,init}) \simeq 26$ eV and the cutoff $k_B T_{e0,cut} \simeq 8$ eV plotted in Fig. 14. This places the cutoff close to the ~10 eV initial temperature of OFI-generated hydrogen plasma columns (B. Miao *et al.* 2024).

Above the cutoff, conductive ionization of the inner shock wall is energetically possible; below it, the coupled heat flux is insufficient. Reducing $\alpha$ decreases the heat flux available to ionize the inner shock wall and raises the cutoff temperature, with $T_{e0,\mathrm{cut}} \propto \alpha^{-1}$ for the other parameters held fixed. The fixed-temperature radius curves in Fig. 14 remain unchanged, but the dotted cutoff curve moves upward as $r_{p,\mathrm{min}}(T_{e0,\mathrm{cut}}) \propto \alpha^{-2}$, and temperature curves below the new cutoff become energetically excluded. Increasing $\alpha$ has the opposite effect; Fig. 14 shows the case $\alpha = 0.5$.

Note that this simple scaling model does not take into account rapid plasma cooling from expansion and conduction (T. R. Clark and Milchberg 1997; L. Feder *et al.* 2020; B. Miao *et al.* 2024; Picksley, Alejo, Shalloo, *et al.* 2020), so that use of $T_{e0}$ in the scalings is optimistic; in a more complete model, the cutoff temperature curve in Fig. 14 would be higher. In addition, we have considered ionization of the inner wall of the shock as a minimum requirement, not ionization through the shock. Thus, all points $(N_g, r_p)$ in Fig. 14 close to the cutoff curve are marginal for shock ionization.

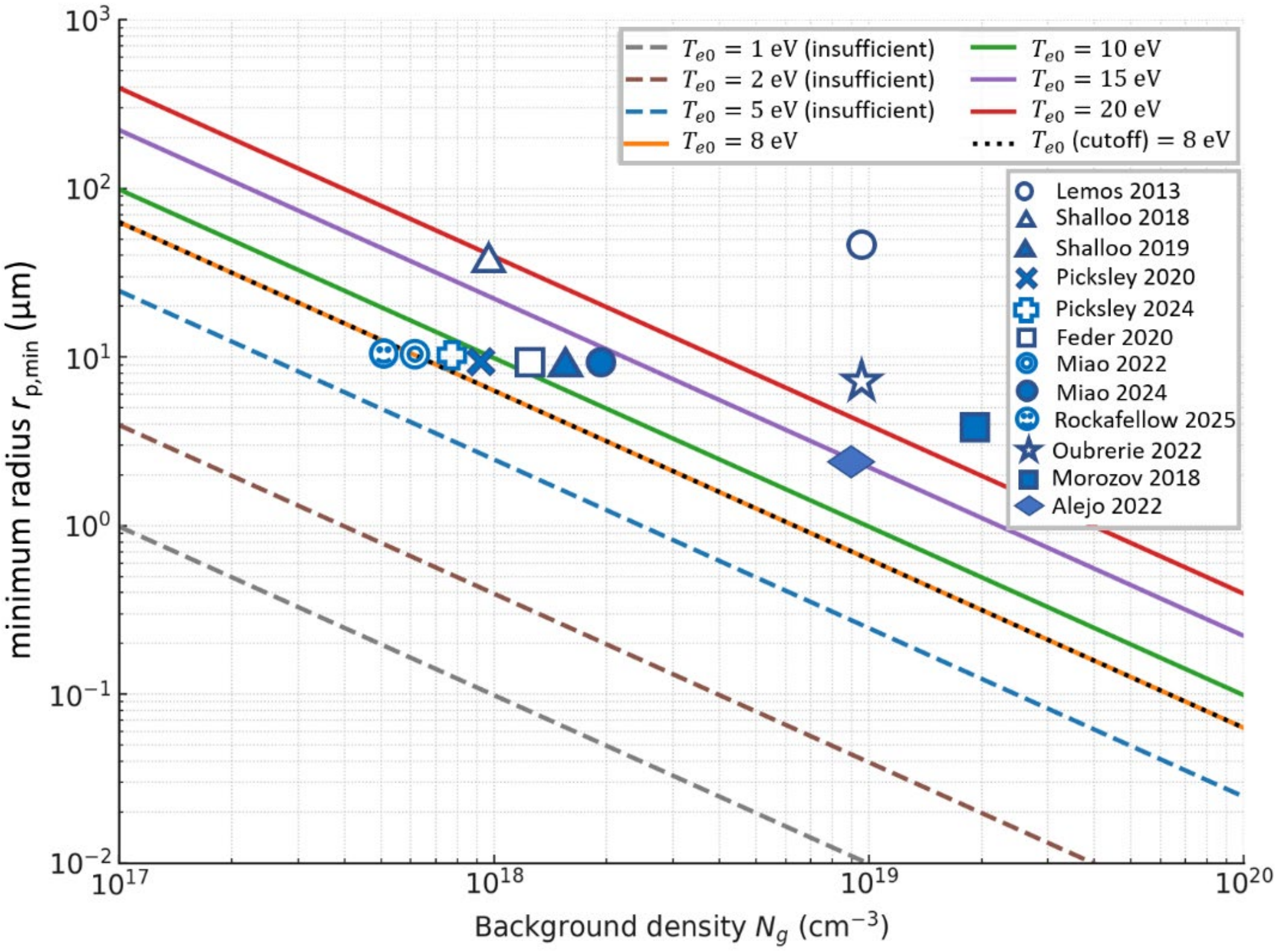

**Figure 14.** Scaling estimate of minimum initial plasma radius $r_{p,min}$ of an impulsively heated hydrogen OFI plasma that permits conductive ionization of the inner wall of the evolving peripheral neutral shock. Solid curves are energetically allowed; dashed curves are not. ($N_g, r_p$) points from several recent experiments are overlaid, where $N_g$ is background gas density and $r_p$ is the initial plasma radius. Experiments near the cutoff curve are marginal for conductive ionization of the inner neutral-shock wall. Adapted from (Rockafellow *et al.* 2026).

## 3. Single-pulse OFI-generated hydrodynamic waveguides

As discussed earlier, OFI-generated hydrodynamic waveguides were first explored in (Lemos, Grismayer, Cardoso, Figueira, *et al.* 2013; Lemos, Grismayer, Cardoso, Geada, *et al.* 2013), where 0.4 ps, λ=800 nm pulses were focused by a spherical lens to generate plasma columns in hydrogen and helium gas jets of neutral density $N_g < 10^{19}$ cm$^{-3}$, with shadowgraphic and interferometric images in Fig. 15(a)(b) showing the initial plasma column, which has an initial radius $r_p$~50 μm. As with hydrodynamic shock expansion of collisionally-heated waveguides (T. R. Clark and Milchberg 1997), an evolving waveguide and cylindrical blast wave were measured (Fig. 15(c)(d)). Waveguide formation and guiding were demonstrated up to 8 mm for moderate intensity (~$10^{15}$ W/cm$^2$) pulses, with on-axis plasma densities $N_{e0} >$ ~$10^{18}$ cm$^{-3}$ (Lemos *et al.* 2018). (R. J. Shalloo *et al.* 2018) then reported the first experimental demonstration of low-density hydrodynamic optical-field-ionized waveguides with on-axis densities relevant to multi-GeV LWFA by starting with lower initial hydrogen density $N_g$~$10^{18}$ $cm^{-3}$. Here, a spherical lens was also employed, generating 4-mm-long initial plasma channels of radius $r_0$~40 μm that hydrodynamically evolved into waveguides with on-axis densities as low as $N_{e0}$~$2 \times 10^{17}$ cm$^{-3}$ (Fig. 15 (e)(f)). In Fig. 14, we overlay the $r_p$, $N_g$ points from these two experiments, and see that both cases are well above the cutoff (dotted line) for ionization of the neutral hydrogen shock, consistent with the measured plasma density profiles.

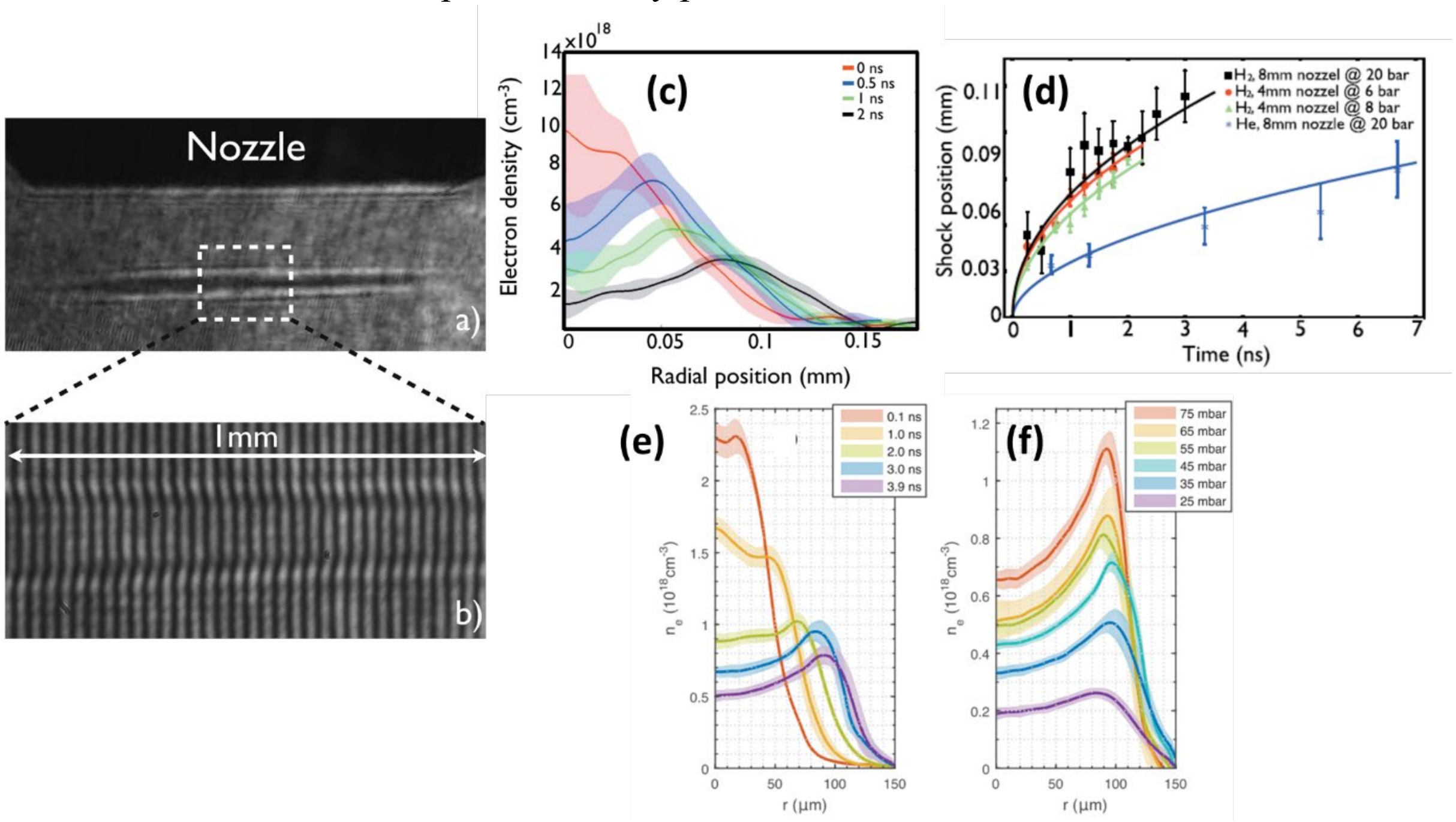

**Figure 15.** Expansion of OFI plasmas generated by lens focusing. Adapted from (Lemos, Grismayer, Cardoso, Figueira, *et al.* 2013): **(a)(b)** Shadowgraphic and interferometric images of the initial lens-generated plasma. **(c)(d)** Evolving plasma density profile and shock propagation structure extracted from interferometry. Adapted from (R. J. Shalloo *et al.* 2018): **(e)(f)** Longitudinal interferometry measurements showing expansion of lens-generated OFI channels for (e) fixed cell pressure and (f) varied cell pressure.

High repetition rate generation of OFI plasma waveguides, using a lens focus in a chamber backfilled with $H_2$, was demonstrated in (Alejo *et al.* 2022), where single-color transverse interferometry measured the guide refractive index profile. Figure 16 shows the stability of guide generation at 0.4 kHz over 6.5 hours, with the gas recovering between shots. Here, $N_g{\sim}9\times10^{18}$ cm$^{-3}$ and $r_p{\sim}3\mu m$, with this point on Fig. 14 sitting above the cutoff line, showing that the predicted likelihood of shock ionization is consistent with the measured plasma density profile.

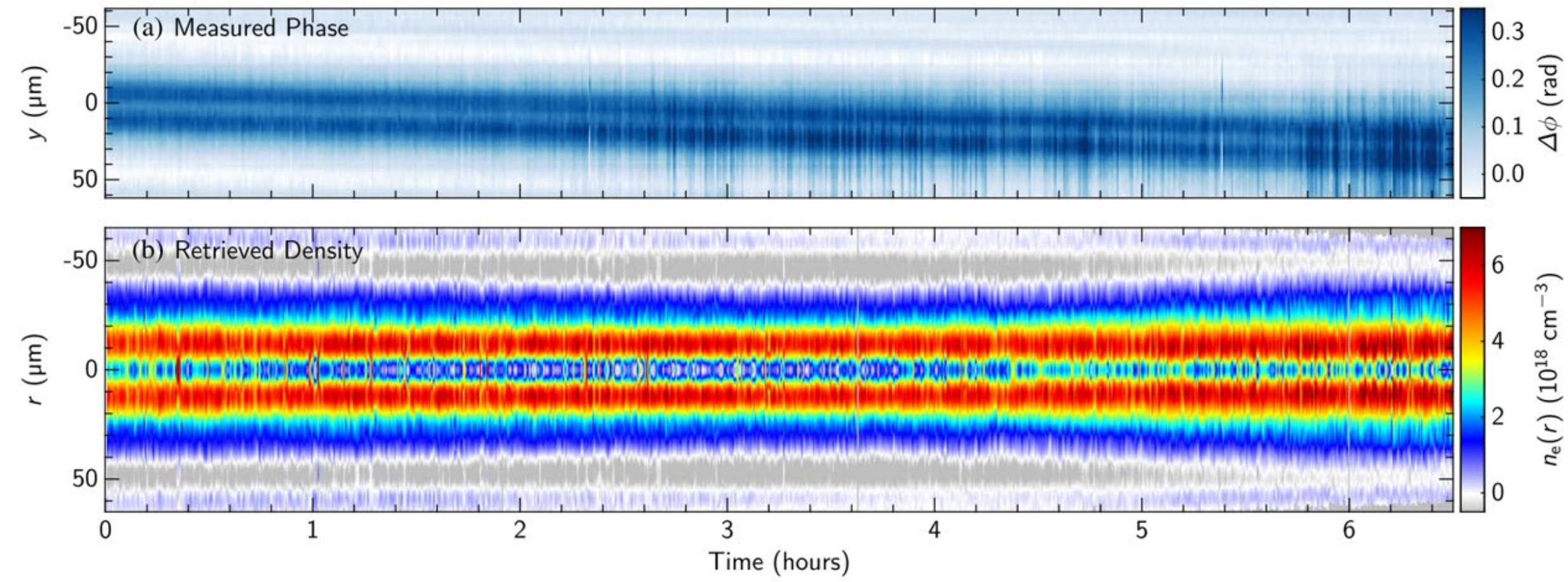

**Figure 16.** Hours-long stability of plasma channels generated at 0.4 kHz by a lens focused to a 5 μm FWHM spot in a chamber backfilled with $H_2$ at $N_g \sim 9 \times 10^{18}\ cm^{-3}$ (adapted from (Alejo *et al.* 2022)). **(a)** Phase measured via single-color transverse interferometry. **(b)** Extracted plasma density.

Later experiments used axicon lenses to make axially extended OFI plasmas with $J_0$ Bessel beams. The central maximum of a $J_0$ beam generally produces a much narrower initial plasma column (Durfee, Lynch, and Milchberg 1995) than a lens for a similar plasma length, and thus a smaller thermal energy per unit length. In (R. J. Shalloo *et al.* 2019), 1.6 cm long channels were generated with a $J_0$ Bessel beam in a hydrogen gas cell for $N_g = 1.6 \times 10^{18}$ cm$^{-3}$ and $r_p \sim 10$μm. This is consistent with the Bessel beam ray approach angle $\gamma = 2.5°$, which gives a $J_0$ central peak radius of $r_0 = 2.405/k \sin\gamma \sim 7$ μm. These initial channels were found to generate weaker plasma shocks (Fig. 17(a)(b)) than in the lens-generated waveguide experiments (Lemos, Grismayer, Cardoso, Figueira, *et al.* 2013; R. J. Shalloo *et al.* 2018). This corresponds to a less confining waveguide, whose leakage attenuation length was calculated in (B. Miao *et al.* 2020) to be $L_{1/e} \sim 0.5$ cm, and is consistent with marginal shock ionization (see Fig. 14). Additional work (R. Shalloo 2018) found that the transmission was higher for high-intensity ($> 10^{16} W/cm^2$) guided pulses than for low-power pulses. The proposed role of additional ionization by the injected pulse is discussed in Sec. III.C.4(ii).

These low density channels were later extended to 10 cm (Picksley, Alejo, Cowley, *et al.* 2020), using a longer Bessel beam focus and a longer hydrogen cell. As in (R. J. Shalloo *et al.* 2019), interferometric measurements showed low contrast shocks (Fig. 17(c)), suggesting similar weak confinement, with reported attenuation lengths $L_{1/e} < 10\ cm$. In this experiment, $N_g \lesssim 10^{18}$ cm$^{-3}$ and $r_p \sim 10$ μm, also conditions for marginal shock ionization (see Fig. 14).

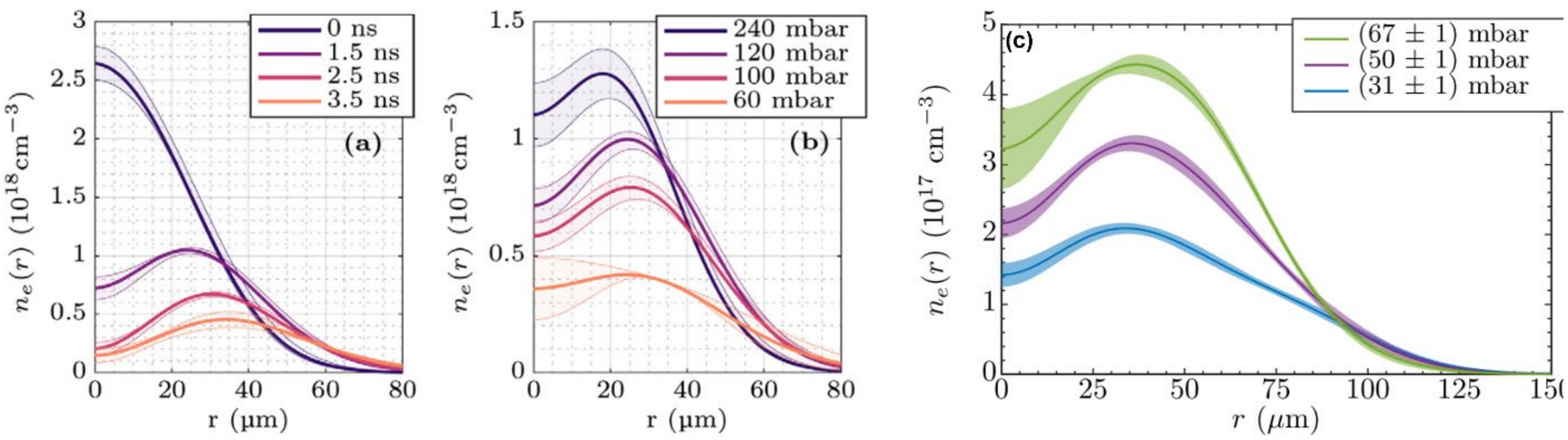

**Figure 17.** Expansion of plasma columns generated through field ionization by a Bessel beam (panels **(a)(b)** adapted from (R. J. Shalloo *et al.* 2019), panel **(c)** adapted from (Picksley, Alejo, Cowley, *et al.* 2020)).

Contemporaneous measurements of OFI waveguide generation using similar Bessel beam geometry ($\gamma = 2.3°$) and $H_2$ gas density employed femtosecond two-color interferometry of fast-evolving plasmas (see Sec. V.A.1) (L. Feder *et al.* 2020; B. Miao *et al.* 2020; B. Miao *et al.* 2024). This revealed the individual contributions of electrons and neutrals to the time-evolving refractive index structure. We note that when the plasma contribution dominates the refractive index, as it does in collisionally heated plasma waveguides, single color interferometry is sufficient (T. R. Clark and Milchberg 1997). For OFI-heated guides, however, possibly incomplete ionization of the neutral shock necessitates the measurement of both plasma and neutral contributions. Figure 18 (from (L. Feder *et al.* 2020)) plots the time-resolved plasma and neutral density profiles as a function of delay after OFI heating of the $H_2$ gas. As seen in Fig. 18(a), the electron density relaxes and widens while maintaining its central peak, simultaneously driving a shock in the peripheral neutral gas. Thermal conduction from the central plasma appears insufficient to ionize the neutral shock, with no plasma waveguide cladding developing. Later measurements under similar gas, laser, and Bessel beam conditions revealed the same behaviour (B. Miao *et al.* 2024). The $(N_g, r_p)$ points for these experiments are plotted in Fig. 14; they are close to the cutoff for neutral shock ionization. Simulations of Bessel beam OFI-induced hydrodynamics in $H_2$ gas (B. Miao *et al.* 2024) are consistent with these measurements: For $N_g = 2 \times 10^{18}$ cm$^{-3}$ and Bessel beam ray axis-approach angle $\gamma = 2.3°$, the simulations give $T_{e0} \sim 10$ eV, initial OFI plasma radius $r_p \sim 10\ \mu m$, and no development of an ionized shock (see Sec. V.B.3, Fig. 38).

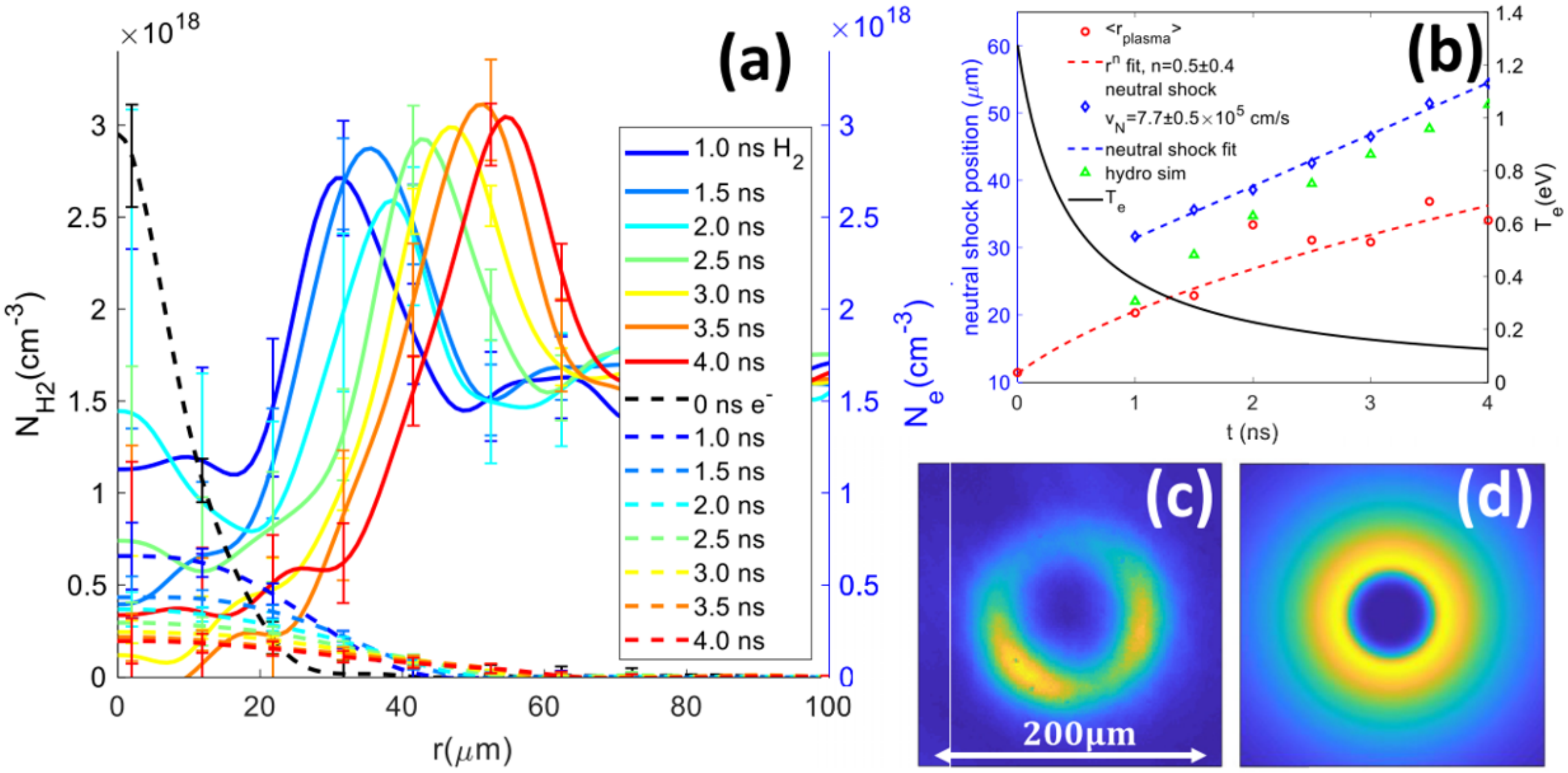

**Figure 18.** Expansion of an OFI generated plasma column and the surrounding neutral shock (adapted from (L. Feder *et al.* 2020)). **(a)** Two-color interferometry showing combined expansion of plasma and neutral shock; there is no evidence of a plasma cladding. **(b)** Plot of measured and simulated neutral shock expansion speed and temperature. **(c)** Measured and **(d)** simulated end-on image of a low-power probe pulse trapped by the neutral $H_2$ annulus.

OFI plasma waveguides have also been generated in the extended focus of an "axiparabola" (Smartsev *et al.* 2019). Like an axicon, an axiparabola focuses the radial annuli of a near-field beam to different longitudinal locations in the far-field, producing a Bessel-like beam with an axially extended focus (see Sec. V.B.2). In (Oubrerie, Leblanc, *et al.* 2022), a 1.5 cm long axiparabola-formed plasma channel in a hydrogen gas jet (with $N_g \sim 10^{19}\ cm^{-3}$ and $r_p \sim 8\ \mu m$) expanded into a plasma waveguide with on-axis plasma density $N_{e0} \sim 1.5 \times 10^{18}$ cm$^{-3}$. Here, the plasma density was $10\,\times$ higher than in (R. J. Shalloo *et al.* 2019; Picksley, Alejo, Cowley, *et al.* 2020), and interferometry showed a higher refractive index contrast between core and cladding, which would lead to lower waveguide leakage. This is consistent with $(N_g, r_p)$ for this experiment plotted in Fig. 14, where the point is well above the cutoff curve for ionization of the neutral shock.

To summarize this section: for the low densities appropriate for multi-GeV LWFA, long, narrow, low-density plasmas generated by single-pulse OFI heating may not evolve into a highly confining plasma waveguide because of weak ionization of the peripheral neutral shock; the waveguide's plasma cladding is either insufficient or completely lacking. This suggests the need for auxiliary methods for generating the plasma cladding. This is in contrast to single-pulse collisionally heated plasma waveguides, whose evolution begins at orders-of-magnitude higher initial density and temperature (see Sec. III.A), with typical $(N_g, r_p)$ points effectively lying significantly beyond any cutoffs in Fig. 14.

### 4. Methods for generating the plasma cladding in low density OFI waveguides

As discussed, Fig. 18(a) shows direct measurement of the evolving plasma core and neutral shock (L. Feder *et al.* 2020). The neutral shock maintains a roughly constant expansion speed (Fig. 18(b)) and amplitude, while the plasma column expansion slows due to cooling, with the on-axis density dropping to ~10% of its initial value. The neutral shock annulus supports a low-intensity ring mode (which an ionized shock would not do), as shown in the guided probe beam image in panel (c) and the propagation simulation in panel (d). The presence of the annular neutral shock in low density OFI-generated plasma columns makes possible two auxiliary methods for producing the plasma waveguide cladding, which we now review.

### (i) Two-Bessel method for core and cladding generation

Here, two Bessel beam pulses are used (B. Miao *et al.* 2020), the first to generate the plasma waveguide core and the second to generate the cladding. The first pulse is a zero order Bessel beam ( $J_0$ ), whose high intensity central maximum generates the waveguide core via OFI, which drives the outward propagating annular shock in the neutrals. The second is a higher order $J_q$ pulse ($q = 8$ or $q = 16$) which is delayed in time so that its high-intensity off-axis ring overlaps with the expanding neutral shock annulus and ionizes it, forming the plasma cladding. In this experiment, the $J_0$ pulse was formed with a reflective axicon. At a variable delay ($\tau_d \approx 1-3$ ns) the second pulse, transmitted through a $q^{th}$ order spiral phase plate to apply a $q\varphi$ azimuthal phase ($0 \leq \varphi < 2\pi$), was focused by the same axicon to form the $J_q$ Bessel beam. Deformable mirror correction of the beam wavefronts in advance of the axicon and waveplate (B. Miao, Feder, *et al.* 2022) enabled the production of beams with high-fidelity to the Bessel functional form over the full depth of focus $L_{focus} \lesssim 30$ cm (see Sec. V.B).

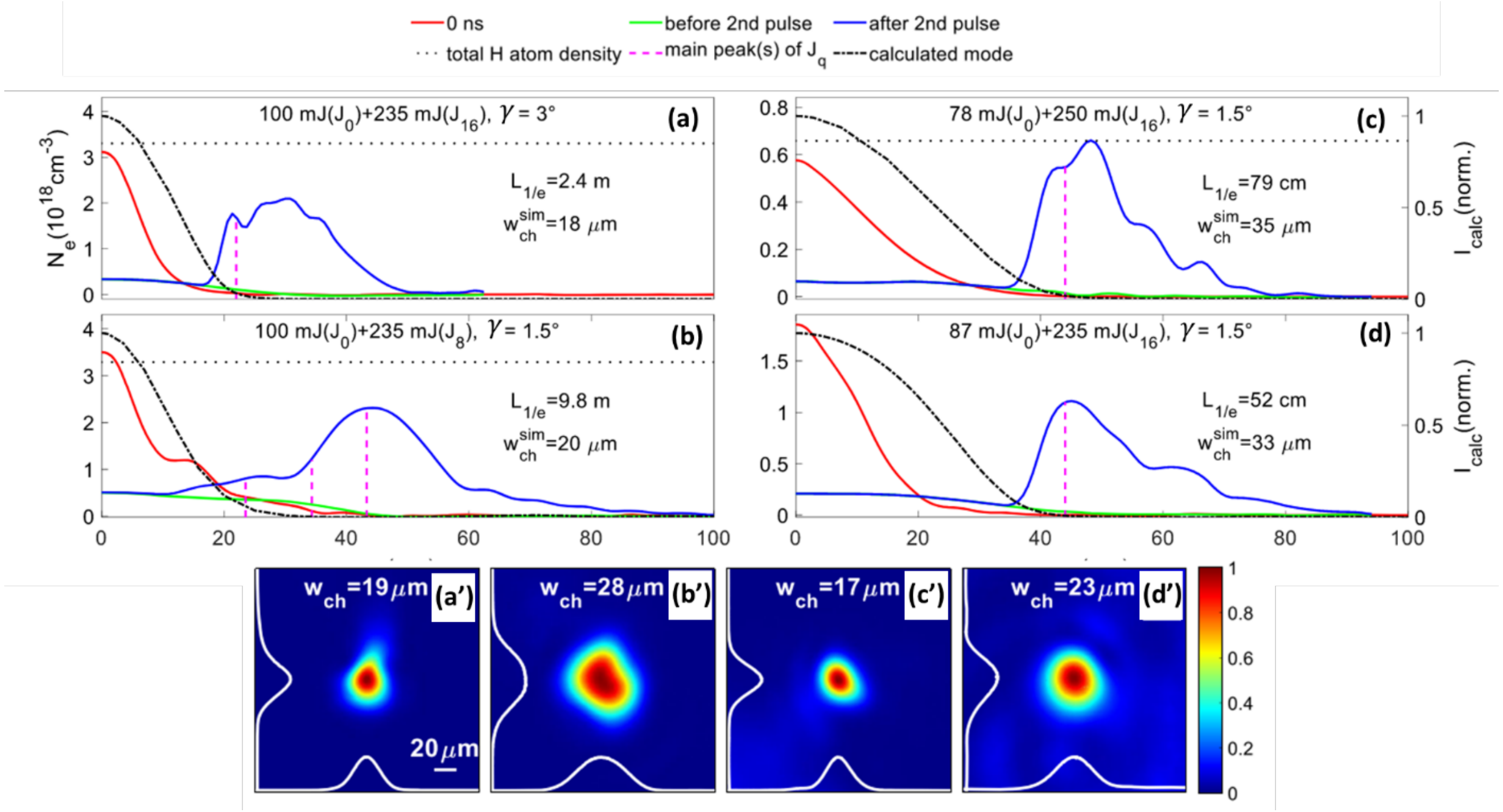

**Figure 19.** Tunable structure of waveguides generated by the two-Bessel method (adapted from (B. Miao *et al.* 2020)). **(a)-(d)** Extracted waveguide parameters for various laser/plasma conditions. **(a′)-(d′)** Guided modes collected for each of the profiles in (a)-(d).

As shown in Fig. 19, extensive control of waveguide parameters was achieved through variation of the initial gas $H_2$ density $N_g$, the delay $\tau_d$ between the $J_0$ and $J_q$ beams, axicon approach angle $\gamma$, and order $q$ of the $J_q$ cladding-forming beam. Characterization of the channel structure with transverse interferometry (panels (a)-(d)) demonstrated full ionization of the waveguides, with modal structure well approximated by a step-index model (see Sec. II.B.3). The high contrast between the core and cladding produced highly confining guides with calculated attenuation lengths (T. R. Clark and Milchberg 2000) as high as $L_{1/e}\sim 9.8$ m, greatly exceeding $L_{guide} = 30$ cm for this case. Figure 19(a')-(d') shows measured end modes corresponding to the various guides in Fig. 19(a)-(d). The guide and mode in panels (d) and (d′) are from guiding of $\sim 10^{17}\ \mathrm{W/cm^2}$ pulses in 5 cm hydrogen plasma waveguides generated in gas sheets from an extended supersonic slit jet (B. Miao *et al.* 2020; B. Miao 2020).

**(ii) Self-waveguiding method for cladding generation**

As first shown in (Morozov *et al.* 2018), a delayed intense short pulse injected along the axis of an expanding Bessel beam-induced OFI plasma column (up to 3.5 mm long in hydrogen) can guide more efficiently than a weaker pulse. At higher intensity, the transverse wings in the leading edge of the pulse more readily ionize the inside walls of the neutral shock and enhance the waveguide cladding, and thereby the guiding efficiency. While experimental density profiles were not extracted, simulations in (Morozov *et al.* 2018) for those experimental conditions ($N_g\sim 2\times 10^{19}\ cm^{-3}$ and $r_p\sim 4\ \mu m$) showed that a single Bessel beam pulse could form only a weak plasma cladding; this is consistent with the $(N_g, r_p)$ point located above the cutoff in Fig. 14. In this case, the very narrow initial plasma was estimated to rapidly cool from ~10 eV to ~1 eV within 100 ps, so the thermal flux available for ionizing the shock is lower than the estimates leading to Fig. 14. This explains the need in (Morozov *et al.* 2018) for an injected intense pulse to further build up the cladding for efficient guiding.

As discussed in Sec. III.C.2, for long, narrow OFI plasmas generated in low density gases ($N_g < \sim 10^{18}\ cm^{-3}$) appropriate for multi-GeV LWFA, a plasma cladding is, at best, marginally generated by a single-pulse Bessel beam: for example, measurements and simulations in (L. Feder *et al.* 2020; B. Miao *et al.* 2024) show no cladding generated at all. Just as in the high density case (Morozov *et al.* 2018), the cladding in low density guides can be formed by ionization of the inside wall of the neutral shock (see Fig. 18(a)) by the transverse wings of the far leading edge of a sufficiently intense injected pulse. This was demonstrated with experiments and simulations in (L. Feder *et al.* 2020) where the process was described as *self-waveguiding,* with the far leading edge of the pulse generating the cladding—on the fly—to strongly confine the rest of the pulse. This is distinct from *relativistic self-guiding* (Sec. I.A), which requires at least an order of magnitude greater peak intensity.

For efficient self-waveguiding, the threshold laser intensity of $\sim 10^{14}\ \mathrm{W/cm^2}$ for OFI of hydrogen (Corkum, Burnett, and Brunel 1989) in the transverse wings of the pulse sets the on-axis intensity threshold as $\sim 10^{17}\ \mathrm{W/cm^2}$ for a guide with $w_{ch}\sim 20$ µm. A simulation using the code FBPIC (Rémi Lehe *et al.* 2016) of the self-waveguiding process is shown in Fig. 20(a), where the leading edge of the pulse ionizes the inside wall of the neutral shock well in advance of the intensity peak. The simulation corresponds to the experimental conditions in (L. Feder *et al.* 2020), where pulses of peak intensity $\sim 10^{17}\ \mathrm{W/cm^2}$ were guided in 10 cm long OFI channels generated by $J_0$ pulses in 10 cm long hydrogen gas jets. For the same initial density profile and higher peak intensity corresponding to $a_0 \sim 1$, Fig. 20(b) shows the leading edge fully ionizing the shock (B. Miao, Shrock, *et al.* 2022). The stronger pulse supplies a larger field through leakage across the developing plasma cladding (see Sec. II.B.3), enabling ionization through the full shock width. The steep field dependence of the ionization rate accounts for the pronounced difference in radial extent between the two cases.

As the cladding is generated early in the pulse envelope, it provides a guiding structure for the main body of the pulse. Its optical confinement is described by the modal theory developed in (Durfee, Lynch, and Milchberg 1994; T. R. Clark and Milchberg 2000), while relativistic and wake-induced modifications of pulse propagation are treated by the reduced models of (P. Sprangle et al. 1992; E. Esarey et al. 2000; Benedetti et al. 2012), as reviewed in Sec. II. Additional simulations in (L. Feder *et al.* 2020) demonstrated that self-waveguiding is not limited to the plasma/neutral hydrodynamic shock structures formed by expansion of OFI plasmas, but is expected for *any* prepared hybrid plasma/neutral or purely neutral density structure which produces a sufficient core-cladding plasma index contrast after ionization. The latter includes gases with on-axis density depressions achieved by non-laser methods.

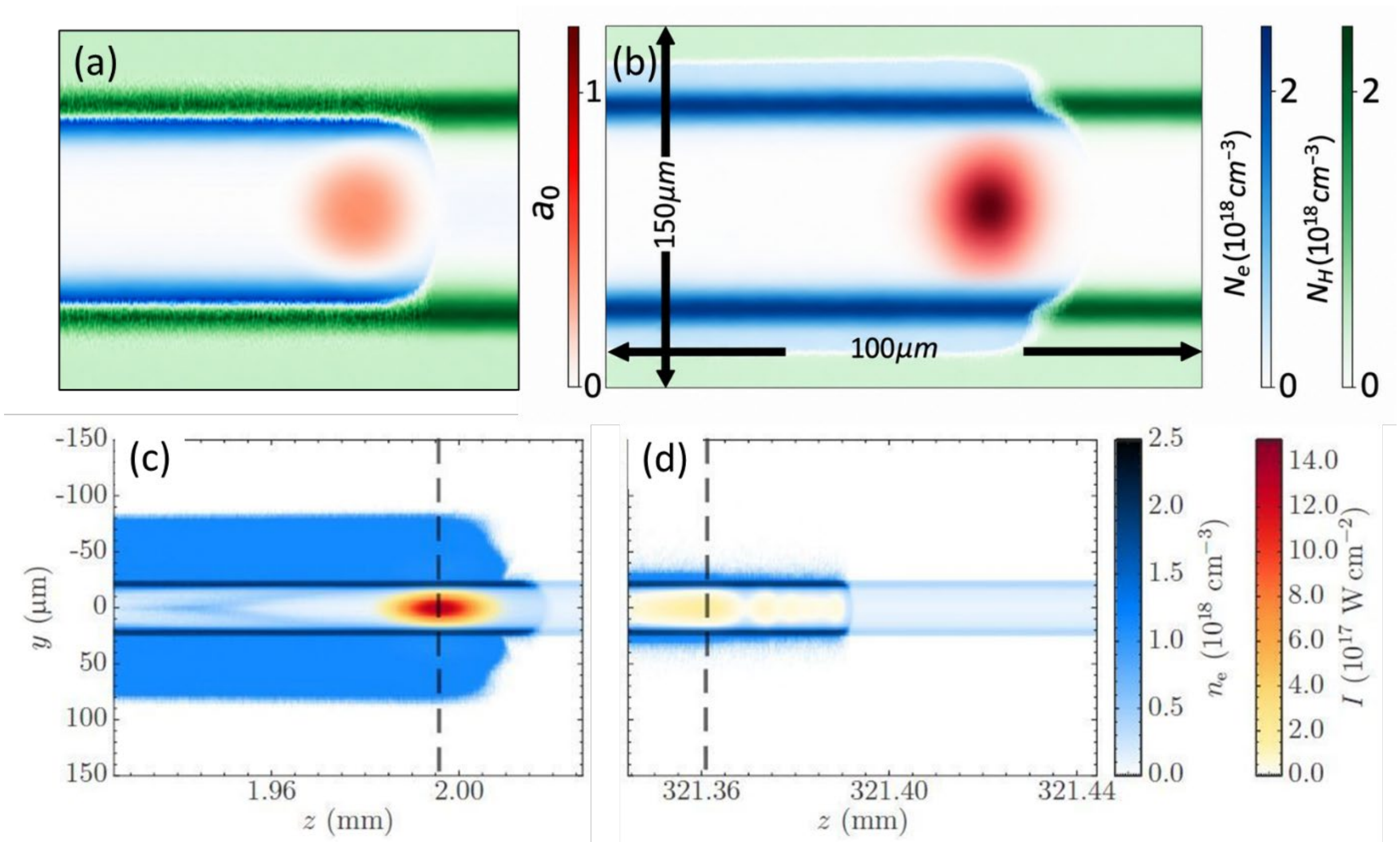

**Figure 20.** Simulations of the self-waveguiding process for the same prepared index structure in hydrogen at **(a)** moderate pulse intensity just sufficient to ionize the neutral shock inner wall (conditions of (L. Feder *et al.* 2020)), and **(b)** for $a_0 \sim 1$, showing full shock ionization. **(c)(d)** Simulations showing electron density and laser pulse profiles early and late in a ~30 cm structure, showing elimination of leakage and ionization outside the guide. ((a) and (b) are adapted from (B. Miao, Shrock, *et al.* 2022); (c) and (d) are adapted from (Picksley, Alejo, Shalloo, *et al.* 2020)).

(Picksley, Alejo, Shalloo, *et al.* 2020) contemporaneously investigated this mechanism, described as "conditioned hydrodynamic optical-field-ionized" (CHOFI) channels. They measured enhanced transmission of the injected pulse and reconstructed electron-density profiles, measured with single-color interferometry, showing a deeper, thicker plasma wall after its passage. An off-axis density exceeding that expected from full ionization of the ambient gas supported the interpretation that compressed peripheral neutral gas supplied part of the additional plasma. Hydrodynamic and particle-in-cell simulations described the formation of such a structure and its ionization by the pulse's leading edge (the "conditioning" pulse), producing a channel that confined the remainder of the pulse, as seen in simulations (Fig. 20(c)(d)) showing the pulse and electron density profile at the beginning and end of propagation. In panel (c), the self-waveguiding pulse is seen to excite a linear wake. The authors noted that this mechanism had played a role in their previous Bessel-OFI waveguide experiments (R. J. Shalloo *et al.* 2019; Picksley, Alejo, Cowley, *et al.* 2020), where the guide profiles measured after single-pulse OFI generation had insufficient cladding for the observed guiding efficiency (see Sec. III.C.3). An earlier qualitative proposal appeared in (R. Shalloo 2018), suggesting, without identifying the presence and role of the neutral gas shock, that ionization of surrounding gas by an injected pulse could strengthen confinement.

Figure 21(a) plots measured guide profiles produced by self-waveguided pulses of increasing energy, demonstrating that the transverse extent of ionization into the neutral shock increases while the near-axis structure is unaffected. Panel (b) plots the waveguide structure for increasing neutral gas density and panel (c) plots the structure for varying delays between the $J_0$ pulse and the self-waveguiding pulse, highlighting the nearly independent control of waveguide core and cladding.

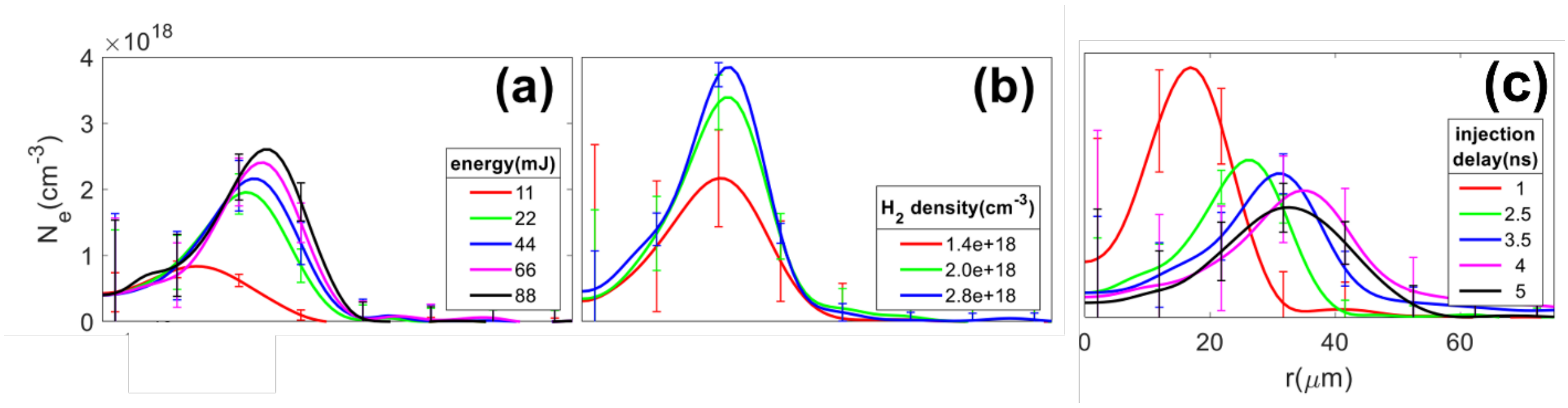

**Figure 21.** Plasma density profiles measured by two-color interferometry ~5 ps after passage of self-waveguided pulse for **(a)** varied self-waveguided pulse energy at fixed injection delay of 2.5 ns and $H_2$ density $1.6 \times 10^{18}\ cm^{-3}$ **(b)** varied $H_2$ gas density at injection delay 2.5 ns. **(c)** varied delay between $J_0$ pulse and injected (self-waveguided) pulse. (Adapted from (L. Feder *et al.* 2020)).

Full-channel resolved images of self-waveguiding are shown in Fig. 22. Figure 22(a) shows a sequence of plasma fluorescence images from pulses of increasing energy injected into the prepared index structure (plasma core with neutral density shock wall) generated in a 10 cm gas jet by a single $J_0$ pulse (L. Feder *et al.* 2020). At the highest energies, it is seen that the pulse "burns through" to the end of the jet. For these channels, pulse front erosion loss, from pre-cladding diffraction and leakage plus energy absorbed by cladding generation, was ~6 mJ/cm. In general, the erosion loss is dominated by diffraction and leakage, with a cladding ionization cost of < 1 mJ for the entire waveguide (L. Feder *et al.* 2020). For the higher intensities required for LWFA ($a_0 >$ ~1), complete ionization of the shock is expected (B. Miao, Shrock, *et al.* 2022). Figure 22(b), for a shorter and denser hydrogen jet, shows a time sequence of interferograms, separated by ~0.5 ps delays, capturing the self-waveguiding pulse propagating from right to left (Morozov *et al.* 2018).

It is clear from these images that there is significantly more plasma after passage of the self-waveguiding pulse, in agreement with the analysis earlier in this section and in Sec. III.C.2.

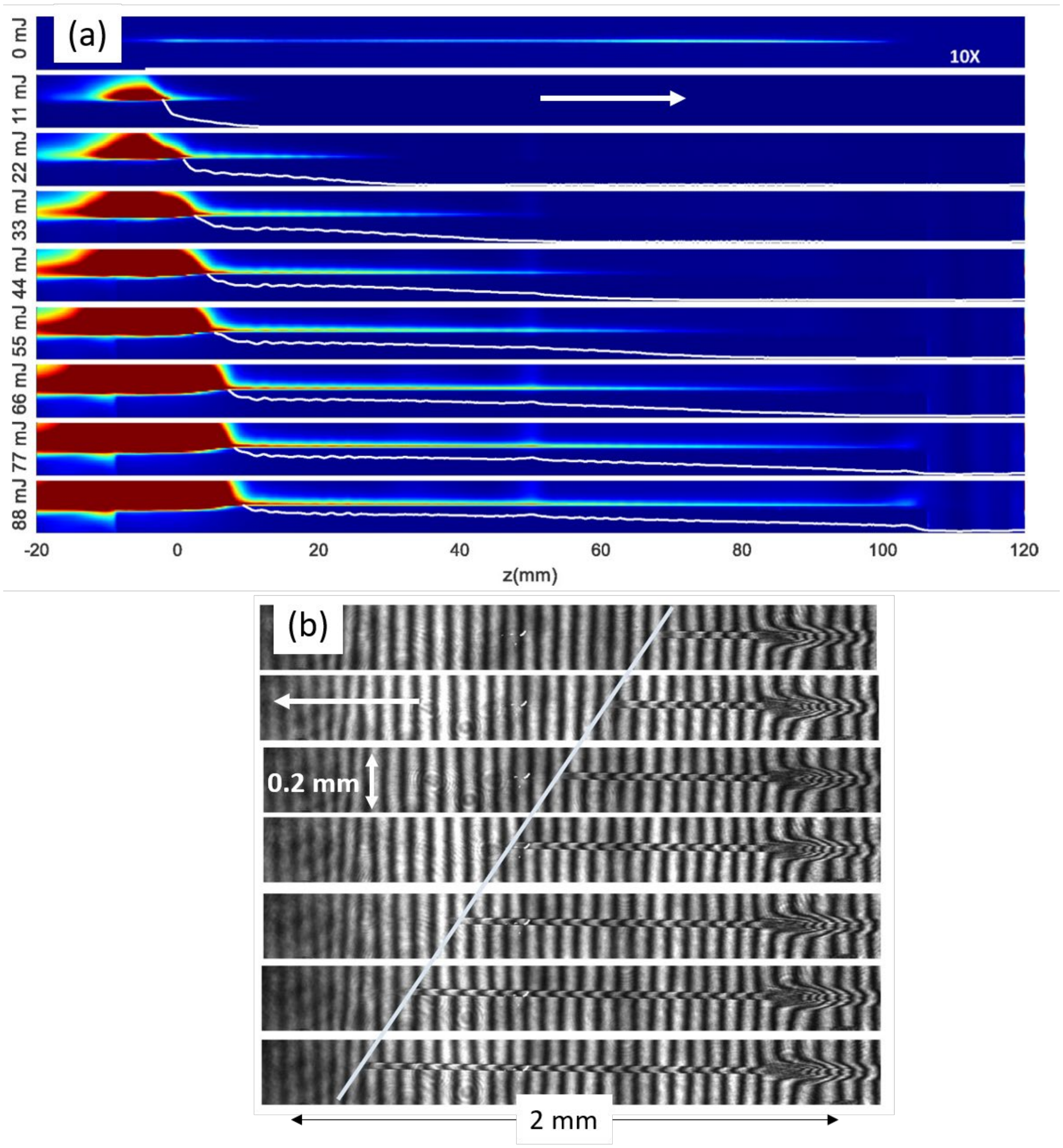

**Figure 22. (a)** Fluorescence from self-waveguiding pulses of increasing energy in a low-density hydrogen gas jet. The pulse injected into the prepared index structure propagates from left to right (adapted from (L. Feder *et al.* 2020)). **(b)** Interferograms of plasma generated by self-waveguiding in a short, dense hydrogen jet vs. probe delay. Here the pulse propagates right to left. The panels are at increasing ~0.5 ps delays from top to bottom (adapted from (Morozov *et al.* 2018)).

Another feature in Fig. 22(a) is the axial modulation of the fluorescence, which corresponds to beating of the $(p, m) = (0,0)$ and $(1,0)$ modes, with the strongest modulations near the guide entrance manifesting as saturated fluorescence. Figure 23 plots the radially integrated fluorescence signal vs. z for different delays between the $J_0$ and self-waveguiding pulses (i.e. different channel sizes). The measured modulation period is in agreement with the beating period calculated from the quasibound modes (L. Feder *et al.* 2020), with the period increasing with delay. The coupling into these modes and consequent beating was also observed in accompanying propagation simulations (L. Feder *et al.* 2020).

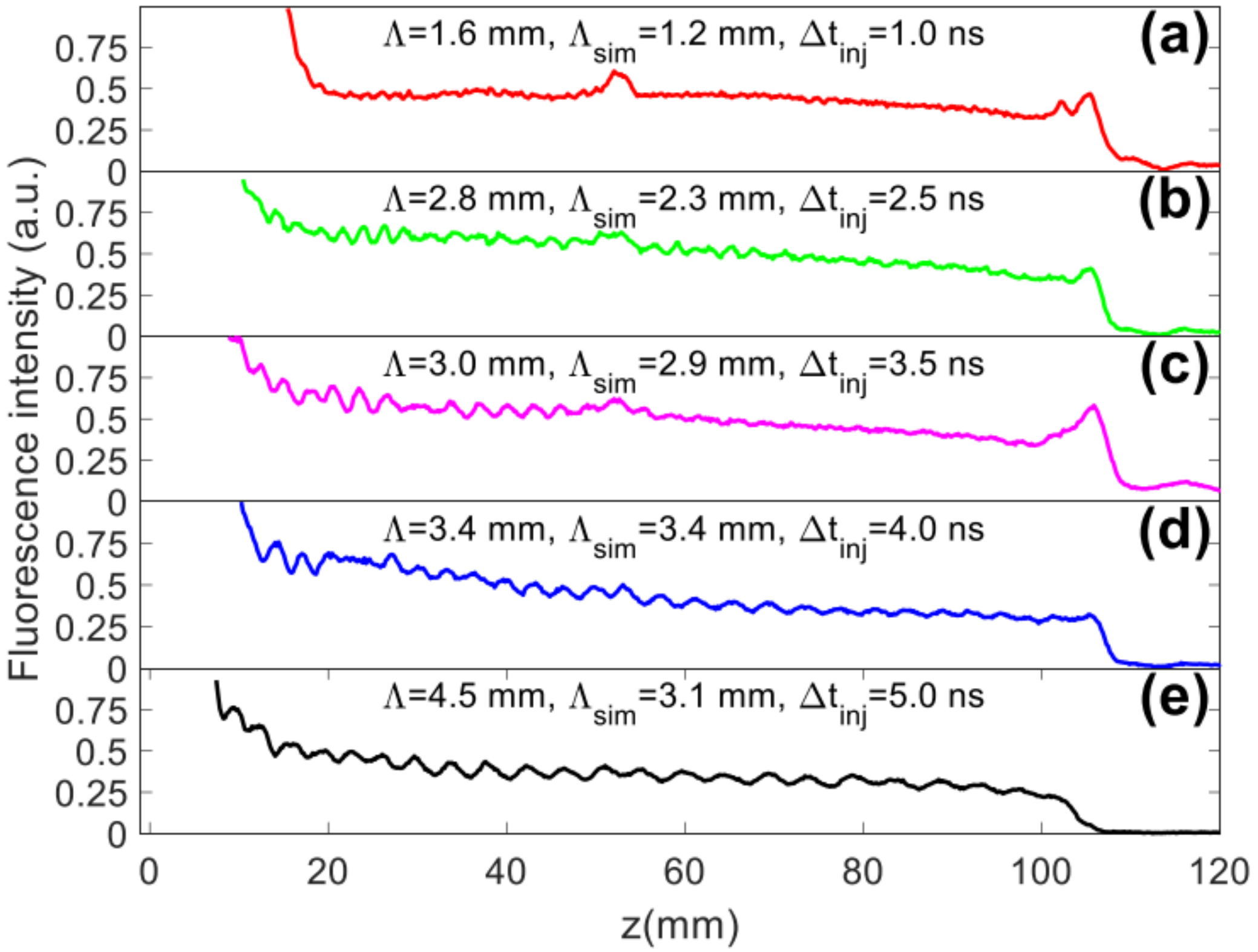

**Figure 23.** Integrated fluorescence signals from self-waveguiding pulses vs. injection delay $\Delta t_{inj}$ from the initiating $J_0$ pulse. The measured and simulated axial modulation periods are $\Lambda$ and $\Lambda_{sim}$ (adapted from (L. Feder *et al.* 2020)).

As a final validation of the self-waveguiding picture, a low-power, frequency doubled probe pulse was coupled into the channel 5 ps after the self-waveguiding pulse (L. Feder *et al.* 2020). As discussed in Sec. II, a unique feature of plasma waveguides is that well-bound mode structure is wavelength independent (Durfee, Lynch, and Milchberg 1995; T. R. Clark and Milchberg 2000). Comparisons of exit mode sizes for the $\lambda = 800$ nm self-waveguiding pulse and $\lambda = 400$ nm probe pulse were observed to closely agree for a range of self-waveguiding pulse energies.

The energy of a self-waveguiding pulse must scale as $w_{ch}^2$ in order to maintain the threshold intensity for OFI at the neutral channel walls. For the 2-Bessel method, the energy required in the $J_q$ beam for cladding generation scales as $w_{ch}^{0.63}$ (B. Miao *et al.* 2020; L. Feder *et al.* 2020; Linus Feder 2021), which is more energy efficient for larger guides. However, for applications such as multi-GeV LWFA using petawatt class lasers, which use meter-scale plasma waveguides, the energy cost for either method is ~1 J, a small fraction of the tens of joules used in the LWFA drive pulse. Because the self-waveguiding method is easier to implement than the 2-Bessel method, it has been used in recent multi-GeV experiments (B. Miao, Shrock, *et al.* 2022; Shrock 2023; Picksley *et al.* 2024; Rockafellow *et al.* 2025).

## IV. CAPILLARY WAVEGUIDES

While the methods discussed in Sec. III rely on laser pulses to produce the plasma waveguide structure, another branch of techniques relies on guiding in hollow dielectric capillary tubes, with and without plasma-generating electrical discharges along their length. In this section we will review these methods.

### A. Hollow dielectric capillary (no discharge)

In the first demonstration of high-power guiding in a capillary tube structure, in this case without a discharge, a 1 TW subpicosecond pulse was guided up to 13 cm in a ~270 μm inner diameter hollow circular-cross-section glass capillary (Jackel *et al.* 1995). As the guided intensity was increased, a sharp increase in throughput was identified when the guiding mechanism shifted from grazing incidence reflection off the capillary wall to plasma wall reflection. In (Borghesi *et al.* 1998), plasma-wall reflection inside of a 100 μm diameter hollow capillary enabled guiding of >10 TW pulses over 10 cm with ~50% transmission.

Both experiments demonstrated multimode guiding due to coupling mismatch. For a hollow dielectric capillary, the modes are Bessel-like hybrid EH modes (Cros *et al.* 2002) (as opposed to the TEM modes of plasma waveguides (T. R. Clark and Milchberg 2000)). For these modes, coupling of an incident Gaussian beam is optimized when the waist satisfies $w_0 = 0.645a$, where $a$ is the capillary inner radius. Under this condition, monomode guiding of >10 TW pulses to $\sim 10^{16}$ W/cm$^2$ was achieved (Dorchies *et al.* 1999). However, laser damage—particularly due to mode beating and misalignment (Dorchies *et al.* 1999)—was found to significantly impede the use of hollow capillaries for pulse intensities $> 10^{16}$ W/cm$^2$. Funnel or cone entrances were also proposed to improve coupling (Andreev *et al.* 2008). A cone was employed in a LWFA experiment (Kitagawa *et al.* 2004), where ~10 J, ~0.6 ps laser pulses to $\sim 10^{18}$ W/cm$^2$ were injected into a 60 μm inner diameter glass capillary. The capillary walls were ionized by a prepulse, generating a glass plasma in which electrons were accelerated up to ~100 MeV in a thermal spectrum. Here, the capillary was destroyed in each laser shot. In general, filling the capillary with a specified gas, as one would need to do for better control of applications, was found to significantly reduce transmission due to ionization induced refraction of the beam (Dorchies *et al.* 1999). This, plus lack of control of the plasma density profile, limited their use in applications requiring extended laser-plasma interaction, particularly LWFA.

At peak laser intensities well below the ionization threshold of fill gases in hollow capillaries ($< \sim 10^{13}$ W/cm$^2$), propagating pulses can undergo extreme spectral broadening via self-phase modulation from the bound electron Kerr effect, generating ultrabroadband spectra (Nisoli, De Silvestri, and Svelto 1996). These pulses are then compressed to only a few optical cycles in duration (<10 fs) using prisms or chirped mirrors (Sartania *et al.* 1997). The generation of few cycle pulses is a rich field (Brabec and Krausz 2000), including applications to attosecond physics and light sources (Krausz and Ivanov 2009) and kHz, few-cycle pulse-driven LWFA (Guénot *et al.* 2017; Salehi *et al.* 2021; Lazzarini *et al.* 2023; Railing *et al.* 2024).

### B. Discharge capillaries

Some of the above issues are solved by employing an electrical discharge along the capillary axis; this generates and heats a preformed plasma within the capillary structure. The fast establishment of a radial pressure equilibrium between the hot center of the discharge and the cool capillary wall forms a plasma density profile with low density in the center and higher density at the wall; a plasma waveguide is formed inside of the capillary. Because of the much higher plasma densities approached at the wall, capillary discharge waveguides are generally much more optically confining than laser-induced hydrodynamic waveguides, with guided modes better described as bound than leaky (see Fig. 1 and discussion).

#### 1. Ablative discharge capillary

When a high voltage discharge is driven along the capillary axis, plasma is formed from discharge-induced ablation of the capillary walls. For capillary diameters <1mm, radial plasma pressure equilibrium is achieved within a few hundred ns, and a concave plasma density profile is established with a plasma density minimum on axis. The first generation and guiding in this type of waveguide was presented in (Ehrlich *et al.* 1996), where a slow discharge at ~10 kV was driven in a 1-cm-long polypropylene cylinder with a 350 μm diameter axial hole between two electrodes. Slow discharges have a current rise time $\tau_r > \sim R_c/c_s$, where $R_c$ is the capillary radius and $c_s$ is the ion sound speed. For the conditions of (Ehrlich *et al.* 1996), $R_c/c_s \sim 50$ ns and $\tau_r \sim 100 - 200$ ns. Guiding of pulses with intensities $> 10^{16}$ W/cm$^2$ was achieved over 11 Rayleigh ranges with on-axis plasma electron densities $> 10^{19}$ cm$^{-3}$. Prior work had demonstrated guiding in 1D plasma waveguides formed by discharges along a 1-cm-long slotted $CF_2$ capillary (Zigler *et al.* 1996). The discharge formed a 1D plasma waveguide with an on-axis density of $10^{19}$ cm$^{-3}$ between the slot walls, which confined a beam in one dimension but not the other.

Ablative discharge capillaries were further improved with the development of a polyethylene 'double' capillary, where discharge plasma was generated in a short (3 mm) region and allowed to expand into a longer (>1 cm) region (of the same capillary diameter) and heated through additional discharge. A schematic is shown in Figure 24(a) (Kaganovich *et al.* 1997). Measurements of the Stark broadening of the H-alpha line in the discharge plasma enabled time-resolved observation of the plasma evolution (Ehrlich *et al.* 1998), and showed that the parabolic plasma profile developed several hundred nanoseconds after the discharge was initiated. An example of these measurements is shown in Fig. 24(b)-(b''') along with propagation simulations and experimental data showing the effect of varying channel parameters on the mode beating frequency (Ehrlich *et al.* 1998) in panel (c) .

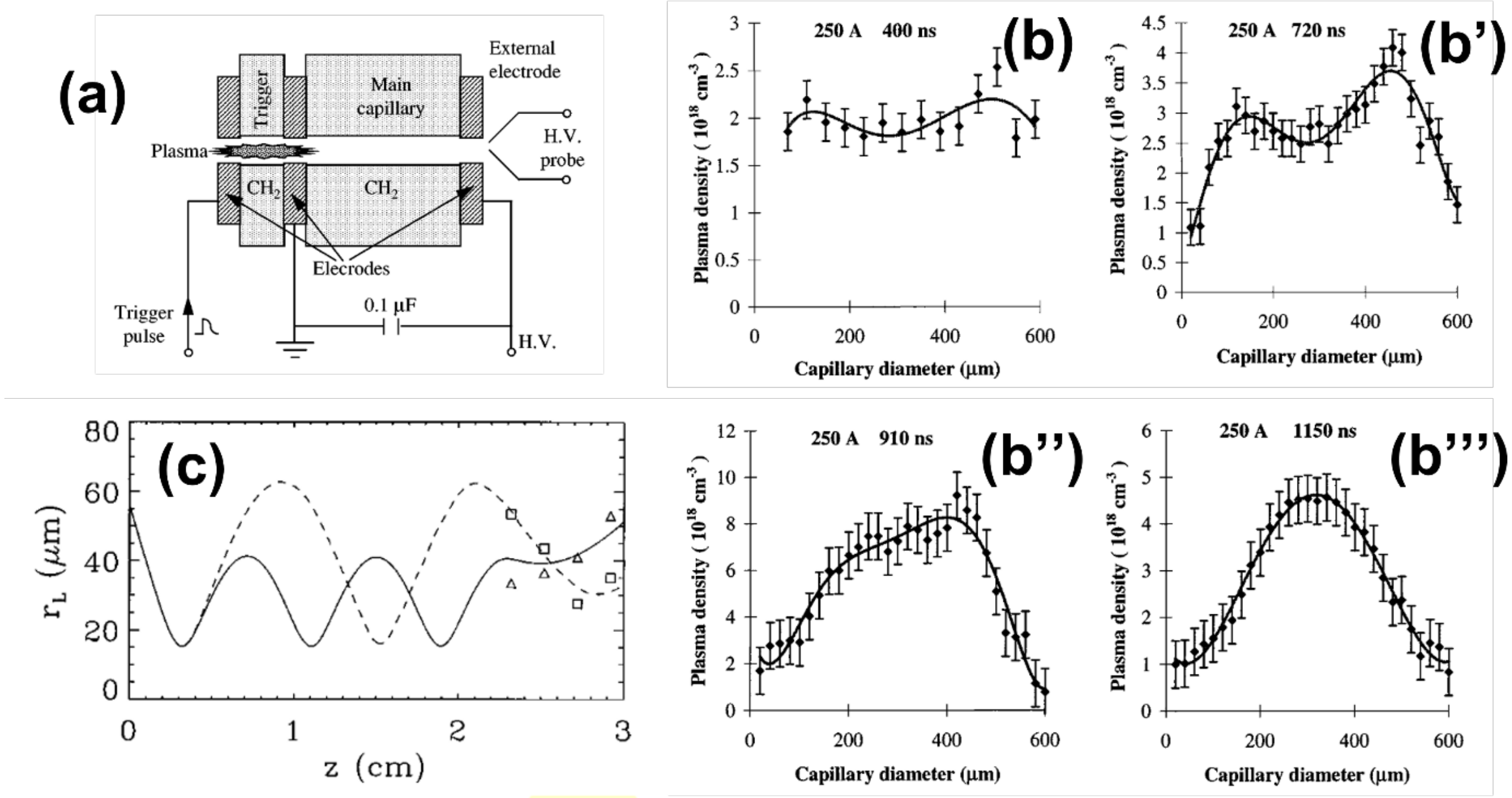

**Figure 24.** Waveguide generation in ablative discharge capillaries (panel (a) adapted from (Kaganovich *et al.* 1997), panels (b)-(c) adapted from (Ehrlich *et al.* 1998)). **(a)** Schematic of 'double capillary'. **(b)-(b‴)** Evolution of plasma inside ablative discharge capillary. **(c)** simulated and measured pulse evolution in waveguides formed via ablative discharge capillary.

The double capillary design was employed to guide $10^{16}\ \mathrm{W/cm^2}$ pulses over 6.6 cm with on-axis density $\sim 5\times10^{18}\ \mathrm{cm^{-3}}$ (Ehrlich *et al.* 1998) and later for pulses with intensities above $10^{17}\ \mathrm{W/cm^2}$ over 2 cm with on-axis densities $> 10^{18}\ \mathrm{cm^{-3}}$ (Kaganovich *et al.* 1999). A modification of the double capillary design employing square geometry enabled more detailed characterization of the plasma evolution with transverse interferometry (see Sec. V.A.1) (Jones *et al.* 2003).

Guiding of picosecond pulses in ablative discharge capillaries of a different design was explored in (Hooker, Spence, and Smith 2000) and various loss mechanisms were examined. Longitudinal interferometry of the channel (see Sec. V.A.1) verified the formation of a concave guiding structure on a few hundred nanosecond timescale (Spence, Burnett, and Hooker 1999). In a similar ablative capillary experiment, an EUV laser was used for longitudinal shadowgraphy of the evolving channel (Marconi *et al.* 2000). Additional work (Spence and Hooker 2000a) indicated that the partial ionization of carbon resulted in ionization-induced defocusing at high laser intensities, limiting the applicability of ablative capillaries for experiments requiring transmission of high-intensity pulses. The reliance on wall ablation for plasma also greatly limits the lifetime of ablative capillaries.

### 2. Gas-filled fast discharge capillary

Discharges in gas-filled capillaries are classified as fast or slow. Fast discharges have a current rise time $\tau_r \ll \sim R_c/c_s$. Here, the inward radial force density $j_z B_\varphi/c$ —from the magnetic field $B_\varphi$ generated by the discharge current $j_z$—dominates the thermal pressure gradient force density in the plasma, giving rise to a "z-pinch" where the plasma is compressed radially inward (Rocca *et al.* 1995; Rocca 1999). For capillary diameters $< \sim 1$ mm and $c_s \sim 10^5 - 10^6$ cm/s for plasmas at a few eV in temperature, $R_c/c_s \sim 100$ ns. Fast capillary discharges with $\tau_r \sim 30$ ns have been employed to generate 3-12 cm plasma columns as the gain medium for soft x-ray lasers (Rocca *et al.* 1994). During the z-pinch implosion, a concave plasma density profile develops before maximum compression on axis. Subsequent experiments suggested that this transient, concave plasma structure assisted in guiding the x-ray laser beam through the plasma (Hosokai *et al.* 1997; Moreno *et al.* 1998). Guiding of high-power optical pulses in this transient structure was first reported in (Hosokai *et al.* 2000), using a peak ~5 kA discharge with $\tau_r \sim 15\ ns$. Accompanying MHD simulations identified formation of an imploding shock and transient channel about 8 ns after the discharge began and lasting for about 500 ps. The discharge profile and simulated electron density are reproduced in Figure 25(a)(b). In (Fauser and Langhoff 2000) it was suggested that a longer-lived transient channel may be formed during the relaxation phase after the implosion has reached the axis and the plasma shock is reflected back toward the capillary wall. High power guiding experiments and longitudinal interferometry of a 5.5 cm-long fast-discharge channel (Luther *et al.* 2004) demonstrated the progression from multimode to single mode guiding during the z-pinch compression phase (Fig. 25(c), panels A-C), followed by beam refraction from the pinched plasma (panel D). This was followed by weak guiding in the relaxing plasma (panel E), which was confirmed to be a leaky guide by additional far field imaging showing the characteristic conical radiating fields (T. R. Clark and Milchberg 2000) described in Sec. II.B. Figure 25(d) plots the total (solid line) and guided (dashed line, using an iris to isolate the guided mode) transmission through the capillary vs. delay from the onset of the discharge. The vertical dashed lines correspond to the labeled panels in Fig. 25(c). Simulations in (Bobrova, Bulanov, Esaulov, *et al.* 2000) further suggested that a quasi-stable equilibrium might be found after the reflected shock has returned to the capillary wall.

Reliance on the transient structure generated by z-pinch implosion imposes significant limitations on the broader application of fast discharge waveguides for guiding high-intensity pulses. Although the use of a fill gas inside the capillary reduces damage from wall ionization as well as the introduction of carbon and other highly multiply ionizable contaminants which reduce waveguide performance, the structure observed in (Hosokai *et al.* 2000) only existed for ~ 500 ps. This limits the tunability of guide parameters and poses difficulties for implementation on the scale of tens of cm. Furthermore, simulations suggest that for some combinations of gas and discharge parameters, the reliance on magnetic pressure in the z-pinch to drive the channel formation opens the door to potential MHD instabilities seeded by wake excitation (Bobrova, Bulanov, Esaulov, *et al.* 2000; Bobrova, Bulanov, Farina, *et al.* 2000). That said, the ability of fast discharge capillaries to deeply ionize Ar and other multiply ionizable gases made them an ideal candidate for short wavelength pumping applications.

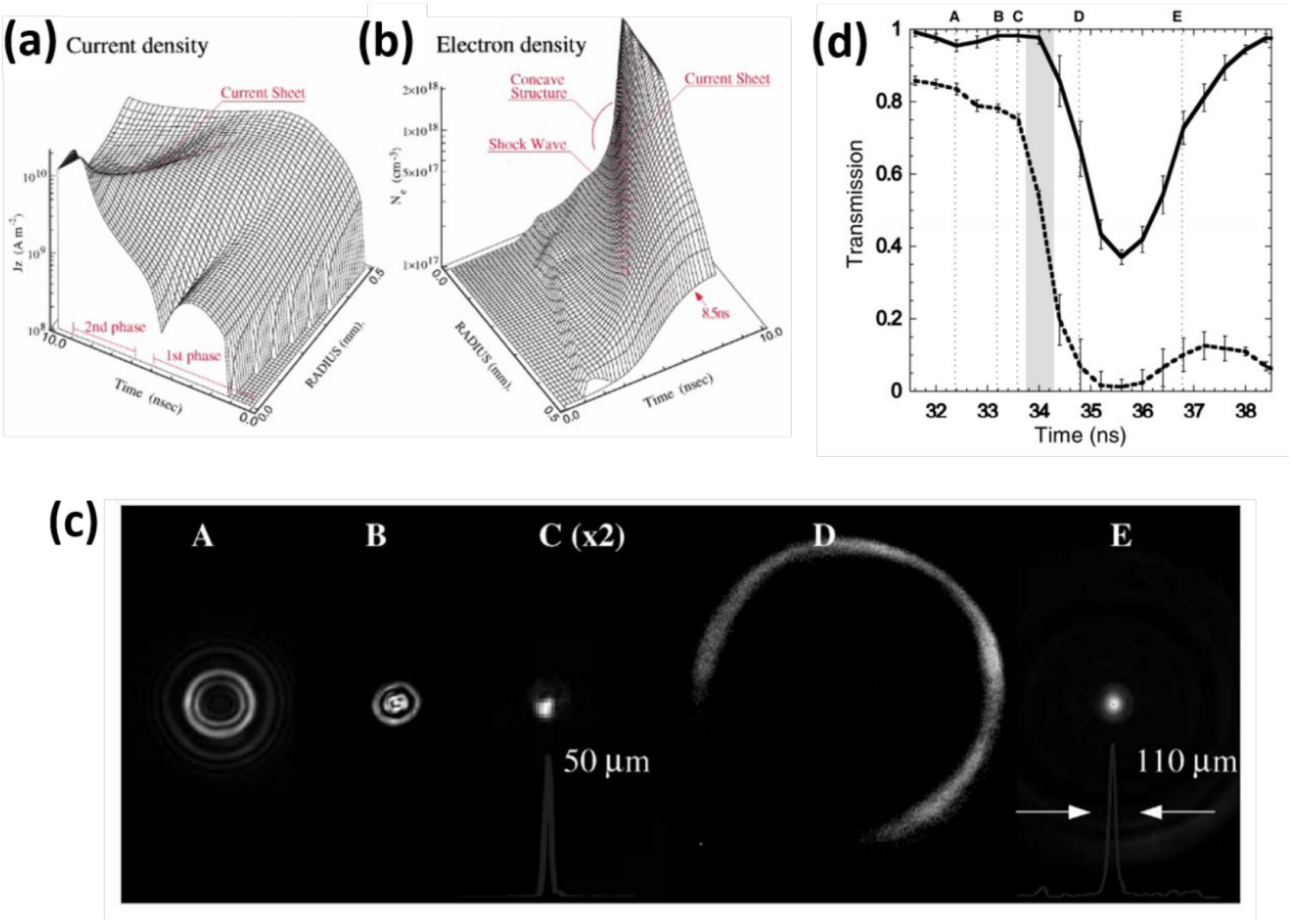

**Figure 25.** Waveguide generation by fast discharge in a gas-filled capillary. Panels (a)(b) are adapted from (Hosokai *et al.* 2000); panels (c)(d) are adapted from (Luther *et al.* 2004). **(a)** Simulated discharge current density distribution. **(b)** Simulated density profile corresponding to the discharge profile in (a). **(c)** Exit intensity profiles during compression (A–C), strong refraction near pinch (D), and post-pinch expansion (E). **(d)** Total transmission (solid curve, iris open) and guided transmission (dashed curve, iris closed) versus time from discharge onset. The vertical dashed lines A–E mark the times of the corresponding images in (c).

### 3. Gas-filled slow discharge capillary

Many of the limitations of ablative and fast gas-filled discharge capillaries are overcome with use of slow discharge gas filled capillaries with $\tau_r > \sim R_c/c_s$, first employed as plasma waveguides in (Spence, Butler, and Hooker 2001; Butler, Spence, and Hooker 2002; Spence, Butler, and Hooker 2003). The slow (typically $\sim 100$ ns $< \tau_r < \sim 1$ μs) and relatively low current discharge ensures that the plasma dynamics are dominated by thermal, rather than magnetic (z-pinching) pressure, while the use of a low-Z fill gas ensures generation of a fully ionized plasma. In these experiments, 67 mbar $H_2$ was introduced into 300 μm internal diameter ceramic (alumina) capillaries 2-5 cm in length. A 1.7 nF capacitor charged to 25 kV drove a $\tau_r \sim 200$ ns discharge with $\sim 500$ A peak current. Longitudinal interferometry measured the formation of a parabolic density profile $\sim 50$ ns after discharge onset, with example profiles (Spence and Hooker 2000b) in Fig. 26(a) showing on-axis densities $> \sim 2 \times 10^{18}$ cm$^{-3}$. Laser energy transmission as a function of delay with respect to discharge onset (Butler, Spence, and Hooker 2002) is plotted in Fig. 26(b) for various fill pressures and capillary lengths, with transmission up to 80% for the 5 cm capillary. It is seen that transmission continues for several hundred nanoseconds even after decay of the discharge current. Corresponding guided exit modes are plotted in Fig. 26(c), which shows a $>$ 500 ns onset to $\sim 80\%$ guiding efficiency.

An important part of the capillary waveguide system is the electrical driver. The current rise time, peak current, and pulse duration determine the breakdown and heating history and therefore the ionization state, radial temperature gradient, and time window over which the guiding profile is reproducible. Trigger jitter sets the shot-to-shot timing uncertainty for laser injection; electrode geometry, circuit and plasma impedance, reflections, and longitudinal current uniformity affect axial reproducibility; and recharge, gas recovery, and wall cooling constrain the repetition rate. (Spence and Hooker 2000b) documented the compact capacitor-discharge circuit used for the early hydrogen guide. (Abuazoum *et al.* 2011) developed a low-jitter solid-state pulsed supply, and (Dyson, Thornton, and Hooker 2016) demonstrated a compact Marx-bank driver. These circuit choices must be considered together with capillary geometry and gas fill when assessing waveguide performance.

Slow gas filled discharge capillaries (referred to as 'discharge capillaries' throughout the rest of this section) have seen widespread use and until recently were the main method for generation of plasma waveguides in high energy gain LWFAs (Leemans *et al.* 2006; Rowlands-Rees *et al.* 2008; Leemans *et al.* 2014; Gonsalves *et al.* 2019; Qin *et al.* 2022; H. Lu *et al.* 2011).

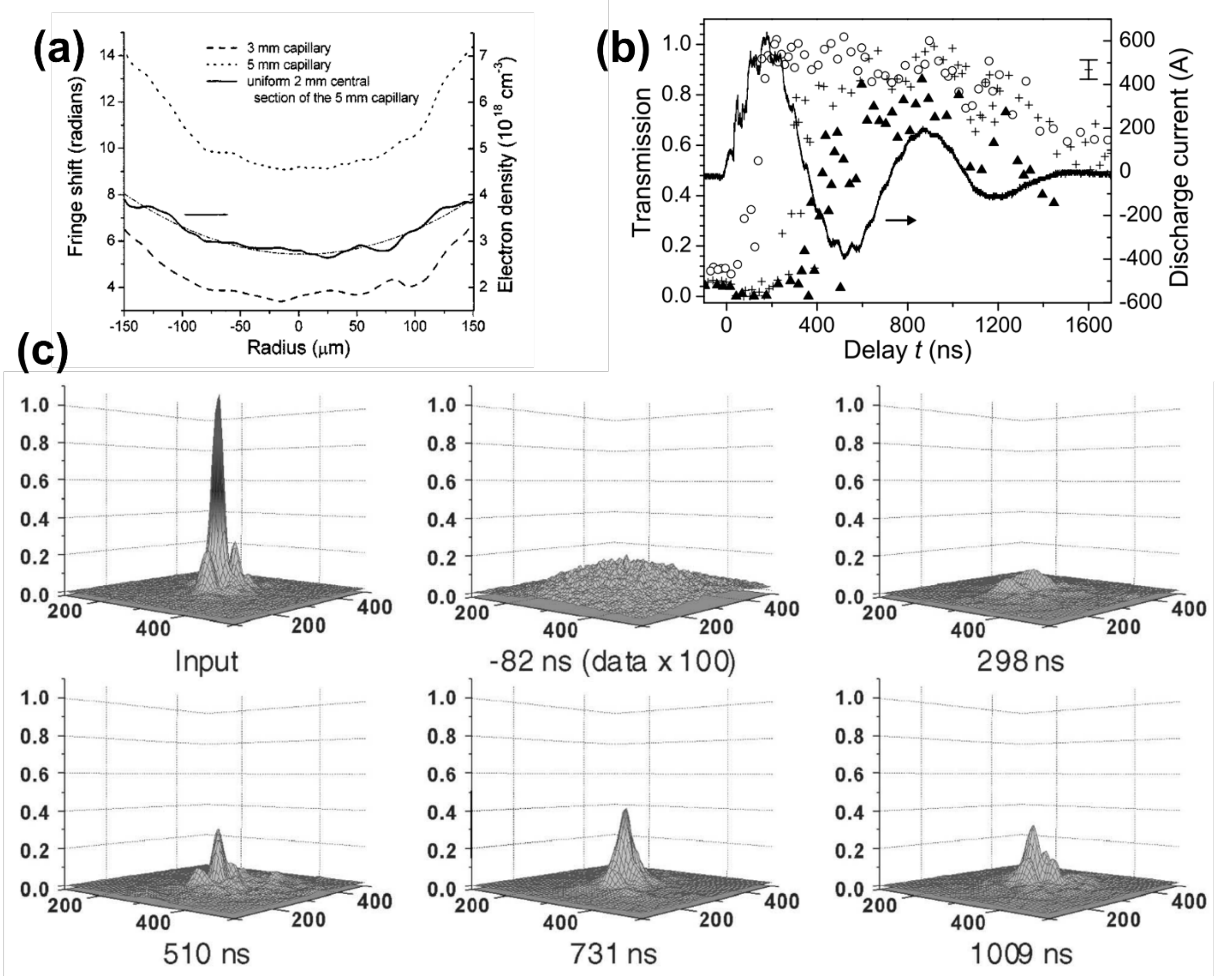

**Figure 26.** Plasma waveguide formation by a slow discharge in a hydrogen-filled capillary (panel (a) adapted from (Spence and Hooker 2000b) and panels (b),(c) adapted from (Butler, Spence, and Hooker 2002)). **(a)** Measured transverse density profiles of plasma waveguides. **(b)** Measured transmission efficiencies vs. delay with respect to onset of discharge current (solid curve). Triangles are for a 5 cm long capillary filled with 330 mbar $H_2$. **(c)** Waveguide exit mode intensity profiles vs. delay with respect to discharge onset for the 330 mbar, 5 cm capillary.

A model of plasma evolution and resultant waveguide properties in discharge capillaries is outlined in (Bobrova *et al.* 2002), which presents 1D MHD simulations modeling the plasma and wall dynamics (the same technique has also been used to model ablative and fast gas filled capillaries (Bobrova *et al.* 1998; Bobrova, Bulanov, Farina, *et al.* 2000)). Employing a discharge of the form $I(t) = I_0 sin(\pi t/\tau_d)$ with parameters comparable to (Spence and Hooker 2000b; Butler, Spence, and Hooker 2002), Bobrova *et al.* identify three characteristic stages of slow discharge capillaries. These are denoted by the dashed lines in Figure 27, which is modified from (Bobrova *et al.* 2002) and shows the simulated evolution of the plasma density, temperature, and ionization state throughout the discharge.

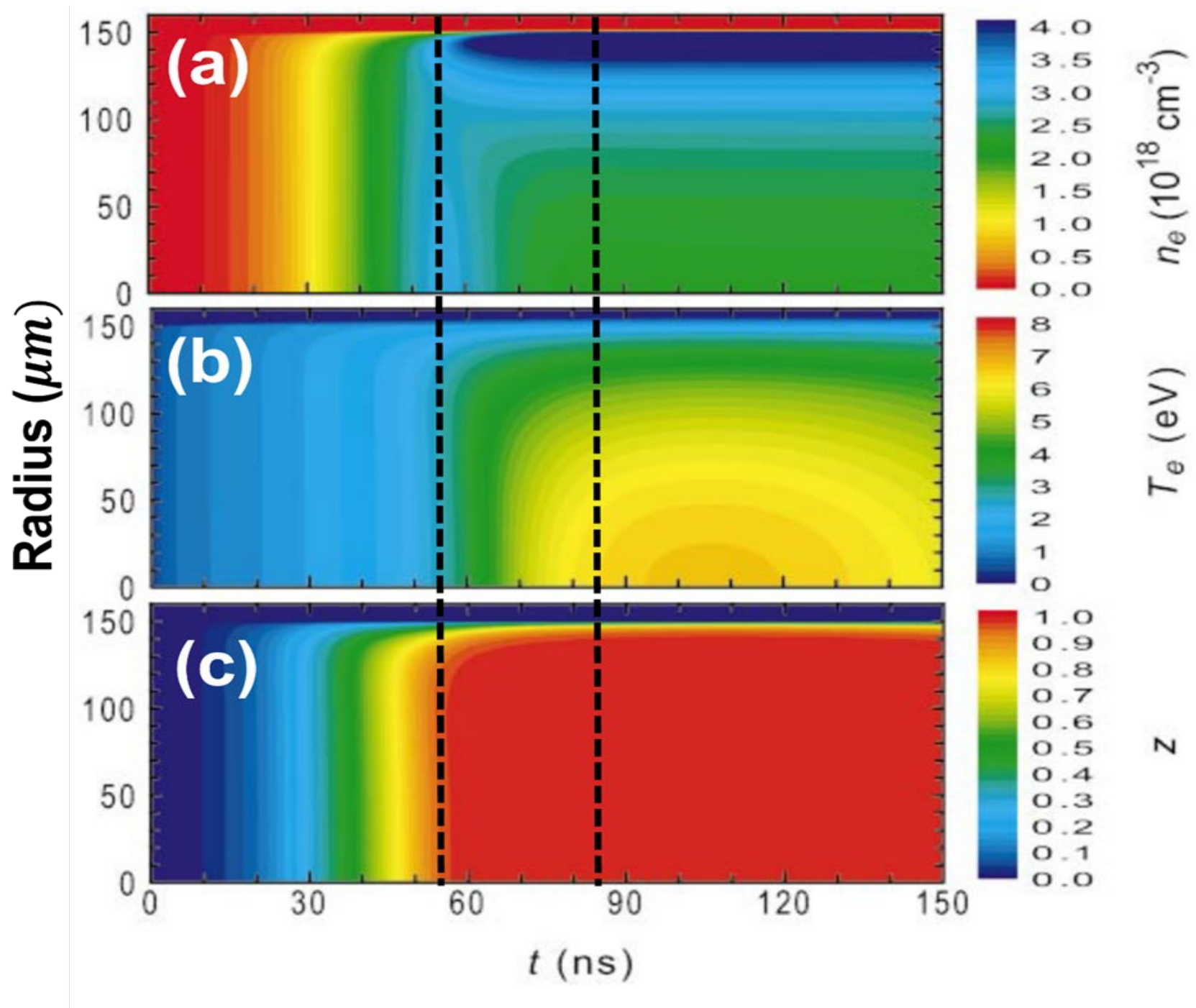

**Figure 27.** Quasi-static model of plasma evolution in hydrogen-filled discharge capillaries (adapted from (Bobrova *et al.* 2002)). **(a)** Evolution of plasma density. **(b)** Evolution of plasma temperature. **(c)** Evolution of ionization state.

During the first stage, after an initial breakdown which enables transmission of the discharge through the gas, the increasing current results in a steady increase of the ionization level (and therefore plasma density) and temperature. During this initial phase, the radial distribution of these parameters remains very homogeneous. This occurs up to the point when the plasma becomes fully ionized (~55 ns for the parameters shown in Fig. 27). Over the next ~30 ns, ohmic heating in the plasma combined with conduction to cool capillary walls results in a new, radially convex temperature distribution. Since the pressure is approximately uniform across the capillary, this also results in a radially concave redistribution of the plasma density forming a guiding structure. Once this redistribution is complete, the plasma reaches a quasi-steady-state equilibrium for the duration of the discharge, forming a long-lived, high-contrast plasma waveguide. The degree to which equilibrium is preserved depends largely on longitudinal plasma flow out of the entrance and exit of the capillary and the structure persists after the discharge has ended until sufficient heat is conducted through the walls for plasma recombination to begin (Bobrova *et al.* 2002).

In conjunction with their presentation of MHD simulations, Bobrova *et al.* developed a simplified quasistatic model (QSM) of the plasma evolution and guiding parameters by considering the balance of ohmic heating in the plasma with conduction to the walls of the alumina capillary. A key finding of this description is that the channels produced by the discharge are parabolic and largely insensitive to the amplitude of the discharge current, with the on-axis density $N_{e0}$ determined only by initial atomic gas density $\mathrm{N}_0$ and ionization state Z in the capillary,

$$N_{e0} \approx 0.7364\, ZN_0\ . \tag{39}$$

For the approximately parabolic radial density profiles predicted by the QSM, and sufficient density contrast to confine the modes of interest, capillary discharge waveguides are well-described by the model of parabolic channels discussed in Sec. II.B.2. Since both the modal structure of parabolic plasma waveguides and the QSM are well-defined, it is straightforward to determine the matched spot size, $w_{ch}$ from only the capillary radius, $R_c$, and the initial capillary density, $N_0$ (Bobrova *et al.* 2002):

$$w_{ch}[\mu m] \approx 1.48 \times 10^5 \frac{\sqrt{R_c[\mu m]}}{(ZN_0[cm^{-3}])^{1/4}} \quad . \tag{40}$$

This unique determination of optical properties in highly parabolic channels enables high-accuracy modeling of guided mode spot size evolution (mode beating) (E Esarey and Leemans 1999; E. Esarey *et al.* 2000) and allows for study of channel properties from analysis of transmitted modes alone (Sec. V.A.2, (Gonsalves *et al.* 2010)).

Analysis in (Bobrova *et al.* 2002) suggested formation of a ~0.8 μm plasma layer at the capillary boundary and heating up to 200 C. This is at odds with a nonlocal thermal equilibrium model (n-LTE) derived in (Broks, Garloff, and Van Der Mullen 2005) which predicts a wall temperature >1000 C (still too low for catastrophic capillary damage to occur). Broks *et al.* model the plasma as a quasi-neutral fluid while the wall is modeled as an insulator. As such, electrons do not directly transfer heat to the wall, but rather through heavy particles in a plasma sheath at the wall-plasma boundary. The wall is also assumed to include absorbed hydrogen which can be released during heating and by ion-wall collisions. As shown in Fig. 28, the n-LTE model also predicts three characteristic phases of plasma evolution, which are similar to those in (Bobrova *et al.* 2002), though the non-equilibrium evolution during phase II slightly differs.

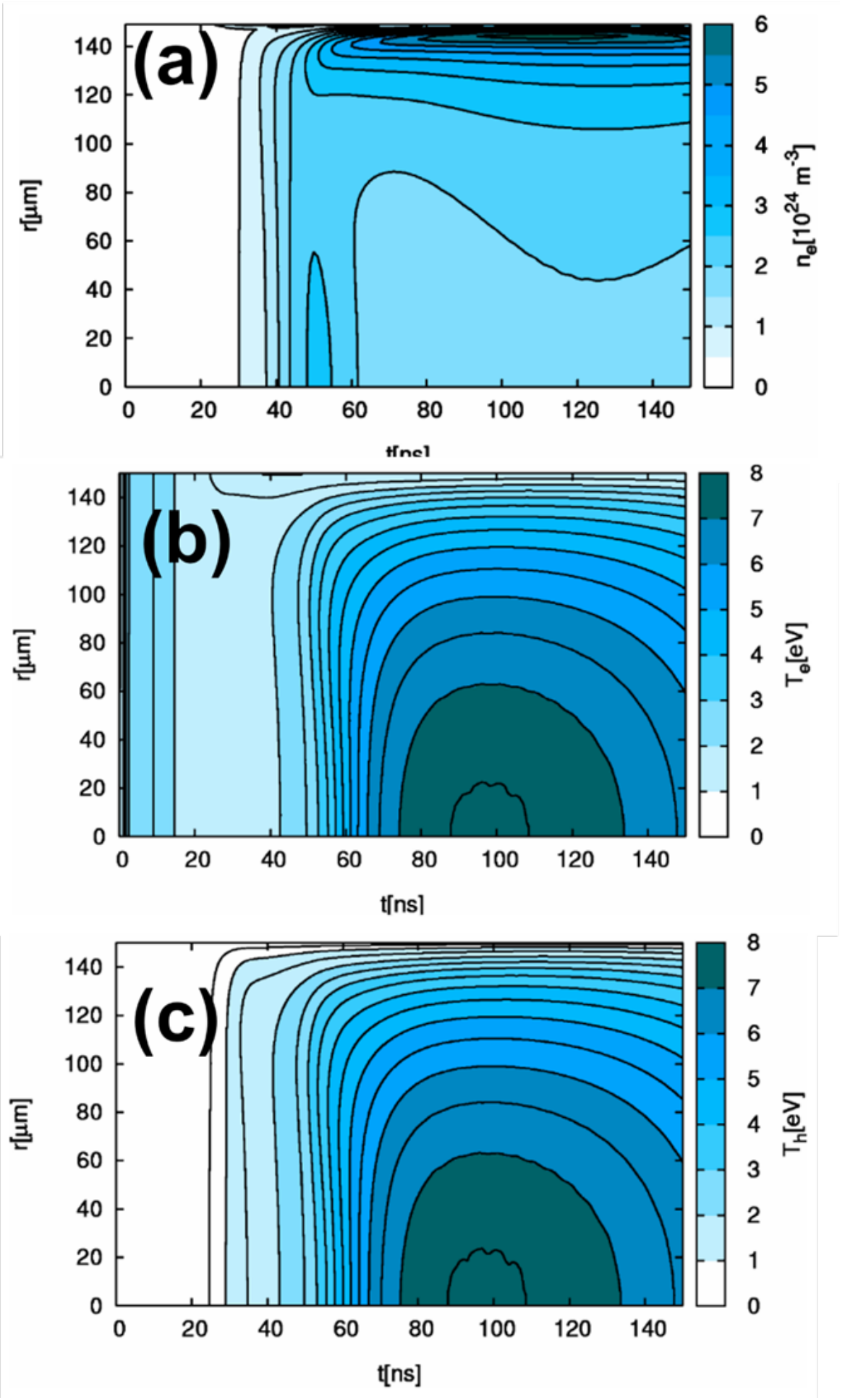

**Figure 28.** Nonlocal thermal equilibrium model of hydrogen-filled capillary discharge waveguide formation (adapted from (Broks, Garloff, and Van Der Mullen 2005)). **(a)** Evolution of plasma density. **(b)** Evolution of electron temperature. **(c)** Evolution of heavy particle temperature.

A parametric study of the model derived empirical formulas for several of the guiding properties, including the fundamental mode size (Broks, Van Dijk, and Van Der Mullen 2006):

$$w_{ch}[m] = 7.3 \times 10^3 \left(N_{0,H_2}[m^{-3}]\right)^{-0.25} (R_c[m])^{0.5625}, \tag{41}$$

where $N_{0,H_2}$ is the initial molecular $H_2$ density. This differs slightly from the finding in Eq. (40), which in the same practical units and assuming full ionization of $H_2$ is (Broks, Van Dijk, and Van Der Mullen 2006)

$$w_{ch}[m] = 4.7 \times 10^3 \left(N_{0,H_2}[m^{-3}]\right)^{-0.25} (R_c[m])^{0.5} \tag{42}$$

Though both models predict parabolic channels with similar $w_{ch}$, they predict different on-axis plasma densities. Transverse interferometry (see Sec. V.A.1) of discharge channels in a square capillary (Broks *et al.* 2007; Gonsalves *et al.* 2007) suggested that the n-LTE model was a better fit to the experiments performed in (Spence, Butler, and Hooker 2001; Butler, Spence, and Hooker 2002). But given the difficulty of directly measuring axial density in cylindrical capillaries (Gonsalves *et al.* 2010), a similar comparison has not been performed in that geometry. Crucially, the models presented in both (Bobrova *et al.* 2002) and (Broks, Garloff, and Van Der Mullen 2005) find dependence on the density in the fundamental mode size of the channel. Thus, for a fixed capillary radius, the optical properties of the waveguide are not independently tunable from the on-axis plasma density. Moreover, the capillary radius cannot be made arbitrarily small, posing difficulties for the implementation of capillary discharge waveguides with low on-axis plasma densities (Sec. IV.B.4).

The ability of discharge capillary-generated channels to guide high intensity pulses in plasmas with densities $\sim 10^{18}\ \text{cm}^{-3}$ opened a new regime of guided LWFA experiments. Capillary discharges were employed in the first demonstration of GeV LWFA in a pre-formed plasma channel (Leemans *et al.* 2006), where $w_0 \sim 25\ \mu\text{m}$ ~20-30 TW pulses accelerated electrons up to ~1 GeV in 3.3 cm capillaries. Examples of the transmission for this waveguide with a lower power (~5 TW) pulse are shown in Figure 29(a). Low-power transmission measurements more directly characterize channel guiding losses (Sec. V.A.2), since at higher intensities the transmitted energy is also reduced by energy transfer to the plasma wake (Shadwick, Schroeder, and Esarey 2009; Shiraishi *et al.* 2013). The drive pulses were not optimally matched to the channels produced in the 190 and 310 μm diameter capillaries and relied on self-injection to couple electrons into the wake. For a 40 TW drive pulse, the guide's input and transmitted modes are shown in Fig. 29(b)(c).

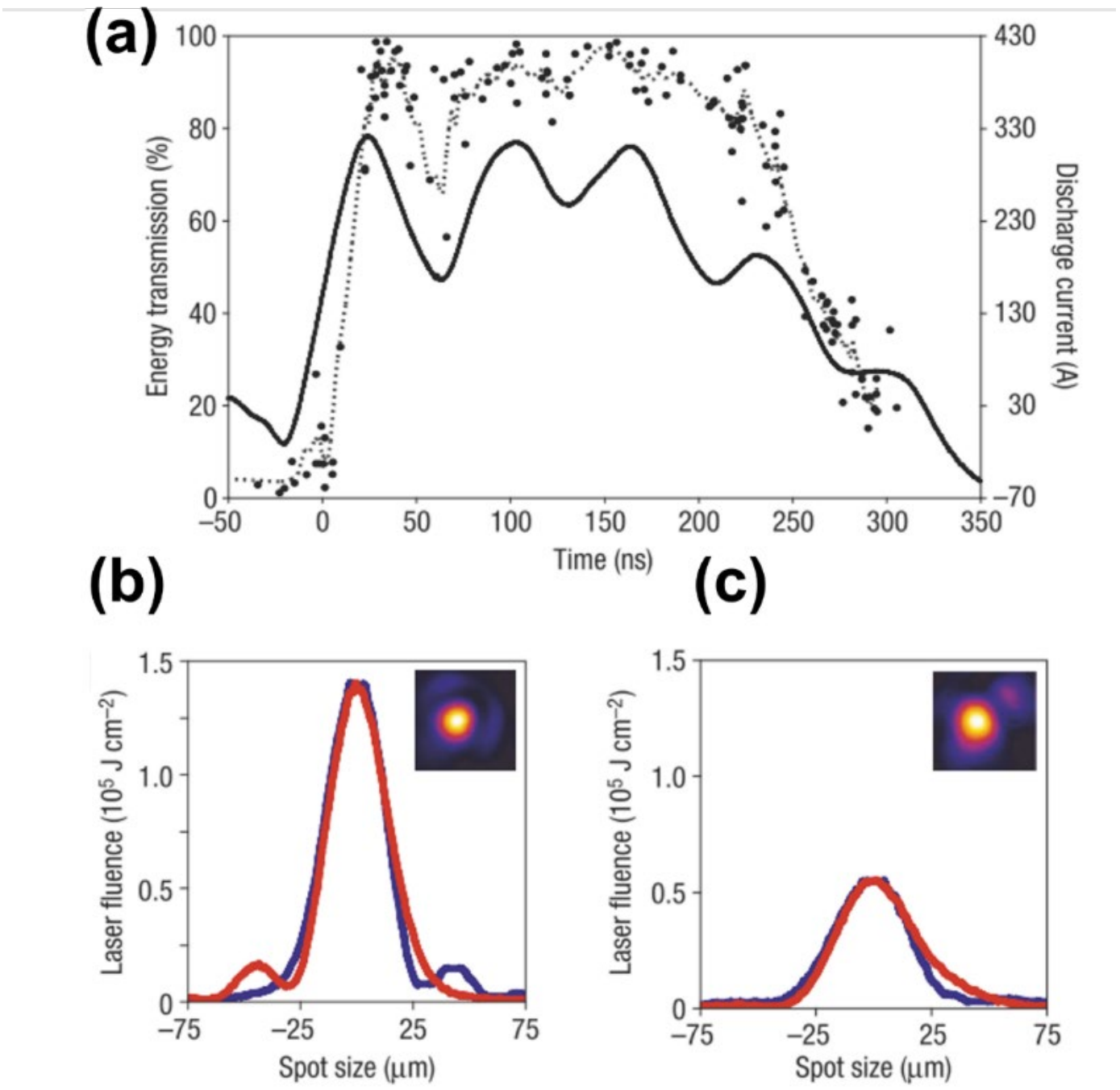

**Figure 29.** High power guiding in 3.3-cm-long gas-filled capillary discharge waveguide (adapted from (Leemans *et al.* 2006)). **(a)** Points**:** Measured transmission vs delay from discharge onset. Solid curve: discharge current. For a 40 TW input pulse: **(b)** Guide input mode **(c)** transmitted mode. The blue (red) curves are horizontal (vertical) lineouts.

GeV-level electron beams were also produced with guided 20 TW pulses (Karsch *et al.* 2007). Other experiments explored the production of 200 MeV beams in a partially ionized channel (Rowlands-Rees *et al.* 2008), and the effects of bubble regime self-focusing (W. Lu *et al.* 2007) on mismatched drive pulse coupling in the production of ~500 MeV electron bunches (Ibbotson *et al.* 2010). A hybrid capillary with a ~1 mm high density region at the entrance was used to generate few-hundred MeV electron bunches with narrow <5% energy spreads due to localized electron injection from wake-velocity manipulation through a combination of density transition (Bulanov *et al.* 1998; Suk *et al.* 2001; Geddes *et al.* 2008; Schmid *et al.* 2010; Hue *et al.* 2023) and self-focusing in the high-density portion of gas (Gonsalves *et al.* 2011). In a gas-filled discharge capillary, longitudinal density tailoring through gas feed lines of different cross sections was demonstrated together with laser transmission above 90% (M. S. Kim *et al.* 2013).

Acceleration of electrons above 4 GeV was achieved in (Leemans *et al.* 2014) where a 0.3 PW pulse with $a_0 = 1.6$, $w_0 = 52\ \mu m$ was guided over 9 cm in a 500 μm discharge capillary with on axis densities as low as $7 \times 10^{17}$ cm$^{-3}$ and mode sizes which varied from ~$60 - 70$ μm over the density range explored. Analysis found significant coupling into higher order modes due to mismatch of the non-Gaussian far field of the high-power focus. Because of the large spot size and low leakage rate inherent in waveguides generated through capillary discharge, beating persisted throughout propagation of the pulse without appreciable group velocity walkoff from the fundamental mode (see Sec. II.D). This beating was particularly pronounced for the larger $w_{ch}$ at lower densities (desirable for higher energy gain in LWFA) and could result in capillary damage due to the concentration of energy outside the nominal mode size of the waveguide (nearer to the capillary walls) (Gonsalves *et al.* 2015).

### 4. Discharge capillary with a heater laser

Laser wakefield acceleration to energies beyond the few GeV range with ≲ PW scale lasers requires plasma densities ~ $10^{17}$ cm$^{-3}$ with $a_0$ ~2 drive pulses over meter-scale acceleration lengths (B. Miao, Shrock, *et al.* 2022; Picksley *et al.* 2024; Rockafellow *et al.* 2025). Development of a TeV-scale laser-driven accelerator envisions the sequential staging of many individual modules with energy gains ~10 GeV in each module (Schroeder *et al.* 2010; Carl A. Lindstrøm 2021), while development of a single 100 GeV stage has been proposed operating at plasma densities ~$10^{16}$ cm$^{-3}$ (Ludwig *et al.* 2025). At current PW-level laser facilities (Y. Wang *et al.* 2017; Jourdain *et al.* 2021; Maksimchuk *et al.* 2025; Nakamura *et al.* 2017), sufficient intensities for LWFA are achievable with focusing geometries producing ≲ 100 μm spot sizes. The scaling in Eqs. (39-42) indicates that a capillary with radius $R_c \approx 70$ μm would be required to guide a 70 μm spot with on-axis density $N_{e0} = 10^{17}$ cm$^{-3}$, and even with large $f/\#$ focusing optics producing 100 μm spots, significant interaction between the high intensity laser pulse (as well as the discharge itself) and the capillary wall is unavoidable, resulting in high wall temperatures as well as potentially catastrophic damage (Bobrova *et al.* 2013) .

In (Bobrova *et al.* 2013), use of an auxiliary few-nanosecond 'heater' laser pulse was proposed as a mechanism for increasing the channel contrast and enabling guiding of smaller spot sizes at lower densities than achievable in a guide formed through capillary discharge alone. The heater pulse is matched so that it is guided in the capillary channel, and the IB heating of the plasma within the heater pulse mode results in hydrodynamic expansion of the plasma at the local sound speed. One-dimensional MHD simulations reproduced in Fig. 30 demonstrate the clear deepening of the channel which persists for much longer than the 1 ns pulse used in those simulations.

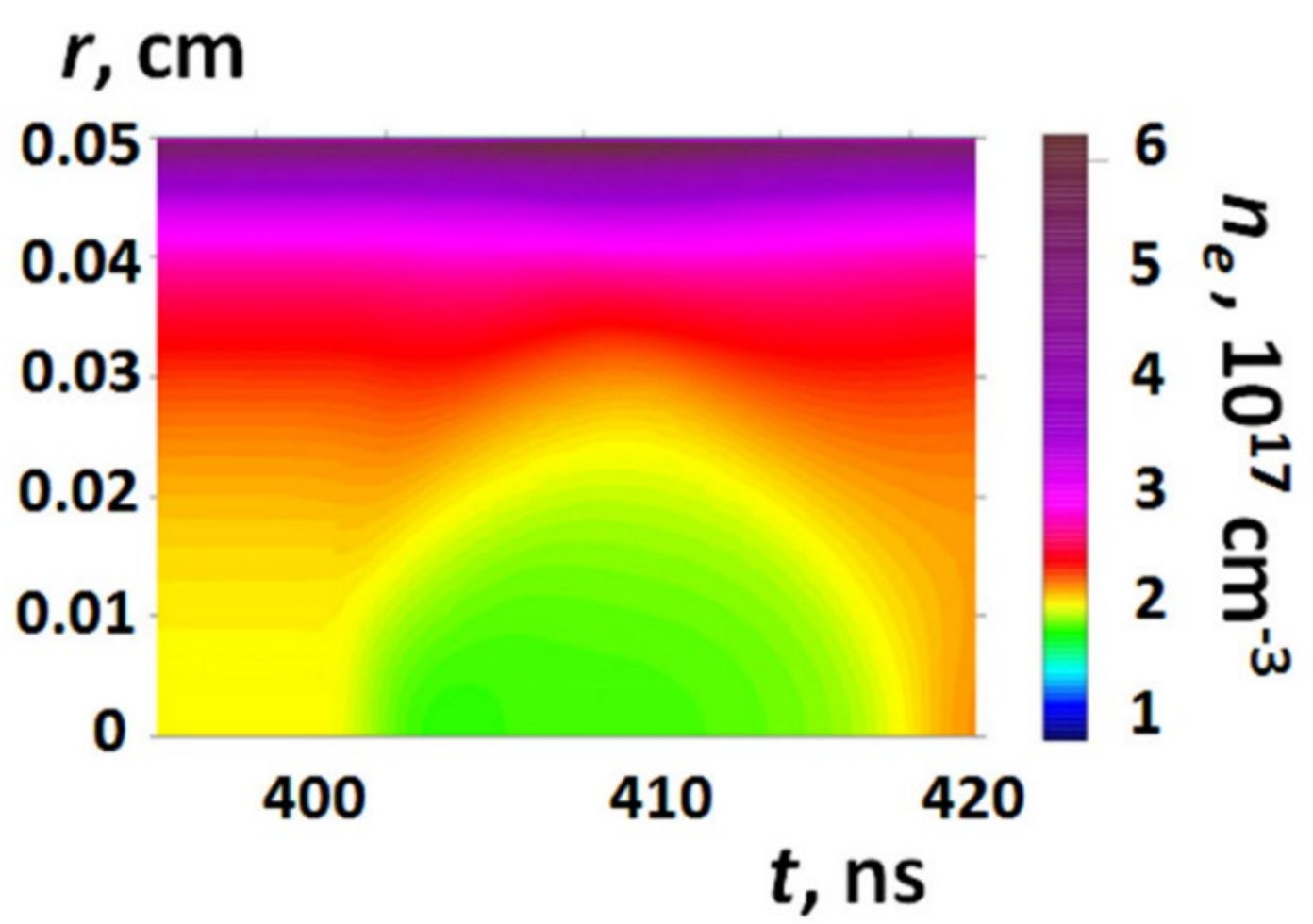

**Figure 30.** Simulated channel formation in a laser-heated discharge capillary (adapted from (Bobrova *et al.* 2013)).

With longer (> 1 ns) heater pulses, different temporal slices of the pulse experience different guiding structures due to the hydrodynamic evolution of the channel which occurs on a ns time scale, effectively generating a more confining guide for later portions of the pulse. This 'self-guiding' enabled generation of a low density ($\sim 2 \times 10^{17}$ cm$^{-3}$) channel with a mode size $w_{ch}$ ~60 $\mu m$ in 800 μm diameter hydrogen filled discharge capillaries (Gonsalves *et al.* 2019). Matched guiding of 31 J, 35 fs (0.85 PW), $a_0 = 2.2$ pulses over 20 cm in these channels resulted in acceleration to 7.8 GeV. The significant effect of the heater pulse on the channel size in those experiments is demonstrated in Fig. 31, where it can be seen that an appropriately timed, ~6 ns heater pulse was able to reduce the spot size by almost 1.5 × at the operating density.

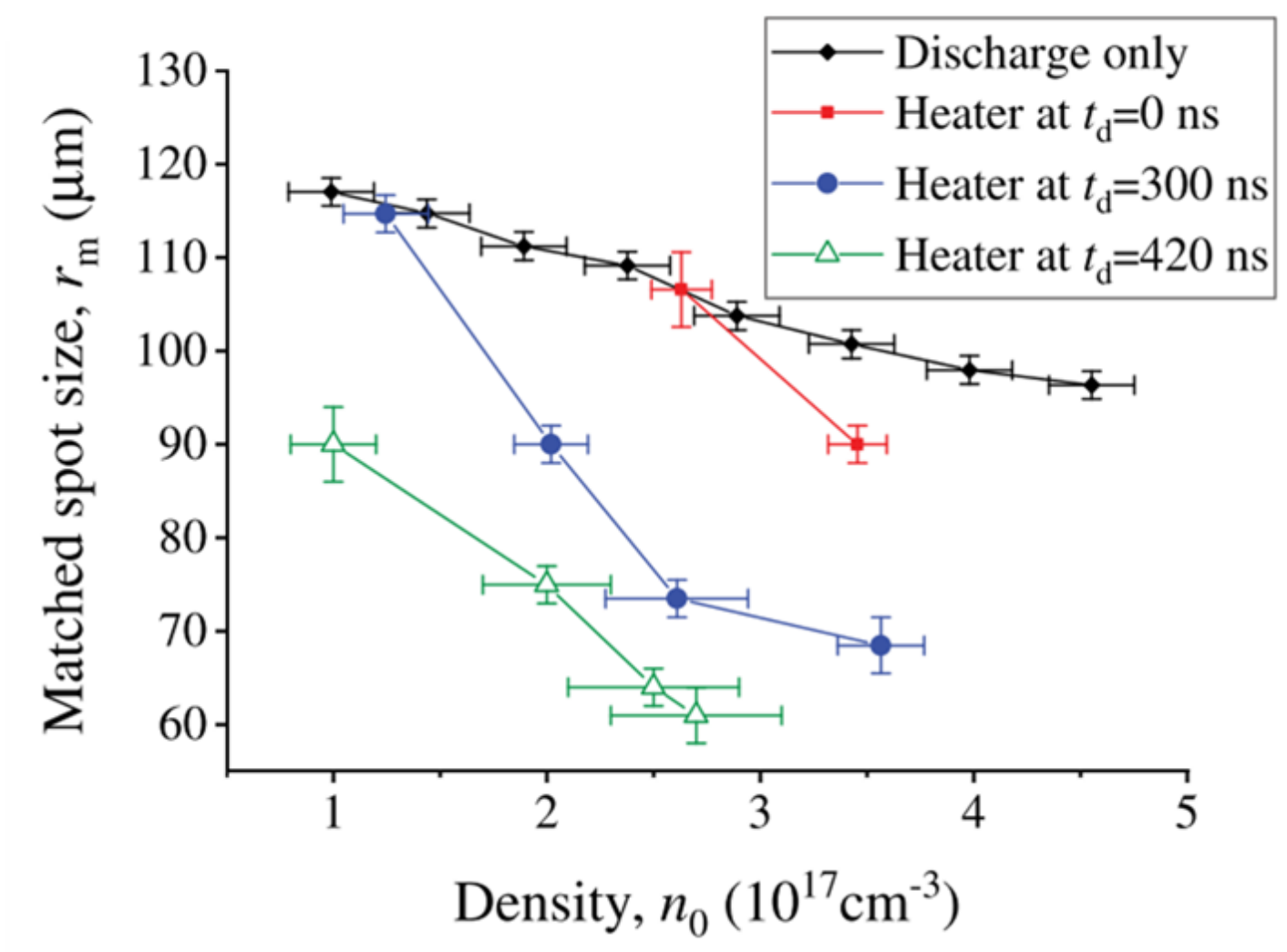

**Figure 31.** Matched spot size for various capillary-heater pulse conditions (adapted from (Gonsalves *et al.* 2019)).

Despite this success, discharge capillaries still face significant obstacles to achieving the lower densities desired for higher energy gain. Longitudinal density upramps can mitigate dephasing by increasing the wake phase velocity (P. Sprangle *et al.* 2001; Rittershofer *et al.* 2010). The tapered discharge capillary discussed above (M. S. Kim *et al.* 2013) showed longitudinal density control, although dephasing mitigation was not demonstrated. Maintaining the desired transverse guiding structure along such a ramp presents an additional challenge: the relationship between density and mode size in Eqs. (39–42) couples longitudinal density tailoring to changes in $w_{ch}$. Density tailoring in gas-fed capillaries operated without a discharge is discussed in Sec. V.C.2. The azimuthal magnetic field of a capillary discharge can also focus electron beams, enabling active plasma lenses (van Tilborg *et al.* 2015; C. A. Lindstrøm *et al.* 2018) and coupling between LWFA stages (Steinke *et al.* 2016).

## V. EXPERIMENTAL AND MODELING CONSIDERATIONS FOR PLASMA WAVEGUIDES

In this section we provide a brief overview of some of the considerations for implementing plasma waveguides in high-intensity laser experiments. This includes diagnostic methods, setups for the experimental realization of plasma waveguides, and modeling of OFI heating and hydrodynamic waveguide formation. We note that an extensive discussion of diagnostics for LWFA experiments is found in (Downer *et al.* 2018).

### A. Diagnostics

#### 1. Spatial interferometry and wavefront sensors

One of the most widely employed methods for characterizing laboratory plasmas is spatial interferometry (Muraoka and Maeda 2000). It is based on the change in phase $\Delta\phi$ of an electromagnetic beam propagating through a medium with space and frequency dependent refractive index shift $\Delta n(\mathbf{r},\omega)$. The phase shift induced in a probe beam along a straight path P along $y$ through the medium is $\Delta\phi(x,z) = k_0 \int_P \Delta n(\mathbf{r},\omega)dy$, where $k_0$ is the vacuum wavenumber of the probe beam and the phase shift is imprinted on the transverse phase front of the beam. Interference of the probe beam with a reference beam (with $\Delta\phi = 0$) produces an interference pattern from which the 2D phase $\Delta\phi(x,z)$ is extracted using various algorithms employing fast Fourier transforms, such as (Takeda, Ina, and Kobayashi 1982). The specific type of interferometer employed depends on the experimental constraints; these include Michelson, Mach-Zehnder, and folded wavefront (shearing) interferometers (Hutchinson 2002). In folded wavefront interferometry, a single probe beam passes through the index structure, with part of the beam picking up the 2D phase shift imprint from the structure and part missing the structure. The beam then enters the interferometer where it is split; the phase shifted portion is then interfered with the unshifted portion.

For the low density, low collisionality plasmas typical of plasma waveguides, $\Delta n(\mathbf{r},\omega) = n(\mathbf{r},\omega) - 1$, where $n(\mathbf{r},\omega) \cong 1 - N_e(\mathbf{r})/2N_{cr} + \Delta n_0(\mathbf{r},\omega)$ from Eq. (6), where $\Delta n_0(\mathbf{r},\omega) = 2\pi\chi(\mathbf{r},\omega)$ is the index contribution of any neutrals present. In general, each dispersively different species contributing to the composite refractive index requires a separate color probe beam for its contribution to be extracted. In $\Delta\phi(x,z)$ measured for plasma waveguides, $z$ is the waveguide axis and $x$ is the transverse coordinate; $y$ is the probe beam direction. In some cases, single-color measurements before and after a pulse can also constrain the initial neutral density if the post-pulse ionization state is known and hydrodynamic redistribution between the measurements is negligible (Picksley, Alejo, Shalloo, *et al.* 2020).

The first interferometric measurement of a plasma waveguide was performed using a 70 ps, λ=532 nm probe pulse and a folded wavefront interferometer (T. R. Clark and Milchberg 1997) to measure the evolution of hydrodynamic plasma waveguides collisionally heated by a 100ps, $\lambda = 1064$ nm $J_0$ Bessel beam. In that experiment, the neutral contribution to the phase shift was negligible compared to the plasma ($\Delta n_0 (N_e/N_{cr})^{-1} \ll 1$), leaving the need for only a single-color interferometric probe beam. The cylindrical symmetry of the Bessel-beam-generated hydrodynamic waveguide ($N_e(\mathbf{r}) = N_e(r)$) was then exploited to enable extraction of $N_e(r,z)$ from the chordally integrated phase shift, $\Delta\phi(x,z)$, using the Abel transform (Montgomery Smith, Keefer, and Sudharsanan 1988; Brill *et al.* 1990; T. R. Clark and Milchberg 1997; Dribinski *et al.* 2002; Hutchinson 2002): $N_e(r,z) = (2N_{cr}/k\pi)\int_r^{r_b} dx\,(x^2 - r^2)^{-1/2}\,(d\Delta\phi(x,z)/dx)$, where $r_b$ is the outer radial boundary of the index structure. In similar hydrodynamic waveguide interferometry experiments since then, Abel inversion has been used to extract radially resolved waveguide index profiles.

As seen in Fig. 32(a)(c′), the phase shift imposed on the transmitted probe beam appears as bending of the interference fringes. In panel (a), the probe is propagated transverse to the waveguide (T. R. Clark and Milchberg 1997), and the map of relative phase shift across the region of interest is extracted in panel (b) by Fourier analysis (Takeda, Ina, and Kobayashi 1982). In the interferogram of panel (c′), corresponding to the fast discharge capillary experiment of (Luther *et al.* 2004), the probe is propagated along the capillary axis. The corresponding plasma density profiles in panel (c) are determined directly from the 2D phase shift map extracted from the interferogram. The use of high repetition rate laser systems for plasma generation and interferometric probing enables averaging over many shots to cancel shot and camera readout noise (Y. H. Chen *et al.* 2007) to achieve measurement of small phase shifts down to the few milliradian level. For spatial interferometry of plasma waveguides, this has enabled measurement of plasma densities $< 10^{17}$ cm$^{-3}$. In practice, one can obtain $\overline{\Delta\phi}(x,z)$ by averaging over thousands of shots, and then extract the refractive index profile via Abel inversion. For axially uniform waveguides, one can average even further by computing the axial average $\langle \overline{\Delta\phi}(x,z) \rangle_z$ before Abel inversion. For non-uniform channels, 2D profiles may be extracted from interferograms by the use of forward fitting LG functions (R. J. Shalloo *et al.* 2019), or by inclusion of azimuthal terms (guided by expected asymmetries) in the Abel inversion (B. Miao *et al.* 2020).

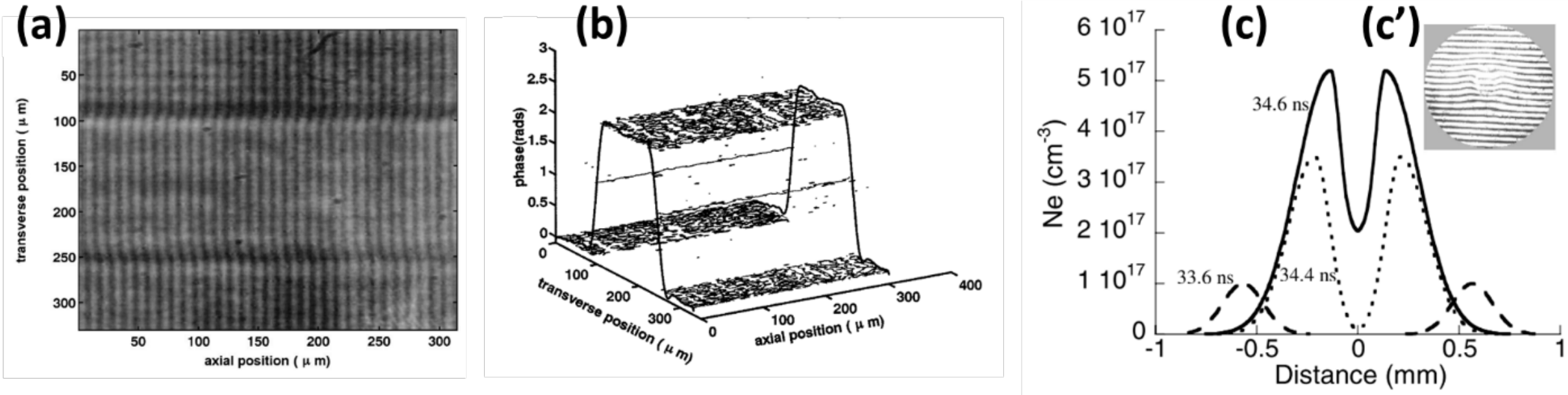

**Figure 32.** Examples of phase extraction from interferometry. **(a)** Raw interferogram from transverse probing of a hydrodynamically generated channel and **(b)** extracted phase map. **(c)** Inferred density from a longitudinal probe and **(c′)** raw interferogram. (Panels (a)(b) adapted from (T. R. Clark and Milchberg 1997), panels (c)(c′) adapted from (Luther *et al.* 2004)).

In later work, spatial interferometry of plasma waveguides was extended to 100 fs duration probe beams, specifically to resolve the fast time evolution of hollow plasma channels generated by higher order $J_5$ beams (Fan *et al.* 2000). In addition to the needed spatial overlap of probe and reference pulses in spatial interferometry, femtosecond probe and reference pulses must accurately overlap in time to maximize interference fringe visibility; this overlap is much less sensitive for the long pulses used in (T. R. Clark and Milchberg 1997).

Two-color interferometry for separating free-electron and neutral-gas contributions to plasma refractivity dates to the early development of optical plasma diagnostics. (McLean and Ramsden 1965) demonstrated two-wavelength separation in shock-heated hydrogen, showing that neglecting neutral gas caused single-wavelength measurements to underestimate the electron density. Early laser implementations followed: (Alcock and Ramsden 1966) probed a laser-produced air spark using a ruby laser's fundamental and second harmonic, while (Gusinow, Gerardo, and Verdeyen 1966) used two He–Ne laser interferometers to measure electron and neutral-atom densities in helium and argon afterglows. Later examples include (Weber and Fulghum 1997), who used a two-color continuous wave probe beam to measure the plasma and neutral contributions in pulsed power plasmas, and (Gonsalves *et al.* 2007), who used ~10 ns, two-color pulses to probe capillary discharge waveguides. The latter experiments required transparent capillaries with rectangular cross sections and optically smooth outer and inner walls to minimize distortion of the transverse probe beam (Jones *et al.* 2003; Gonsalves *et al.* 2007; Broks *et al.* 2007). While the plane surfaces minimized probe beam distortion, the non-circular geometry was not ideal for plasma waveguide generation, and the inner walls were roughened by the discharges. (Point *et al.* 2014) also used ~10 ns, two-color interferometry to probe femtosecond filament-induced HV electric discharges.

As discussed in Sec. III, two-color spatial interferometry with femtosecond probe pulses directly extracts the separate electron and neutral density profiles in fast-evolving OFI plasmas (B. Miao *et al.* 2020; L. Feder *et al.* 2020; B. Miao *et al.* 2024). In one version of the scheme (B. Miao *et al.* 2024), the two-color probe beam (λ=400 nm and λ=800 nm) propagated collinearly through the index structure, and then each color was directed by a beamsplitter into its own folded wavefront interferometer. In these low-density waveguide experiments, generating $\overline{\Delta\phi_{400nm}}(x,z)$ and $\overline{\Delta\phi_{800nm}}(x,z)$ by averaging over thousands of shots is essential for extracting the low-density plasma and neutral contributions.

For plasma or gas profiles that are very short or low density, longitudinal probing (along z) enables direct extraction of the transverse index variation $\Delta n(\mathbf{r}_\perp,\omega) = \Delta n(x,y,\omega)$, without the need for Abel inversion or its assumption of cylindrical symmetry (Luther et al. 2004; Wahlstrand et al. 2014; R. J. Shalloo et al. 2018; R. J. Shalloo et al. 2019). However, to avoid excessive probe wavefront distortion, either the sample length $L$ or the maximum transverse gradient of the refractive index ($|\nabla_t n|_{max}$) needs to be small enough so that $L < (2n\delta_{min}/|\nabla_t n|_{max})^{1/2}$ , where $\delta_{min}$ is the transverse size of the minimum desirable and resolvable feature. For example, for the OFI guide measured (with transverse interferometry) in Fig. 18(a), taking $|\nabla_t n|_{max} \sim \Delta r_{shock}^{-1} |\Delta N_e/2N_{cr}| \sim 0.6\ \mathrm{cm}^{-1}$ and $\delta_{min} \sim \Delta r_{shock} \sim 10\ \mu\mathrm{m}$, where $\Delta r_{shock}$ is the shock width, indicates that longitudinal interferometry is appropriate only for $L_{guide} < \sim 1\ mm$. While this is much shorter than practical low density OFI guides, it remains useful for proof-of-principle experiments on very short sections.

For conventional cylindrical or optically inaccessible discharge capillaries, the difficulty of transverse probe access generally favors longitudinal interferometry, where extracted central plasma density is longitudinally averaged over the probe path. A technique using longitudinal interferometry in the spectral domain was reported in (Daniels *et al.* 2015), where the probe passed along the capillary plasma axis and the reference passed outside it. This approach relies on the known frequency dependence of the laser group velocity to extract plasma density from the spectral interferogram. Later work (van Tilborg *et al.* 2018) built on this technique by copropagating two pulses of different central wavenumbers ($k_0$ and $2k_0$) through the waveguide, doubling the $k_0$ pulse, and then interfering the two $2k_0$ pulses to extract the axial average on-axis plasma density. Another approach for measuring the longitudinal density variation used a translatable mask to select axial regions to probe via longitudinal interferometry (R. Li *et al.* 2025).

Wavefront sensors measure probe-beam phase profiles without interferometer optics, at the expense of spatial resolution or field of view (Baker *et al.* 2002; Plateau *et al.* 2010). A Shack–Hartmann sensor reconstructs the phase from local wavefront slopes, measured through focal-spot displacements behind a microlens array. Its spatial resolution is limited by the microlens size and spacing, typically at least ten times the $\sim 10$ μm camera pixel spacing available for interferometry. Interferometric resolution is instead limited by the imaging aperture and fringe spacing (Plateau *et al.* 2010). Magnifying the probed region improves wavefront-sensor resolution but reduces the field of view. Both techniques can yield ambiguous results when phase shifts vary by $\Delta\phi \gtrsim 2\pi$ across small transverse regions of the probe beam.

## 2. Guided beam imaging and analysis

The most straightforward tool for assessing guiding and transverse mode structure is end mode imaging: direct imaging of the exit of the waveguide where the guided beam emerges, first employed in (Durfee and Milchberg 1993). For low-power guiding of only the lowest order (0,0) mode, this enables direct measurement of the fundamental mode size $w_{ch}$. For multimode guiding, the imaged intensity profile at the exit may be complicated and dependent on the phase of beating between transverse modes of different amplitudes. These are often unknown, but it was demonstrated in (Gonsalves *et al.* 2010) that measurement of the shifts in the position of the exit mode centroid, due to systematically varied coupling offsets of a lower power guided pulse, could be employed to deduce $w_{ch}$. Key to this technique is knowledge of the mode structure of the waveguide.

The transmission efficiency is estimated from the ratio of the spatially integrated exit and entrance fluence images. The injected mode and the exit mode can be equivalent-plane imaged or relay imaged after reflection from calibrated attenuators moved into the beam. Some configurations enable direct imaging of high intensity focal spots using sufficient beam attenuation, with the measurements becoming more difficult at ultra-high laser intensity (Nakamura *et al.* 2017). For nonrelativistic pulses $a_0 << 1$, the efficiency ratio accounts for coupling losses at the entrance, and leakage and erosion losses during propagation in the guide. If the transverse mode structure of the guide is known, then images of the injected pulse focus at the guide entrance can be used to estimate coupling efficiencies into the various transverse modes, as discussed in Sec. II.C. However, at the guide exit, except for the case of very few modes, it is difficult to determine from imaging alone which guided modes have survived. If guiding is clearly in the (0,0) mode, then transmission measurements for different waveguide propagation distances can be fit to an exponential decay curve to find $L_{1/e}$ (Picksley, Alejo, Cowley, *et al.* 2020). For waveguides generated in long gas jets, the guide length can be adjusted by limiting the length of the gas sheet, and the exit mode imaged as a function of length (Picksley *et al.* 2024). In (Gonsalves *et al.* 2020), the output mode of a low-power probe was measured as a function of delay after auxiliary laser heating of a discharge capillary, resolving a reduced guided spot size and enhanced fluence.

For high intensity pulses ($a_0 \sim 1$ and higher), significant laser energy can be deposited into plasma wake excitation, and simple transmission measurements no longer reflect guiding losses alone. The leading portion of the pulse propagates in its self-generated plasma density depression and experiences red-shifting, $\Delta\omega \sim (k_0 v_g z/2N_{cr})\, \partial N_{e0}/\partial\xi < 0$, where $\xi = v_g t - z$ is a local coordinate that increases towards the back of the pulse. Guided pulse spectral evolution and laser-wake energy transfer dynamics in the case of incomplete modal walkoff were simulated in (Shiraishi *et al.* 2013); these enabled estimates of the strength of wakefields excited in capillary discharge waveguides. Red-shifted exit spectra were also measured in optically generated plasma waveguides (B. Miao, Shrock, *et al.* 2022; Picksley *et al.* 2024). Blue-shifted components in the transmitted spectrum often result from ionization (Wood, Siders, and Downer 1991).

Recent analysis of relativistic pulse propagation in LWFA shows that the complex evolution of $a_0 > 1$ pulses propagating in meter-scale plasma waveguides creates new demands for broadband or spectrally resolved exit-mode diagnostics (Shrock *et al.* 2024). For pulses of a few tens of femtoseconds, such as the 40-fs example in Sec. II.D.1, and channels with $w_{ch} \lesssim 50\ \mu m$ and propagation lengths $L_{guide} \gtrsim 20\ cm$, group velocity walkoff of higher order modes causes them to increasingly lag and temporally separate behind the (0,0) mode, reducing their temporal overlap and suppressing linear mode beating. The required distance depends on pulse duration and the mode pair; nonlinear stage II beating can persist, as discussed in Sec. II.E and Fig. 7. Nevertheless, an end mode image will capture and integrate all modes that survive without appreciable leakage to the end of the guide, obscuring the characteristic features of multimode propagation. At high intensity, the different excited modes experience different spectral evolution, with the fundamental mode, as the primary driver of the plasma wake, being preferentially red-shifted. For the long propagation lengths employed in (B. Miao, Shrock, *et al.* 2022; Picksley *et al.* 2024), this red-shifting is so significant that most of the energy coupled into the fundamental mode at $\lambda = 0.8\ \mu m$ shifts to wavelengths well beyond $\lambda = 1\ \mu m$, where the sensitivity of standard CCD/CMOS sensor chips is minimal, effectively rendering the fundamental exit mode undetectable by visible or near-IR imaging diagnostics. As a result, the less red-shifted higher order mode structure may appear more pronounced in exit mode images. For example, if a particular higher order mode is preferentially excited due to pointing fluctuations (see Sec. II.C), then it could dominate the exit mode image. This effect has been described in simulations in (Shrock 2023), where a straightforward experiment is proposed to measure this mode-dependent spectral evolution. Thus, while conventional end-mode imaging remains useful, quantitative modal-energy partition in this regime may require broadband or spectrally resolved collection.

Another technique for imaging high-intensity propagation in plasma waveguides was presented in the first papers demonstrating them (Durfee and Milchberg 1993; Durfee, Lynch, and Milchberg 1994; Durfee, Lynch, and Milchberg 1995). Here, Thomson side-scattering was used to image the guided light as it propagated down the waveguide, with additional scattering bursts observed as the injected laser pulse entered and exited the waveguide. Imaging of the scattering was done through narrowband interference filters to eliminate the contribution of plasma fluorescence to the images. For leaky modes, the reduction in side scattering with propagation distance enabled the direct measurement of $L_{1/e}$.

### 3. Measurement of longitudinal gas density profiles in meter-scale gas jets

The implementation of meter-scale plasma waveguides requires the characterization of meter-scale gas jets (B. Miao *et al.* 2025) to determine the longitudinal variation of gas density. A fluorescence technique first presented in (B. Miao, Shrock, *et al.* 2022; Shrock *et al.* 2022) characterized meter-scale target gas profiles in (B. Miao *et al.* 2024; Picksley *et al.* 2024; Rockafellow *et al.* 2025; Šišma *et al.* 2026). The technique (illustrated in Fig. 33) relies on the recombination fluorescence of hydrogen plasmas generated by $J_0$ Bessel beams.

First, a longitudinally resolved map of fluorescence intensity is obtained by imaging the long, $J_0$-beam-induced plasma generated in the target chamber backfilled to a set of known pressures (with the jet in place). The fluorescence signal at a given axial ($z$) location is a function of the local neutral gas density and Bessel beam intensity. Thus, an image of $J_0$-induced fluorescence from the jet's pulsed gas sheet (with the chamber evacuated) can be interpolated to determine the longitudinally-resolved neutral gas density profile above the jet. Sample fluorescence images from the pulsed gas sheet and from a uniform backfill are shown in Fig. 33(a)(b). Panel (c) plots the transversely integrated fluorescence signal for varying backfill pressure and for the pulsed gas sheet (black line). The $z$-dependent pulsed gas density profile is then extracted, with sample profiles plotted in Sec. V.C.1 (Fig. 39(a)).

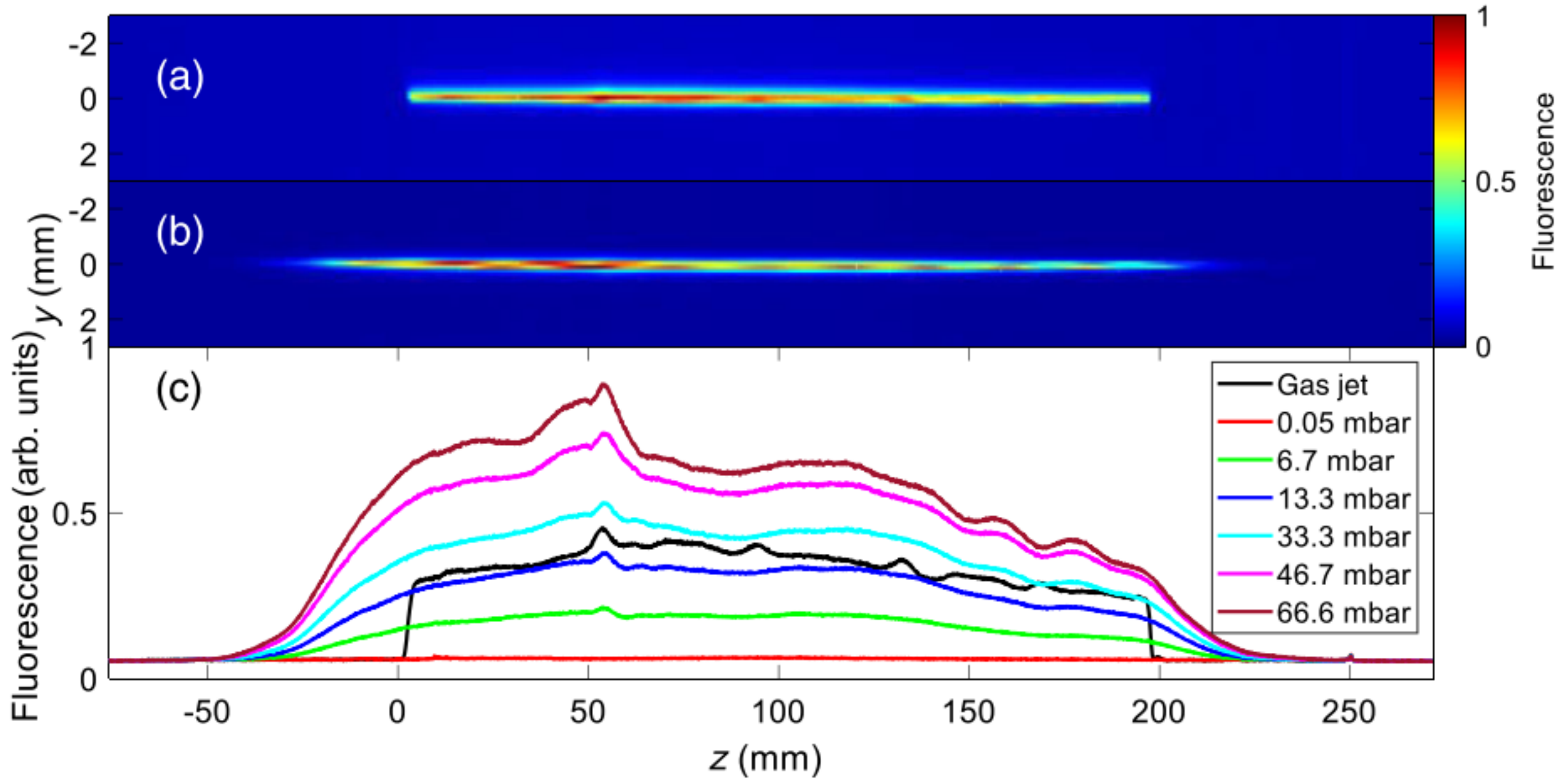

**Figure 33.** Fluorescence technique for longitudinal characterization of meter-scale gas jets (adapted from (B. Miao, Shrock, *et al.* 2022)). **(a)** Example of hydrogen plasma fluorescence signal in the pulsed gas sheet above the jet. **(b)** Example of hydrogen plasma fluorescence in backfill gas above the jet. **(c)** Radially integrated backfill fluorescence signals and gas jet fluorescence signal.

Importantly, the accuracy of this method is independent of longitudinal variations in the Bessel beam intensity profile. This technique has been applied to hydrogen and helium plasmas (Shrock *et al.* 2022; Šišma *et al.* 2026) and can be employed for other working gases with suitable emission lines accessible by an imaging system. While Bessel beam refraction by the plasma limits the technique to densities below the effective critical density $N_{e0} < N_{cr}\, sin^2\, \gamma$ (B. Miao *et al.* 2024), this is still a wide parameter space. For example, for a Bessel beam of radius $R_b = 2.5$ cm and desired waveguide length $L_{guide} = 1\ m$ (B. Miao *et al.* 2025), one needs $\gamma \sim 1.5°$ (see Sec. V.B.1 below), for which the initial plasma density must be $N_{e0} < 1.2 \times 10^{18}$ cm$^{-3}$, corresponding to fully ionized $H_2$ gas at density $N_g < 10^{18}$ cm$^{-3}$. Given the $\sim 10\times$ reduction of plasma density by subsequent hydrodynamic expansion (L. Feder *et al.* 2020), this initial gas density is appropriate for accessing the desired range of $N_{e0} \sim 10^{17}$ cm$^{-3}$ for multi-GeV LWFA.

## B. Bessel and Bessel-like beams

Bessel beams (Durnin 1987; Durnin, Miceli, and Eberly 1987; Vasara, Turunen, and Friberg 1989) and Bessel-like beams (Davidson, Friesem, and Hasman 1991; Sochacki, Kołodziejczyk, *et al.* 1992; Popov *et al.* 1998; Dharmavarapu, Bhattacharya, and Juodkazis 2018; Smartsev *et al.* 2019; Tripathi *et al.* 2025) have been commonly employed in the formation of hydrodynamic plasma waveguides since their first demonstration in (Durfee and Milchberg 1993). The Bessel beam is a class of solution to the Helmholtz wave equation (see Sec. II.B, Eq. (9)). For the case of a uniform background medium of index $n_0$, one solution is the $q^{th}$ order Bessel beam of Eq. (12). Note that this beam is not physically realizable, as it requires an infinite aperture and infinite energy. A realizable, finite energy beam, such as a Bessel-Gauss beam passed through a finite aperture, is typically also called a "Bessel beam". This is because the near-axis transverse intensity profile is practically indistinguishable from that of an infinite aperture Bessel beam.

### 1. Generation of Bessel beams

In the experiments discussed in this review, Bessel beams were typically generated with optical elements called axicons, introduced in (McLeod 1954), in which they were discussed in terms of ray optics. In a wave optics description, the most common axicon geometry (the "conical axicon") applies a linear radial phase shift $\Phi(\rho) = -k_\perp \rho = -k\rho \sin\gamma$ to the flat phase fronts of an incident beam, where $\rho$ is the radial coordinate at the axicon input surface. This generates an output optical beam with a conical phase front whose rays approach the optical axis at angle $\gamma$, with perpendicular wavenumber magnitude and parallel wavenumber $k_\perp = k \sin\gamma$ and $\beta = k_z = k\cos\gamma$. The self-interference of the conical wave gives rise to the Bessel function field dependence of Eq. (12), as illustrated in Fig. 34(a). Conical axicons applied to the generation of plasma waveguides have included transmissive glass cones, which apply $\Phi(\rho)$ via refraction (Durfee and Milchberg 1993; Gaul et al. 2000; R. J. Shalloo et al. 2019; Morozov et al. 2018), conical mirrors applying $\Phi(\rho)$ via reflection (B. Miao *et al.* 2020; Šišma *et al.* 2026), or "ring gratings", which apply $\Phi(\rho)$ via diffraction (B. Miao, Shrock, *et al.* 2022; Shrock 2023). The latter is the element illustrated in Fig. 34(a). To generate Bessel beams with $q = |m| \geq 1$, a spiral phase plate with azimuthal phase $\Phi(\varphi) = m\varphi$ $(0 \leq \varphi \leq 2\pi)$ is centered on the beam axis in advance of the axicon (Fan *et al.* 2000; S. Gessner *et al.* 2016; S. J. Gessner 2016; B. Miao 2020; B. Miao *et al.* 2020), as shown in Fig. 34(a).

If an $n^{th}$ order super-Gaussian field, $A(\rho) = A_0 e^{-(\rho/R_b)^n} e^{im\varphi}$, is incident on a conical axicon (after a spiral phase plate if $|m| \geq 1$ is desired), the near-axis field

$$E(r, \varphi, z) = E_0 (z \tan\gamma / R_b)^{1/2} e^{-(z \tan\gamma/R_b)^n} e^{i\beta z} J_{|m|}(k_\perp r) e^{im\varphi} \quad (43)$$

is generated over its depth of focus

$$L_f \approx \frac{R_b}{\tan\gamma} , \quad (44)$$

the approximate axial distance over which the transverse form of the field is Bessel-like and invariant. In Eqs. (43) and (44), $r$ is the radial coordinate in the focal volume, $z$ is distance on the optical axis from the axicon center, and $L_f$ is determined by the mapping of the thin annulus at $\rho = R_b$ on the axicon input face to the beam axis via $z = \rho/tan\,\gamma$. Note that for Eq. (43) as written, the total power at fixed $z$ is proportional to $\int_0^R |E(r, \varphi, z)|^2 2\pi r dr$, which diverges as $R \to \infty$. In practice, either the effective beam radius or the axicon imposes a finite aperture. Assuming that the input beam radius is much smaller than the axicon radius, and defining an effective aperture radius $\rho_{max}$ by $e^{-(\rho_{max}/R_b)^n} = \varepsilon \ll 1$, the power integral is truncated at $R(z) = \rho_{max} - z\tan\gamma$. When reasonable axial uniformity of $I(r, z)$ is needed for a given input beam profile and axicon, the simplest way to achieve it is to axially overfill the desired interaction region. For example, for gas jets one would use $L_f \sim$ several $L_{jet}$.

In the early hydrodynamic plasma waveguide experiments, which used collisional ionization and heating (Durfee and Milchberg 1993; Gaul *et al.* 2000), conical glass refractive axicons were used (Durfee and Milchberg 1993; Sheng *et al.* 2005) because the Bessel beam-forming pulses were of modest peak power ($\sim 10^9 - 10^{10}$ W) and there was negligible nonlinear phase accumulation in the glass. The much higher peak power pulses needed to generate long OFI waveguides (up to $\sim 10^{13}$ W) are best suited for either reflective axicons (conical mirrors) or very thin diffractive axicons (G. J. Swanson 1991; Erteza 1995; O'Shea *et al.* 2009). Recently, 0.5 mm thick diffractive transmission axicons (ring gratings) have been employed to simplify experimental geometries (B. Miao, Shrock, *et al.* 2022; Shrock *et al.* 2024; Rockafellow *et al.* 2025) so that the Bessel and LWFA drive beams co-propagate rather than counter propagate as needed with a reflective axicon. Off-axis axicons may also be employed to simplify reflective geometries; an off-axis reflective axicon has recently been used in OFI plasma waveguide experiments at ELI-Beamlines (Šišma *et al.* 2026). A schematic of Bessel beam generation using a diffractive transmission axicon (and an optional phase plate for higher order $J_q$ beams ($q \geq 1$)) is shown in Fig. 34(a). Higher order Bessel beams may also be formed using kinoforms, which imprint both radial and azimuthal phase and have been employed to generate hollow plasma channels (Fan *et al.* 2000; S. J. Gessner 2016). Figures 34(b) and (c) are photos of the experiment in (B. Miao, Shrock, *et al.* 2022) showing plasma generation and waveguide injection with a 280 TW LWFA drive pulse. The injected pulse undergoes self-waveguiding in the index structure (plasma core and neutral density shock walls) prepared by a $J_0$ Bessel beam pulse. Coupling of the drive pulse to the waveguide used a drilled mirror with an on-axis hole of radius $a_h$, where typically $a_h \ll R_b$. Here, the Bessel beam depth of focus is shortened to $L_f = (R_b - a_h)/\tan\gamma$.

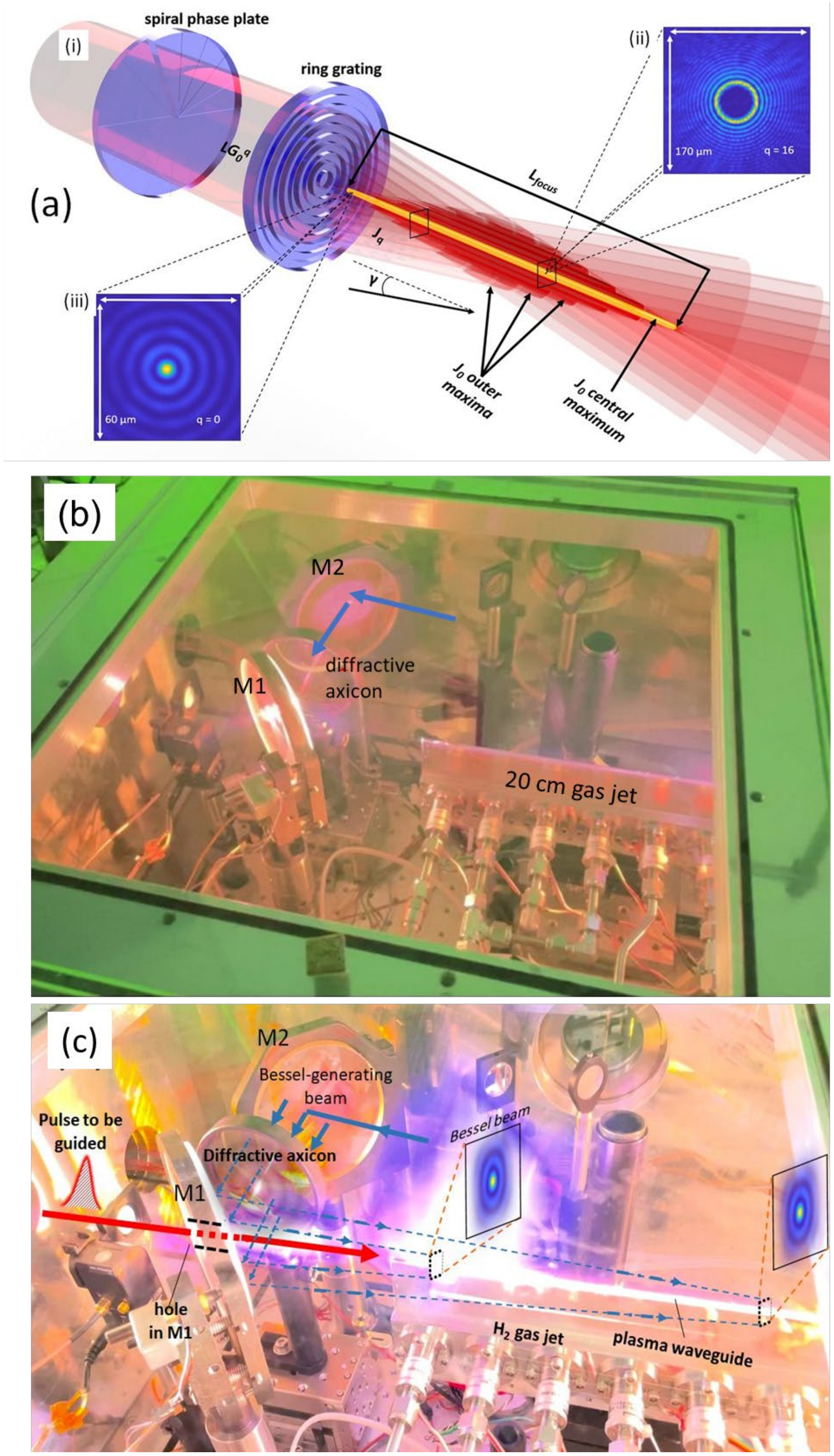

**Figure 34.** Meter-scale plasma generation by high-fidelity Bessel beams (panel (a) adapted from (Shrock *et al.* 2022), panels (b) and (c) adapted from (Bo Miao, Shrock, and Milchberg 2023)). **(a)** (i) Schematic of Bessel beam formation. A collimated, finite aperture beam passes through a spiral phase plate (for $q \geq 1$ case) and diffractive axicon (ring grating) to form a $J_q$ Bessel beam. The radial phase applied by the ring grating causes each annulus of the incident beam to approach the transmission axis at an angle $\gamma$, generating the quasi-uniform, extended Bessel beam focus of length $L_f \approx R_b/tan\,\gamma$. (ii), (iii) Imaged intensity profiles of $J_0$ and $J_{16}$ beams generated by this approach. **(b)** photo of Bessel beam-induced OFI heating of $H_2$ gas sheet from 20 cm gas jet. Weak H-α fluorescence is seen above the nozzle. The beam entering the diffractive axicon is reflected from mirror M2, and the exiting beam reflects from M1 to form the Bessel beam. **(c)** Same as above, except with 280 TW LWFA drive pulse injected after 2.5 ns delay into the OFI waveguide through a central hole of radius $a_h$ in M1. Here, the Bessel focus is of length $L_f \approx (R_b - a_h)/tan\,\gamma$. The Bessel beam rays are shown as blue dashed lines at an angle $\gamma$ with respect to the optical axis. (Photos in (b) and (c) courtesy of Reed Hollinger, Colorado State Univ.).

Maximizing fidelity of the real beam over the full length $L_f$ to the Bessel functional form in Eq. (43) near the optical axis is especially important for generation of cylindrically symmetric meter-scale waveguides. Because each near-field beam ring at radial position $\rho$ at the axicon entrance is mapped to a different $z$-location along the focal line, Bessel beams are particularly susceptible to wavefront aberrations, which manifest as increasing z-dependent beam distortion, resulting in plasma channels shorter than $L_f$. For standard beams such as Gaussian beams, in-situ wavefront correction with a deformable mirror is now a common tool for optimizing focal spot quality; optimization methods directly employing SLMs have also been used for a range of beam types (Wulff *et al.* 2006; Čižmár, Mazilu, and Dholakia 2010; Houzet *et al.* 2016). However, those methods are not readily applicable to the extended depth of field of a Bessel beam focus produced by a fixed optic. Ref. (B. Miao, Feder, *et al.* 2022) reported a deformable-mirror-based method for reconstruction of the near-field wavefront only from intensity measurements collected along the focal line of a $J_{q\geq 0}$ Bessel beam. This method produces a beam with high fidelity to the Bessel functional form over the full length $L_f$, as seen in Fig. 35 for $q = 0$ and $q = 16$, and has been crucial for generating meter-scale plasma waveguides used for multi-GeV electron acceleration (B. Miao, Shrock, *et al.* 2022; Shrock *et al.* 2024; Rockafellow *et al.* 2025).

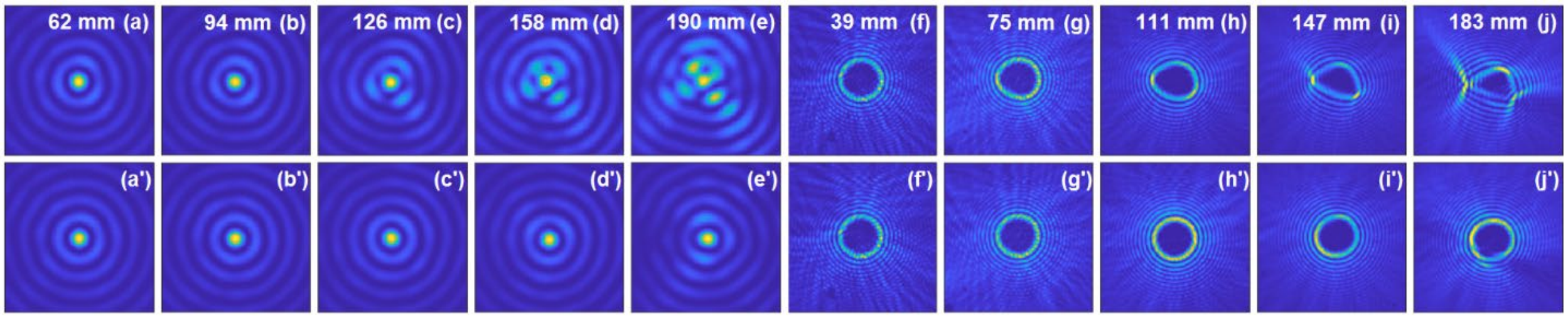

**Figure 35.** Examples of corrected Bessel beams applied to OFI plasma waveguide generation, from (Shrock *et al.* 2022). **(a)-(e)** and **(a′)-(e′)** show the uncorrected and corrected intensity profiles for a $J_0$ Bessel beam formed with a reflective axicon vs. position along the focal line. The window size is 59μm × 59μm. **(f)-(j)** and **(f′)-(j′)** show the same for a $J_{16}$ beam. The window size is 168μm × 168μm.

## 2. Generation of Bessel-like beams

Beyond ensuring fidelity to the Bessel functional form, controlling the axial dependence of peak intensity is desirable for longitudinal tailoring of plasma waveguides. The radial amplitude of the incident beam and the phase imposed by the axicon provide complementary controls: the phase determines where each incident annulus contributes along the focal line, while the amplitude determines the energy supplied by that annulus.

For a generalized axicon with radial phase $\Phi(\rho)$ (Sochacki, Kołodziejczyk, *et al.* 1992; Čižmár and Dholakia 2009), the local radial wavenumber is $k_r(\rho) = d\Phi/d\rho < 0$ for inward-directed rays. In the paraxial approximation, an annulus at radius $\rho$ contributes near $z = -k\rho/k_r(\rho) \simeq \rho/\tan\gamma(\rho)$. Extending the stationary-phase treatment of (Friberg 1996; Fan *et al.* 2000) gives the near-axis field

$$E(r, \phi, z) \propto A(\rho) \left[\frac{\rho|k_r(\rho)|}{|D(\rho)|}\right]^{1/2} J_q(|k_r(\rho)|r) e^{iq\phi} e^{i\Psi(z)} e^{ikr^2/(2z)} \tag{45}$$

where $D(\rho) = 1 - \rho k_r'(\rho)/k_r(\rho)$ and $\Psi(z) = kz + \Phi(\rho) + k\rho^2/(2z) + (\pi/4)\mathrm{sgn}D(\rho)$. Here $k_r{}' = dk_r/d\rho$, and $\rho = \rho(z)$ is the contributing annulus. Equation (45) assumes a single isolated stationary annulus with $D \neq 0$, away from aperture edges. Note that the phase factors do not affect the intensity profile. The effective focal length, $L_f = \max[\rho/\tan\gamma(\rho)] - \min[\rho/\tan\gamma(\rho)]$, is determined by the illuminated radial interval on the axicon.

The conical axicon of Sec. V.B.1 is the linear-phase case, $\Phi(\rho) = -k\sin\gamma\,\rho$, with constant $k_r$ and $D = 1$. Equation (45) then recovers the intensity envelope of Eq. (43): the axial intensity envelope is proportional to $\rho|A(\rho)|^2$, evaluated at $\rho \simeq z\tan\gamma$. Figure 36 illustrates the effects of illumination and aperture for $q = 0$, comparing apertured plane-wave and apertured and unapertured Gaussian input profiles.

A varying radial phase changes both this mapping and the local transverse scale. For example, a "logarithmic axicon" with $\Phi(\rho) = -\Phi_0\ln(\rho/\rho_0)$ has $k_r = -\Phi_0/\rho$ and $D = 2$. Thus, for $q = 0$ and uniform illumination $A(\rho) = A_0$, Eq. (45) gives constant on-axis intensity. Finite-aperture diffraction produces axial oscillations, which can be strongly suppressed by smoothly apodizing the inner and outer edges of an annular logarithmic axicon, yielding a nearly uniform plateau with smooth entrance and exit regions (Jaroszewicz *et al.* 1993). A diffractive log axicon was first demonstrated by (Popov *et al.* 1998), followed by a refractive log axicon (Golub *et al.* 2010).

Controlling the axial dependence of peak intensity using refractive or reflective axicons is difficult owing to the fabrication tolerances required for the needed surface figure, leaving diffractive optics as a practical option (Popov *et al.* 1998; Dharmavarapu, Bhattacharya, and Juodkazis 2018). Recently, a Bessel-like beam from a diffractive logarithmic axicon (Tripathi *et al.* 2025; Sloss *et al.* 2026) with $\Phi(\rho) = -\Phi_0\ln(a + b\rho^2)$ formed a 30-cm-long plasma waveguide with an integrated funnel-shaped entrance. The entrance funnel, for improving injected laser coupling, was formed by the faster hydrodynamic expansion induced by the wider Bessel central maximum and resulting greater OFI heating per length of channel (see Eq. (37)) at small $z$.

Other generalized axicons include the 'axilens' (Davidson, Friesem, and Hasman 1991; Sochacki, Bará, *et al.* 1992), which combines the transverse focal energy concentration of a spherical lens with the extended depth of focus of an axicon, and the 'axiparabola' (Smartsev *et al.* 2019; Oubrerie, Andriyash, *et al.* 2022) which combines the functions of an off-axis parabolic mirror and an axicon. The axiparabola maps incident annuli to a prescribed finite focal interval, allowing the axial energy distribution to be tailored through its surface profile and illumination. Localizing this interval to the gas target can reduce energy deposition outside the interaction region. For the axiparabola designed to give an axially uniform $I(r = 0, z)$, an incident ray at radius $\rho$ from an incident beam profile $A(\rho) = A_0$ is reflected and focused to $z = f(\rho) = f_0 + L_f(\rho/R)^2$ along the focal line. Here, $f_0$ is the start of the focal line, $L_f$ is the depth of focus, and $R$ is the radius of the incident beam. A scheme for shaping smooth or structured on-axis quasi-Bessel intensity profiles produced by axiparabola focusing was developed by (Touguet *et al.* 2026). The localized extended focus of an axiparabola has been used to generate OFI plasma waveguides for controlled LWFA (Sec. III.C.3; (Oubrerie, Leblanc, *et al.* 2022; Lahaye *et al.* 2025)), including longitudinally density-ramped guides formed in a tilted gas jet (Lahaye *et al.* 2026).

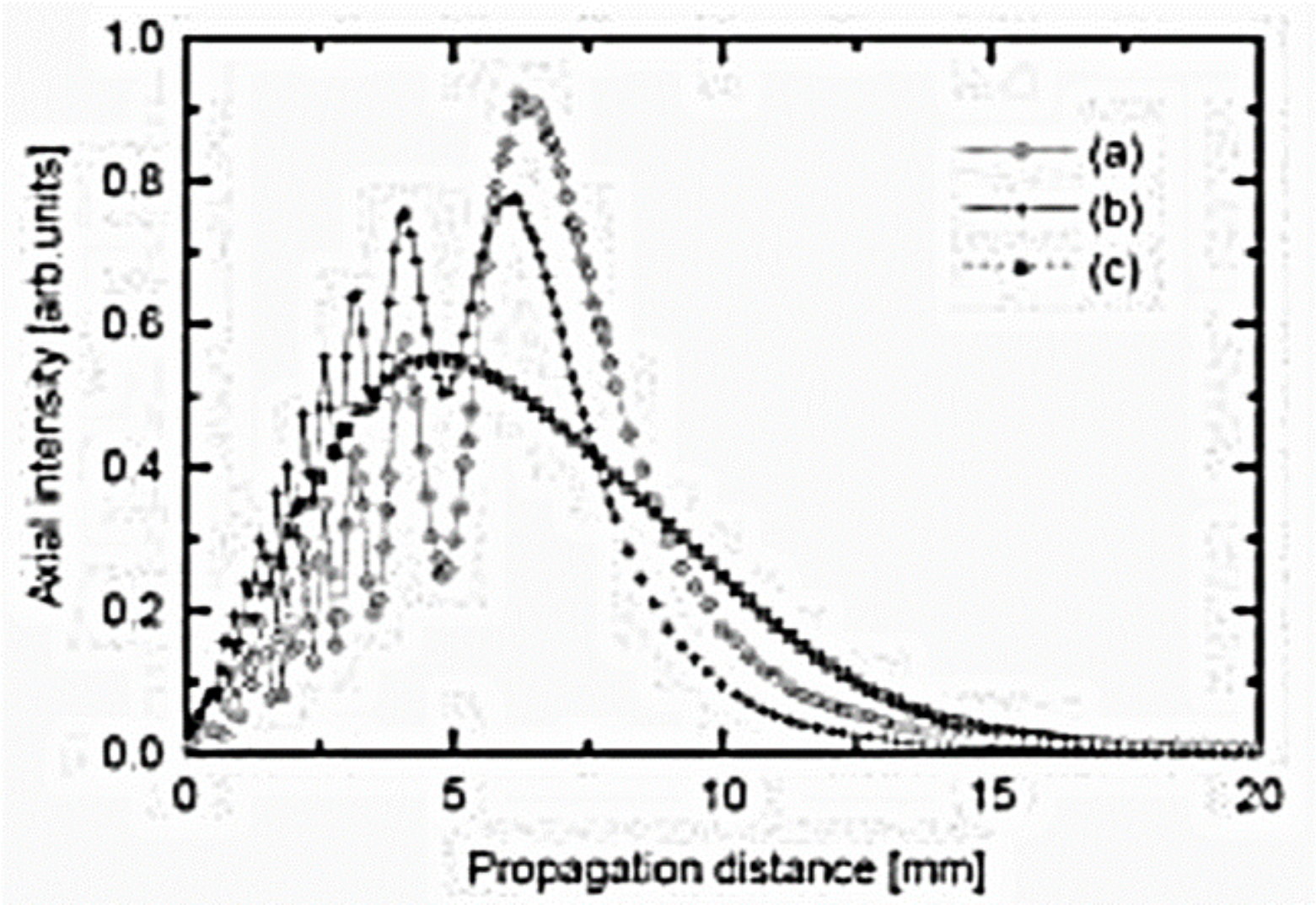

**Figure 36.** Comparison of on-axis intensity $I(r = 0, z)$ for different near field profiles focused by an axicon (adapted from (McGloin and Dholakia 2005)). (a) Plane wave incident on axicon with finite aperture, (b) Gaussian beam incident on axicon with finite aperture, (c) Gaussian beam incident on axicon with infinite aperture.

### 3. Propagation of intense Bessel beams in ionizing gases and initiation of plasma waveguides

With laser-based generation of plasma waveguides increasingly using Bessel or Bessel-like beams, the beam-plasma interaction must be considered, especially for meter-scale waveguides. For such guides, the rays of a standard Bessel beam must approach the optical axis at a small angle $\gamma$. For example, for a beam with radius $R_b = 2$ cm incident on an axicon for a desired 50-cm-long plasma ($\approx L_f$), the required ray approach angle is $\gamma = \tan^{-1}(R_b/L_f) \sim 2.3°$, approximately the Bessel ray angle in Figs. 34(b) and (c). For these conditions, the effective critical electron density for total internal reflection and exclusion of grazing incidence Bessel beam rays (directed from neutral gas to plasma) is $N_{cr,eff} = N_{cr} \, sin^2 \gamma \sim 2.7 \times 10^{18} \, cm^{-3}$ at $\lambda_0 = 800$ nm, with strong ray refraction occurring even for $N_e < N_{cr,eff}$. The optimal waveguide conditions for multi-GeV LWFA are meter-scale guide lengths and central densities $N_{e0} \sim 10^{17}$ cm$^{-3}$, reached after the $\sim 10 \times$ density drop (L. Feder *et al.* 2020; B. Miao *et al.* 2024) from hydrodynamic expansion of OFI plasma at initial electron density $\sim 10^{18}$ cm$^{-3}$. As the latter is generated from fully ionized $H_2$ gas at $N_g \sim 10^{18}$ cm$^{-3}$, it is clear that generation of meter-scale OFI plasmas with Bessel beams requires self-consistent propagation simulations to accurately model the gas ionization and heating.

This analysis was performed in recent work benchmarking simulations against measurements of OFI hydrodynamic waveguides (B. Miao *et al.* 2024). Self-consistent propagation simulations of Bessel beam-induced OFI heating were used to determine the 3D electron density and temperature profiles needed as initial conditions for hydrocode simulations of waveguide evolution (discussed in Sec. V.D) using the code SPARC (Gordon *et al.* 2006). The simulated channels agreed well with experimental measurements of the electron and neutral density profiles obtained through two-color interferometry. Both profiles are needed for accurately modeling the physics of OFI-based plasma waveguide formation discussed in Secs. III.C.1 and III.C.2.

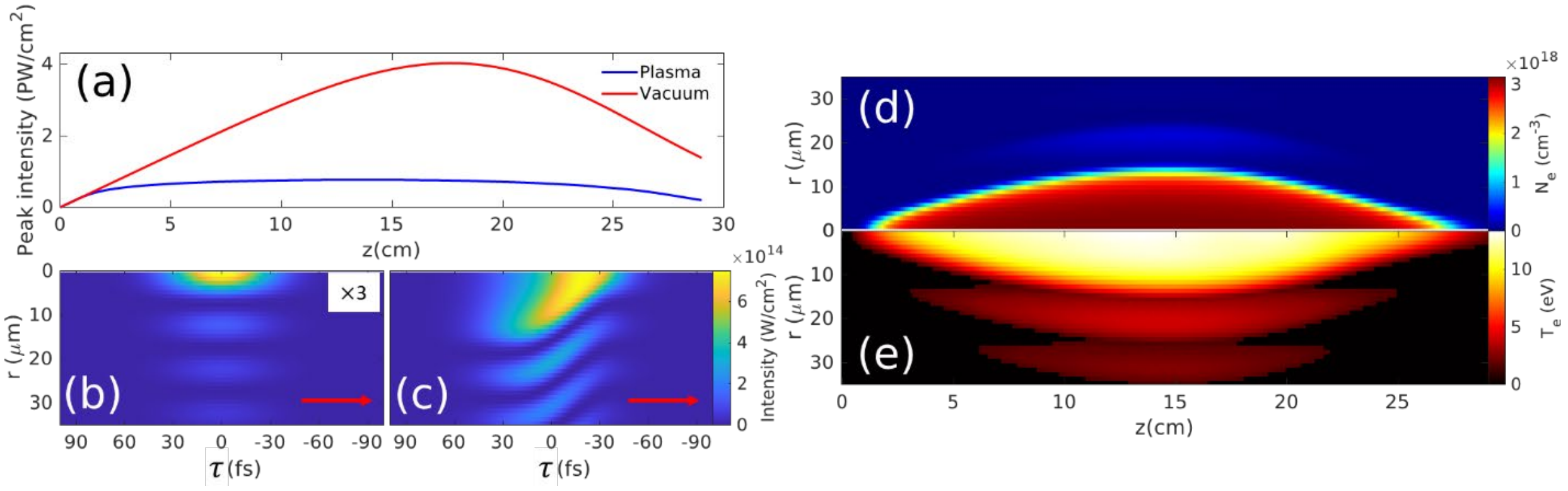

**Figure 37.** Simulation results of OFI plasma generation and heating in hydrogen using the nonlinear propagation code YAPPE (L. Feder *et al.* 2020; B. Miao *et al.* 2024) for linearly polarized (LP) Bessel beam pulses (40 mJ, 50 fs FWHM, 4$^{\text{th}}$ order super-Gaussian profile with $1/e$ field radius $w_0 = 1\ cm,\ \gamma = 2.3°$). Adapted from (B. Miao *et al.* 2024). **(a)** Peak on-axis instantaneous laser intensity with (blue) and without (red) $H_2$ gas present**. (b)** Bessel beam intensity envelope (× 3) at $z = 0.9$ cm. Red arrow indicates propagation direction. **(c)** Bessel beam intensity envelope at $z = 15$ cm. $\tau = \xi/v_g = t - z/v_g$ is the temporal coordinate in the moving window. **(d)** Electron density profile and **(e)** electron temperature profile after passage of Bessel beam pulse.

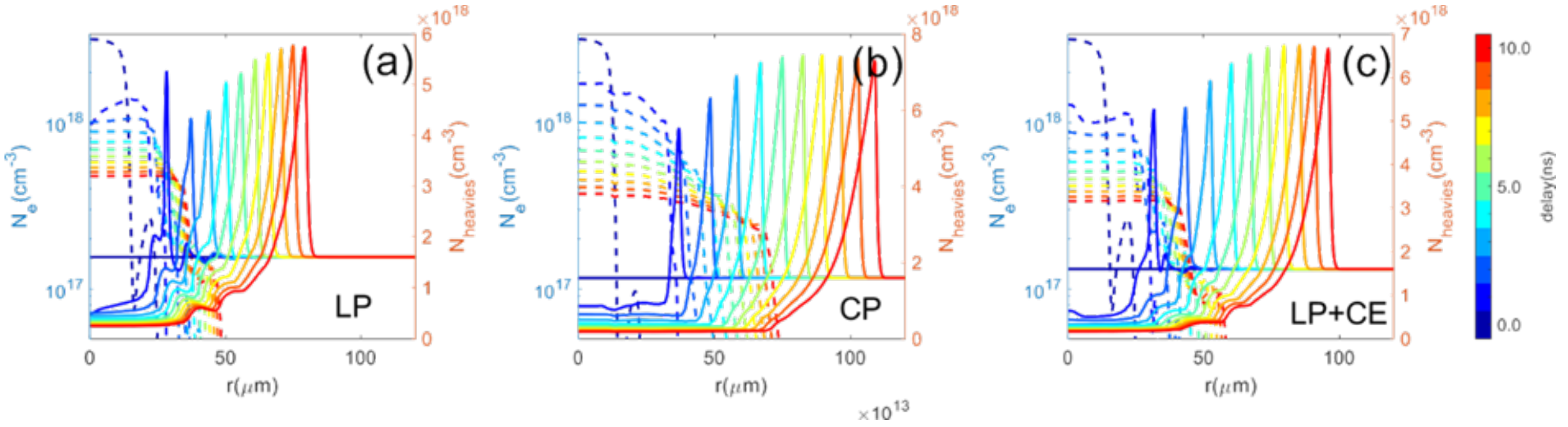

**Figure 38.** SPARC (Gordon *et al.* 2006) hydrodynamic simulation result using YAPPE output for **(a)** linear polarization and **(b)** circular polarization as initial conditions. **(c)** LP+CE (linear polarization and Coulomb explosion-imposed ion temperature of $T_i = 3$ eV). All panels adapted from (B. Miao *et al.* 2024).

Figure 37 shows a simulation of intense linearly polarized (LP) Bessel beam ionization and heating of a 30 cm length of $H_2$ gas, using the code YAPPE (B. Miao *et al.* 2020; L. Feder *et al.* 2020), which implements the unidirectional pulse propagation algorithm (Kolesik and Moloney 2004). The red curve in (a) and the intensity profile in (b) are for the undistorted Bessel beam. The blue curve in (a) and panel (c) show the significant effect of refractive distortion of the Bessel beam. A useful consequence of the refraction is that the peak intensity in (a) is flattened along z, lending itself to more plasma uniformity along the beam axis. Electron density and temperature profiles are plotted in Fig. 37(d) and (e), providing the initial plasma pressure profile used as an input to the SPARC hydrodynamic simulations (Figure 38 (a)). As noted in the discussion of Sec. III.C.3, the expanding electron density profiles remain peaked on axis, with no development of an ionized shock to serve as a plasma cladding, in agreement with two-color interferometry measurements (L. Feder *et al.* 2020; B. Miao *et al.* 2024). A similar YAPPE simulation was run for circular polarization (CP), with the resulting electron density and temperature profiles used as initial conditions for SPARC in Fig. 38(b). The simulations predict that CP-heated plasma expands significantly faster than LP; however, the measurements show that LP and CP heated plasmas expand similarly. A likely explanation is that LP-favored Coulomb explosion of the intermediate species $H_2^+$ instantaneously boosts the ion temperature and drives faster expansion to be competitive with CP (B. Miao *et al.* 2024).

### C. Gas jets, gas cells, and discharge capillaries

The choice of method for plasma waveguide generation is informed by a combination of desired waveguide parameters, diagnostic access, and ease of implementation. The waveguides primarily discussed in this review are laser driven (laser ionization of gas plumes from gas jets or quasistatic fills in gas cells) or electrically driven (discharge capillaries). In this section, we highlight specific implementations of these methods.

#### 1. Axially extended supersonic slit nozzles

As discussed earlier, a major application of plasma waveguides is LWFA, which sets specific requirements on waveguide properties. For example, in order to reach $\sim 10$ GeV energy gain in a single LWFA stage without dephasing of the acceleration process (E. Esarey, Schroeder, and Leemans 2009), a laser pulse must maintain its normalized vector potential $a_0 > 1$ by optical guiding in a low-density meter-scale plasma waveguide with on-axis density $N_e \sim 10^{17} cm^{-3}$ (B. Miao, Shrock, *et al.* 2022). This imposes demanding requirements on a plasma target: it should be easily accessible by experimental diagnostics, free standing to avoid laser and plasma damage of nearby materials, capable of high repetition rate operation, and programmable in density profile. Well-known plasma targets such as gas cells (Osterhoff *et al.* 2008; Audet *et al.* 2018; R. J. Shalloo *et al.* 2019) and capillary discharges (Butler, Spence, and Hooker 2002; Leemans *et al.* 2014; Gonsalves *et al.* 2019) cannot meet all of these requirements. Free-standing gas jets, however, can provide a useful combination of optical access, profile programmability, and resistance to nearby material damage, with tradeoffs in gas load and flow uniformity. The dependence of gas-density profiles on nozzle geometry and gas species has been examined experimentally and numerically in shorter supersonic and subsonic targets for LWFA (Lorenz *et al.* 2019), illustrating how nozzle design can control density uniformity and gradients.

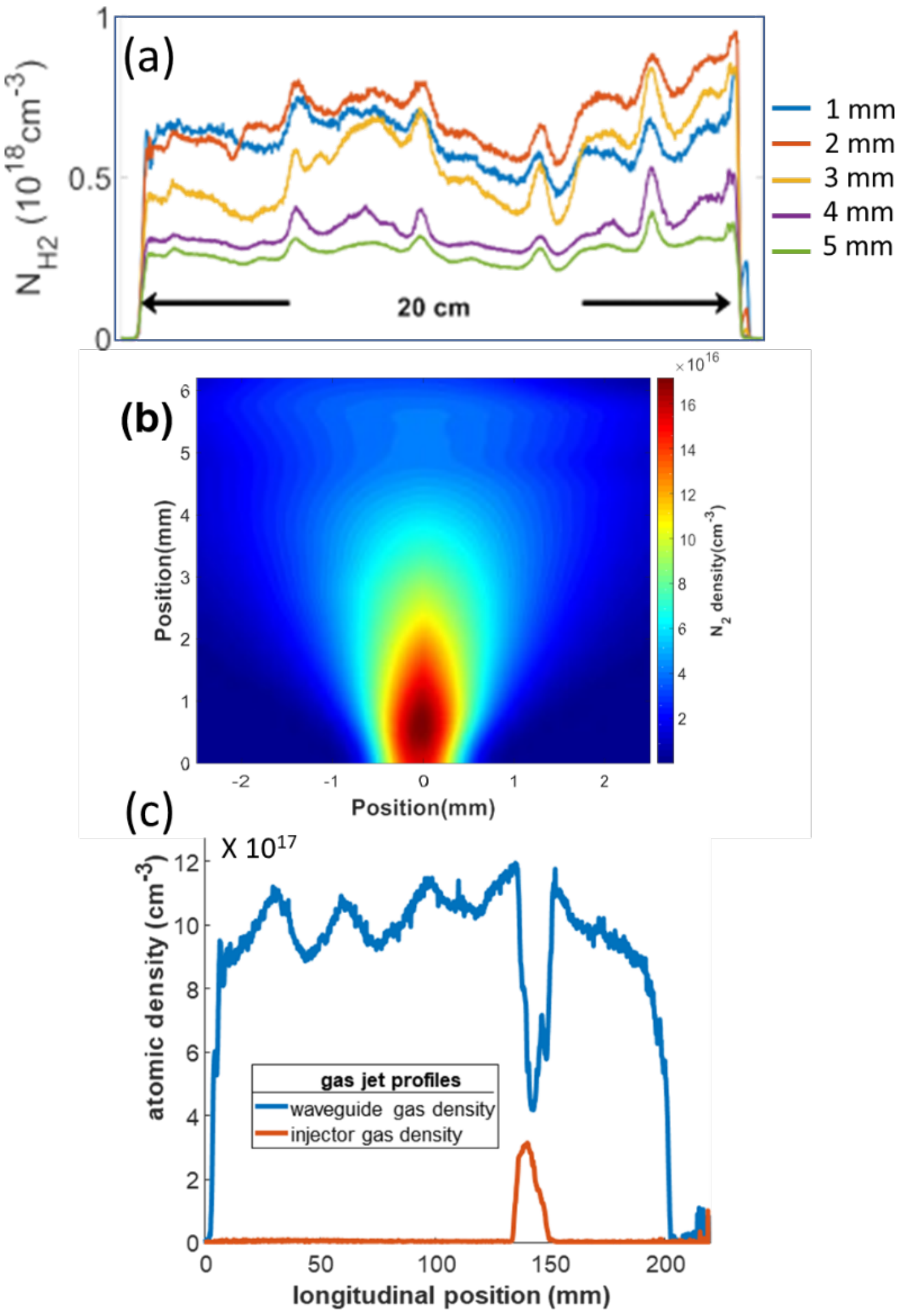

**Figure 39.** Long, low density gas jets (panels (a)-(b) adapted from (Shrock *et al.* 2022), panel (c) adapted from (Shrock *et al.* 2024)) **(a)** Longitudinally-resolved $H_2$ density profiles of a 20 cm long jet, at various heights above the jet nozzle, extracted using fluorescence diagnostic (Sec. V.A.3). **(b)** Transverse $H_2$ density profile from longitudinal interferometry of the 20 cm jet used in (B. Miao, Shrock, *et al.* 2022). **(c)** Example of longitudinal gas density profiles with use of a localized $N_2$ injector located 14 cm from the waveguide entrance (z = 0).

A free-standing plasma waveguide can be initiated by a Bessel beam focus in the gas plume above the jet nozzle orifice, and the full length of the waveguide is accessible for diagnostics. Figure 34(b) shows a 20 cm $H_2$ jet OFI-heated by a $J_0$ Bessel beam, followed after ~2.5 ns delay by injection and self-waveguiding of a 280 TW LWFA drive pulse. If the jet is supersonic, the edges of the gas flow are sharpened, and a specific gas density operating point can be located even further from the orifice, avoiding interaction of the wings of the Bessel beam focus with the nozzle (Shrock 2023). The attainable repetition rate depends on available vacuum pumping speed, valve and gas-flow recovery, and the laser repetition rate. All of these features have been implemented in the jets used in (B. Miao *et al.* 2020; L. Feder *et al.* 2020; B. Miao, Shrock, *et al.* 2022; Picksley *et al.* 2024; B. Miao *et al.* 2025; Rockafellow *et al.* 2025). In each of these experiments, the jet design is similar: high-pressure gas backs multiple solenoid valves which feed a small stabilizing/mixing reservoir beneath a narrow throat. The flow becomes supersonic as the gas moves into the wider nozzle region above the throat. All of these jets have used $H_2$ as the working gas because it is fully ionized at the relatively low OFI threshold of $\sim 10^{14}\ \mathrm{W/cm^2}$. Doping of the working gas with a few percent $N_2$ enabled ionization injection (Rowlands-Rees *et al.* 2008; Pak *et al.* 2010; McGuffey *et al.* 2010; Clayton *et al.* 2010; M. Chen *et al.* 2012).

Figure 39 shows results for the supersonic (Mach 4), 20 cm long jet used in the LWFA experiments of (Shrock *et al.* 2024). Longitudinal $H_2$ density profiles vs. height above the jet nozzle are plotted in panel (a) (Sec. V.A.3 describes this fluorescence-based measurement), panel (b) plots the transverse $H_2$ density profile determined by longitudinal interferometry, and panel (c) plots longitudinal density profiles from a modified version of the jet that axially localized the flow of dopant $N_2$ gas. The localized-dopant profile enabled controlled ionization injection, producing repeatable quasi-monoenergetic 2.5 GeV bunches. These results also suggested that use of localized higher Z gases could be used to locally modify $w_{ch}$ (due to the slower expansion speed of the neutral shock), altering injection dynamics by squeezing the guided mode to a smaller size. A complementary proposal uses an optically generated annular plasma to produce a localized waveguide constriction (R. J. Shalloo *et al.* 2024). Simulations showed that wake expansion as the drive pulse leaves the constriction enables localized electron injection and subsequent acceleration to 1.1 GeV. Earlier experiments used plasma density transitions to trigger injection in plasma waveguides, producing 1.1- and 1.2-GeV beams (Oubrerie, Leblanc, *et al.* 2022; Picksley *et al.* 2023).

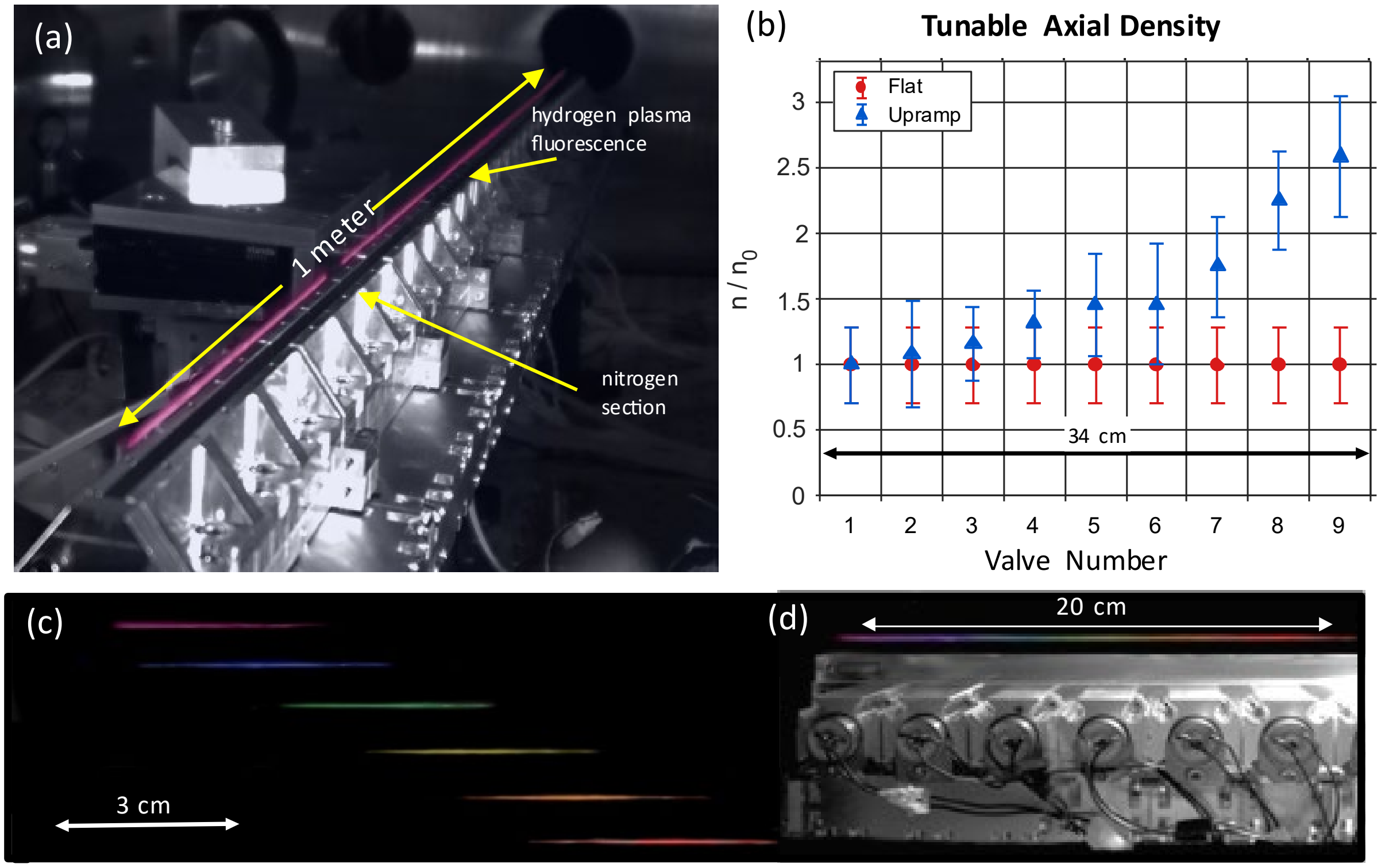

**Figure 40. (a)** A 1-meter-long plasma generated in a modular supersonic gas jet by a 50 fs J0 Bessel beam. Each of 9 modules is comprised of 3 sections. An injector valve in one of the 27 jet sections is backed by nitrogen, with hydrogen backing elsewhere. The plasma fluorescence is imaged through a H-α line filter, where the nitrogen section appears as a gap. **(b)** Demonstration of axial gas density tunability in a 34-cm-long Maryland helium jet using longitudinal interferometry measurements of a representative jet section at different backing pressures and valve opening times. Red points: identical settings across all sections, illustrating a flat density profile. Blue points: individually tailored settings illustrating a customizable longitudinal density upramp. Error bars are the $\pm 1$ standard deviation over 10 jet pulses. **(c)** Bessel OFI-induced hydrogen recombination fluorescence from each of six valves on a sectional gas jet, with superimposed colors to differentiate each valve. **(d)** Hydrogen fluorescence from all 6 valves fired together showing smoothed profile, with superimposed colors identifying contribution from each valve. Adapted from (Rockafellow *et al.* 2026)

To date, the RMS longitudinal variation in jet density is typically $< \sim 25\%$ (Fig. 39(a)) due to various imperfections in jet fabrication. This variation has had little impact on the waveguide mode structure since OFI plasma heating largely depends on the single atom interaction (B. Miao *et al.* 2024) and the plasma density difference between the waveguide core and cladding has been sufficiently large. Nevertheless, the axial variation may affect electron bunch quality in a LWFA, as discussed in (B. Miao, Shrock, *et al.* 2022).

Multi-valve supersonic jets of this type, up to 30 cm in length, have been used to achieve 10-GeV-class operation. A LBNL-Maryland collaboration (Picksley *et al.* 2024) generated quasi-monoenergetic peaks up to 9.2 GeV with charge extending beyond 10 GeV with the BELLA petawatt laser, and in similar experiments (Rockafellow *et al.* 2025) produced electron beams with few-pC spectral features to 10 GeV and beyond with ~ nC-level charge in the 1-7 GeV range. In (Picksley *et al.* 2024), $N_2$ dopant for ionization injection was restricted to the first 12 cm of the channel to suppress injection and acceleration of electron bunches later in the guide. Recently, longitudinal density ramping, with the goal of LWFA dephasing mitigation, has been applied to a 30 cm gas jet of this type by adjusting the throat width and tilting the nozzle (R. Li *et al.* 2025). A flexible approach to ramping is controlling the density along the extended jet via gas valve backing pressure and timing (B. Miao *et al.* 2025). Recently, a higher pressure, single-valve-driven 20-cm-long helium jet has been employed for OFI-waveguide-based multi-GeV LWFA experiments (Šišma *et al.* 2026).

For jets beyond $\sim 30 - 40$ cm in length, a modular design is advantageous: fabrication tolerances can be maintained over shorter sections, and well-characterized sections can be linked to construct jets of arbitrary length. Recently, a 1-m-long hydrogen plasma has been generated by a Bessel-like beam (from a log axicon (Tripathi *et al.* 2025)) in a 1-m-long Mach 5 jet (see Fig. 40), constructed from nine ~11-cm-long modules, each with three sub-sections (27 valves total). The jet design enabled control of gas flow and species localization via valve-resolved gas species, backing pressure, and timing (B. Miao *et al.* 2025). A version of this jet presented in (Rockafellow *et al.* 2026) enabled a 34 cm density-upramped plasma waveguide to achieve quasi-monoenergetic electron energies of 9.5 GeV (Rockafellow 2026).

### 2. Gas cells

The use of elongated gas cells (Picksley, Alejo, Cowley, *et al.* 2020) in which to generate long plasma waveguides requires laser entrance and exit apertures. The entrance aperture is needed for access by the waveguide initiating pulse (typically a Bessel beam) and the intense drive pulse; the exit aperture enables imaging of the guided pulse and passage of any products of the interaction such as accelerated electron bunches or short wavelength light. The inevitable gas leakage from the apertures into the enclosing vacuum chamber sets up aperture-sized density ramps that extend from inside the gas cell into the chamber, limiting the sharpness of gas onset and the steady state density of the gas. Simple uniformly filled gas cells offer limited independent control over longitudinal density variation, and their apertures are susceptible to damage and widening by the waveguide generating pulse, the guided high intensity pulse, and contact with the plasma.

Capillaries can also serve as gas cells without an electrical discharge, with gas supplied through side inlets and escaping through the open ends. The flowing gas establishes the target density profile, and the incident laser pulse produces the plasma and relativistically self-guides in it. (Osterhoff *et al.* 2008) used a sapphire gas cell to obtain reproducible gas profiles and stable electron beams with energies up to ∼200 MeV. Such gas cells also support sustained accelerator operation. Using a continuously supplied sapphire capillary gas cell, (Maier *et al.* 2020) demonstrated day-long LWFA operation at 1 Hz, producing 100,000 consecutive electron bunches with peak energy 368 MeV with 2.4% rms variation; the target sustained more than 200,000 shots before replacement. In addition, longitudinal tailoring is possible through the gas-feed geometry and pressures: (J. Kim *et al.* 2021) demonstrated a density upramp in a 7-mm-long sapphire capillary using two independently supplied gas inlets of unequal width. Using this design, (Nam *et al.* 2024) demonstrated dephasing mitigation, increasing electron bunch energy by $> 2\times$ to 260 MeV. Such designs provide control of the longitudinal gas distribution without requiring a discharge-generated guiding channel.

Specialized design of gas cells to measure specific waveguide properties has also been reported. In (Picksley, Alejo, Cowley, *et al.* 2020), a gas cell with adjustable length (also see (Brandi *et al.* 2016)) was employed to characterize the attenuation lengths of OFI-generated channels. A new adjustable design of hybrid gas cell and supersonic jet proposed in (C. Aniculaesei *et al.* 2018) for high energy gain LWFA was employed in (Picksley *et al.* 2023) to demonstrate >1 GeV LWFA by self-waveguiding pulses.

### 3. Capillary discharges

Capillary discharges have been extensively discussed in Sec. IV. Here we briefly summarize several augmentations that improve the function of discharge capillary plasma waveguides. In order to overcome the constraints on $w_{ch}$ imposed by a fixed capillary wall radius, (K. K. Swanson *et al.* 2021) demonstrated a cryogenically formed discharge capillary waveguide. A controllable wall radius was achieved by flowing $N_2O$ through the liquid nitrogen-cooled capillary. This enabled tuning of the $w_{ch}$ to match an injected pulse. In another design, water cooling and current profiles designed to mitigate erosion during the discharge were employed (Gonsalves *et al.* 2016) to demonstrate capillary discharge waveguide formation at kHz repetition rates. Hybrid gas-jet/capillary targets provide additional control of electron injection through an entrance density transition (Gonsalves *et al.* 2011), as discussed in Sec. IV.B.3. An auxiliary nanosecond laser pulse can also heat the discharge plasma and modify its transverse density profile, as discussed in Sec. IV.B.4.

## D. Hydrodynamic modeling of waveguide formation

Direct characterization of the complete waveguide plasma and optical properties is not always achievable—particularly for the meter-scale waveguides discussed in Sec. III.C.4, where diagnostic access along the full length can be difficult. Thus, high-fidelity modeling of waveguide formation is an essential tool. As discussed in Secs. IV.B.3 and V.B.3, accurate simulation of waveguide formation requires careful treatment of boundary conditions and incorporation of multi-physics modeling. Here we briefly review computational approaches applied to modeling the various plasma waveguides covered in this review.

A key challenge for modeling laser-generated waveguides is the disparate timescales among ultrafast laser heating, electron thermal transport in the gas/plasma, and plasma hydrodynamics.

Early modeling of plasma waveguides generated by collisional heating with 100ps laser pulses (Durfee and Milchberg 1993; Durfee, Lynch, and Milchberg 1994; Durfee, Lynch, and Milchberg 1995; T. R. Clark and Milchberg 1997) employed a single fluid Lagrangian hydrodynamics code to model plasma waveguide generation and evolution, efficiently computing shock dynamics. This code did not model the laser interaction self-consistently. With the later development of a Helmholtz equation solver for generalized Bessel beams in media with complex refractive index profiles (T. R. Clark and Milchberg 1998a), a Lagrangian hydrocode was subsequently developed incorporating self-consistent Bessel beam gas/plasma interaction. This was applied to experiments where the 100 ps Bessel beam pulses were long enough to resonantly side-couple to the evolving plasma waveguides as they were generated (Fan, Parra, and Milchberg 2000). This code was employed along with the 1D MHD code HELIOS (MacFarlane, Golovkin, and Woodruff 2006) in (R. J. Shalloo *et al.* 2018) to model expansion of an ultrashort pulse OFI plasma generated by a lens. The simulated shock expansion was found to agree with both Sedov-Taylor blast wave theory (Zel'dovich and Raizer 1967) and measured cylindrical shock expansion.

For ultrashort-pulse OFI waveguides, the separation between laser heating and hydrodynamic expansion allows the modeling to proceed in stages. A calculation of ionization and heating first supplies the spatial density and temperature profiles needed to initialize the hydrodynamic evolution. The Eulerian code SPARC (Gordon *et al.* 2006) was employed by (Morozov *et al.* 2018; B. Miao *et al.* 2024) to model this evolution. SPARC treats the electron, ion, and neutral fluids separately, allowing their distinct responses to be followed through the expanding plasma and shock region. In (B. Miao *et al.* 2024), the initial conditions for meter-scale OFI waveguides were supplied by YAPPE (L. Feder *et al.* 2020), which computes self-consistent ultrashort-pulse Bessel beam propagation and the resulting ionization and temperature profiles. (Picksley, Alejo, Shalloo, *et al.* 2020) used the single-fluid, multi-temperature FLASH code (Fryxell *et al.* 2000), initialized with ionization and heating calculated from a prescribed laser field.

(Mewes *et al.* 2023) developed a complementary framework combining plane-wave WarpX PIC calculations (Vay *et al.* 2021) of ionization and electron heating with a single-fluid hydrodynamic model implemented in (*COMSOL Multiphysics*). The calculated electron densities and temperatures were tabulated versus intensity and mapped onto a prescribed spatial laser profile to initialize the expansion. Comparison with interferometric measurements tested the predicted channel evolution, and the calculations explored how gas density, laser pulse energy, and expansion delay could be used to tune the matched spot size. Both this initialization and that used by (Picksley, Alejo, Shalloo, *et al.* 2020) neglect ionization-induced refraction of the channel-forming pulse.

To extend these calculations from waveguide formation to intense-pulse propagation, the hydrodynamically evolved profiles at the chosen injection delay can be transferred to a propagation or PIC code such as WarpX. For self-waveguiding, this transfer must retain the neutral gas distribution as well as the plasma profile, so that further ionization by the injected pulse and the resulting changes in guiding can be followed. The injected laser field can be specified using LASY (Thévenet *et al.* 2025), which constructs pulse profiles from measurements, analytical expressions, or simulation output and exports them in the openPMD format supported by WarpX. This sequence connects the channel-forming pulse, hydrodynamic expansion, and subsequent guided interaction, while allowing each stage to be modeled on its appropriate timescale.

Modeling capillary discharge waveguides faces its own complexities. In particular, plasma behaviour at the capillary walls is difficult to model, but plays an important role in thermal transport (Broks, Van Dijk, and Van Der Mullen 2006; Broks *et al.* 2007). Two approaches (Bobrova *et al.* 2002; Broks, Garloff, and Van Der Mullen 2005) to modeling waveguide formation in capillary discharges are discussed in Sec. IV.B.3.

## VI. CONCLUSIONS

With the increasing number of petawatt class laser facilities coming online worldwide (J. W. Yoon *et al.* 2021; ICUIL 2024), plasma waveguides are key to laser-driven generation of secondary sources of photons and charged particles, among these facilities' main applications. These applications require extended high intensity pulse propagation over distances much greater than the characteristic diffractive spreading distance, or Rayleigh range. The high intensities of interest are well beyond the optical field ionization thresholds for atoms, so one must consider mitigation of diffraction in a plasma.

In this article, we have reviewed the physics of plasma waveguides, pre-formed optical fibers made of plasma, and their development and use. Plasma waveguides have transverse mode structure that enables extended and controlled high intensity interaction with the medium. Compared with relativistic self-guiding in plasma, preformed plasma waveguides can significantly reduce the laser energy requirements for a particular nonlinear process and offer control of axial and transverse laser field profiles. This is important for excitation of desirable nonlinear responses such as plasma wakes for electron acceleration.

After discussing the general principles of guiding in plasma waveguides, including linear and nonlinear propagation, we have emphasized the two types of waveguides most in use: laser-produced hydrodynamic plasma waveguides and discharge capillary plasma waveguides. Recent results with laser-produced waveguides, specifically ones generated by optical field ionization by short pulse Bessel and Bessel-like beams in elongated supersonic gas jets, are particularly promising. These experiments demonstrate flexible shot-to-shot optical and gas jet control of transverse and longitudinal waveguide profiles and therefore mode structure. These waveguides are free-standing, located away from the jet nozzle, eliminating both laser-induced and plasma-induced damage to material structures, and enabling both active and passive diagnostic access from all directions. Gas species composition and location can be controlled by separately backed gas injectors. A recent modular version of these supersonic jets has demonstrated Bessel-beam-induced generation of 1-meter-long hydrogen plasmas.

We have also discussed simulations of plasma waveguide generation and pulse propagation. High performance simulations, including hydrocodes, particle-in-cell codes and spectrally and carrier resolved nonparaxial laser propagation codes will continue to be indispensable tools for designing and interpreting high intensity guiding experiments. With high intensity guiding over meter lengths and even longer on the horizon, computational needs will only increase. Boosted frame simulations like (Vay *et al.* 2021) will play a major role.

Going forward, several developments in laser-generated plasma waveguides appear especially promising. Axial density ramps for dephasing mitigation were proposed (Katsouleas 1986; P. Sprangle *et al.* 2001; Pukhov and Kostyukov 2008; Rittershofer *et al.* 2010; Wen *et al.* 2010; Döpp *et al.* 2016) and demonstrated in short targets (Jaehoon Kim, Ju, and Yoo 2011; Guillaume *et al.* 2015; Gustafsson *et al.* 2024; Nam *et al.* 2024; Lahaye *et al.* 2026). Gas jets offer particular flexibility for control of the axial density distribution. Recently, (Lahaye *et al.* 2026) demonstrated dephasing mitigation in a density-ramped OFI waveguide, increasing electron energies by 40% to 1.6 GeV over a ~2 cm acceleration distance. To extend this principle to tens-of-cm-long waveguides, gas jets have been designed that can introduce a density upramp by controlling the throat size and tilting the jet (R. Li *et al.* 2025) or tuning the backing pressure and timing of cm-scale sections of a 1-meter-long nozzle (B. Miao *et al.* 2025). Extending controlled density ramps throughout multi-meter-scale waveguides remains a route toward 100 GeV (Ludwig *et al.* 2025). Generalized axicons are now a basic element for producing Bessel and Bessel-like beams for OFI-based plasma generation (Smartsev et al. 2019; B. Miao et al. 2020; B. Miao, Shrock, et al. 2022; Šišma et al. 2026); recent versions have been customized to sculpt waveguide entrance funnels for improved drive-laser coupling (Tripathi *et al.* 2025). Future generalized-axicon designs can exploit OFI to create axially varying structures for quasi-phase matching of short-wavelength generation (Zepf *et al.* 2007), terahertz generation (Antonsen, Palastro, and Milchberg 2007), quasi-phase-matched laser acceleration (York *et al.* 2008; Layer *et al.* 2007; S. J. Yoon, Palastro, and Milchberg 2014), and guided-mode control for electron injection in LWFA. These applications also require improved drive-laser pointing (Amodio *et al.* 2025) and mode stability, since pointing fluctuations can excite unwanted higher-order modes or cause shot-to-shot variation of the lowest-order mode (B. Miao, Shrock, *et al.* 2022).

Recent advances in plasma waveguides have transformed the length scales over which relativistically intense laser–matter interactions can be studied and exploited. The demonstration of meter-scale, laser-generated waveguides extends high-intensity interaction lengths by orders of magnitude beyond those achieved just a few years ago, opening new possibilities for laser wakefield acceleration at substantially higher electron energies. Progress in controlling waveguide formation and structure is making it possible to tailor the plasma environment to the demands of a particular interaction. These developments will benefit not only particle acceleration, but also nonlinear frequency conversion, short-wavelength and terahertz radiation generation, and other processes that require sustained, intense optical fields. The convergence of long interaction lengths and greater control over the guiding structure is opening a new regime of relativistic nonlinear optics, with implications for fundamental physics and for advanced sources of radiation and energetic particles.

## ACKNOWLEDGMENTS

The authors thank Linus Feder, Andrew Goffin, Scott Hancock, Matthew Le, Frederica Liu, Lucas Railing, Ari Sloss, Nishchal Tripathi, Mayank Gupta, and Marly Young for useful discussions and technical assistance, Jens Osterhoff for providing valuable technical feedback, and Mike Downer for suggesting this article. This work was supported by the U.S. Department of Energy (DE-SC0015516 and LaserNetUS DE-SC0019076/FWP#SCW1668), the National Science Foundation (PHY2010511), and the Defense Advanced Research Projects Agency (DARPA) under the Muons for Science and Security Program. Simulations at Maryland used DoD high performance computing support provided through the Office of Naval Research (N00014-20-1-2233) and the Air Force Office of Scientific Research (FA9550-21-1-0405). WarpX is primarily funded by the US DOE Exascale Computing Project, with contributors at LBNL, LLNL, CEA-LIDYL, SLAC, DESY, CERN, and TAE Technologies.

## REFERENCES

Abuazoum, S., S. M. Wiggins, R. C. Issac, G. H. Welsh, G. Vieux, M. Ganciu, and D. A. Jaroszynski, 2011, "A High Voltage Pulsed Power Supply for Capillary Discharge Waveguide Applications," Rev Sci Instrum **82**, 063505.

Agrawal, Govind, 2019, *Nonlinear Fiber Optics* (Academic Press).

Albert, F., N. Lemos, J. L. Shaw, B. B. Pollock, C. Goyon, W. Schumaker, A. M. Saunders, et al., 2017, "Observation of Betatron X-Ray Radiation in a Self-Modulated Laser Wakefield Accelerator Driven with Picosecond Laser Pulses," Phys Rev Lett (American Physical Society) **118**, 134801.

Albert, F., A. G.R. Thomas, S. P.D. Mangles, S. Banerjee, S. Corde, A. Flacco, M. Litos, et al., 2014, "Laser Wakefield Accelerator Based Light Sources: Potential Applications and Requirements," Plasma Phys. Control. Fusion **56**, 084015.

Alcock, A. J., and S. A. Ramsden, 1966, "Two Wavelength Interferometry of a Laser-Induced Spark in Air," Appl. Phys. Lett. **8**, 187–188.

Alejo, A., J. Cowley, A. Picksley, R. Walczak, and S. M. Hooker, 2022, "Demonstration of Kilohertz Operation of Hydrodynamic Optical-Field-Ionized Plasma Channels," Phys. Rev. Accel. Beams (American Physical Society) **25**, 011301.

Alexeev, I., T. M. Antonsen, K. Y. Kim, and H. M. Milchberg, 2003, "Self-Focusing of Intense Laser Pulses in a Clustered Gas," Phys Rev Lett (American Physical Society) **90**, 103402.

Ambat, M. V., J. L. Shaw, J. J. Pigeon, K. G. Miller, T. T. Simpson, D. H. Froula, and J. P. Palastro, 2023, "Programmable-Trajectory Ultrafast Flying Focus Pulses," Opt. Express (Optica Publishing Group) **31**, 31354–31368.

Amherd, Noel A., and George C. Vlases, 1974, "Trapping and Absorption of an Axially Directed CO2 Laser Beam by a θ-Pinch Plasma," Appl. Phys. Lett. **24**, 93–95.

Amodio, Alessio, Dan Wang, Curtis Berger, Hai-En Tsai, Samuel K. Barber, Jeroen van Tilborg, Alexander Picksley, et al., 2025, "Pointing Stabilization of a 1 Hz High-Power Laser via Machine Learning," High Power Laser Sci. Eng. **13**, e35.

Andreev, Nikolay E., Brigitte Cros, Gilles Maynard, Patrick Mora, and Franck Wojda, 2008, "Coupling Efficiency of Intense Laser Pulses to Capillary Tubes for Laser Wakefield Acceleration," IEEE Trans. Plasma Sci. **36**, 1746–1750.

Aniculaesei, C., Hyung Taek Kim, Byung Ju Yoo, Kyung Hwan Oh, and Chang Hee Nam, 2018, "Novel Gas Target for Laser Wakefield Accelerators," Rev. Sci. Instrum. **89**, 1–5.

Aniculaesei, Constantin, Thanh Ha, Samuel Yoffe, Lance Labun, Stephen Milton, Edward McCary, Michael M. Spinks, et al., 2023, "The Acceleration of a High-Charge Electron Bunch to 10 GeV in a 10-Cm Nanoparticle-Assisted Wakefield Accelerator," Matter Radiat. Extrem. **9**, 014001.

Antonsen, Thomas M., Jr., John Palastro, and Howard M. Milchberg, 2007, "Excitation of Terahertz Radiation by Laser Pulses in Nonuniform Plasma Channels," Phys. Plasmas **14**, 033107.

Antonsen, Thomas M., and Patrick Mora, 1995, "Leaky Channel Stabilization of Intense Laser Pulses in Tenuous Plasmas," Phys. Rev. Lett. **74**, 4440–4443.

Arber, T. D., K. Bennett, C. S. Brady, A. Lawrence-Douglas, M. G. Ramsay, N. J. Sircombe, P. Gillies, et al., 2015, "Contemporary Particle-in-Cell Approach to Laser-Plasma Modelling," Plasma Phys. Control. Fusion (IOP Publishing) **57**, 113001.

Arrowsmith, C. D., K. G. Miller, M. V. Ambat, S.-W. Bahk, I. A. Begishev, J. Bromage, S. Bucht, et al., 2026, "Dephasingless Laser Wakefield Acceleration of Electrons Using a Flying Focus," Nat. Phys. (Nature Publishing Group) **22**, 1513–1518.

Askar'yan, G A, and N M Tarasova, 1974, "Passage of Accelerated Particles and Quanta through a Medium along a Reduced-Density Channel Produced by a Laser Beam," JTEP Lett **20**.

Audet, T. L., P. Lee, G. Maynard, S. Dobosz Dufrénoy, A. Maitrallain, M. Bougeard, P. Monot, and B. Cros, 2018, "Gas Cell Density Characterization for Laser Wakefield Acceleration," Nucl. Instrum. Methods Phys. Res. Sect. Accel. Spectrometers Detect. Assoc. Equip. (North-Holland) **909**, 383–386.

Augst, S., D. Strickland, D. D. Meyerhofer, S. L. Chin, and J. H. Eberly, 1989, "Tunneling Ionization of Noble Gases in a High-Intensity Laser Field," Phys. Rev. Lett. **63**, 2212–2215.

Baker, K. L., J. Brase, M. Kartz, S. S. Olivier, B. Sawvel, and J. Tucker, 2002, "The Use of a Shack-Hartmann Wave Front Sensor for Electron Density Characterization of High Density Plasmas," Rev. Sci. Instrum. **73**, 3784.

Benedetti, C., F. Rossi, C. B. Schroeder, E. Esarey, and W. P. Leemans, 2015, "Pulse Evolution and Plasma-Wave Phase Velocity in Channel-Guided Laser-Plasma Accelerators," Phys. Rev. E (American Physical Society) **92**, 023109.

Benedetti, C., C. B. Schroeder, E. Esarey, and W. P. Leemans, 2012, "Quasi-Matched Propagation of Ultra-Short, Intense Laser Pulses in Plasma Channels," Phys. Plasmas **19**, 053101.

Bergé, L., S. Skupin, R. Nuter, J. Kasparian, and J. P. Wolf, 2007, "Ultrashort Filaments of Light in Weakly Ionized, Optically Transparent Media," Rep. Prog. Phys. (IOP Publishing) **70**, 1633.

Birdsall, C. K., and A. D. Langdon, 1991, *Plasma Physics via Computer Simulation* (CRC Press, Boca Raton).

Bobrova, N. A., S. V. Bulanov, A. A. Esaulov, and P. V. Sasorov, 2000, "Capillary Discharges for Guiding of Laser Pulses," Plasma Phys. Rep. **26**, 10–20.

Bobrova, N. A., S. V. Bulanov, D. Farina, R. Pozzoli, T. L. Razinkova, J. I. Sakai, P. V. Sasorov, and I. V. Sokolov, 2000, "MHD Simulations of Plasma Dynamics in Pinch Discharges in Capillary Plasmas," Laser Part. Beams **18**, 623–638.

Bobrova, N. A., S. V. Bulanov, D. Farina, R. Pozzoli, T. L. Razinkova, J. Sakai, and P. V. Sasorov, 1998, "Dissipative MHD Simulation of Capillary Plasmas for Guiding of Intense Ultrashort Laser Pulses," J Phys Soc Jpn **67**, 3437–3442.

Bobrova, N. A., A. A. Esaulov, J. I. Sakai, P. V. Sasorov, D. J. Spence, A. Butler, S. M. Hooker, and S. V. Bulanov, 2002, "Simulations of a Hydrogen-Filled Capillary Discharge Waveguide," Phys. Rev. E - Stat. Phys. Plasmas Fluids Relat. Interdiscip. Top. **65**, 1–11.

Bobrova, N. A., P. V. Sasorov, C. Benedetti, S. S. Bulanov, C. G.R. Geddes, C. B. Schroeder, E. Esarey, and W. P. Leemans, 2013, "Laser-Heater Assisted Plasma Channel Formation in Capillary Discharge Waveguides," Phys. Plasmas **20**, 020703.

Borghesi, M., A. J. Mackinnon, R. Gaillard, O. Willi, and A. A. Offenberger, 1998, "Guiding of a 10-TW Picosecond Laser Pulse through Hollow Capillary Tubes," Phys. Rev. E - Stat. Phys. Plasmas Fluids Relat. Interdiscip. Top. **57**, 4899–4902.

Boyd, Robert, 2008, *Nonlinear Optics* (Elsevier).

Brabec, Thomas, and Ferenc Krausz, 2000, "Intense Few-Cycle Laser Fields: Frontiers of Nonlinear Optics," Rev. Mod. Phys. **72**, 545–591.

Braginskii, S. I., 1965, "Transport Processes in a Plasma," in *Reviews of Plasma Physics, Volume 1*, edited by M. A. Leontovich (Consultants Bureau, New York), p. 205.

Brandi, F., F. Giammanco, F. Conti, F. Sylla, G. Lambert, and L. A. Gizzi, 2016, "Note: Real-Time Monitoring via Second-Harmonic Interferometry of a Flow Gas Cell for Laser Wakefield Acceleration," Rev. Sci. Instrum. **87**, 4–7.

Brill, B., B. Arad, M. Kishenevsky, A. Ludmirsky, and A. Zigler, 1990, "Density Measurement of Dense Capillary Discharge Plasma Using Soft X-Ray Backlighting," J. Phys. Appl. Phys. **23**, 1064–1068.

Broks, B. H. P., W. Van Dijk, J. J. A. W. van der Mullen, A. J. Gonsalves, T. P. Rowlands-Rees, and S. M. Hooker, 2007, "Modeling of a Square Pulsed Capillary Discharge Waveguide for Interferometry Measurements," Phys. Plasmas **14**, 023501.

Broks, B. H.P., K. Garloff, and J. J.A.M. Van Der Mullen, 2005, "Nonlocal-Thermal-Equilibrium Model of a Pulsed Capillary Discharge Waveguide," Phys. Rev. E - Stat. Nonlinear Soft Matter Phys. **71**, 1–10.

Broks, B. H.P., W. Van Dijk, and J. J.A.M. Van Der Mullen, 2006, "Parameter Study of a Pulsed Capillary Discharge Waveguide," J. Phys. Appl. Phys. **39**, 2377–2383.

Bulanov, S., N. Naumova, F. Pegoraro, and J. Sakai, 1998, “Particle Injection into the Wave Acceleration Phase Due to Nonlinear Wake Wave Breaking,” Phys. Rev. E - Stat. Phys. Plasmas Fluids Relat. Interdiscip. Top. **58**, R5257–R5260.

Butler, A., A. J. Gonsalves, C. M. McKenna, D. J. Spence, S. M. Hooker, S. Sebban, T. Mocek, I. Bettaibi, and B. Cros, 2003, “Demonstration of a Collisionally Excited Optical-Field-Ionization Xuv Laser Driven in a Plasma Waveguide,” Phys. Rev. Lett. **91**, 1–4.

Butler, A., D. J. Spence, and S. M. Hooker, 2002, “Guiding of High-Intensity Laser Pulses with a Hydrogen-Filled Capillary Discharge Waveguide,” Phys. Rev. Lett. **89**, 185003.

C. Caizergues, S. Smartsev, V. Malka, and C. Thaury, 2020, “Phase-Locked Laser-Wakefield Electron Acceleration,” Nat. Photonics **14**, 475.

Canagasabey, Albert, Costantino Corbari, Alexey V Gladyshev, Flavien Liegeois, Sebastien Guillemet, Yves Hernandez, Mikhail V Yashkov, et al., 2009, “High-Average-Power Second-Harmonic Generation from Periodically Poled Silica Fibers,” Opt. Lett. **34**, 2483–2485.

Chen, M., E. Esarey, C. B. Schroeder, C. G.R. Geddes, and W. P. Leemans, 2012, “Theory of Ionization-Induced Trapping in Laser-Plasma Accelerators,” Phys. Plasmas **19**, 033101.

Chen, Y. H., S. Varma, A. York, and H. M. Milchberg, 2007, “Single-Shot, Space- and Time-Resolved Measurement of Rotational Wavepacket Revivals in H2, D2, N2, O2, and N2O,” Opt. Express **15**, 11341–11357.

Ching, E. S. C., P. T. Leung, A. Maassen van den Brink, W. M. Suen, S. S. Tong, and K. Young, 1998, “Quasinormal-Mode Expansion for Waves in Open Systems,” Rev. Mod. Phys. (American Physical Society) **70**, 1545–1554.

Chou, M.-C., P.-H. Lin, C.-A. Lin, J.-Y. Lin, J. Wang, and S.-Y. Chen, 2007, “Dramatic Enhancement of Optical-Field-Ionization Collisional-Excitation X-Ray Lasing by an Optically Preformed Plasma Waveguide,” Phys. Rev. Lett. (American Physical Society) **99**, 063904.

Chu, T. K., and L. C. Johnson, 1975, “Measurement of the Development and Evolution of Shock Waves in a Laser-Induced Gas Breakdown Plasma,” Phys. Fluids **18**, 1460–1466.

Čižmár, Tomáš, and Kishan Dholakia, 2009, “Tunable Bessel Light Modes: Engineering the Axial Propagation,” Opt. Express (Optica Publishing Group) **17**, 15558–15570.

Čižmár, Tomáš, Michael Mazilu, and Kishan Dholakia, 2010, “In Situ Wavefront Correction and Its Application to Micromanipulation,” Nat. Photonics **4**, 388–394.

Clark, T. R., and H. M. Milchberg, 1997, “Time- and Space-Resolved Density Evolution of the Plasma Waveguide,” Phys. Rev. Lett. **78**, 2373–2376.

———, 1998a, “Frequency Selective Tunnel Coupling to the Plasma Fiber,” Phys. Rev. Lett. **81**, 357–360.

———, 1998b, “Laser-Driven Implosion of a Cylindrical Plasma,” Phys. Rev. E - Stat. Phys. Plasmas Fluids Relat. Interdiscip. Top. **57**, 3417–3422.

———, 2000, “Optical Mode Structure of the Plasma Waveguide,” Phys. Rev. E **61**, 1954–1965.

Clark, Thomas R., Jr., 1998, “The Hydrodynamics and Optical Mode Structure of the Plasma Waveguide,” Ph.D. dissertation, (University of Maryland, College Park).

Clayton, C. E., J. E. Ralph, F. Albert, R. A. Fonseca, S. H. Glenzer, C. Joshi, W. Lu, et al., 2010, “Self-Guided Laser Wakefield Acceleration beyond 1 GeV Using Ionization-Induced Injection,” Phys. Rev. Lett. **105**, 105003.

*COMSOL Multiphysics* (COMSOL AB, Stockholm, Sweden), https://www.comsol.com/.

Cooley, J. H., J. Z. Wu, T. M. Antonsen, K. Y. Kim, I. Alexeev, J. Fan, and H. M. Milchberg, 2003, “Effective Coupling of Ultra-Intense Laser Pulses to Funnel-Mouthed Plasma Waveguides,” OSA Trends Opt. Photonics Ser. **88**, 1042–1043.

Corde, S., K. Ta Phuoc, G. Lambert, R. Fitour, V. Malka, A. Rousse, A. Beck, and E. Lefebvre, 2013, “Femtosecond x Rays from Laser-Plasma Accelerators,” Rev. Mod. Phys. (American Physical Society) **85**, 1–48.

Corkum, P. B., N. H. Burnett, and F. Brunel, 1989, “Above-Threshold Ionization in the Long-Wavelength Limit,” Phys Rev Lett (American Physical Society) **62**, 1259–1262.

Couairon, A., and A. Mysyrowicz, 2007, “Femtosecond Filamentation in Transparent Media,” Phys. Rep. (North-Holland) **441**, 47–189.

Cowie, L. L., and C. F. McKee, 1977, “The Evaporation of Spherical Clouds in a Hot Gas. I - Classical and Saturated Mass Loss Rates,” Astrophys. J. **211**, 135.

Cros, B., C. Courtois, G. Matthieussent, A. Di Bernardo, D. Batani, N. Andreev, and S. Kuznetsov, 2002, “Eigenmodes for Capillary Tubes with Dielectric Walls and Ultraintense Laser Pulse Guiding,” Phys. Rev. E - Stat. Phys. Plasmas Fluids Relat. Interdiscip. Top. **65**, 1–7.

Daniels, J., J. Van Tilborg, A. J. Gonsalves, C. B. Schroeder, C. Benedetti, E. Esarey, and W. P. Leemans, 2015, “Plasma Density Diagnostic for Capillary-Discharge Based Plasma Channels,” Phys. Plasmas (American Institute of Physics Inc.) **22**.

Davidson, N., A. A. Friesem, and E. Hasman, 1991, “Holographic Axilens: High Resolution and Long Focal Depth,” Opt. Lett. (Optica Publishing Group) **16**, 523–525.

Dawson, J. M., A Hertzberg, R. E. Kidder, H. G. Ahlstrom, and L. E. Steinhauer, 1971, *Plasma Physics and Controled Nuclear Fusion Research 1971*, edited by R. Schenin and J. W. Weil (Madison), p. 673.

Decker, C. D., W. B. Mori, K.-C. Tzeng, and T. Katsouleas, 1996, “The Evolution of Ultra-intense, Short-pulse Lasers in Underdense Plasmas,” Phys. Plasmas **3**, 2047–2056.

Depresseux, A., E. Oliva, J. Gautier, F. Tissandier, J. Nejdl, M. Kozlova, G. Maynard, et al., 2015, “Table-Top Femtosecond Soft X-Ray Laser by Collisional Ionization Gating,” Nat. Photonics (Nature Publishing Group) **9**, 817–821.

Dharmavarapu, Raghu, Shanti Bhattacharya, and Saulius Juodkazis, 2018, “Diffractive Optics for Axial Intensity Shaping of Bessel Beams,” J. Opt. U. K. (IOP Publishing) **20**.

Dimitrov, D. A., R. E. Giacone, D. L. Bruhwiler, R. Busby, J. R. Cary, C. G. R. Geddes, E. Esarey, and W. P. Leemans, 2007, “Coupling of Laser Energy into Plasma Channels,” Phys Plasmas **14**, 043105.

Ditmire, T., T. Donnelly, A. M. Rubenchik, R. W. Falcone, and M. D. Perry, 1996, “Interaction of Intense Laser Pulses with Atomic Clusters,” Phys. Rev. A (American Physical Society) **53**, 3379–3402.

Ditmire, T., R. A. Smith, and M. H. R. Hutchinson, 1998, “Plasma Waveguide Formation in Predissociated Clustering Gases,” Opt. Lett. **23**, 322.

Ditmire, T., R. A. Smith, J. W.G. Tisch, and M. H.R. Hutchinson, 1997, “High Intensity Laser Absorption by Gases of Atomic Clusters,” Phys. Rev. Lett. **78**, 3121–3124.

Djordjević, B. Z., C. Benedetti, C. B. Schroeder, E. Esarey, and W. P. Leemans, 2018, “Filtering Higher-Order Laser Modes Using Leaky Plasma Channels,” Phys. Plasmas **25**, 013103.

Döpp, A., E. Guillaume, C. Thaury, A. Lifschitz, K. Ta Phuoc, and V. Malka, 2016, “Energy Boost in Laser Wakefield Accelerators Using Sharp Density Transitions,” Phys. Plasmas **23**, 056702.

Dorchies, F., J. R. Marquès, B. Cros, G. Matthieussent, C. Courtois, T. Vélikoroussov, P. Audebert, et al., 1999, “Monomode Guiding of 1016W/Cm2 Laser Pulses over 100 Rayleigh Lengths in Hollow Capillary Dielectric Tubes,” Phys. Rev. Lett. **82**, 4655–4658.

Downer, M. C., R. Zgadzaj, A. Debus, U. Schramm, and M. C. Kaluza, 2018, “Diagnostics for Plasma-Based Electron Accelerators,” Rev. Mod. Phys. (American Physical Society) **90**, 35002.

Drake, R Paul, 2018, *High-Energy-Density Physics*, Graduate Texts in Physics (Springer International Publishing, Cham), http://link.springer.com/10.1007/978-3-319-67711-8.

Dribinski, Vladimir, Alexei Ossadtchi, Vladimir A. Mandelshtam, and Hanna Reisler, 2002, “Reconstruction of Abel-Transformable Images: The Gaussian Basis-Set Expansion Abel Transform Method,” Rev. Sci. Instrum. **73**, 2634.

Dudley, John M., Goëry Genty, and Stéphane Coen, 2006, “Supercontinuum Generation in Photonic Crystal Fiber,” Rev. Mod. Phys. **78**, 1135–1184.

Durfee, C. G., F. Lynch, and H. M. Milchberg, 1994, “Mode Properties of a Plasma Waveguide for Intense Laser Pulses,” Opt. Lett. **19**, 1937.

Durfee, C. G., J. Lynch, and H. M. Milchberg, 1995, “Development of a Plasma Waveguide for High-Intensity Laser Pulses,” Phys. Rev. E **51**, 2368–2389.

Durfee, C. G., and H. M. Milchberg, 1993, “Light Pipe for High Intensity Laser Pulses,” Phys. Rev. Lett. **71**, 2409–2412.

Durnin, J., 1987, “Exact Solutions for Nondiffracting Beams I The Scalar Theory,” J. Opt. Soc. Am. A **4**, 651.

Durnin, J., J. Miceli, and J. H. Eberly, 1987, “Diffraction-Free Beams,” Phys. Rev. Lett. **58**, 1499–1501.

Dyson, A. E., C. Thornton, and S. M. Hooker, 2016, “A Compact, Low Cost Marx Bank for Generating Capillary Discharge Plasmas,” Rev Sci Instrum **87**, 093302.

Ehrlich, Y., C. Cohen, D. Kaganovich, A. Zigler, R. F. Hubbard, P. Sprangle, and E. Esarey, 1998, “Guiding and Damping of High-Intensity Laser Pulses in Long Plasma Channels,” J. Opt. Soc. Am. B **15**, 2416.

Ehrlich, Y., C. Cohen, A. Zigler, J. Krall, P. Sprangle, and E. Esarey, 1996, “Guiding of High Intensity Laser Pulses in Straight and Curved Plasma Channel Experiments,” Phys. Rev. Lett. **77**, 4186–4189.

Erteza, Ireena A., 1995, “Diffraction Efficiency Analysis for Multi-Level Diffractive Optical Elements” (Sandia National Laboratories, Albuquerque, NM).

Esarey, E, and W P Leemans, 1999, “Nonparaxial Propagation of Ultrashort Laser Pulses in Plasma Channels,” Phys. Rev. E **59**, 1082–1095.

Esarey, E., C. B. Schroeder, and W. P. Leemans, 2009, “Physics of Laser-Driven Plasma-Based Electron Accelerators,” Rev. Mod. Phys. **81**, 1229–1285.

Esarey, E., C. B. Schroeder, B. A. Shadwick, J. S. Wurtele, and W. P. Leemans, 2000, “Nonlinear Theory of Nonparaxial Laser Pulse Propagation in Plasma Channels,” Phys. Rev. Lett. (American Physical Society) **84**, 3081–3084.

Esarey, E., P. Sprangle, J. Krall, A. Ting, and G. Joyce, 1993, “Optically Guided Laser Wake-Field Acceleration,” Phys Fluids B **5**, 2690–2697.

Esarey, Eric, Jonathan Krall, and Phillip Sprangle, 1994, “Envelope Analysis of Intense Laser Pulse Self-Modulation in Plasmas,” Phys. Rev. Lett. **72**, 2887–2890.

Fabbro, Rémy, Claire Max, and Edouard Fabre, 1985, “Planar Laser-driven Ablation: Effect of Inhibited Electron Thermal Conduction,” Phys. Fluids **28**, 1463–1481.

Fan, J., T. R. Clark, and H. M. Milchberg, 1998, “Generation of a Plasma Waveguide in an Elongated, High Repetition Rate Gas Jet,” Appl. Phys. Lett. **73**, 3064–3066.

Fan, J., E. Parra, I. Alexeev, K. Y. Kim, H. M. Milchberg, L. Ya Margolin, and L. N. Pyatnitskii, 2000, “Tubular Plasma Generation with a High-Power Hollow Bessel Beam,” Phys. Rev. E - Stat. Phys. Plasmas Fluids Relat. Interdiscip. Top. **62**, 7603–7606.

Fan, J., E. Parra, K. Y. Kim, I. Alexeev, H. M. Milchberg, J. Cooley, and T. M. Antonsen, 2002, “Resonant Self-Trapping of High Intensity Bessel Beams in Underdense Plasmas,” Phys. Rev. E **65**, 056408.

Fan, J., E. Parra, and H. M. Milchberg, 2000, "Resonant Self-Trapping and Absorption of Intense Bessel Beams," Phys. Rev. Lett. **84**, 3085–3088.

Fauser, C., and H. Langhoff, 2000, "Focussing of Laser Beams by Means of a Z-Pinch Formed Plasma Guiding System," Appl. Phys. B Lasers Opt. **71**, 607–609.

Feder, L., B. Miao, J. E. Shrock, A. Goffin, and H. M. Milchberg, 2020, "Self-Waveguiding of Relativistic Laser Pulses in Neutral Gas Channels," Phys. Rev. Res. (American Physical Society) **2**, 43173.

Feder, Linus, 2021, "Laser Wakefield Acceleration in Optical Field Ionized Plasma Waveguides," http://hdl.handle.net/1903/28898.

Fonseca, R. A., L. O. Silva, F. S. Tsung, V. K. Decyk, W. Lu, C. Ren, W. B. Mori, et al., 2002, "OSIRIS: A Three-Dimensional, Fully Relativistic Particle in Cell Code for Modeling Plasma Based Accelerators," in *Computational Science — ICCS 2002*, edited by Peter M. A. Sloot, Alfons G. Hoekstra, C. J. Kenneth Tan, and Jack J. Dongarra (Springer Berlin Heidelberg, Berlin, Heidelberg), pp. 342–351.

Friberg, Ari T., 1996, "Stationary-Phase Analysis of Generalized Axicons," JOSA A (Optica Publishing Group) **13**, 743–750.

Froula, Dustin H., David Turnbull, Andrew S. Davies, Terrance J. Kessler, Dan Haberberger, John P. Palastro, Seung-Whan Bahk, et al., 2018, "Spatiotemporal Control of Laser Intensity," Nat. Photonics (Nature Publishing Group) **12**, 262–265.

Fryxell, B., K. Olson, P. Ricker, F. X. Timmes, M. Zingale, D. Q. Lamb, P. MacNeice, R. Rosner, J. W. Truran, and H. Tufo, 2000, "FLASH: An Adaptive Mesh Hydrodynamics Code for Modeling Astrophysical Thermonuclear Flashes," Astrophys. J. Suppl. Ser. (The American Astronomical Society) **131**, 273–334.

Gaul, E. W., S. P. Le Blanc, A. R. Rundquist, R. Zgadzaj, H. Langhoff, and M. C. Downer, 2000, "Production and Characterization of a Fully Ionized He Plasma Channel," Appl. Phys. Lett. **77**, 4112–4114.

Geddes, C. G.R., K. Nakamura, G. R. Plateau, Cs Toth, E. Cormier-Michel, E. Esarey, C. B. Schroeder, J. R. Cary, and W. P. Leemans, 2008, "Plasma-Density-Gradient Injection of Low Absolute-Momentum-Spread Electron Bunches," Phys. Rev. Lett. **100**, 215004.

Geddes, C. G.R., Cs. Toth, J. Van Tilborg, E. Esarey, C. B. Schroeder, D. L. Bruhwiler, C. Nieter, J. Cary, and W. P. Leemans, 2004, "High-Quality Electron Beams from a Laser Wakefield Accelerator Using Plasma-Channel Guiding," Nature **431**, 2–5.

Geddes, C. G.R., Cs Tóth, J. Van Tilborg, E. Esarey, C. B. Schroeder, D. Bruhwiler, C. Nieter, J. Cary, and W. P. Leemans, 2005, "Production of High-Quality Electron Bunches by Dephasing and Beam Loading in Channeled and Unchanneled Laser Plasma Accelerators," Phys. Plasmas **12**, 1–10.

Geltner, I., Y. Ping, and S. Suckewer, 2003, “Guiding and Transmission of Ultrashort High-Intensity Laser Pulses in Discharge-Ablated LiF Microcapillaries,” J Opt Soc Am B **20**, 616–620.

Gessner, Spencer, Erik Adli, James M. Allen, Weiming An, Christine I. Clarke, Chris E. Clayton, Sebastien Corde, et al., 2016, “Demonstration of a Positron Beam-Driven Hollow Channel Plasma Wakefield Accelerator,” Nat. Commun. **7**, 5–10.

Gessner, Spencer J., 2016, “Demonstration of the Hollow Channel Plasma Wakefield Accelerator,” SLAC-R-1073, 1340170, http://www.osti.gov/servlets/purl/1340170/.

Giacone, R. E., D. A. Dimitrov, D. L. Bruhwiler, P. Messmer, J. R. Cary, C. G. R. Geddes, E. Esarey, and W. P. Leemans, 2005, *Simulations of Laser Pulse Coupling and Transmission Efficiency in Plasma Channels,* (IEEE, Knoxville, Tennessee), pp. 2179–2181, https://proceedings.jacow.org/p05/PAPERS/TPAE031.PDF.

Goffin, A, I Larkin, A Tartaro, A Schweinsberg, A Valenzuela, E W Rosenthal, and H M Milchberg, 2023, “Optical Guiding in 50-Meter-Scale Air Waveguides,” Phys Rev X (American Physical Society) **13**, 11006.

Goffin, A., A. Tartaro, and H. M. Milchberg, 2023, “Quasi-Steady-State Air Waveguide,” Optica (Optica Publishing Group) **10**, 505.

Golub, Ilya, Brahim Chebbi, Dagan Shaw, and Dariusz Nowacki, 2010, “Characterization of a Refractive Logarithmic Axicon,” Opt. Lett. **35**, 2828.

Gonsalves, A. J., F. Liu, N. A. Bobrova, P. V. Sasorov, C. Pieronek, J. Daniels, S. Antipov, et al., 2016, “Demonstration of a High Repetition Rate Capillary Discharge Waveguide,” J. Appl. Phys. **119**, http://dx.doi.org/10.1063/1.4940121.

Gonsalves, A. J., K. Nakamura, C. Benedetti, C. V. Pieronek, S. Steinke, J. H. Bin, S. S. Bulanov, et al., 2020, “Laser-Heated Capillary Discharge Plasma Waveguides for Electron Acceleration to 8 GeV,” Phys. Plasmas **27**, 053102.

Gonsalves, A. J., K. Nakamura, J. Daniels, C. Benedetti, C. Pieronek, T. C.H. De Raadt, S. Steinke, et al., 2019, “Petawatt Laser Guiding and Electron Beam Acceleration to 8 GeV in a Laser-Heated Capillary Discharge Waveguide,” Phys. Rev. Lett. (American Physical Society) **122**, 84801.

Gonsalves, A. J., K. Nakamura, J. Daniels, H. S. Mao, C. Benedetti, C. B. Schroeder, Cs Tóth, et al., 2015, “Generation and Pointing Stabilization of Multi-GeV Electron Beams from a Laser Plasma Accelerator Driven in a Pre-Formed Plasma Waveguide,” Phys. Plasmas **22**, http://dx.doi.org/10.1063/1.4919278.

Gonsalves, A. J., K. Nakamura, C. Lin, J. Osterhoff, S. Shiraishi, C. B. Schroeder, C. G. R. Geddes, Cs Tóth, E. Esarey, and W. P. Leemans, 2010, “Plasma Channel Diagnostic Based on Laser Centroid Oscillations,” Phys. Plasmas **17**, 056706.

Gonsalves, A. J., K. Nakamura, C. Lin, D. Panasenko, S. Shiraishi, T. Sokollik, C. Benedetti, et al., 2011, "Tunable Laser Plasma Accelerator Based on Longitudinal Density Tailoring," Nat. Phys. (Nature Publishing Group) **7**, 862–866.

Gonsalves, A. J., T. P. Rowlands-Rees, B. H. P. Broks, J. J. A. M. van der Mullen, and S. M. Hooker, 2007, "Transverse Interferometry of a Hydrogen-Filled Capillary Discharge Waveguide," Phys. Rev. Lett. (American Physical Society) **98**, 025002.

Gordon, D, P Sprangle, S Slinker, R Fernsler, and M Lampe, 2006, "SPARC—A Simulation Model for Electrical Charges," NRL Memo. Rep., 6706–6790.

Granzow, N, S P Stark, M A Schmidt, A S Tverjanovich, L Wondraczek, P StJ Russell, W J Wadsworth, et al., 2011, "Supercontinuum Generation in Chalcogenide-Silica Step-Index Fibers," Opt. Express **19**, 21003–21010.

Guénot, D., D. Gustas, A. Vernier, B. Beaurepaire, F. Böhle, M. Bocoum, M. Lozano, et al., 2017, "Relativistic Electron Beams Driven by kHz Single-Cycle Light Pulses," Nat. Photonics (Nature Publishing Group) **11**, 293–296.

Guillaume, E., A. Döpp, C. Thaury, K. Ta Phuoc, A. Lifschitz, G. Grittani, J. P. Goddet, et al., 2015, "Electron Rephasing in a Laser-Wakefield Accelerator," Phys. Rev. Lett. **115**, 155002.

Gusinow, M. A., J. B. Gerardo, and J. T. Verdeyen, 1966, "Investigation of Electronic Recombination in Helium and Argon Afterglow Plasmas by Means of Laser Interferometric Measurements," Phys. Rev. **149**, 91–96.

Gustafsson, Cornelia, Erik Löfquist, Kristoffer Svendsen, Andrea Angella, Anders Persson, and Olle Lundh, 2024, "Combined Plasma Lens and Rephasing Stage for a Laser Wakefield Accelerator," Sci. Rep. (Nature Publishing Group) **14**, 26286, https://www.nature.com/articles/s41598-024-78143-6.

Hagena, O. F., and W. Obert, 1972, "Cluster Formation in Expanding Supersonic Jets: Effect of Pressure, Temperature, Nozzle Size, and Test Gas," J. Chem. Phys. **56**, 1793–1802.

Heckl, O. H., C. R.E. Baer, C. Kränkel, S. V. Marchese, F. Schapper, M. Holler, T. Südmeyer, et al., 2009, "High Harmonic Generation in a Gas-Filled Hollow-Core Photonic Crystal Fiber," Appl. Phys. B Lasers Opt. **97**, 369–373.

Hine, G. A., A. J. Goers, L. Feder, J. A. Elle, S. J. Yoon, and H. M. Milchberg, 2016, "Generation of Axially Modulated Plasma Waveguides Using a Spatial Light Modulator," Opt. Lett. (Optica Publishing Group) **41**, 3427.

Hojbota, Calin Ioan, Mohammad Mirzaie, Do Yeon Kim, Tae Gyu Pak, Mohammad Rezaei-Pandari, Vishwa Bandhu Pathak, Jong Ho Jeon, et al., 2023, "High-Energy Betatron Source Driven by a 4-PW Laser with Applications to Non-Destructive Imaging," Eur. Phys. J. A **59**, 247.

Hooker, S. M., D. J. Spence, and R. A. Smith, 2000, "Guiding of High-Intensity Picosecond Laser Pulses in a Discharge-Ablated Capillary Waveguide," J. Opt. Soc. Am. B **17**, 90.

Hosokai, Tomonao, Masaki Kando, Hideki Dewa, Hideyuki Kotaki, Syuji Kondo, Noboru Hasegawa, Kazuhisa Nakajima, and Kazuhiko Horioka, 2000, “Optical Guidance of Terrawatt Laser Pulses by the Implosion Phase of a Fast Z-Pinch Discharge in a Gas-Filled Capillary,” Opt. Lett. **25**, 10.

Hosokai, Tomonao, Mitsuo Nakajima, Takayuki Aoki, Masao Ogawa, and Kazuhiko Horioka, 1997, “Correlation between Soft X-Ray Emission and Dynamics of Fast Capillary Discharges,” Jpn. J. Appl. Phys. Part 1 Regul. Pap. Short Notes Rev. Pap. **36**, 2327–2335.

Houard, Aurélien, Pierre Walch, Thomas Produit, Victor Moreno, Benoit Mahieu, Antonio Sunjerga, Clemens Herkommer, et al., 2023, “Laser-Guided Lightning,” Nat. Photonics (Nature Research) **17**, 231–235.

Houzet, J., N. Faure, M. Larochette, A.-C. Brulez, S. Benayoun, and C. Mauclair, 2016, “Ultrafast Laser Spatial Beam Shaping Based on Zernike Polynomials for Surface Processing,” Opt. Express **24**, 6542.

Huang, C., V. K. Decyk, C. Ren, M. Zhou, W. Lu, W. B. Mori, J. H. Cooley, T. M. Antonsen, and T. Katsouleas, 2006, “QUICKPIC: A Highly Efficient Particle-in-Cell Code for Modeling Wakefield Acceleration in Plasmas,” J. Comput. Phys. **217**, 658–679.

Hue, Céline S., Yang Wan, Eitan Y. Levine, and Victor Malka, 2023, “Control of Electron Beam Current, Charge, and Energy Spread Using Density Downramp Injection in Laser Wakefield Accelerators,” Matter Radiat. Extrem. (AIP Publishing, LLC) **8**, https://doi.org/10.1063/5.0126293.

Hutchinson, I. H., 2002, *Principles of Plasma Diagnostics* (Cambridge University Press), 2nd ed., https://www.cambridge.org/core/product/identifier/9780511613630/type/book.

Ibbotson, T. P.A., N. Bourgeois, T. P. Rowlands-Rees, L. S. Caballero, S. I. Bajlekov, P. A. Walker, S. Kneip, et al., 2010, “Laser-Wakefield Acceleration of Electron Beams in a Low Density Plasma Channel,” Phys. Rev. Spec. Top. - Accel. Beams **13**, 8–11.

ICUIL, 2024, “Https://Www.Icuil.Org/.”

Jackel, S., R. Burris, J. Grun, A. Ting, C. Manka, K. Evans, and J. Kosakowskii, 1995, “Channeling of Terawatt Laser Pulses by Use of Hollow Waveguides,” Opt. Lett. **20**, 1086.

Jaroszewicz, Zbigniew, Jacek Sochacki, Andrzej Kołodziejczyk, and Leszek Rafał Staronski, 1993, “Apodized Annular-Aperture Logarithmic Axicon: Smoothness and Uniformity of Intensity Distributions,” Opt. Lett. (Optica Publishing Group) **18**, 1893–1895.

Jhajj, N., I. Larkin, E. W. Rosenthal, S. Zahedpour, J. K. Wahlstrand, and H. M. Milchberg, 2016, “Spatiotemporal Optical Vortices,” Phys. Rev. X (American Physical Society) **6**.

Jhajj, N., E. W. Rosenthal, R. Birnbaum, J. K. Wahlstrand, and H. M. Milchberg, 2014, “Demonstration of Long-Lived High-Power Optical Waveguides in Air,” Phys. Rev. X **4**, 011027.

J.J. Pigeon, P. Franke, M. Lim Pac Chong, J. Katz, R. Boni, C. Dorrer, J.P. Palastro, and D.H. Froula, 2024, “Ultrabroadband Flying Focus Using an Axiparabola Echelon Pair,” Opt. Express **32**, 576.

Jones, T. G., A. Ting, D. Kaganovich, C. I. Moore, and P. Sprangle, 2003, “Spatially Resolved Interferometric Measurement of a Discharge Capillary Plasma Channel,” Phys. Plasmas **10**, 4504–4512.

Jourdain, N., U. Chaulagain, M. Havlík, D. Kramer, D. Kumar, I. Majerová, V. T. Tikhonchuk, G. Korn, and S. Weber, 2021, “The L4n Laser Beamline of the P3-Installation: Towards High-Repetition Rate High-Energy Density Physics at ELI-Beamlines,” Matter Radiat. Extrem. **6**.

Kaganovich, D., P. V. Sasorov, Y. Ehrlich, C. Cohen, and A. Zigler, 1997, “Investigations of Double Capillary Discharge Scheme for Production of Wave Guide in Plasma,” Appl. Phys. Lett. **71**, 2925–2927.

Kaganovich, D., A. Ting, C. I. Moore, A. Zigler, H. R. Burris, Y. Ehrlich, R. Hubbard, and P. Sprangle, 1999, “High Efficiency Guiding of Terawatt Subpicosecond Laser Pulses in a Capillary Discharge Plasma Channel,” Phys. Rev. E - Stat. Phys. Plasmas Fluids Relat. Interdiscip. Top. **59**, R4769–R4772.

Karsch, S., J. Osterhoff, A. Popp, T. P. Rowlands-Rees, Zs Major, M. Fuchs, B. Marx, et al., 2007, “GeV-Scale Electron Acceleration in a Gas-Filled Capillary Discharge Waveguide,” New J. Phys. **9**.

Katsouleas, T., 1986, “Physical Mechanisms in the Plasma Wake-Field Accelerator,” Phys. Rev. A **33**, 2056–2064, https://link.aps.org/doi/10.1103/PhysRevA.33.2056.

Kiani, Leily, Tong Zhou, Seung-Whan Bahk, Jake Bromage, David Bruhwiler, E. Michael Campbell, Zenghu Chang, et al., 2023, “High Average Power Ultrafast Laser Technologies for Driving Future Advanced Accelerators,” J. Instrum. (IOP Publishing) **18**, T08006.

Kim, H. T., K. H. Pae, H. J. Cha, I. J. Kim, T. J. Yu, J. H. Sung, S. K. Lee, T. M. Jeong, and J. Lee, 2013, “Enhancement of Electron Energy to the Multi-GeV Regime by a Dual-Stage Laser-Wakefield Accelerator Pumped by Petawatt Laser Pulses,” Phys Rev Lett **111**, 165002.

Kim, H. T., V. B. Pathak, K. H. Pae, A. Lifschitz, F. Sylla, J. H. Shin, C. Hojbota, et al., 2017, “Stable Multi-GeV Electron Accelerator Driven by Waveform-Controlled PW Laser Pulses,” Sci Rep **7**, 10203.

Kim, J., V. L. J. Phung, K. Roh, M. Kim, K. Kang, and H. Suk, 2021, “Development of a Density-Tapered Capillary Gas Cell for Laser Wakefield Acceleration,” Rev. Sci. Instrum. **92**, 023511, https://doi.org/10.1063/5.0009632.

Kim, Jaehoon, Kim Geun Ju, and Seung Hoon Yoo, 2011, “Energy Enhancement Using an Upward Density Ramp in Laser Wakefield Acceleration,” J. Korean Phys. Soc. (The Korean Physical Society) **59**, 3166–3170,

https://www.kci.go.kr/kciportal/ci/sereArticleSearch/ciSereArtiView.kci?sereArticleSearchBean.artiId=ART001602714.

Kim, K. Y., I. Alexeev, J. Fan, E. Parra, and H. M. Milchberg, 2002, “Plasma Waveguides: Addition of End Funnels and Generation in Clustered Gases,” AIP Conf. Proc. **647**, 646–653.

Kim, K. Y., I. Alexeev, E. Parra, and H. M. Milchberg, 2003, “Time-Resolved Explosion of Intense-Laser-Heated Clusters,” Phys Rev Lett (American Physical Society) **90**, 023401.

Kim, Ki-Yong, 2003, “Measurement of Ultrafast Dynamics in the Interaction of Intense Laser Pulses with Gases, Atomic Clusters, and Plasmas,” *ProQuest Dissertations and Theses*, Ph.D., (University of Maryland, College Park, United States -- Maryland), https://www.proquest.com/dissertations-theses/measurement-ultrafast-dynamics-interaction/docview/305321403/se-2?accountid=14696.

Kim, M. S., D. G. Jang, T. H. Lee, I. H. Nam, I. W. Lee, and H. Suk, 2013, “Characteristics of a Tapered Capillary Plasma Waveguide for Laser Wakefield Acceleration,” Appl Phys Lett **102**, 204103.

Kimura, W. D., H. M. Milchberg, P. Muggli, X. Li, and W. B. Mori, 2011, “Hollow Plasma Channel for Positron Plasma Wakefield Acceleration,” Phys. Rev. Spec. Top. - Accel. Beams **14**, 1–11.

Kiselev, S., A. Pukhov, and I. Kostyukov, 2004, “X-Ray Generation in Strongly Nonlinear Plasma Waves,” Phys. Rev. Lett. **93**, 1–4.

Kitagawa, Yoneyoshi, Yasuhiko Sentoku, Shin Akamatsu, Wataru Sakamoto, Ryosuke Kodama, Kazuo A. Tanaka, Ken Azumi, et al., 2004, “Electron Acceleration in an Ultraintense-Laser-Illuminated Capillary,” Phys. Rev. Lett. **92**, 18–21.

Kling, Matthias F, Carmen S Menoni, Cameron Geddes, Almantas Galvanauskas, Felicie Albert, Leily Kiani, Michael Chini, et al., 2024, “Roadmap on Basic Research Needs for Laser Technology,” J. Opt. (IOP Publishing) **27**, 013002.

Kolesik, M., and J. V. Moloney, 2004, “Nonlinear Optical Pulse Propagation Simulation: From Maxwell’s to Unidirectional Equations,” Phys. Rev. E - Stat. Phys. Plasmas Fluids Relat. Interdiscip. Top. **70**, 036604.

Krall, J., E. Esarey, P. Sprangle, and G. Joyce, 1994, “Propagation of Radius-Tailored Laser Pulses over Extended Distances in a Uniform Plasma,” Phys. Plasmas **1**, 1738–1743.

Kramida, A., Yu. Ralchenko, J. Reader, and NIST ASD Team, 2024, “NIST Atomic Spectra Database (Version 5.12),” https://nist.gov.

Krausz, Ferenc, and Misha Ivanov, 2009, “Attosecond Physics,” Rev. Mod. Phys. **81**, 163–234.

Kumarappan, V., K. Y. Kim, and H. M. Milchberg, 2005, “Guiding of Intense Laser Pulses in Plasma Waveguides Produced from Efficient, Femtosecond End-Pumped Heating of Clustered Gases,” Phys. Rev. Lett. **94**, 205004.

Lahaye, R., I. A. Andriyash, J. Gautier, O. Kononenko, A. Leblanc, J.-P. Goddet, A. Tafzi, and C. Thaury, 2026, “Experimental Demonstration of Dephasing Reduction in an Optically Guided Laser-Plasma Accelerator,” Phys Rev Res. **8**, 023274.

Lahaye, R., K. Oubrerie, O. Kononenko, J. Gautier, I. A. Andriyash, and C. Thaury, 2025, “Quasimonoenergetic Multi-GeV Electron Acceleration in a Plasma Waveguide,” Phys Rev Accel Beams **28**, 091301.

Layer, B. D., A. York, T. M. Antonsen, S. Varma, Y. H. Chen, Y. Leng, and H. M. Milchberg, 2007, “Ultrahigh-Intensity Optical Slow-Wave Structure,” Phys. Rev. Lett. **99**, 035001.

Layer, B. D., A. G. York, S. Varma, Y.-H. Chen, and H. M. Milchberg, 2009, “Periodic Index-Modulated Plasma Waveguide,” Opt. Express **17**, 4263–4267.

Lazzarini, C. M., G. M. Grittani, P. Valenta, I. Zymak, R. Antipenkov, U. Chaulagain, L. V. N. Goncalves, et al., 2023, “50 MeV Electron Beams Accelerated by a Terawatt Scalable kHz Laser,” Phys. Plasmas **31**, 030703–9.

Le, M. S., G. A. Hine, A. Goffin, J. P. Palastro, and H. M. Milchberg, 2024, “Self-Focused Pulse Propagation Is Mediated by Spatiotemporal Optical Vortices,” Phys. Rev. Lett. (American Physical Society) **133**.

Leemans, W. P., A. J. Gonsalves, H. S. Mao, K. Nakamura, C. Benedetti, C. B. Schroeder, Cs Tóth, et al., 2014, “Multi-Gev Electron Beams from Capillary-Discharge-Guided Subpetawatt Laser Pulses in the Self-Trapping Regime,” Phys. Rev. Lett. **113**, 245002.

Leemans, W. P., B. Nagler, A. J. Gonsalves, Cs Tóth, K. Nakamura, C. G. R. Geddes, E. Esarey, C. B. Schroeder, and S. M. Hooker, 2006, “GeV Electron Beams from a Centimetre-Scale Accelerator,” Nat. Phys. **2**, 696–699.

Lehe, R., B. Miao, J. E. Shrock, and B. Hidding, 2025, “Summary of Working Group 1: Laser-Driven Plasma Wakefield Acceleration,” Nucl. Instrum. Methods Phys. Res. Sect. Accel. Spectrometers Detect. Assoc. Equip. **1072**, 170133.

Lehe, Rémi, Manuel Kirchen, Igor A. Andriyash, Brendan B. Godfrey, and Jean Luc Vay, 2016, “A Spectral, Quasi-Cylindrical and Dispersion-Free Particle-In-Cell Algorithm,” Comput. Phys. Commun. (Elsevier B.V.) **203**, 66–82.

Lemos, N., L. Cardoso, J. Geada, G. Figueira, F. Albert, and J. M. Dias, 2018, “Guiding of Laser Pulses in Plasma Waveguides Created by Linearly-Polarized Femtosecond Laser Pulses,” Sci. Rep. (Springer US) **8**, 3165.

Lemos, N., T. Grismayer, L. Cardoso, G. Figueira, R. Issac, D. A. Jaroszynski, and J. M. Dias, 2013, “Plasma Expansion into a Waveguide Created by a Linearly Polarized Femtosecond Laser Pulse,” Phys. Plasmas **20**, 063102.

Lemos, N., T. Grismayer, L. Cardoso, J. Geada, G. Figueira, and J. M. Dias, 2013, “Effects of Laser Polarization in the Expansion of Plasma Waveguides,” Phys. Plasmas **20**.

Leung, P. T., S. Y. Liu, and K. Young, 1994, “Completeness and Orthogonality of Quasinormal Modes in Leaky Optical Cavities,” Phys. Rev. A **49**, 3057–3067.

Li, R., A. Picksley, C. Benedetti, F. Filippi, J. Stackhouse, L. Fan-Chiang, H. E. Tsai, et al., 2025, “Longitudinal Tapering in Gas Jets for Increased Efficiency of 10-GeV Class Laser Plasma Accelerators,” Rev. Sci. Instrum. **96**, 043306.

Li, X F, A L Huillier, M Ferray, L A Lompre, and G Mainfray, 1989, “Multiple-Harmonic Generation in Rare Gases at High Laser Intensity,” Phys. Rev. A **39**, 5751–5761.

Liberman, Aaron, Anton Golovanov, Slava Smartsev, Anda-Maria Talposi, Sheroy Tata, and Victor Malka, 2026, “First Electron Acceleration in a Tunable-Velocity Laser Wakefield,” Phys. Rev. Res. **8**, L022001.

Lin, Shao-Chi, 1954, “Cylindrical Shock Waves Produced by Instantaneous Energy Release,” J. Appl. Phys. (AIP Publishing) **25**, 54–57.

Lindstrøm, C. A., E. Adli, G. Boyle, R. Corsini, A. E. Dyson, W. Farabolini, S. M. Hooker, et al., 2018, “Emittance Preservation in an Aberration-Free Active Plasma Lens,” Phys. Rev. Lett. (American Physical Society) **121**, 194801.

Lindstrøm, Carl A., 2021, “Staging of Plasma-Wakefield Accelerators,” Phys. Rev. Accel. Beams (American Physical Society) **24**, 14801.

Liu, Jinjun, Edcel J. Salumbides, Urs Hollenstein, Jeroen C. J. Koelemeij, Kjeld S. E. Eikema, Wim Ubachs, and Frédéric Merkt, 2009, “Determination of the Ionization and Dissociation Energies of the Hydrogen Molecule,” J. Chem. Phys. **130**, 174306.

Lorenz, S., G. Grittani, E. Chacon-Golcher, C. M. Lazzarini, J. Limpouch, F. Nawaz, M. Nevrkla, L. Vilanova, and T. Levato, 2019, “Characterization of Supersonic and Subsonic Gas Targets for Laser Wakefield Electron Acceleration Experiments,” Matter Radiat. Extrem. **4**, 015401.

Lu, Haiyang, Mingwei Liu, Wentao Wang, Cheng Wang, Jiansheng Liu, Aihua Deng, Jiancai Xu, et al., 2011, “Laser Wakefield Acceleration of Electron Beams beyond 1 GeV from an Ablative Capillary Discharge Waveguide,” Appl. Phys. Lett. **99**, 091502.

Lu, W., M. Tzoufras, C. Joshi, F. S. Tsung, W. B. Mori, J. Vieira, R. A. Fonseca, and L. O. Silva, 2007, “Generating Multi-GeV Electron Bunches Using Single Stage Laser Wakefield Acceleration in a 3D Nonlinear Regime,” Phys. Rev. Spec. Top. - Accel. Beams **10**, 061301.

Ludwig, J. D., S. C. Wilks, A. J. Kemp, G. J. Williams, N. Lemos, E. Rockafellow, B. Miao, et al., 2025, “Laser Based 100 GeV Electron Acceleration Scheme for Muon Production,” Sci. Rep. **15**, 25902.

Luther, B. M., Y. Wang, M. C. Marconi, J. L.A. Chilla, M. A. Larotonda, and J. J. Rocca, 2004, “Guiding of Intense Laser Beams in Highly Ionized Plasma Columns Generated by a Fast Capillary Discharge,” Phys. Rev. Lett. **92**, 1–4.

MacFarlane, J. J., I. E. Golovkin, and P. R. Woodruff, 2006, "HELIOS-CR – A 1-D Radiation-Magnetohydrodynamics Code with Inline Atomic Kinetics Modeling," J. Quant. Spectrosc. Radiat. Transf. (Pergamon) **99**, 381–397.

Maier, Andreas R., Niels M. Delbos, Timo Eichner, Lars Hübner, Sören Jalas, Laurids Jeppe, Spencer W. Jolly, et al., 2020, "Decoding Sources of Energy Variability in a Laser-Plasma Accelerator," Phys. Rev. X **10**, 031039.

Maksimchuk, A., J. Nees, B. Hou, R. Anthony, J. Bae, F. Bayer, M. Burger, et al., 2025, "The ZEUS Multi-Petawatt Laser System," Phys. Plasmas **32**, 103107.

Marconi, M. C., C. H. Moreno, J. J. Rocca, V. N. Shlyaptsev, and A. L. Osterheld, 2000, "Dynamics of a Microcapillary Discharge Plasma Using a Soft X-Ray Laser Backlighter," Phys. Rev. E - Stat. Phys. Plasmas Fluids Relat. Interdiscip. Top. **62**, 7209–7218.

Marcuse, D., 1977, "Loss Analysis of Single-Mode Fiber Splices," Bell Syst. Tech. J. (John Wiley & Sons, Ltd) **56**, 703–718.

Matthews, D. L., P. L. Hagelstein, M. D. Rosen, M. J. Eckart, N. M. Ceglio, A. U. Hazi, H. Medecki, et al., 1985, "Demonstration of a Soft X-Ray Amplifier," Phys. Rev. Lett. (American Physical Society) **54**, 110–113.

McGloin, D., and K Dholakia, 2005, "Bessel Beams: Diffraction in a New Light," Contemp. Phys. **46**, 15–28.

McGuffey, C., A. G.R. Thomas, W. Schumaker, T. Matsuoka, V. Chvykov, F. J. Dollar, G. Kalintchenko, et al., 2010, "Ionization Induced Trapping in a Laser Wakefield Accelerator," Phys. Rev. Lett. **104**, 025004.

McLean, E. A., and S. A. Ramsden, 1965, "Optical Interferometric and Spectroscopic Measurements of Electron Density in a Plasma," Phys. Rev. **140**, A1122.

McLeod, John H., 1954, "The Axicon: A New Type of Optical Element," J. Opt. Soc. Am. **44**, 592.

McPherson, A., G. Gibson, H. Jara, U. Johann, T. S. Luk, I. A. McIntyre, K. Boyer, and C. K. Rhodes, 1987, "Studies of Multiphoton Production of Vacuum-Ultraviolet Radiation in the Rare Gases," JOSA B (Optica Publishing Group) **4**, 595–601.

Mewes, S. M., G. J. Boyle, A. Ferran Pousa, R. J. Shalloo, J. Osterhoff, C. Arran, L. Corner, R. Walczak, S. M. Hooker, and M. Thévenet, 2023, "Demonstration of Tunability of HOFI Waveguides via Start-to-End Simulations," Phys. Rev. Res. (American Physical Society) **5**, 033112.

Miao, B., 2020, "Laser Wakefield Accelerator Experiments: Coherent Injection Radiation and Optical Field Ionization Based Plasma Waveguides" (University of Maryland, College Park).

Miao, B., L. Feder, J. E. Shrock, A. Goffin, and H. M. Milchberg, 2020, "Optical Guiding in Meter-Scale Plasma Waveguides," Phys. Rev. Lett. **125**, 74801.

Miao, B., L. Feder, J. E. Shrock, and H. M. Milchberg, 2022, “Phase Front Retrieval and Correction of Bessel Beams,” Opt. Express **30**, 11360.

Miao, B., E. Rockafellow, J. E. Shrock, S. W. Hancock, D. Gordon, and H. M. Milchberg, 2024, “Benchmarking of Hydrodynamic Plasma Waveguides for Multi-GeV Laser-Driven Electron Acceleration,” Phys Rev Accel Beams (American Physical Society) **27**, 081302.

Miao, B., J. E. Shrock, L. Feder, R. C. Hollinger, J. Morrison, R. Nedbailo, A. Picksley, et al., 2022, “Multi-GeV Electron Bunches from an All-Optical Laser Wakefield Accelerator,” Phys. Rev. X (American Physical Society) **12**, 31038.

Miao, B., J. E. Shrock, E. Rockafellow, A. J. Sloss, and H. M. Milchberg, 2025, “Meter-Scale Supersonic Gas Jets for Multi-GeV Laser-Plasma Accelerators,” Rev. Sci. Instrum. (American Institute of Physics) **96**, 043003.

Miao, Bo, Jaron Shrock, and Howard Milchberg, 2023, “Electrons See the Guiding Light,” Phys. Today (American Institute of Physics) **76**, 54–55.

Milchberg, H. M., Y.-H. Chen, Y.-H. Cheng, N. Jhajj, J. P. Palastro, E. W. Rosenthal, S. Varma, J. K. Wahlstrand, and S. Zahedpour, 2014, “The Extreme Nonlinear Optics of Gases and Femtosecond Optical Filamentation,” Phys. Plasmas (AIP Publishing) **21**, 100901.

Milchberg, H. M., C. G. Durfee, and J. Lynch, 1995, “Application of a Plasma Waveguide to Soft-x-Ray Lasers,” J. Opt. Soc. Am. B (Optica Publishing Group) **12**, 731–737.

Milchberg, H. M., C. G. Durfee, and T. J. McIlrath, 1995, “High-Order Frequency Conversion in the Plasma Waveguide,” Phys. Rev. Lett. **75**, 2494–2497.

Milchberg, H. M., S. J. McNaught, and E. Parra, 2001, “Plasma Hydrodynamics of the Intense Laser-Cluster Interaction,” Phys. Rev. E - Stat. Phys. Plasmas Fluids Relat. Interdiscip. Top. **64**, 7.

Milchberg, H.M, K.Y Kim, V Kumarappan, B.D Layer, and H Sheng, 2006, “Clustered Gases as a Medium for Efficient Plasma Waveguide Generation,” Philos. Trans. R. Soc. Math. Phys. Eng. Sci. **364**, 647–661.

Miller, Kyle G., Jacob R. Pierce, Manfred V. Ambat, Jessica L. Shaw, Kale Weichman, Warren B. Mori, Dustin H. Froula, and John P. Palastro, 2023, “Dephasingless Laser Wakefield Acceleration in the Bubble Regime,” Sci. Rep. **13**, 21306.

Mlejnek, M, E M Wright, and J V Moloney, 1998, “Dynamic Spatial Replenishment of Femtosecond Pulses Propagating in Air,” Opt. Lett. **23**, 382–384.

Molen, G. M., M. Kristiansen, and M. O. Hagler, 1973, “CO2 Laser Beam Refraction in a Linear Discharge Plasma,” Appl. Phys. Lett. **23**, 601–603.

Montgomery Smith, L., Dennis R. Keefer, and S. I. Sudharsanan, 1988, “Abel Inversion Using Transform Techniques,” J. Quant. Spectrosc. Radiat. Transf. **39**, 367–373.

Mora, Patrick, and Jr. Antonsen Thomas M., 1997, “Kinetic Modeling of Intense, Short Laser Pulses Propagating in Tenuous Plasmas,” Phys. Plasmas **4**, 217–229.

Moreno, C. H., M. C. Marconi, V. N. Shlyaptsev, B. R. Benware, C. D. Macchietto, J. L.A. Chilla, J. J. Rocca, and A. L. Osterheld, 1998, “Two-Dimensional near-Field and Far-Field Imaging of a Ne-like Ar Capillary Discharge Table-Top Soft-x-Ray Laser,” Phys. Rev. - At. Mol. Opt. Phys. **58**, 1509–1514.

Morozov, A., A. Goltsov, Q. Chen, M. Scully, and S. Suckewer, 2018, “Ionization Assisted Self-Guiding of Femtosecond Laser Pulses,” Phys. Plasmas **25**, 053110.

Muraoka, K., and M. Maeda, 2000, *Laser-Aided Diagnostics of Plasmas and Gases* (Taylor & Francis Group, LLC, Boca Raton).

Nakamura, Kei, Hann-Shin Mao, Anthony J. Gonsalves, Henri Vincenti, Daniel E. Mittelberger, Joost Daniels, Arturo Magana, Csaba Toth, and Wim P. Leemans, 2017, “Diagnostics, Control and Performance Parameters for the BELLA High Repetition Rate Petawatt Class Laser,” IEEE J. Quantum Electron. **53**, 1–21.

Nam, Inhyuk, Minseok Kim, Myung Hoon Cho, Dogeun Jang, MinSup Hur, and Hyyong Suk, 2024, “Enhancement of Electron Energies in the Laser Wakefield Acceleration Using a Density Up-Ramp Capillary Gas-Cell,” Phys. Scr. **99**, 015603, https://iopscience.iop.org/article/10.1088/1402-4896/ad1452.

Nikitin, S. P., I. Alexeev, J. Fan, and H. M. Milchberg, 1999, “High Efficiency Coupling and Guiding of Intense Femtosecond Laser Pulses in Preformed Plasma Channels in an Elongated Gas Jet,” Phys. Rev. E - Stat. Phys. Plasmas Fluids Relat. Interdiscip. Top. **59**, R3839–R3842.

Nikitin, S. P., T. M. Antonsen, T. R. Clark, Yuelin Li, and H. M. Milchberg, 1997, “Guiding of Intense Femtosecond Pulses in Preformed Plasma Channels,” Opt. Lett. **22**, 1787.

Nisoli, M., S. De Silvestri, and O. Svelto, 1996, “Generation of High Energy 10 Fs Pulses by a New Pulse Compression Technique,” Appl. Phys. Lett. **68**, 2793–2795.

O’Shea, Donald C., Thomas J. Suleski, Alan D. Kathman, and Dennis W. Prather, 2009, *Diffractive Optics: Design, Fabrication, and Test*, *Diffractive Optics: Design, Fabrication, and Test* (SPIE Press, Bellingham, Washington).

Osterhoff, J., A. Popp, Zs Major, B. Marx, T. P. Rowlands-Rees, M. Fuchs, M. Geissler, et al., 2008, “Generation of Stable, Low-Divergence Electron Beams by Laser-Wakefield Acceleration in a Steady-State-Flow Gas Cell,” Phys. Rev. Lett. **101**, 085002.

Oubrerie, Kosta, Igor A Andriyash, Ronan Lahaye, Slava Smartsev, Victor Malka, and Cédric Thaury, 2022, “Axiparabola: A New Tool for High-Intensity Optics,” J. Opt. **24**, 045503.

Oubrerie, Kosta, Adrien Leblanc, Olena Kononenko, Ronan Lahaye, Igor A. Andriyash, Julien Gautier, Jean-Philippe Goddet, et al., 2022, “Controlled Acceleration of GeV Electron Beams in an All-Optical Plasma Waveguide,” Light Sci Appl (Springer US) **11**, 180.

Pak, A., K. A. Marsh, S. F. Martins, W. Lu, W. B. Mori, and C. Joshi, 2010, “Injection and Trapping of Tunnel-Ionized Electrons into Laser-Produced Wakes,” Phys. Rev. Lett. **104**, 025003.

Palastro, J. P., J. L. Shaw, P. Franke, D. Ramsey, T. T. Simpson, and D. H. Froula, 2020, “Dephasingless Laser Wakefield Acceleration,” Phys. Rev. Lett. (American Physical Society) **124**, 134802.

Picksley, A., A. Alejo, J. Cowley, N. Bourgeois, L. Corner, L. Feder, J. Holloway, et al., 2020, “Guiding of High-Intensity Laser Pulses in 100-Mm-Long Hydrodynamic Optical-Field-Ionized Plasma Channels,” Phys. Rev. Accel. Beams (American Physical Society) **23**, 81303.

Picksley, A., A. Alejo, R. J. Shalloo, C. Arran, A. Von Boetticher, L. Corner, J. A. Holloway, et al., 2020, “Meter-Scale Conditioned Hydrodynamic Optical-Field-Ionized Plasma Channels,” Phys. Rev. E **102**, 53201.

Picksley, A., J. Chappell, E. Archer, N. Bourgeois, J. Cowley, D. R. Emerson, L. Feder, et al., 2023, “All-Optical GeV Electron Bunch Generation in a Laser-Plasma Accelerator via Truncated-Channel Injection,” Phys. Rev. Lett. (American Physical Society) **131**, 245001.

Picksley, A., J. Stackhouse, C. Benedetti, K. Nakamura, H.-E. Tsai, R. Li, B. Miao, et al., 2024, “Matched Guiding and Controlled Injection in Dark-Current-Free, 10-GeV-Class, Channel-Guided Laser-Plasma Accelerators,” Phys. Rev. Lett. (American Physical Society) **133**, 255001.

Plateau, G. R., N. H. Matlis, C. G.R. Geddes, A. J. Gonsalves, S. Shiraishi, C. Lin, R. A. Van Mourik, and W. P. Leemans, 2010, “Wavefront-Sensor-Based Electron Density Measurements for Laser-Plasma Accelerators,” Rev. Sci. Instrum. **81**, 0833108.

Point, G., Y. Brelet, L. Arantchouk, J. Carbonnel, B. Prade, A. Mysyrowicz, and A. Houard, 2014, “Two-Color Interferometer for the Study of Laser Filamentation Triggered Electric Discharges in Air,” Rev Sci Instrum **85**, 123101.

Poling, Bruce E., John M. Prausnitz, and John P. O’Connell, 2020, *Properties of Gases and Liquids, Fifth Edition*, McGraw-Hill’s AccessEngineering (McGraw-Hill Education, New York, N.Y), Fifth edition.

Popov, Sergei Yu, Ari T. Friberg, Marko Honkanen, Jari Lautanen, Jari Turunen, and Bernd Schnabel, 1998, “Apodized Annular-Aperture Diffractive Axicons Fabricated by Continuous-Path-Control Electron Beam Lithography,” Opt. Commun. **154**, 359–367.

Pukhov, A., and I. Kostyukov, 2008, “Control of Laser-Wakefield Acceleration by the Plasma-Density Profile,” Phys. Rev. E **77**, 025401, https://link.aps.org/doi/10.1103/PhysRevE.77.025401.

Qin, Zhiyong, Wentao Li, Jiaqi Liu, Jiansheng Liu, Wentao Wang, Changhai Yu, Zhijun Zhang, et al., 2022, “Multi-GeV Cascaded Laser Wakefield Acceleration in a Hybrid Capillary Discharge Waveguide,” New J. Phys. (IOP Publishing) **24**, 073048.

Railing, L. M., M. S. Le, C. M. Lazzarini, and H. M. Milchberg, 2024, "Loss-Free Shaping of Few-Cycle Terawatt Laser Pulses," Opt Lett (Optica Publishing Group) **49**, 1433–1436.

Rainville, Alexander, Mathew Whittlesey, Christopher Pasquale, Yanwen Jing, Mingshu Chen, Siyun Chen, Hanzhang Pei, et al., 2024, "Near-Complete Extraction of Maximum Stored Energy from Large-Core Fibers Using Coherent Pulse Stacking Amplification of Femtosecond Pulses," Optica (Optica Publishing Group) **11**, 1540–1548.

Raĭzer, I͡U. P., 1997, *Gas Discharge Physics* (Springer, Berlin).

Regan, S. P., R. Epstein, V. N. Goncharov, I. V. Igumenshchev, D. Li, P. B. Radha, H. Sawada, et al., 2007, "Laser Absorption, Mass Ablation Rate, and Shock Heating in Direct-Drive Inertial Confinement Fusiona)," Phys. Plasmas **14**, 056305.

Rittershofer, W., C. B. Schroeder, E. Esarey, F. J. Grüner, and W. P. Leemans, 2010, "Tapered Plasma Channels to Phase-Lock Accelerating and Focusing Forces in Laser-Plasma Accelerators," Phys. Plasmas **17**, 063104, https://pubs.aip.org/pop/article/17/6/063104/262666/Tapered-plasma-channels-to-phase-lock-accelerating.

Rocca, J. J., 1999, "Table-Top Soft x-Ray Lasers," Rev. Sci. Instrum. **70**, 3799–3827.

Rocca, J. J., V. Shlyaptsev, F. G. Tomasel, O. D. Cortzar, D. Hartshorn, and J. L.A. Chilla, 1994, "Demonstration of a Discharge Pumped Table-Top Soft-x-Ray Laser," Phys. Rev. Lett. **73**, 2192–2195.

Rocca, J. J., F. G. Tomasel, M. C. Marconi, V. N. Shlyaptsev, J. L. A. Chilla, B. T. Szapiro, and G. Giudice, 1995, "Discharge-pumped Soft-x-ray Laser in Neon-like Argon," Phys. Plasmas **2**, 2547–2554.

Rockafellow, E, 2026, "Dephasing Mitigation of Multi-GeV Electron Beams in a Ramped Plasma Waveguide," *Indico Global (Indico)*, Accessed October 29, https://indico.global/event/14577/contributions/157916/.

Rockafellow, E., B. Miao, J. E. Shrock, A. Sloss, M. S. Le, S. W. Hancock, S. Zahedpour, et al., 2025, "Development of a High Charge 10 GeV Laser Electron Accelerator," Phys. Plasmas (American Institute of Physics) **32**, 053102.

Rockafellow, E, A J Sloss, J E Shrock, B Miao, N Tripathi, S W Hancock, D F Gordon, and H M Milchberg, 2026, "Low-Density Hydrodynamic Plasma Waveguides: Cladding Generation and Gas Jet Development," Plasma Phys. Control. Fusion (IOP Publishing) **68**, 095002.

Rosenthal, E. W., N. Jhajj, J. K. Wahlstrand, and H. M. Milchberg, 2014, "Collection of Remote Optical Signals by Air Waveguides," Optica (Optical Society of American (OSA)) **1**, 5–9.

Rousse, Antoine, Kim Ta Phuoc, Rahul Shah, Alexander Pukhov, Eric Lefebvre, Victor Malka, Sergey Kiselev, et al., 2004, "Production of a keV X-Ray Beam from Synchrotron Radiation in Relativistic Laser-Plasma Interaction," Phys. Rev. Lett. **93**, 1–4.

Rowlands-Rees, T. P., C. Kamperidis, S. Kneip, A. J. Gonsalves, S. P.D. Mangles, J. G. Gallacher, E. Brunetti, et al., 2008, “Laser-Driven Acceleration of Electrons in a Partially Ionized Plasma Channel,” Phys. Rev. Lett. **100**, 1–4.

Rundquist, Andy, Charles G Durfee, Zenghu Chang, Catherine Herne, Sterling Backus, Margaret M Murnane, and Henry C Kapteyn, 1998, “Phase-Matched Generation of Coherent Soft X-Rays,” Science **280**, 1412–1415.

Russell, Philip, 2003, “Photonic Crystal Fibers,” Science **299**, 358–362.

Sainte-Marie, A., O. Gobert, and F. Quéré, 2017, “Controlling the Velocity of Ultrashort Light Pulses in Vacuum through Spatio-Temporal Couplings,” Optica **4**, 1298–1304.

Salehi, F., M. Le, L. Railing, M. Kolesik, and H. M. Milchberg, 2021, “Laser-Accelerated, Low-Divergence 15-MeV Quasimonoenergetic Electron Bunches at 1 kHz,” Phys. Rev. X **11**, 021055.

Sartania, S., Z. Cheng, M. Lesner, G. Tempea, Ch Spielmann, F. Krausz, and K. Ferencz, 1997, “Generation of 0.1-TW, 5-Fs Optical Pulses at a 1 kHz Repetition Rate,” Opt. Lett. **22**, 1562–1564.

Schaffer, Chris B, André Brodeur, and Eric Mazur, 2001, “Measurement Science and Technology Laser-Induced Breakdown and Damage in Bulk Transparent Materials Induced by Tightly Focused Femtosecond Laser Pulses,” Meas Sci Technol **12**, 1784–1794.

Schmid, K., A. Buck, C. M.S. Sears, J. M. Mikhailova, R. Tautz, D. Herrmann, M. Geissler, F. Krausz, and L. Veisz, 2010, “Density-Transition Based Electron Injector for Laser Driven Wakefield Accelerators,” Phys. Rev. Spec. Top. - Accel. Beams **13**, 091301.

Schmidt, G, and W Horton, 1985, “Self-Focusing of Laser Beams in the Beat-Wave Accelerator,” Comments Plasma Phiscs Control. Fusion **9**, 85–90.

Schroeder, C. B., C. Benedetti, E. Esarey, J. van Tilborg, and W. P. Leemans, 2011, “Group Velocity and Pulse Lengthening of Mismatched Laser Pulses in Plasma Channels,” Phys. Plasmas **18**, 083103.

Schroeder, C. B., E. Esarey, C. G.R. Geddes, C. Benedetti, and W. P. Leemans, 2010, “Physics Considerations for Laser-Plasma Linear Colliders,” Phys. Rev. Spec. Top. - Accel. Beams **13**, 101301.

Shadwick, B. A., C. B. Schroeder, and E. Esarey, 2009, “Nonlinear Laser Energy Depletion in Laser-Plasma Accelerators,” Phys. Plasmas **16**, 056704.

Shalloo, R, 2018, “Hydrodynamic Optical-Field-Ionized Plasma Waveguides for Laser Plasma Accelerators,” PhD Thesis, (University of Oxford).

Shalloo, R. J., C. Arran, L. Corner, J. Holloway, J. Jonnerby, R. Walczak, H. M. Milchberg, and S. M. Hooker, 2018, “Hydrodynamic Optical-Field-Ionized Plasma Channels,” Phys. Rev. E (American Physical Society) **97**, 053203.

Shalloo, R. J., C. Arran, A. Picksley, A. Von Boetticher, L. Corner, J. Holloway, G. Hine, et al., 2019, "Low-Density Hydrodynamic Optical-Field-Ionized Plasma Channels Generated with an Axicon Lens," Phys. Rev. Accel. Beams (American Physical Society) **22**, 41302.

Shalloo, R. J., A. Ferran Pousa, M. Mewes, S. Jalas, M. Kirchen, R. D'Arcy, J. Osterhoff, K. Põder, and M. Thévenet, 2024, "Controlled Injection in a Laser Plasma Accelerator via an Optically Generated Waveguide Constriction," https://arxiv.org/abs/2410.15937.

Sheng, H., K. Y. Kim, V. Kumarappan, B. D. Layer, and H. M. Milchberg, 2005, "Plasma Waveguides Efficiently Generated by Bessel Beams in Elongated Cluster Gas Jets," Phys. Rev. E - Stat. Nonlinear Soft Matter Phys. **72**, 036411.

Shiraishi, S., C. Benedetti, A. J. Gonsalves, K. Nakamura, B. H. Shaw, T. Sokollik, J. Van Tilborg, et al., 2013, "Laser Red Shifting Based Characterization of Wakefield Excitation in a Laser-Plasma Accelerator," Phys. Plasmas **20**, 063103.

Shrock, J. E., 2023, "Multi-GeV Laser Wakefield Acceleration in Optically Generated Plasma Waveguides" (University of Maryland, College Park), https://doi.org/10.13016/qijs-yx3f.

Shrock, J. E., B. Miao, L. Feder, and H. M. Milchberg, 2022, "Meter-Scale Plasma Waveguides for Multi-GeV Laser Wakefield Acceleration," Phys. Plasmas (AIP Publishing LLC) **29**, 073101.

Shrock, J. E., E. Rockafellow, B. Miao, M. Le, R. C. Hollinger, S. Wang, T. Wang, et al., 2024, "Guided Mode Evolution and Ionization Injection in Meter-Scale Multi-GeV Laser Wakefield Accelerators," Phys. Rev. Lett. (American Physical Society) **133**, 045002.

Šišma, Jiří, Michal Nevrkla, Filip Vitha, Sebastian Lorenz, Illia Zymak, Alžběta Špádová, Andrea Kollárová, et al., 2026, "High-Repetition-Rate, All-Reflective Optical Guiding and Electron Acceleration in Helium Using an off-Axis Axicon," High Power Laser Sci. Eng. **14**, e52.

Skrodzki, P. J., M. Burger, and I. Jovanovic, 2017, "Transition of Femtosecond-Filament-Solid Interactions from Single to Multiple Filament Regime," Sci. Rep. (Nature Publishing Group) **7**, 12740.

Sloss, A.J. et al., 2026, "Effects of Laser-Plasma Waveguide Coupling on Accelerated Electron Bunch Stability and Spatial Profiles," *Indico Global (Indico)*, Accessed October 29, https://indico.global/event/14577/contributions/157908/.

Smartsev, Slava, Clément Caizergues, Kosta Oubrerie, Julien Gautier, Jean-Philippe Goddet, Amar Tafzi, Kim Ta Phuoc, Victor Malka, and Cédric Thaury, 2019, "Axiparabola: A Long-Focal-Depth, High-Resolution Mirror for Broadband High-Intensity Lasers," Opt. Lett. **44**, 3414.

Snyder, Allan W., and John D. Love, 1984, *Optical Waveguide Theory*, *Optical Waveguide Theory* (Chapman and Hall, London).

Sochacki, J., S. Bará, Z. Jaroszewicz, and A. Kołodziejczyk, 1992, "Phase Retardation of the Uniform-Intensity Axilens," Opt. Lett. (Optica Publishing Group) **17**, 7–9.

Sochacki, J., A. Kołodziejczyk, Z. Jaroszewicz, and S. Bará, 1992, “Nonparaxial Design of Generalized Axicons,” Appl. Opt. **31**, 5326.

Spence, D. J., P. D. S. Burnett, and S. M. Hooker, 1999, “Measurement of the Electron-Density Profile in a Discharge-Ablated Capillary Waveguide,” Opt. Lett. **24**, 993.

Spence, D. J., A. Butler, and S. M. Hooker, 2001, “First Demonstration of Guiding of High-Intensity Laser Pulses in a Hydrogen-Filled Capillary Discharge Waveguide,” J. Phys. B At. Mol. Opt. Phys. **34**, 4103–4112.

———, 2003, “Gas-Filled Capillary Discharge Waveguides,” J. Opt. Soc. Am. B **20**, 138.

Spence, D. J., and S. M. Hooker, 2000a, “Simulations of the Propagation of High-Intensity Laser Pulses in Discharge-Ablated Capillary Waveguides,” Pac. Rim Conf. Lasers Electro-Opt. CLEO - Tech. Dig. **17**, 293–294.

———, 2000b, “Investigation of a Hydrogen Plasma Waveguide,” Phys Rev E (American Physical Society) **63**, 015401.

Spitzer, Lyman, 2006, *Physics of Fully Ionized Gases* (Dover Publ, Mineola, NY), 2. rev. ed.

Sprangle, P., E. Esarey, J. Krall, and G. Joyce, 1992, “Propagation and Guiding of Intense Laser Pulses in Plasmas,” Phys. Rev. Lett. **69**, 2200.

Sprangle, P., B. Hafizi, and J. R. Peñano, 2000, “Laser Pulse Modulation Instabilities in Plasma Channels,” Phys Rev E (American Physical Society) **61**, 4381–4393.

Sprangle, P., B. Hafizi, J. R. Peñano, R. F. Hubbard, A. Ting, C. I. Moore, D. F. Gordon, A. Zigler, D. Kaganovich, and T. M. Antonsen, 2001, “Wakefield Generation and GeV Acceleration in Tapered Plasma Channels,” Phys. Rev. E **63**, 564051–5640511.

Sprangle, Phillip, Jonathan Krall, and Eric Esarey, 1994, “Hose-Modulation Instability of Laser Pulses in Plasmas,” Phys. Rev. Lett. **73**, 3544–3547.

Stark, H., A. Klenke, M. Benner, J. Buldt, and J. Limpert, 2023, “32 mJ, 158 Fs Pulses at 20 kHz Repetition Rate by Spatiotemporal Coherent Combination of a Fiber Laser System,” in *Laser Congress 2023 (ASSL, LAC)* (Optica Publishing Group, Tacoma, Washington), p. ATh3A.5, https://opg.optica.org/abstract.cfm?URI=ASSL-2023-ATh3A.5.

Steingrube, D. S., E. Schulz, T. Binhammer, M. B. Gaarde, A. Couairon, U. Morgner, and M. Kovačev, 2011, “High-Order Harmonic Generation Directly from a Filament,” New J. Phys. (IOP Publishing) **13**, 043022.

Steinhauer, L. C., 1970, “Laser Heating of a Stationary Plasma,” https://ntrs.nasa.gov/citations/19710027808.

Steinhauer, Loren C., and Harlow G. Ahlstrom, 1971a, “Propagation of Coherent Radiation in a Cylindrical Plasma Column,” Phys. Fluids **14**, 1109–1114.

———, 1971b, “One-Dimensional Laser Heating of a Stationary Plasma,” Phys. Fluids **14**, 81–93.

Steinke, S., J. van Tilborg, C. Benedetti, C. G. R. Geddes, C. B. Schroeder, J. Daniels, K. K. Swanson, et al., 2016, “Multistage Coupling of Independent Laser-Plasma Accelerators,” Nature (Nature Publishing Group) **530**, 190–193.

Suckewer, S., C. H. Skinner, H. Milchberg, C. Keane, and D. Voorhees, 1985, “Amplification of Stimulated Soft X-Ray Emission in a Confined Plasma Column,” Phys. Rev. Lett. (American Physical Society) **55**, 1753–1756.

Suk, H., N. Barov, J. B. Rosenzweig, and E. Esarey, 2001, “Plasma Electron Trapping and Acceleration in a Plasma Wake Field Using a Density Transition,” Phys. Rev. Lett. **86**, 1011–1014.

Sun, Guo-Zheng, Edward Ott, Y. C. Lee, and Parvez Guzdar, 1987, “Self-Focusing of Short Intense Pulses in Plasmas,” Phys. Fluids **30**, 526.

Swanson, G J, 1991, “Binary Optics Technology : Theoretical Limits on the Diffraction Efficiency of Multilevel Diffractive Optical Elements” (Massachusetts Insititute of Technology, Lincoln Lab, Lexington, Ma), Vol. 914.

Swanson, K. K., A. J. Gonsalves, H. S. Mao, T. Sipla, S. S. Bulanov, N. A. Bobrova, P. V. Sasorov, et al., 2021, “Cryogenically Formed Discharge Waveguide,” Phys. Rev. Accel. Beams (American Physical Society) **24**, 91301.

Tajima, T., and J. M. Dawson, 1979, “Laser Electron Accelerator,” Phys. Rev. Lett. **43**, 267–270.

Takeda, Mitsuo, Hideki Ina, and Seiji Kobayashi, 1982, “Fourier-Transform Method of Fringe-Pattern Analysis for Computer-Based Topography and Inteferometry.,” J. Opt. Soc. Am. **72**, 156–160.

Thévenet, M, I A Andriyash, L Fedeli, A Ferran Pousa, A Huebl, S Jalas, M Kirchen, et al., 2025, “LASY: LAser Manipulations Made eaSY,” J. Phys. Conf. Ser. (IOP Publishing) **3124**, 012014.

Tosa, V., E. Takahashi, Y. Nabekawa, and K. Midorikawa, 2003, “Generation of High-Order Harmonics in a Self-Guided Beam,” Phys. Rev. - At. Mol. Opt. Phys. **67**, 4.

Touguet, Jérôme, Igor Andriyash, Ronan Lahaye, Guillaume Chapelant, Julien Gautier, Lucas Rovige, and Cédric Thaury, 2026, “Advanced Shaping of Quasi-Bessel Beams for High-Intensity Applications,” High Power Laser Sci. Eng., 1–10.

Tripathi, N., B. Miao, A. Sloss, E. Rockafellow, J. E. Shrock, S. W. Hancock, and H. M. Milchberg, 2025, “Longitudinal Shaping of Plasma Waveguides Using Diffractive Axicons for Laser Wakefield Acceleration,” Opt. Lett. (Optica Publishing Group) **50**, 3441.

van Tilborg, J., A. J. Gonsalves, E. H. Esarey, C. B. Schroeder, and W. P. Leemans, 2018, "Density Characterization of Discharged Gas-Filled Capillaries through Common-Path Two-Color Spectral-Domain Interferometry," Opt. Lett. **43**, 2776.

van Tilborg, J., S. Steinke, C. G. R. Geddes, N. H. Matlis, B. H. Shaw, A. J. Gonsalves, J. V. Huijts, et al., 2015, "Active Plasma Lensing for Relativistic Laser-Plasma-Accelerated Electron Beams," Phys. Rev. Lett. (American Physical Society) **115**, 184802.

Vasara, Antti, Jari Turunen, and Ari T. Friberg, 1989, "Realization of General Nondiffracting Beams with Computer-Generated Holograms," J. Opt. Soc. Am. A **6**, 1748.

Vay, J. L., 2007, "Noninvariance of Space- and Time-Scale Ranges under a Lorentz Transformation and the Implications for the Study of Relativistic Interactions," Phys. Rev. Lett. **98**, 130405.

Vay, J. L., A. Huebl, A. Almgren, L. D. Amorim, J. Bell, L. Fedeli, L. Ge, et al., 2021, "Modeling of a Chain of Three Plasma Accelerator Stages with the WarpX Electromagnetic PIC Code on GPUs," Phys. Plasmas **28**, 023105.

Volfbeyn, P., E. Esarey, and W. P. Leemans, 1999, "Guiding of Laser Pulses in Plasma Channels Created by the Ignitor-Heater Technique," Phys. Plasmas **6**, 2269–2277.

Wahlstrand, J. K., Y. H. Cheng, and H. M. Milchberg, 2012, "Absolute Measurement of the Transient Optical Nonlinearity in N 2, O 2, N 2O, and Ar," Phys. Rev. - At. Mol. Opt. Phys. **85**.

Wahlstrand, J. K., N. Jhajj, E. W. Rosenthal, S. Zahedpour, and H. M. Milchberg, 2014, "Direct Imaging of the Acoustic Waves Generated by Femtosecond Filaments in Air," Opt. Lett. (Optica Publishing Group) **39**, 1290–1293.

Wang, Xiaoming, Rafal Zgadzaj, Neil Fazel, Zhengyan Li, S. A. Yi, Xi Zhang, Watson Henderson, et al., 2013, "Quasi-Monoenergetic Laser-Plasma Acceleration of Electrons to 2 GeV," Nat. Commun. (Nature Publishing Group) **4**, 1988.

Wang, Yong, Shoujun Wang, Alex Rockwood, Bradley M. Luther, Reed Hollinger, Alden Curtis, Chase Calvi, Carmen S. Menoni, and Jorge J. Rocca, 2017, "0 . 85 PW Laser Operation at 3 . 3 Hz and Second-Harmonic Beamline," Opt. Lett. **42**, 3828–3831.

Weber, B. V., and S. F. Fulghum, 1997, "A High Sensitivity Two-Color Interferometer for Pulsed Power Plasmas," Rev Sci Instrum **68**, 1227–1232.

Wen, Meng, Baifei Shen, Xiaomei Zhang, Fengchao Wang, Zhangying Jin, Liangliang Ji, Wenpeng Wang, Jiancai Xu, and Kazuhisa Nakajima, 2010, "Controlled Electron Acceleration in the Bubble Regime by Optimizing Plasma Density," New J. Phys. **12**, 045010, https://doi.org/10.1088/1367-2630/12/4/045010.

Winterfeldt, Carsten, Christian Spielmann, and Gustav Gerber, 2008, "Colloquium: Optimal Control of High-Harmonic Generation," Rev. Mod. Phys. **80**, 117–140.

Wood, Wm. M., C. W. Siders, and M. C. Downer, 1991, "Measurement of Femtosecond Ionization Dynamics of Atmospheric Density Gases by Spectral Blueshifting," Phys Rev Lett (American Physical Society) **67**, 3523–3526.

Wu, Jianzhou, James H. Cooley, Thomas M. Antonsen, and Howard M. Milchberg, 2005, "Effective Coupling of Ultraintense Laser Pulse to Funnel-Mouthed Plasma Waveguides," Phys. Plasmas **12**, 1–8.

Wulff, Kurt D., Daniel G. Cole, Robert L. Clark, Roberto DiLeonardo, Jonathan Leach, Jon Cooper, Graham Gibson, and Miles J. Padgett, 2006, "Aberration Correction in Holographic Optical Tweezers," Opt. Express **14**, 4170.

Xia, Yuhui, Zhenan Wang, Ziyao Tang, Jianghao Hu, Xinyang Liu, Letian Liu, Laifu Man, et al., 2026, "Damping Dynamics of the Centroid Oscillation of a Relativistic Laser Pulse in a Plasma Channel," Phys. Rev. Appl. (American Physical Society) **26**, 034076.

Yariv, Amnon, 1989, *Quantum Electronics* (Wiley).

Yoon, Jin Woo, Yeong Gyu Kim, Il Woo Choi, Jae Hee Sung, Hwang Woon Lee, Seong Ku Lee, and Chang Hee Nam, 2021, "Realization of Laser Intensity over 10 23 W/Cm 2," Optica (Optica Publishing Group) **8**, 630.

Yoon, S. J., A. J. Goers, G. A. Hine, J. D. Magill, J. A. Elle, Y.-H. Chen, and H. M. Milchberg, 2013, "Shock Formation in Supersonic Cluster Jets and Its Effect on Axially Modulated Laser-Produced Plasma Waveguides," Opt. Express (Optica Publishing Group) **21**, 15878.

Yoon, S. J., J. P. Palastro, and H. M. Milchberg, 2014, "Quasi-Phase-Matched Laser Wakefield Acceleration," Phys. Rev. Lett. (American Physical Society) **112**, 134803.

York, A. G., H. M. Milchberg, J. P. Palastro, and T. M. Antonsen, 2008, "Direct Acceleration of Electrons in a Corrugated Plasma Waveguide," Phys. Rev. Lett. **100**.

Zel'dovich, Ya. B., and Yu. P. Raizer, 1967, *Physics of Shock Waves and High-Temperature Hydrodynamic Phenomena*, *New York: Academic Press*, https://ui.adsabs.harvard.edu/abs/1967pswh.book.....Z.

Zepf, M., B. Dromey, M. Landreman, P. Foster, and S. M. Hooker, 2007, "Bright Quasi-Phase-Matched Soft-X-Ray Harmonic Radiation from Argon Ions," Phys. Rev. Lett. **99**.

Zhou, Tong, John Ruppe, Cheng Zhu, I-Ning Hu, John Nees, and Almantas Galvanauskas, 2015, "Coherent Pulse Stacking Amplification Using Low-Finesse Gires-Tournois Interferometers," Opt. Express (Optica Publishing Group) **23**, 7442.

Zigler, A., Y. Ehrlich, C. Cohen, J. Krall, and P. Sprangle, 1996, "Optical Guiding of High-Intensity Laser Pulses in a Long Plasma Channel Formed by a Slow Capillary Discharge," J. Opt. Soc. Am. B **13**, 68.